\documentclass[10pt]{article}
\usepackage{eqsection,latexsym,epsf,cite,epsfig,amssymb}

\begin{document}

\title{\bf Critical Phenomena and Renormalization-Group Theory}
\author{
  \\
  {\bf  Andrea Pelissetto}              \\[-0.cm]
  {\it Dipartimento di Fisica and INFN -- Sezione di Roma I}    \\[-0.cm]
  {\it Universit\`a degli Studi di Roma ``La Sapienza"}        \\[-0.cm]
  {\it I-00185 Roma, ITALIA}          \\[-0.cm]
  { E-mail: {\tt Andrea.Pelissetto@roma1.infn.it}}   \\[-0.cm]
  \\ 
  {\bf  Ettore Vicari}              \\[-0.cm]
  {\it Dipartimento di Fisica and INFN -- Sezione di Pisa}    \\[-0.cm]
  {\it Universit\`a degli Studi di Pisa}        \\[-0.cm]
  {\it I-56127 Pisa, ITALIA}          \\[-0.cm]
  { E-mail: {\tt Ettore.Vicari@df.unipi.it}}   \\[-0.cm]
  {\protect\makebox[5in]{\quad}}  % To force authors' names to be written
                                  %   vertically, one above another.
                                  % (\author seems to put them side-by-side
                                  %   if there is room.)
  \\[1cm]
{\Large April  2002} \\[1cm]
{\bf \Large To appear in Physics Reports}  \\
}
\vspace{0.5cm}

\date{}

%\vspace{0.2cm}

\maketitle

\onecolumn
\begin{abstract}

{\large

We review results concerning the critical behavior
of spin systems at equilibrium.
We consider the Ising and the general O($N$)-symmetric universality
classes, including the $N\rightarrow 0$ limit that describes the
critical behavior of self-avoiding walks.
For each of them, we review 
the estimates of the critical exponents, of the equation of state, 
of several amplitude ratios, and of the two-point function of the order
parameter. We report results in three and two dimensions.
We discuss the crossover phenomena that are observed in 
this class of systems. In particular, we review the 
field-theoretical and numerical studies of systems with medium-range
interactions.
 
Moreover, we consider several examples of magnetic and structural
phase transitions, which are described by more complex 
Landau-Ginzburg-Wilson Hamiltonians, such as
$N$-component systems 
with cubic anisotropy, 
O($N$)-symmetric systems in the presence 
of quenched disorder, frustrated spin systems with 
noncollinear or canted order, and 
finally, a class of systems described by the tetragonal Landau-Ginzburg-Wilson 
Hamiltonian with three quartic couplings.
The results for the tetragonal Hamiltonian are original, in particular
we present the six-loop perturbative series for 
the $\beta$-functions. Finally, we consider a Hamiltonian with symmetry
O$(n_1)\oplus$O$(n_2)$ that is relevant for the description of multicritical 
phenomena.

\vskip 1truecm
}
\end{abstract}
%\twocolumn

%\clearpage

\newcommand{\be}{\begin{equation}}
\newcommand{\ee}{\end{equation}}
\newcommand{\bea}{\begin{eqnarray}}
\newcommand{\eea}{\end{eqnarray}}
\newcommand{\<}{\langle}
\renewcommand{\>}{\rangle}

%\ltapprox and \gtapprox produce > and < signs with twiddle underneath
\def\spose#1{\hbox to 0pt{#1\hss}}
\def\ltapprox{\mathrel{\spose{\lower 3pt\hbox{$\mathchar"218$}}
 \raise 2.0pt\hbox{$\mathchar"13C$}}}
\def\gtapprox{\mathrel{\spose{\lower 3pt\hbox{$\mathchar"218$}}
 \raise 2.0pt\hbox{$\mathchar"13E$}}}

\def\bsigma{\mbox{\protect\boldmath $\sigma$}}
\def\btau{\mbox{\protect\boldmath $\tau$}}
\def\bphi{\mbox{\protect\boldmath $\varphi$}}
\def\bz{\mbox{\protect\boldmath $z$}}
\def\bw{\mbox{\protect\boldmath $w$}}
\def\hatp{\hat p}
\def\hatl{\hat l}
\def\smfrac#1#2{{\textstyle\frac{#1}{#2}}}
\def\case#1#2{{\textstyle\frac{#1}{#2}}}

\def\msbar{ {\overline{\hbox{\scriptsize MS}}} }
\def\normalmsbar{ {\overline{\hbox{\normalsize MS}}} }

\newcommand{\R}{\hbox{{\rm I}\kern-.2em\hbox{\rm R}}}
\newcommand{\N}{\hbox{{\rm I}\kern-.2em\hbox{\rm N}}}

\newcommand{\reff}[1]{(\ref{#1})}

\tableofcontents
\clearpage

\section*{Plan of the review}

The main issue of this review is the critical behavior of spin 
systems at equilibrium. 

In Sec. 1 we introduce the notations and the basic
renormalization-group results for the critical exponents,
the equation of state, and the two-point function of the order
parameter, which are used throughout 
the paper.

In Sec. 2 we outline the most important methods that are used 
in the study of equilibrium spin systems: high-temperature expansions,
Monte Carlo methods, and field-theoretical me\-th\-ods. 
It is not a comprehensive review of these techniques; the purpose is to 
present the most efficient methods and to discuss their possible 
sources of error.

In the following sections we focus on specific systems and
universality classes. 
Sec. 3 is dedicated to the Ising universality class in three and two 
dimensions. Secs. 4 and 5 consider the three-dimensional 
$XY$ and Heisenberg universality classes respectively.
In Sec. 6 we discuss the three-dimensional O($N$) universality
classes with $N\geq 4$,  with special 
emphasis on the physically relevant cases $N=4$ and $N=5$.
Secs. 7 and 8 are devoted to the special 
critical behaviors of the two-dimensional models 
with continuous O($N$) symmetry,
i.e. the Kosterlitz-Thouless transition, which occurs in the $XY$ model, and
the peculiar exponential behavior characterizing the zero-temperature
critical limit of the  O($N$) vector model with $N\geq 3$.
Finally,
in Sec. 9 we discuss the limit $N\to 0$ that describes the asymptotic 
properties of self-avoiding walks and of polymers in dilute
solutions and
in the good-solvent regime. For each of these models, we review 
the estimates of the critical exponents, of the equation of state, 
of several universal amplitude ratios, 
and of the two-point function of the order parameter. 
 
In Sec. 10 we discuss the crossover phenomena that are observed in 
this class of systems. In particular, we review the 
field-theoretic and numerical studies of systems with medium-range
interactions. 

In Sec. 11 we consider several examples of magnetic and structural
phase transitions, which are described by more complex 
Landau-Ginzburg-Wilson Hamiltonians. 
We present field-theoretical results and we compare them 
with other theoretical and experimental estimates. 
In Sec. \ref{lsec-cubic} we discuss $N$-component systems 
with cubic anisotropy, and in particular the stability of the 
O$(N)$-symmetric fixed point in the presence of cubic perturbations. 
In Sec. \ref{lsec-random} we consider  O($N$)-symmetric systems in the presence 
of quen\-ched disorder, focusing on the randomly dilute Ising model that shows 
a different type of critical behavior. In Sec. \ref{lsec-frustrated} 
we discuss the critical behavior of frustrated spin systems with 
noncollinear or canted order.
In Sec. \ref{lsec-tetragonal} we discuss a class 
of systems described by the tetragonal Landau-Ginzburg-Wilson 
Hamiltonian with three quartic couplings. 
This section contains original results, in particular
the six-loop perturbative series of the $\beta$-functions.
Finally, in Sec. \ref{LGW-multicritical} we consider a Hamiltonian
with symmetry O$(n_1)\oplus$O$(n_2)$, which is relevant for the 
description of multicritical phenomena.
\bigskip

\noindent
{\bf Acknowledgements}

\bigskip

We thank Tomeu All\'es,
Pasquale Calabrese, Massimo Campostrini, Sergio Caracciolo, 
Jos\'e Carmona, Michele
Caselle, Serena Causo, Alessio Celi, Robert Edwards, 
Martin Hasenbusch, 
Gustavo Mana, 
Victor Mart\'{\i}n-Mayor, Tereza Mendes, Andrea Montanari, 
Paolo Rossi, Alan Sokal,
for collaborating with us on some of the issues
considered in this review.

\clearpage

\section{The theory of critical phenomena} \label{sec-1}

\subsection{Introduction} \label{sec-1.1}

The theory of critical phenomena has quite a long history. In the XIX century 
Andrews \cite{Andrews_1869} discovered a peculiar point in the 
$P-T$ plane of carbon dioxide,
 where the properties of the liquid and of the vapor
become indistinguishable and the system shows critical opalescence: It 
was the first observation of a critical point. Thirty years 
later, Pierre Curie \cite{Curie_1895} discovered the 
ferromagnetic transition in iron and realized the similarities of the 
two phenomena. However, a quantitative theory was still to come. 
Landau \cite{Landau_37} was the first one proposing a general framework that 
provided a unified explanation of these phenomena. His model, 
which corresponds to the mean-field approximation,
gave a good qualitative description of the transitions in fluids and 
magnets. However, Onsager's solution \cite{Onsager_44} 
of the two-dimensional Ising model \cite{Ising_25} and Guggenheim's 
results on the coexistence curve of simple fluids \cite{Guggenheim-45}
showed that Landau's model is not quantitatively correct. 
In the early 60's the modern notations were introduced by 
Fisher \cite{Fisher-63}.
Several scaling relations among critical exponents  were derived
\cite{EF-63,Widom-64,GFSE-64}, and a scaling form for the equation
of state was proposed \cite{Widom-65,DH-65,PP-64-66}. 
A more general framework was introduced by 
Kadanoff \cite{Kadanoff-66}. However, a satisfactory understanding 
was reached only when the scaling ideas were 
reconsidered in the general renormalization-group (RG) framework 
by Wilson 
\cite{Wilson-71,Wilson-71a,WK-74}.
Within the new framework, it was possible to explain the critical
behavior of most of the systems and their universal features; 
for instance, why fluids and 
uniaxial antiferromagnets behave quantitatively in an identical way at the 
critical point.

Since then, critical phenomena have been the object of extensive studies and 
many new ideas have been developed in order to understand the critical 
behavior of increasingly complex systems. Moreover, the concepts that first
appeared in con\-den\-sed-matter physics have been applied 
to different
areas of physics, such as high-energy physics, and even outside, e.g., to 
computer science, biology, economics, and social sciences.

In high-energy physics, the RG theory of
critical phenomena provides the natural framework
for defining quantum field theories 
at a nonperturbative level, i.e., beyond perturbation theory 
(see, e.g., Ref.~\cite{Zinn-Justin-book}).
For example, the Euclidean lattice formulation 
of gauge theories proposed by Wilson \cite{Wilson-74,Wilson-75} 
provides a nonperturbative definition
of quantum chromodynamics (QCD), the theory that is supposed to 
describe the strong interactions in subnuclear physics.
QCD is obtained as the 
critical zero-temperature (zero-bare-coupling)
limit of appropriate four-dimensional lattice models and 
may therefore be considered as a particular 
four-dimensional universality class, characterized by a peculiar 
exponential critical behavior (see, e.g.,
Refs.~\cite{Creutz-book,Mo-Mu-book,Zinn-Justin-book,It-Dr-book}).
Wilson's formulation represented a breakthrough
in the study of QCD, because it lends itself to
nonperturbative computations using
statistical-mechanics techniques, for instance by means of 
Monte Carlo simulations (see, e.g., Ref.~\cite{CJR-83}).

\begin{figure}[t]
%\vskip 1truecm
\begin{tabular}{cc}
\psfig{width=6truecm,angle=0,file=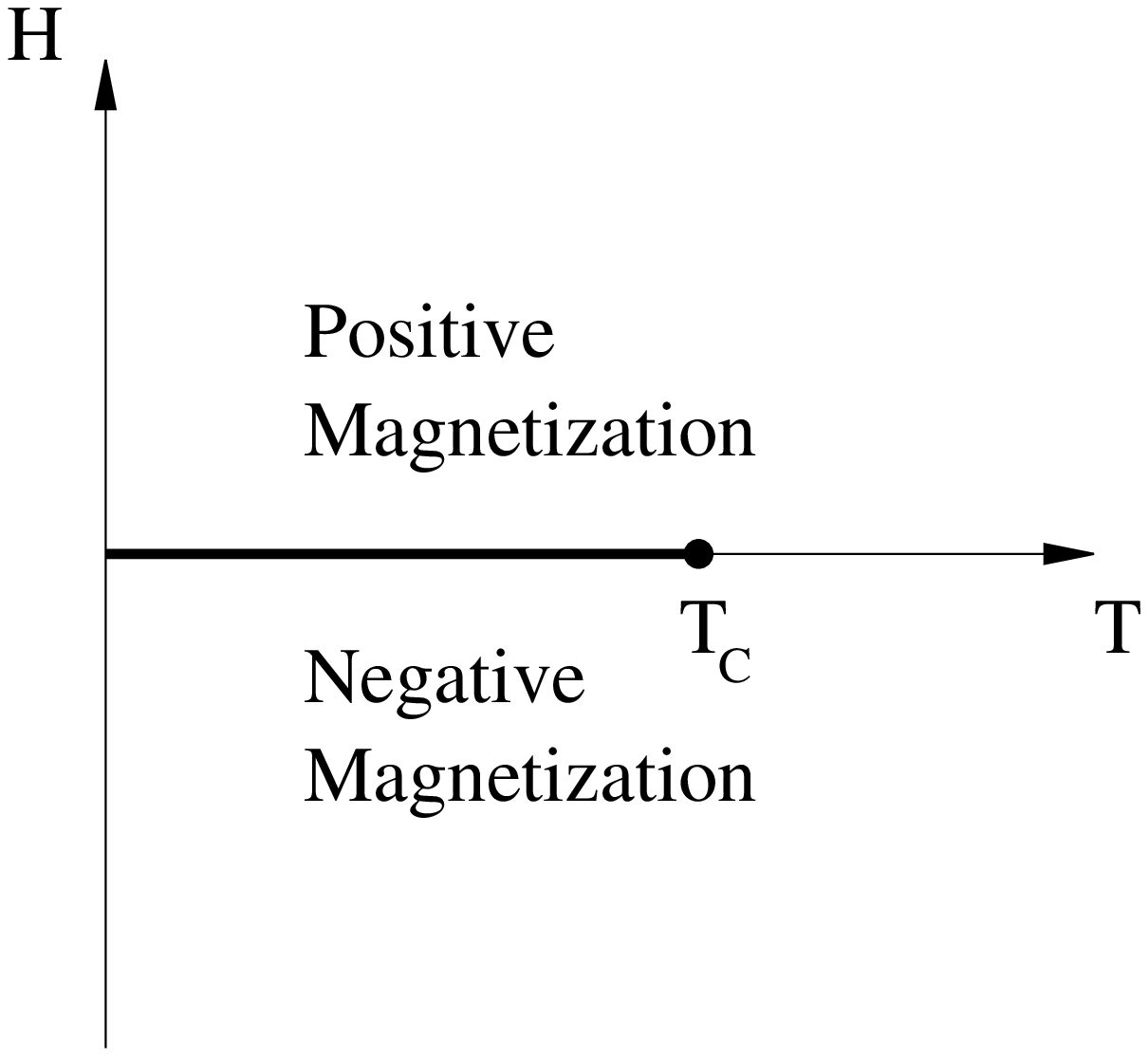} &
\hskip 1truecm
\psfig{width=6truecm,angle=0,file=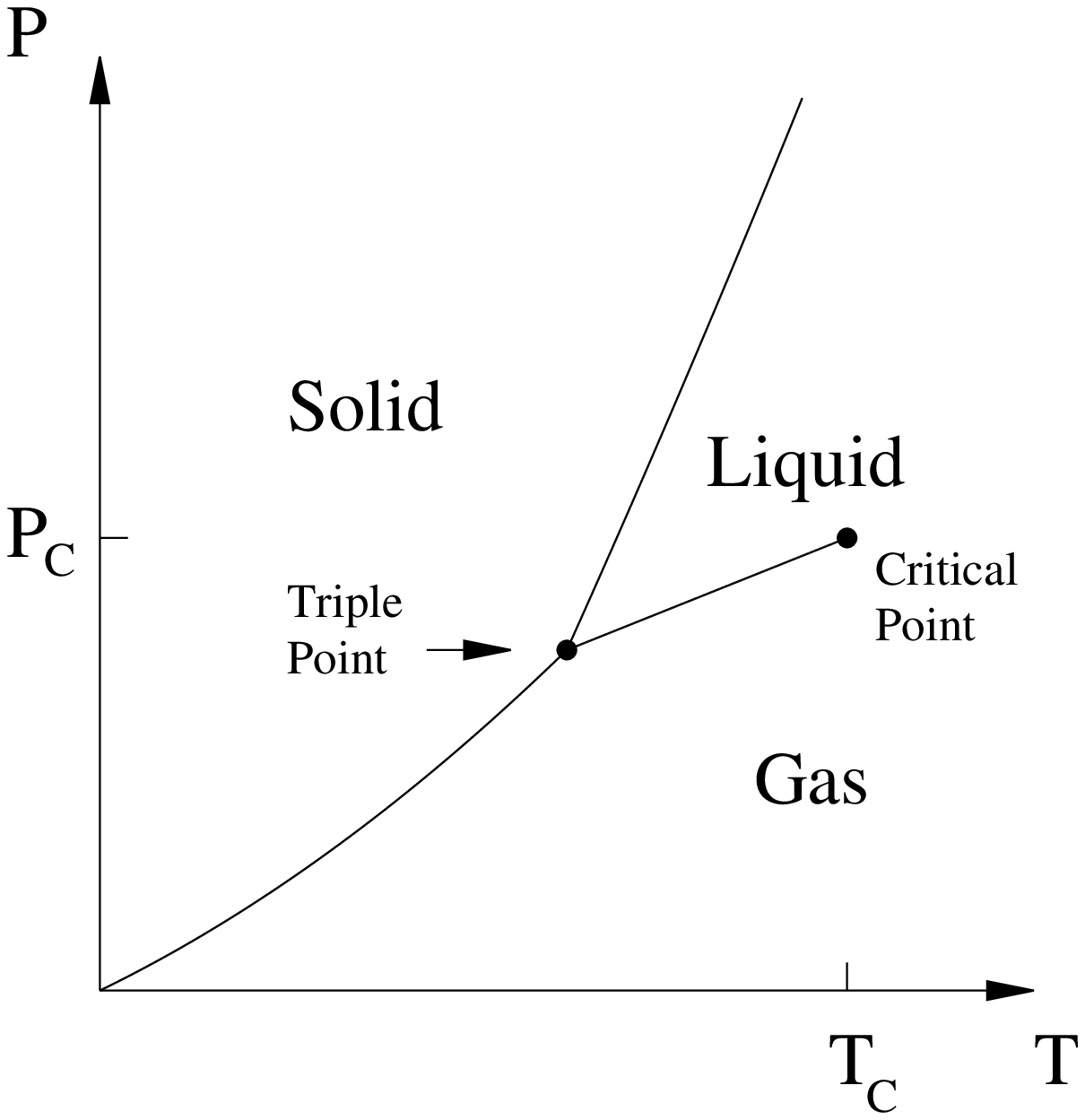} \\
\end{tabular}
\vskip -0.5truecm
\caption{The phase diagram of a magnetic system (left) and of a simple
fluid (right).}
\label{figHT}
\end{figure}

The prototype of models with a continuous phase transition 
is the celebrated Ising
model \cite{Ising_25}. It is defined on a regular lattice with 
Hamiltonian
\be
{\cal H} = - J \sum_{\<ij\>} s_i s_j - H \sum_i s_i,
\ee
where $s_i = \pm1$, and the first sum is extended over all near\-est-neighbor 
pairs $\<ij\>$. The partition function is defined by
\be
Z = \sum_{\{s_i\}} e^{-{\cal H}/T}.
\ee
The Ising model provides a simplified description of a uniaxial 
magnet in which the spins align along a specific direction. The phase
diagram of this system is well known, see Fig. \ref{figHT}. 
%
%% \begin{figure}[t]
%% \vskip 1truecm
%% \centerline{\psfig{width=6truecm,angle=0,file=figPT.ps}}
%% \vskip 0truecm
%% \caption{The phase diagram of a simple fluid.}
%% \label{figPT}
%% \end{figure}
%
For zero magnetic field,
there is a paramagnetic phase for $T>T_c$ and a ferromagnetic phase 
for $T<T_c$, separated by a critical point at $T=T_c$. 
Near the critical point long-range correlations develop,
and the large-scale behavior of the system can be studied
using the RG theory.

The Ising model can easily be mapped into a lattice gas. 
Consider the Hamiltonian
\be
{\cal H} = - 4 J \sum_{\<ij\>} \rho_i \rho_j - \, \mu \sum_i \rho_i ,
\ee
where $\rho_i = 0,1$ depending if the site is empty or occupied, and 
$\mu$ is the chemical potential. If we define $s_i = 2 \rho_i - 1$, 
we reobtain the Ising-model Hamiltonian with $H = 2 q J + \mu/2$, where 
$q$ is the coordination number of the lattice. Thus,
for $\mu = - 4 q J$, there is an equivalent transition separating 
the gas phase for $T>T_c$ from a liquid phase for $T<T_c$. 

\begin{table*}[t]
\caption{Relation between fluid and magnetic quantities. Here 
$\cal F$ and ${\cal A}$ are respectively the Gibbs and the Helmholtz
free energy, $C_P$ and $C_V$ the isobaric and isochoric specific 
heats, $C_M$ and $C_H$ the specific heats at fixed magnetization 
and magnetic field, $\kappa_T$ the isothermic compressibility,
and $\chi$ the magnetic susceptibility. $\rho_c$ and $\mu_c$ are 
the values of the density and of the chemical potential 
at the critical point.}
\label{magneti-fluidi}
\begin{center}
\begin{tabular}{cc}
\hline
& \\[-3.5mm]
FLUID  & MAGNET \\[0.5mm] \hline
& \\[-3mm]
density: $\rho-\rho_c$    & \qquad magnetization $M$ \\
chemical potential: $\mu - \mu_c$  & \qquad magnetic field $H$ \\[2mm]
\hline 
& \\[-1mm]
$C_P = - T\displaystyle\left({\partial^2 {\cal F}\over \partial T^2}\right)_P$
  & 
$C_H = - T\displaystyle\left({\partial^2 {\cal F}\over \partial T^2}\right)_H$ 
\\[5mm]
$C_V = - T \displaystyle\left({\partial^2 {\cal A}\over 
                               \partial T^2}\right)_V$  & 
$C_M = - T \displaystyle\left({\partial^2 {\cal A}\over 
                               \partial T^2}\right)_M$   
\\[5mm]
${\kappa}_T = \displaystyle{1\over \rho} \left({\partial \rho \over 
                               \partial P}\right)_T = 
           - {1\over V} \displaystyle\left({\partial^2 {\cal F}\over \partial
                                           P^2}\right)_T$  &
$\chi = \displaystyle\left({\partial M\over \partial H}\right)_T = 
           - \left({\partial^2 {\cal F}\over \partial H^2}\right)_T$ 
\\[5mm]
\hline
\end{tabular}
\end{center}
\end{table*}

The lattice gas is a crude approximation of a real fluid.
Nonetheless, the universality of the behavior around a continuous
phase-transition point implies that certain quantities, e.g., critical exponents,
some amplitude ratios, scaling functions, and so on, are identical in a real
fluid and in a lattice gas, and hence in the Ising model. Thus,
the study of the Ising model provides {\em exact} predictions for
the critical behavior of real fluids,
and in general for all transitions belonging to the Ising universality
class, whose essential features are a scalar order parameter and
effective short-range interactions.

In the following, we will use a magnetic ``language." 
In Table \ref{magneti-fluidi} we write down 
the correspondences between fluid and magnetic quantities.
The quantity that corresponds  to the magnetic field 
is the chemical potential. However, such a quantity is not 
easily accessible experimentally, and thus one uses the pressure 
as second thermodynamic variable. The phase diagram of a real fluid
is shown in Fig. \ref{figHT} (right). 
The low-temperature line (in boldface) appearing in the magnetic phase diagram
corresponds to the 
liquid-gas transition line between the triple and the critical point. Of course,
this description is only  valid in a neighborhood of the 
critical point. In magnetic systems 
there is a symmetry $M\to -M$, $H\to -H$ that is absent
in fluids. As a consequence, although the leading 
critical behavior is identical, fluids show
subleading corrections that are not present in magnets.

The generalization of the Ising model to systems with an $N$-vector
order parameter and O($N$) symmetry provides other
physically interesting universality classes describing several critical
phenomena in nature, such as some ferromagnetic transitions, the superfluid
transition of $^4$He, the critical behavior of polymers, etc.

This review will mostly focus on the critical behavior 
of $N$-vector models at equilibrium. This issue
has been amply reviewed in the literature, see, e.g.,
Refs. \cite{Fisher-74,Fisher-98,Cardy-book,Zinn-Justin-book,It-Dr-book,Parisibook,Ma-book,KS-01}.
Other reviews can be found in the Domb-Green-Lebowitz book series.
We will mainly discuss the recent developments.
Other systems, described by more complex Landau-Ginzburg-Wilson
Hamiltonians, will be considered in the last section.

\subsection{The models and the basic thermodynamic quantities} \label{sec-1.2}

In this review we mainly deal 
with systems whose critical behavior can
be described by the Heisenberg Hamiltonian
(in Sec. \ref{LGWHa} we will consider some more 
general theories that can be studied with similar techniques). More precisely,
we consider a regular lattice, $N$-vector unit spins defined at the sites of 
the lattice, and the Hamiltonian\footnote{Note that here and in the following
our definitions differ by powers of the temperature from the standard 
thermodynamic definitions. It should be easy for the reader to reinsert
these factors any time they are needed. See Sec. 2.1 of Ref. \cite{PHA-91}
for a discussion of the units.}
\be
{\cal H} = - \beta \sum_{\<ij\>} \vec{s}_i\cdot \vec{s}_j - 
   \sum_i \vec{H}\cdot \vec{s}_i,
\label{NvectorHamiltonian}
\ee
where the summation is extended over all lattice nearest-neighbor pairs 
$\<ij\>$, and $\beta$ is the inverse temperature. This model represents 
the natural generalization of the Ising model, which corresponds to the 
case $N=1$. One may also consider more general Hamiltonians
of the form
\be
{\cal H} = - \beta \sum_{\<ij\>} \vec{\phi}_i\cdot \vec{\phi}_j +
    \sum_i V(\phi_i) -
   \sum_i \vec{H}\cdot \vec{\phi}_i,
\label{generalHamiltonian}
\ee
where $\vec{\phi}_i$ is an $N$-dimensional vector and $V(x)$ is a generic
potential such that 
\be
\int_{-\infty}^\infty e^{b x^2 - V(x)} < + \infty
\ee
for all real $b$. A particular case is the $\phi^4$ 
Hamiltonian 
\begin{equation}
{\cal H} = - \beta \sum_{\<ij\>} \vec{\phi}_i\cdot \vec{\phi}_j +
   \sum_i \left[\lambda (\vec{\phi}_i^{\ 2} - 1)^2 + \phi_i^2\right] -
   \sum_i \vec{H}\cdot \vec{\phi}_i,
\label{latticephi4}
\end{equation}
which is the lattice discretization of the continuum theory
\begin{equation}
{\cal H} = \int d^dx \Bigl\{ \case{1}{2}
    \partial_\mu \vec{\varphi}(x)\cdot \partial_\mu \vec{\varphi}(x) + 
   \case{1}{2} {r} \,\vec{\varphi}(x)\cdot \vec{\varphi}(x)  
      +\case{1}{4!}u \left[ \vec{\varphi}(x)\cdot \vec{\varphi}(x) \right]^2 
       - \vec{H} \cdot\vec{\varphi}(x)\Bigr\},
\end{equation}
where, in the case of a hypercubic lattice, 
\begin{equation}
\varphi = \beta^{1/2} \phi, \qquad 
r = {2-4\lambda\over \beta} - 2d,\qquad
u = {4!\ \lambda\over \beta^2}.
\end{equation}
The partition function is given by 
\be
Z(H,T) = \int \left[\prod_i d\mu(\vec{\phi}_i)\right] \, e^{-{\cal H}},
\ee
where $d\mu(\vec{\phi}) = d^N\! \phi$ when $\phi$ is an unconstrained vector 
and $d\mu(\vec{s}) = d^N\! s\, \delta(s^2 - 1)$ for the Heisenberg Hamiltonian.
We will only consider the classical case, i.e., our spins
will always be classical fields and not quantum operators.

As usual, we introduce the Gibbs free-energy density
\begin{equation}
{\cal F}(H,T) = - {1\over V} \log Z(H,T),
\end{equation}
and the related Helmholtz free-energy density
\begin{equation}
{\cal A}(M,T) = \vec{M}\cdot\vec{H} + {\cal F}(H,T),
\end{equation}
where $V$ is the volume.  Here $\vec{M}$ is the magnetization density 
defined by
\be
\vec{M} = - \left( {\partial {\cal F}\over \partial \vec{H}}\right)_T.
\ee
General arguments of thermal and 
mechanical stability imply $C_P\ge 0$, $C_V\ge 0$, and 
$\kappa_T \ge 0$, and also $C_H\ge 0$, $C_M\ge 0$, 
and\footnote{Note that it is not generically true that the 
magnetic susceptibility is positive. For instance, in diamagnets $\chi< 0$.} 
$\chi\ge 0$. 
These results allow us to prove the convexity\footnote{
We remind the reader that a function $f(x)$ is {\em convex} if 
$f(a x + b y) \le a f(x) + b f(y)$ for all $x,y$, $0\le a,b\le 1$
with $a+b = 1$. If the opposite inequality holds, the function
is {\em concave}.} properties of the 
free energy, for instance using H\"older's inequality
\cite{Griffiths-66}. 
The positivity of the specific heats at constant 
magnetic field and magnetization and of the susceptibility implies
that the Gibbs free energy is concave in $T$ and $H$,
and the Helmholtz free energy is concave in $T$ and convex in $M$.

We consider several thermodynamic quantities:
\begin{enumerate}
\item The magnetic susceptibility $\chi$:
\be
\chi = - {\partial^2 {\cal F}\over \partial\vec{H}\cdot\partial\vec{H}}
   = \sum_x \left[ \< \vec{\phi}_x\cdot \vec{\phi}_0 \> - 
                 \< \vec{\phi}_0 \>^2 \right].
\ee
For vector systems one may  also define 
\be
\chi^{ab} = - {\partial^2 {\cal F}\over \partial H^a \partial H^b},
\ee
and, if $H^a = H\delta^{a 1}$, the 
transverse and longitudinal susceptibilities 
\begin{equation}
\chi_L = \chi^{11}, \qquad
\chi_T = {1\over (N-1)} (\chi - \chi_L).
\end{equation}
Note that $\chi^{ab} = \delta^{ab}\chi_T$ for $a,b\not=1$, because of the 
residual $O(N-1)$ invariance.
\item The $2n$-point connected correlation function $\chi_{2n}$ 
at zero momentum:
\be
\chi_{2n} = - {\partial^{2n} {\cal F}\over 
              (\partial\vec{H}\cdot\partial\vec{H})^n}.
\ee
For the Ising model in the low-temperature phase one should also consider 
odd derivatives of the Gibbs free energy $\chi_{2n+1}$.
\item The specific heat at fixed magnetic field and at fixed magnetization:
\begin{equation}
C_H = - T \left({\partial^2 {\cal F}\over \partial T^2}\right)_H , \qquad
C_M = - T \left({\partial^2 {\cal A}\over \partial T^2}\right)_M . 
\end{equation}
\item The two-point correlation function:
\be
G(x) = \< \vec{\phi}_x\cdot \vec{\phi}_0 \> - 
                 \< \vec{\phi}_0 \>^2 ,
\ee
whose zero-momentum component is the magnetic susceptibility, i.e.
$\chi = \sum_x G(x)$. In the low-temperature phase, for vector models,
one distinguishes transverse and longitudinal contributions. 
If $H^a = H \delta^{a1}$ we define
\begin{eqnarray}
G_L(x) = \< {\phi}_x^1\,  {\phi}_0^1 \> - M^2, \qquad
G_T(x) = \< {\phi}_x^a\,  {\phi}_0^a \>, 
\end{eqnarray}
where $a\not=1$ is not summed over.
\item The exponential or true correlation length (inverse mass gap) 
\be
\xi_{\rm gap} = - \lim\sup_{|x|\to \infty} {|x|\over \log G(x)}\; .
\label{xigap}
\ee
\item The second-moment correlation length
\be
\xi = \left[ {1\over 2d}\, {\sum_x |x|^2 G(x)\over 
                     \sum_x G(x)}\right]^{1/2}.
\label{xism}
\ee
\end{enumerate}

\subsection{Critical indices and scaling relations} \label{sec-1.3}

In three dimensions and  for $H=0$, 
the Hamiltonian (\ref{generalHamiltonian})
displays a low-temperature magnetized phase separated 
from a paramagnetic phase by a critical point. 
The transition may be either first-order or continuous, depending on the 
potential $V(\phi)$.
The continuous transitions are generically characterized by a
nontrivial power-law critical behavior controlled by 
two relevant quantities, the temperature and the external field.
Specific choices of the parameters may lead to multicritical transitions.
For instance, tricritical transitions 
require the additional tuning of one parameter in the potential;
in three dimensions they have mean-field exponents with logarithmic
corrections. We shall not consider them here.
The interested reader should consult Ref. \cite{Lawrie-Sarbach-DG}. 

In two dimensions a power-law critical behavior
is observed only for $N<2$. 
For $N=2$ the systems show a Kosterlitz-Thouless 
transition \cite{KT-73} with a different scaling behavior. 
This is described  in Sec. \ref{XYd2}. 
For $N\ge 3$ there is no finite-temperature 
phase transition and correlations are finite for all temperatures $T\not=0$,
diverging for $T\to 0$.
These systems are discussed in Sec. \ref{sec-5.2}.
In this section 
we confine ourselves to the ``standard" 
critical behavior characterized by power laws.

When the reduced temperature 
\begin{equation}
t\equiv {T-T_c\over T_c} = {\beta_c - \beta\over \beta}
\end{equation}
goes to zero and the magnetic field vanishes, all quantities show 
power-law singularities. It is customary to consider three 
different trajectories in the $(t,H)$ plane.
\begin{itemize}
\item The high-temperature phase at 
zero field: $t>0$ and $H=0$. For $t\to 0$ we 
have 
\begin{equation}
T_c C_H \approx A^+\ t^{-\alpha}, \label{spheath} \\
\end{equation}
for the specific heat, and
\begin{eqnarray}
&\chi \approx N C^+\ t^{-\gamma}, 
\qquad &
\chi_{2n} \approx R_{n,N} C^+_{2n}\ t^{-\gamma_{2n}}, \\
&\xi \approx f^+\ t^{-\nu}, \qquad
&\xi_{\rm gap} \approx f^+_{\rm gap}\ t^{-\nu},
\end{eqnarray}
where $R_{n,N} = {N(N+2)\ldots (N+2n - 2)/(2n-1)!!}$
(note that $R_{n,1} = 1$).
In this phase  the magnetization vanishes.
\item The coexistence curve: $t<0$ and $H=0$. In this case 
we should distinguish scalar systems ($N=1$) from vector systems $(N\ge 2)$. 
Indeed, on the coexistence line
vector systems show Goldstone excitations and the 
two-point function at zero momentum diverges.
The\-re\-fo\-re, $\chi$, $\chi_{2n}$, $\xi$, and $\xi_{\rm gap}$ are 
infinite at the coexistence curve, i.e. for $|H|\to 0$ and any $t< 0$. 
We define 
\begin{eqnarray}
T_c C_H &\approx& A^-\ (-t)^{-\alpha'}, \label{spheatl} \\
|M| &\approx& B (-t)^\beta,
\label{scalings-lowT-2M}
\end{eqnarray}
and for a scalar theory 
\begin{eqnarray}
&\chi \approx C^-\ (-t)^{-\gamma'}, \qquad
&\chi_{n} \approx C^-_{n}\ (-t)^{-\gamma_{n}'},
\label{scalings-lowT-2} \\
&\xi \approx f^-\ (-t)^{-\nu'}, \qquad
&\xi_{\rm gap} \approx f^-_{\rm gap}\ (-t)^{-\nu'}.
\label{scalings-lowT-1}
\end{eqnarray}
In the case of vector models,
a transverse correlation length \cite{Ferer-74} 
\begin{equation}
\xi_{\rm T}\approx f^-_{\rm T} (-t)^{-\nu'}
\end{equation}
is defined from the stiffness constant $\rho_s$ (see Eq. \reff{def-xiT}
below  for the definition). 
In the case of the Ising model, another interesting 
quantity is  the 
interface tension $\sigma$,  which, for $t\to 0^-$, behaves as 
\begin{equation}
\sigma = \sigma_0 (-t)^\mu.
\end{equation}
\item The critical isotherm $t=0$. For $|H|\to 0$ we have
\begin{eqnarray}
\vec{M}   \approx B^c \vec{H}\ |H|^{(1-\delta)/\delta}, \qquad 
\xi       \approx f^c |H|^{-\nu_c}, \qquad 
\xi_{\rm gap} \approx f^c_{\rm gap} |H|^{-\nu_c}.
\end{eqnarray}
The scaling of the $n$-point connected correlation functions is easily 
obtained from that of $\vec{M}$ by taking derivatives with respect to $\vec{H}$.
For instance, we have
\begin{eqnarray}
\chi \approx C^c |H|^{(1-\delta)/\delta}, \qquad
\chi_L \approx C_L^c |H|^{(1-\delta)/\delta}, 
\end{eqnarray}
where
\begin{eqnarray}
C^c = {B^c\over \delta} (1 + N\delta - \delta), \qquad
C_L^c = {B^c\over \delta}.
\end{eqnarray}
\end{itemize}

Moreover,  one introduces the  exponent $\eta$ to describe the behavior of the 
two-point function at the critical point $T=T_c$, $H=0$, i.e.,
\be
G(x) \sim {1\over |x|^{d-2+\eta}} \; .
\ee
The critical exponent 
$\eta$ measures the deviations from a purely Gaussian behavior.

The exponents that we have introduced are not independent. Indeed,  
RG predicts several relations among them. First,
the exponents in the high-temperature phase and on the 
coexistence curve are identical, i.e. 
\begin{eqnarray}
\alpha = \alpha', \qquad  \nu=\nu',\qquad
\gamma = \gamma', \qquad 
\gamma_{2n} = \gamma_{2n}', 
\end{eqnarray} 
Second, the following relations hold:
\begin{eqnarray}
&&\alpha + 2\beta + \gamma = 2, \qquad 
2 - \alpha = \beta (\delta + 1), 
\qquad \beta\delta\nu_c  =  \nu, \nonumber \\
&&\gamma = \nu (2-\eta),\qquad
 \gamma_{2n} = \gamma + 2 (n-1) \Delta_{\rm gap}, 
\end{eqnarray}
where $\Delta_{\rm gap}$ is the ``gap" exponent, 
which controls the radius of the disk in the 
complex-temperature plane without zeroes, i.e. the gap, 
of the partition function (Yang-Lee theorem). Below the 
upper critical dimension, i.e. for $d<4$, also the following 
``hyperscaling" relations are supposed to be valid: 
\begin{eqnarray}
2 - \alpha = d \nu,\qquad
2 \Delta_{\rm gap} = d \nu + \gamma.
\label{hyperscaling-relations}
\end{eqnarray}
Moreover, 
the exponent $\mu$ related to the interface tension in
the Ising model 
satisfies the hyperscaling relation \cite{Widom-65b}
$\mu = (d-1)\nu$.
Using the scaling and hyperscaling relations,
one also obtains
\begin{eqnarray}
\delta = { d +2 - \eta\over d-2+\eta},\qquad
\beta = {\nu\over 2} \left( d-2+\eta\right).
\end{eqnarray}
For $d>4$ the hyperscaling relations do not hold,
and the critical exponents assume the mean-field
values:
\begin{eqnarray}
\gamma = 1 ,\quad
\nu = {1\over 2} ,\quad
\eta = 0 ,\quad
\alpha = 0 ,\quad
\beta = {1\over 2} ,\quad
\delta = 3 ,\quad
\mu = {3\over 2} .
\end{eqnarray}
In the following we only consider the case $d<4$ and thus
we use the hyperscaling relations.

It is important to remark that
the scaling behavior of the specific heat 
given above, cf. Eqs.~(\ref{spheath}) and (\ref{spheatl}), 
is correct only if $\alpha>0$. If $\alpha<0$ 
the analytic background cannot be neglected,
and the critical behavior is 
\be
T_c C_H \approx A^\pm |t|^{-\alpha} + B
\ee
for $\alpha > -1$. 
The amplitudes $A^\pm$ are positive (resp. negative) for $\alpha > 0$
(resp. $\alpha<0$),
see, e.g., Ref.~\cite{CPV-02}.
This fact is confirmed by the critical behavior of all known systems
in the universality classes corresponding to $N$-vector models.
Moreover, there are interesting cases, for instance 
the two-dimensional Ising model, in which the specific heat diverges 
logarithmically, 
\be
T_c C_H \approx - A^\pm \log |t|.
\label{CH-log-def}
\ee

The critical behaviors reported in this section 
are valid asym\-ptotically close to the 
critical point. Scaling corrections
are controlled by a universal exponent $\omega$, which is related to the 
RG dimension of the leading irrelevant operator. 
For $H=0$, both in the high- and in the low-temperature phase,
the scaling corrections are of order $|t|^\Delta$ with $\Delta = \omega \nu$, 
while on the critical isotherm they are of order 
$|H|^{\Delta_c}$ with $\Delta_c = \omega \nu_c$. 

The critical exponents are universal in the sense that 
they have the same value for all systems belonging to a given 
universality class. The amplitudes instead are not universal 
and depend on the microscopic parameters,
and therefore on the particular system considered.  Nonetheless, RG
predicts that some combinations are universal. Several universal 
amplitude ratios are reported in 
Table~\ref{notationsur}. Those involving 
the amplitudes of the susceptibilities and of the correlation lengths 
on the coexistence curve, and 
the amplitude of the interface tension are defined only for 
a scalar theory (Ising universality class).

\begin{table*}
\caption{
Definitions of several universal amplitude ratios. Those involving 
$C^-$, $C^-_n$, $f^-$, $f^-_{\rm gap}$, and $\sigma_0$ are defined only for 
$N=1$.
}
\label{notationsur}
\begin{center}
\begin{tabular}{ll}
\hline 
& \\[-3mm]
\multicolumn{2}{c}{\large{Universal Amplitude Ratios}}\\ 
\hline
& \\[-1mm]
$U_0\equiv A^+/A^-$ & $U_2\equiv C^+/C^-$\\ [1.5mm]
$U_4\equiv C^+_4/C^-_4$ & $R_4^+\equiv - C_4^+B^2/(C^+)^3$ \\ [1.5mm]
$R_c^+\equiv \alpha A^+C^+/B^2$ & $R_c^-\equiv \alpha A^- C^-/B^2$ \\ [1.5mm]
$R_4^-\equiv C_4^-B^2/(C^-)^3$ & 
    $R_\chi\equiv C^+ B^{\delta-1}/(B^c)^\delta$ \\ [1.5mm]
$v_3\equiv - C_3^-B/(C^-)^2$  &  $v_4 \equiv -C_4^-B^2/(C^-)^3 + 3
v_3^2 $ \\ [1.5mm]
%$F_0^\infty\equiv (C^+)^{(3\delta-1)/2}  (B^c)^{-\delta}(-C_4^+)^{(1-\delta)/2} $ 
%&  $f_0^\infty \equiv  R_\chi^{-1}$\\ [1.5mm] 
$g_4^+\equiv -C_4^+/[ (C^+)^2 (f^+)^d]$ $\qquad\qquad$ &  
    $w^2\equiv C^- /[ B^2 (f^-)^d]$ \\ [1.5mm]
$U_\xi\equiv f^+/f^-$ & 
   $U_{\xi_{\rm gap}}\equiv f^+_{\rm gap}/f^-_{\rm gap}$ \\ [1.5mm]  
$Q^+ \equiv \alpha A^+ (f^+)^d$&$ Q^- \equiv \alpha A^- (f^-)^d $ \\ [1.5mm]
$R^+_\xi\equiv (Q^+)^{1/d}$&$Q^+_\xi\equiv f^+_{\rm gap}/f^+$\\ [1.5mm]
$Q^-_\xi\equiv f^-_{\rm gap}/f^-$&$Q^c_\xi\equiv f^c_{\rm gap}/f^c$\\ [1.5mm]
$Q_c \equiv B^2(f^+)^d/C^+$&$ Q_2\equiv (f^c/f^+)^{2-\eta} C^+/C^c$ \\ [1.5mm]
$R_\sigma \equiv \sigma_0 (f^-)^{d-1}$ & 
$R_\sigma^+\equiv \sigma_0 (f^+)^{d-1}$ \\ [1.5mm]
$ P_m \equiv { T_p^\beta B/B^c} $ & 
$ R_p \equiv { C^+/C_p} $ \\ [1.5mm]
\hline
\end{tabular}
\end{center}
\end{table*}

We also consider another trajectory in the $(T,H)$ plane,
the crossover or pseudocritical line $t_{\rm max}(H)$,
which is defined as the reduced temperature for which
the longitudinal magnetic susceptibility $\chi_L(t,H)$
has a maximum at $|H|$ fixed. RG predicts
\begin{eqnarray}
t_{\rm max}(H) = T_p |H|^{1/(\gamma+\beta)},\qquad\qquad
\chi_L(t_{\rm max},H)= C_p t_{\rm max}^{-\gamma}.
\label{chilcr}
\end{eqnarray}
Some related universal amplitude ratios are defined
in Table~\ref{notationsur}.

Finally, we mention the relation between 
ferromagnetic and antiferromagnetic models on bipartite lattices,
such as simple cubic and bcc lattices. 
In an antiferromagnetic 
model the relevant critical quantities are the staggered ones. For instance,
on a cubic lattice the staggered susceptibility 
is given by
\be
  \chi_{\rm stagg} = \sum_x (-1)^{p(x)} \< \sigma_0 \sigma_x\>,
\ee
where $p(x) = {\rm mod}\ (x_1 + \ldots + x_d,2)$ is the parity of $x$. 
One may easily prove that $\chi_{\rm stagg} = \chi_{\rm ferro}$,
where $\chi_{\rm ferro}$ is the ordinary susceptibility in the 
ferromagnetic model. The critical behavior of the staggered quantities
is identical to the critical behavior of the zero-momentum quantities in 
the ferromagnetic model. The critical behavior of the usual
thermodynamic quantities in antiferromagnets
is different, although still related to that of the ferromagnetic
model. For instance, the susceptibility behaves as \cite{Fisher-62}
\be
\chi \approx c_0 + c_1 t + \ldots + b_0 |t|^{1-\alpha} + \ldots
\ee
Higher-order moments of the two-point function, i.e. $\sum_x |x|^n G(x)$,
show a similar behavior \cite{CPRV-02}.

\subsection{Rigorous results for $N=1$} \label{sec-1.4}

Several rigorous results have been obtained for spin systems 
with $N=1$ and $N\rightarrow 0$ (as we shall see, in the limit
$N\rightarrow 0$ spin models can be mapped 
into walk models, see Sec. \ref{n0}). 
We report here only the most relevant ones for $N=1$ and refer the reader to 
Refs. \cite{Baker-book,Fernandez-etal_book,Madras-Slade_93} 
for a detailed presentation of the subject. Most of the results deal with the 
general ferromagnetic Hamiltonian
\be
{\cal H} = - \sum_{i<j} K_{ij} \phi_i\phi_j - \sum_i h_i \phi_i,
\ee
where $K_{ij}$, $h_i$ are arbitrary positive numbers, and the first  
sum is extended over all lattice pairs. The partition function 
is given by
\be 
Z = \int \prod_i \ \left[d\phi_i \; F(\phi_i)\right] \; e^{-{\cal H}},
\ee
where $F(x)$ is an even function satisfying
\be
\int_{-\infty}^\infty dx \, F(x)\, e^{bx^2} < +\infty,
\ee
for all real $b$. For this class of Hamiltonians the following results 
have been obtained:
\begin{itemize}
\item Fisher \cite{Fisher_69} proved the  inequality:
$\gamma \le (2 - \eta) \nu$.
\item Sokal \cite{Sokal_82} proved that 
$\chi \le {\rm const} (1 + \xi^2)$ for $T>T_c$, implying
$\gamma \le 2 \nu$.
If we assume the scaling relation $\gamma = (2 - \eta)\nu$,
then $\eta \nu \ge 0$, and $\eta \ge 0$ since $\nu > 0$.
\item The following Buckingham-Gunton inequalities have been proved
\cite{Buckingham-Gunton_67,Fisher_69}:
\begin{eqnarray}
2 - \eta \le  {d \gamma'\over 2\beta + \gamma'} \le \; {d
\gamma'\over 2 - \alpha'},\qquad
2 - \eta \le  d {\delta - 1\over \delta + 1}.
\end{eqnarray}
\item Sokal \cite{Sokal_81} proved that 
$d \nu' \ge \gamma' + 2 \beta \ge 2 - \alpha'$.
\end{itemize}
Moreover, 
for the $\phi^4$ theory it has been shown that
\cite{Glimm-Jaffe_77,Baker_75} $\gamma \ge 1$,
from which one may derive $\nu \ge 1/2$ using 
$\gamma \le 2 \nu$.

Several additional results have been proved for $d > 4$, showing that 
the exponents have mean-field values. In particular, Aizenman 
\cite{Aizenman_81_82}  proved that $\gamma = 1$, and then
\cite{Aizenman-Fernandez_86} that $\beta = 1/2$ and $\delta = 3$ 
for all $d \ge 5$. Moreover, for the zero-momentum four-point coupling
$g_4$ defined by $g_4 \equiv - \chi_4/(\chi^2 \xi^d)$, 
the inequality \cite{Aizenman_81_82}  
\be
0 \le g_4 \le {{\rm const}\over \xi^{d-4}} \to 0
\ee
holds, implying the absence of scattering (triviality) 
above four dimensions. 
For $d=4$ the RG results \cite{Larkin-Khmelnitskii_69,WR-73,BLZ-73}
\begin{eqnarray}
\chi(t) \sim t^{-1} |\log t|^{1/3}, \qquad
\xi_{\rm gap}(t) \sim  t^{-1/2} |\log t|^{1/6}, \qquad
g_4 \sim {1\over n_0 + |\log t|},
\end{eqnarray}
where $n_0$ is a positive constant, 
have been proved for the weakly coupled $\phi^4$ theory 
\cite{Hara_87,Hara-Tasaki_87}.

\subsection{Scaling behavior of the free energy and of the 
equation of state} \label{sec-1.5}

\subsubsection{Renormalization-group scaling} \label{sec-1.5.1}

According to RG, the Gibbs free energy obeys 
a general scaling law. Indeed, we can write it in terms of 
the nonlinear scaling fields associated with the RG
eigenoperators at the fixed point. If $u_i$ are the scaling fields---they
are analytic functions of $t$, $H$, and of any parameter appearing in the 
Hamiltonian---we have
\be
{\cal F}(H,t) = {\cal F}_{\rm reg}(H,t) + 
                {\cal F}_{\rm sing}(u_1,u_2,\ldots,u_n,\ldots),
\label{Gsing-RG-1}
\ee
where ${\cal F}_{\rm reg}(H,t)$ is an analytic (also at the critical point) 
function of $H$ and $t$ which is usually called background or 
bulk contribution. The function ${\cal F}_{\rm sing}$ obeys a scaling law of 
the form \cite{Wegner-76}:
\begin{eqnarray}
{\cal F}_{\rm sing}(u_1,u_2,\ldots,u_n,\ldots) = 
b^{-d} {\cal F}_{\rm sing}(b^{y_1} u_1,b^{y_2} u_2,\ldots,
   b^{y_n} u_n,\ldots),
\label{Gsing-RGscaling}
\end{eqnarray}
where $b$ is any positive number and $y_n$ are the 
RG dimensions of the scaling fields.\footnote{This 
is the generic scaling form. However, in certain specific
cases, the behavior is more complex with the appearance of logarithmic 
terms. This may be due to resonances between the RG eigenvalues, to
the presence of marginal operators, etc., see Ref. \cite{Wegner-76}.
The simplest example that shows such a behavior is 
the two-dimensional Ising model, see, e.g., Ref.~\cite{Wegner-76}.}
In the models that we 
consider, there are two relevant fields with $y_i>0$, 
and an infinite set of irrelevant fields with $y_i < 0$. The 
relevant scaling fields are associated with the temperature and the 
magnetic field. We assume that they correspond to $u_1$ and $u_2$.
Then $u_1\sim t$ and $u_2\sim |H|$ for $t,|H|\to 0$. If 
we fix 
$b$ by requiring $b^{y_1} |u_1| = 1$ in Eq. \reff{Gsing-RGscaling}, 
we obtain
\begin{eqnarray} 
{\cal F}_{\rm sing}(u_1,u_2,\ldots,u_n,\ldots) = |u_1|^{d/y_1} 
  {\cal F}_{\rm sing}({\rm sign}\, u_1, u_2 |u_1|^{-y_2/ y_1} ,
  \ldots, u_n |u_1|^{-y_n/ y_1}, \ldots). 
\label{Fsing-RG}
\end{eqnarray}
For $n> 2$, $y_i < 0$, so that 
$u_n |u_1|^{-y_n/ y_1}\to0$ for $t\to 0$. Thus,
provided that ${\cal F}_{\rm sing}$ is finite and nonvanishing in this 
limit,\footnote{
This is expected to be true below the upper critical dimension, 
but not above it \cite{Fisher-74-Temple}. 
The breakdown of this hypothesis causes a breakdown
of the hyperscaling relations, and allows the recovery of the mean-field
exponents for all dimensions above the upper critical one.} 
we can rewrite 
\begin{eqnarray} 
 {\cal F}_{\rm sing}(u_1,u_2,\ldots,u_n,\ldots) \approx 
  |t|^{d/y_1}
  {\cal F}_{\rm sing}({\rm sign}\, t, |H| |t|^{-y_2/ y_1} ,
  0,0,\dots).
\label{Gsing-RG-2}
\end{eqnarray}
Using Eqs. \reff{Gsing-RG-1} and \reff{Gsing-RG-2}, we obtain all scaling 
and hyperscaling relations provided that we identify 
\be
y_1 = {1\over \nu}, \qquad\qquad y_2 = {\beta + \gamma\over \nu}.
\ee
Note that the scaling part of the free energy is expressed in terms 
of two different functions, depending on the sign of $t$. However, 
since the free energy is 
analytic along the critical isotherm for $H\not=0$, the two functions are 
analytically related. 

It is possible to avoid the introduction of two different functions
by fixing $b$ so that
$b^{y_2} |u_2| = 1$. This allows us to write, for $t\to 0$ and 
$|H|\to 0$,
\begin{eqnarray}
 {\cal F}_{\rm sing}(u_1,u_2,\ldots,u_n,\ldots) \approx 
   |H|^{d/y_2} {\cal F}_{\rm sing}(t |H|^{y_1/y_2},1,
   0,0,\ldots),
\end{eqnarray}
where we have used the fact that the free energy does not depend on the
direction of $\vec{H}$.

In conclusion, for $|t|\to 0$, $|H|\to 0$, 
$t |H|^{-1/(\beta + \gamma)}$ fixed, we have
\begin{eqnarray}
{\cal F}(H,t) - {\cal F}_{\rm reg}(H,t) 
   \approx  |t|^{d\nu} \widehat{\cal F}_{1,\pm} (H |t|^{-\beta - \gamma}) 
=|H|^{d\nu/(\beta+\gamma)} \widehat{\cal F}_2(t |H|^{-1/(\beta + \gamma)}),
\label{calF-limitescaling}
\end{eqnarray}
where $\widehat{\cal F}_{1,\pm}$ apply for $\pm t > 0$ respectively.
Note that $d\nu = 2 - \alpha$ and $d\nu/(\beta + \gamma) = 1 + 1/\delta$.
The functions $\widehat{\cal F}_{1,\pm}$ and $\widehat{\cal F}_2$ are universal 
apart from trivial rescalings. Eq. \reff{calF-limitescaling} is valid 
in the critical limit. Two types of corrections are expected: 
analytic corrections due to the fact that $u_1$ and $u_2$ are analytic functions
of $t$ and $|H|$, and nonanalytic ones due to the irrelevant 
operators. The leading nonanalytic correction is 
of order $|u_1|^{-y_3/y_1} \sim t^\Delta$, 
or $|u_2|^{-y_3/y_2} \sim |H|^{\Delta_c}$, where we have 
identified $y_3 = - \omega$, $\Delta = \omega\nu$, 
$\Delta_c = \omega \nu_c$.

The Helmholtz free energy obeys similar laws.
In the critical limit, 
for $t\to 0$, $|M|\to 0$, and $t|M|^{-1/\beta}$ fixed, it can be
written as 
\begin{eqnarray}
\Delta {\cal A} = {\cal A}(M,t) - {\cal A}_{\rm reg}(M,t) \approx 
    |t|^{d\nu} \widehat{\cal A}_{1,\pm} (|M| |t|^{-\beta}) = 
   |M|^{\delta+1} \widehat{\cal A}_2(t |M|^{-1/\beta}),
\label{scaling-calF}
\end{eqnarray}
where ${\cal A}_{\rm reg}(M,t)$ is a regular background contribution and
$\widehat{\cal A}_{1,\pm}$ apply for $\pm t > 0$ respectively.
The functions $\widehat{\cal A}_{1,\pm}$ and $\widehat{\cal A}_2$ are universal 
apart from trivial rescalings.
The equation of state is then given by
\begin{equation}
\vec{H} = {\partial{\cal A}\over \partial \vec{M}}. 
\end{equation}

\subsubsection{Normalized free energy and related  quantities} \label{sec-1.5.2}

In this section we define some universal functions related to
$\widehat{\cal A}_{1,+}$ and $\widehat{\cal A}_2$. The 
function $\widehat{\cal A}_{1,-}$ for Ising systems will be discussed
in Sec. \ref{sec-1.5.4}.

The function $\widehat{\cal A}_{1,+} (|M| t^{-\beta})$ may be written 
in terms of a universal function $A_{1}(z)$ normalized in the high-temperature 
phase. The analyticity of the free energy outside the 
critical point and the coexistence curve (Griffiths' analyticity) implies that 
$\widehat{\cal A}_{1,+} (|M| t^{-\beta})$ has a regular expansion in powers of 
$|M|^2 t^{-2\beta}$. We introduce a new variable 
\be 
z \equiv b_1 |M| t^{-\beta},
\label{def-zvariable}
\ee
and write
\be
\widehat{\cal A}_{1,+} (|M| t^{-\beta}) = a_{10} + a_{11} A_1(z),
\ee
where the constants are fixed by requiring that
\be
A_1(z) = {z^2\over 2} + {z^4\over 4!} + O(z^6).
\label{normalization-A1}
\ee
The constants $a_{10}$, $a_{11}$, and $b_1$ can be expressed in terms 
of amplitudes that have been already introduced, i.e.,
\begin{eqnarray}
a_{10} = - {A^+\over (2 - \alpha)(1-\alpha)},  \qquad
a_{11} = - {(C^+)^2\over C^+_4}, \qquad
b_1 = \left[-{C^+_4\over (C^+)^3}\right]^{1/2}.
\label{def-abconstant}
\end{eqnarray}
They are not universal 
since they are normalization factors. On the other hand, 
the ratio $a_{11}/a_{10}$ and the function $A_1(z)$ are universal. 
It is worth mentioning that the ratio $a_{11}/a_{10}$ can be computed from the 
function $A_1(z)$ alone. Indeed, given the function $A_1(z)$, 
there is a unique constant $c$ such that $t^{2-\alpha} (c + A_1(z))$ 
is analytic on the critical isotherm. Such a constant is 
the ratio $a_{10}/a_{11}$.

The function $\widehat{\cal A}_2 (t |M|^{-1/\beta})$ is usually normalized 
imposing two conditions, respectively at the coexistence curve 
and on the critical isotherm. We introduce 
\be 
x \equiv B^{1/\beta}\, t |M|^{-1/\beta},
\label{def-xvariable}
\ee
where $B$ is the amplitude of the magnetization, so that $x = -1$ corresponds
to the coexistence curve. Then, we define
\be
\widehat{\cal A}_2 (t |M|^{-1/\beta}) = a_{20} A_2(x),
\label{normalizzazione-calA2}
\ee
requiring $A_2(0) = 1$. This fixes the constant $a_{20}$:
\be
a_{20} = {(B^c)^{-\delta}\over \delta + 1}.
\label{def-a20constant}
\ee
Again, $a_{20}$ is nonuniversal while $A_2(x)$ is universal.

The functions $A_1(z)$ and $A_2(x)$ are related:
\be
A_1(z) = - {a_{10}\over a_{11}} + B^{\delta+1} {a_{20}\over a_{11}}\ 
  x^{-d\nu}\ A_2(x).
\ee
The scaling equation of state can be written as 
\begin{eqnarray}
\vec{H} =
          a_{11} b_1 {\vec{M}\over |M|} t^{\beta\delta} F(z) 
         = (B^c)^{-\delta} \vec{M} |M|^{\delta - 1} f(x),
\label{eq-stato_a}
\end{eqnarray}
and
\begin{eqnarray}
F(z) \equiv {A_1'}(z)
\qquad\qquad
f(x) \equiv A_2(x) - {x\over d\nu} {A_2'}(x)
\label{eq-stato} 
\end{eqnarray}
Note that $f(0) = 1$, since $A_2(0) = 1$, and $f(-1)=0$ since 
$x=-1$ corresponds to the coexistence curve.

By solving Eq. \reff{eq-stato} with the appropriate 
boundary conditions, it is possible to reobtain  
the free energy. This is trivial in the case of $A_1(z)$. 
In the case of $A_2(x)$ we have
\begin{eqnarray}
A_2(x) = f(x) + {xf'(0)\over 1-\alpha}  - |x|^{2-\alpha} 
  \int^x_0 dy\ |y|^{\alpha-2} \left[f'(y) - f'(0)\right]
\end{eqnarray}
for $\alpha > 0$, see, e.g., Ref.~\cite{BHK-75}.
For $-1 < \alpha < 0$ one needs to perform 
an additional subtraction within the integral \cite{BHK-75}.

It is useful to define universal functions starting from the Gibbs free 
energy, cf. Eq.~(\ref{calF-limitescaling}). We introduce a variable 
\begin{equation}
y \equiv \left({B\over B^c}\right)^{1/\beta} t |H|^{-1/(\beta+\gamma)},
\end{equation}
and define
\begin{equation}
\widehat{\cal F}_2(t |H|^{-1/(\beta + \gamma)}) = 
 - {\delta B^c\over {\delta + 1}}\, G(y),
\end{equation}
so that $G(0)=1$.
The equation of state can now be written as 
\begin{equation}
\vec{M} = B^c \vec{H}\, |H|^{(1-\delta)/\delta} E(y),\qquad
E(y) = G(y) - {y\over d\nu} {\partial G\over \partial y}\; .
\end{equation}
Clearly, $E(y)$ and $f(x)$ are related: 
\begin{eqnarray}
E(y) = f(x)^{-1/\delta} , \qquad
y = x f(x)^{-1/(\beta + \gamma)}.
\label{rel-funw-funx}
\end{eqnarray}
Finally, we introduce a scaling function associated with the 
longitudinal susceptibility, by writing 
\begin{equation}
\chi_L = B^c |H|^{1/\delta-1} D(y),
\end{equation}
where 
\begin{eqnarray}
D(y) = {1\over \delta} 
   \left[E(y) - {y\over \beta} E'(y)\right] 
    = {\beta f(x)^{1-1/\delta} \over 
         \beta\delta f(x) - x f'(x)}.
\label{defDw}
\end{eqnarray}
The function $D(y)$ has a maximum at $y=y_{\rm max}$ corresponding to 
the crossover line defined in Sec. \ref{sec-1.3}. 
We can relate 
$y_{\rm max}$ and $D(y_{\rm max})$ to the amplitude ratios 
$P_m$ and $R_p$ defined
along the crossover line, see Table~\ref{notationsur},
\begin{eqnarray}
y_{\rm max} = \left(P_m\right)^{1/\beta}, \qquad
D(y_{\rm max}) = R_p^{-1} P_m^{1-\delta} R_\chi.
\end{eqnarray}

\subsubsection{Expansion of the equation of state} \label{sec-1.5.3}

The free energy is analytic in the $(T,H)$ plane outside the 
critical point and the coexistence curve.
As a consequence, the functions $A_1(z)$ and 
$F(z)$ have a regular expansion in powers of $z$, with the appropriate 
symmetry under $z\to -z$. The expansion of $F(z)$ can be written as
\begin{equation}
F(z) =\, 
z + \case{1}{6}z^3 + \sum_{n=3} {r_{2n}\over(2n-1)!} z^{2n-1}.
\label{Fzdef}
\end{equation}
The constants $r_{2n}$ can be computed in terms of the $2n$-point
functions $\chi_{2n}$ for $H=0$ and $t\to 0^+$. Explicitly
\begin{eqnarray}
&&r_6 = 10 - {C^+_6 C^+\over (C^+_4)^2}, \nonumber \\
&&r_8 = 280 - 56 {C^+_6 C^+\over (C^+_4)^2}
        + {C^+_8 (C^+)^2\over (C^+_4)^3},\nonumber\\
&&r_{10} =
15400 - 4620 {C^+_6  (C^+)\over (C^+_4)^2} 
   + 126 {(C^+_6)^2  (C^+)^2\over (C^+_4)^4}   
      + 120 {C^+_8  (C^+)^2\over (C^+_4)^3}
  - {C^+_{10}  (C^+)^3\over (C^+_4)^4},
\label{r2jgreen}
\end{eqnarray}
etc. The coefficients $r_{2n}$ are related to the $2n$-point
renormalized coupling constants 
$g_{2n}^+ \equiv\, r_{2n}\, (g_4^+)^{n-1}$.
Griffiths' analyticity implies that  ${\cal A}(M,t)$ has also a regular 
expansion in powers of $t$ for $|M|$ fixed. 
As a consequence, $F(z)$ has the  following
large-$z$ expansion
\begin{equation}
F(z) = z^\delta \sum_{k=0} F^{\infty}_k z^{-k/\beta}.
\label{asyFz}
\end{equation}
The constant $F_0^\infty$ can be expressed in terms of universal
amplitude ratios, using the asymptotic behavior of the magnetization
along the critical isotherm. One obtains
\begin{eqnarray}
F_0^\infty &=& (\delta+1) {a_{20}\over a_{11}} b_1^{-\delta - 1} 
= R_\chi \left(R_4^+\right)^{(1-\delta)/2},
\label{f0inf}
\end{eqnarray}
where $R_\chi$ and $R_4^+$ are
defined in Table~\ref{notationsur}.
The functions $f(x)$ and $F(z)$ are related: 
\begin{eqnarray}
z^{-\delta} F(z) = F_0^\infty f(x),  \qquad
z = z_0 x^{-\beta}, 
\label{relazioneF-f} 
\end{eqnarray}
where
\begin{equation}
z_0^2 = b_1^2 B^2 = R_4^+.
\label{z0}
\end{equation}
Griffiths' analyticity implies that $f(x)$ is regular everywhere for $x>-1$. 
The regularity of $F(z)$ for $z\to 0$ implies 
a large-$x$ expansion of the form
\begin{equation}
f(x) = x^\gamma \sum_{n=0}^\infty f_n^\infty x^{-2n\beta}.
\label{largexfx}
\end{equation}
The coefficients $f_n^\infty$ can be expressed in terms of $r_{2n}$ 
using Eq. (\ref{Fzdef}),
\begin{equation}
f_n^\infty \; =\; z_0^{2n+1-\delta} {r_{2n+2}\over F_0^\infty (2 n + 1)!},
\end{equation}
where $r_2 = r_4 = 1$. In particular, using 
Eqs. (\ref{f0inf}) and (\ref{z0}), one finds
$f_0^\infty = R_\chi^{-1}$.
The function $f(x)$ has a regular expansion in powers of $x$,
\be
f(x) = 1 + \sum_{n=1}^\infty f_n^0 x^n,
\label{expansionfx-xeq0}
\ee
where the coefficients are related to those appearing in Eq. (\ref{asyFz}):
\be
  f_n^0 = {F_n^\infty \over F_0^\infty} z_0^{-n/\beta}.
\ee
Using Eqs.~(\ref{rel-funw-funx}) and (\ref{defDw}) and the above-presented
results, one can derive the expansion of $E(y)$ and $D(y)$ for 
$y\to +\infty$ and $y\to 0$. For $y\to +\infty$, we have
\begin{eqnarray}
E(y) = R_\chi y^{-\gamma} \left[1 + O(y^{-2\beta\delta})\right], \qquad
D(y) = R_\chi y^{-\gamma} \left[1 + O(y^{-2\beta\delta})\right].
\end{eqnarray}

\subsubsection{The behavior at the coexistence curve for 
scalar systems} \label{sec-1.5.4}

For a scalar theory, 
the free energy ${\cal A}(M,t)$ admits a power-series expansion\footnote{
Note that we are not claiming that the free energy is analytic 
on the coexistence curve. Indeed, essential singularities are 
expected \cite{Langer-67,Fisher-67,Andreev-64,FF-70,Isakov-84}. 
Thus, the expansion \reff{expansion-calF-coex} 
should be intended as a formal power series.}
near the coexistence curve, i.e.\ for $t<0$ and $H=0$.  If
$M_0=\lim_{H\to 0^+} M(H)$, for $M>M_0$ (i.e.\ for $H\geq 0$) we can write
\begin{equation}
{\cal A}(M,t) = 
\sum_{j=0} {1\over j!} a_j(t) (M-M_0)^j,
\label{expansion-calF-coex}
\end{equation}
with $a_1(t) = 0$.
This implies an expansion of the  form 
\be
{\cal A}_{\rm sing}(M,t) = 
  (-t)^{2-\alpha} \left[{a}_0 + b_2 Q(u)\right]
\ee
for the singular part of the free energy,
where 
\begin{eqnarray}
{a}_0 = - {A^-\over (\alpha - 1)(\alpha-2)},\qquad
b_2 = {B^2 \over C^-},\qquad
u = B^{-1} M (-t)^{-\beta},
\end{eqnarray}
and $Q(u)$ is normalized so that
\be
Q(u) = \smfrac{1}{2} (u-1)^2 + \sum_{j=3} {v_j\over j!} (u - 1)^j.
\label{sviluppoBu}
\ee
The function $Q(u)$ is universal, as well as the ratio $b_2/a_0$.
The universal constants $v_j$ can be related to critical ratios of the 
correlation functions $\chi_n$. Some explicit formulae are reported in 
Table \ref{notationsur}. 
The constants $v_j$ are related to the 
low-temperature zero-momentum coupling constants
$g^-_n \equiv  v_n w^{n-2}$, where $w^2$ is defined in Table~\ref{notationsur}.
The relation between $f(x)$ and $Q(u)$ is
\begin{eqnarray}
f(x) = b_0 
  u^{-\delta} {dQ(u)\over du}, \qquad
x = -u^{-1/\beta}, 
\end{eqnarray}
where  $b_0 = {U_2/R_\chi}$.
At the coexistence curve,  i.e.  for  $x\to -1$,
\be
f(x) = f_1^{\rm coex} (x + 1) + f_2^{\rm coex} (x+1)^2 + O((x+1)^3),
\label{coexcfx}
\ee
where $f_1^{\rm coex}  = b_0 \beta$.

Using Eqs.~(\ref{rel-funw-funx}) and (\ref{defDw}) and the above-presented
results, we can derive the expansion of $E(y)$ and $D(y)$ for 
$y\to -\infty$,
\begin{eqnarray}
   E(y) \approx (-y)^\beta \left[1 + O( (-y)^{-\beta\delta})\right], \qquad
   D(y) \approx {1\over b_0} 
      (-y)^{-\gamma} \left[1 + O( (-y)^{-\beta\delta})\right].
\end{eqnarray}

\subsubsection{The behavior at the coexistence curve 
for vector systems} \label{sec-1.5.5}

Since the free energy ${\cal A}$ is a function of $|M|$, we have
\begin{eqnarray}
\chi_T = {|M|\over |H|}  ,\qquad
\chi_L = {\partial |M|\over \partial |H|}.
\label{chitl}
\end{eqnarray}
The leading behavior of $\chi_L$ at the coexistence curve can be derived
from the behavior of $f(x)$ for $x\to -1$. The presence of the 
Goldstone singularities drastically changes the behavior of 
$f(x)$ with respect to the scalar case. 
In the vector case the singularity is controlled
by the zero-temperature infrared-stable Gaussian fixed
point~\cite{BW-73,BZ-76,Lawrie-81}. This implies that
\begin{equation}
f(x) \approx  c_f \,(1+x)^{2/(d-2)} 
\label{fxcc} 
\end{equation}
for $x\rightarrow -1$. Therefore
\begin{eqnarray}
\chi_L^{-1} = \delta {|H|\over |M|} - 
      {(B^c)^{-\delta}\over \beta} |M|^{\delta - 1}\,  x f'(x) 
    \propto (-t)^{\beta \delta(d-2)/2 - \beta}\,  |H|^{(4-d)/2}
\end{eqnarray}
near the coexistence curve, showing that
$\chi_L$ diverges as $|H|\to 0$. 

The nature of the corrections to the behavior (\ref{fxcc}) is less
clear. Setting $v\equiv 1+x$ and $y\equiv |H|\ |M|^{-\delta}$, 
it has been conjectured that 
$v$ has  a double expansion in powers
of $y$ and $y^{(d-2)/2}$ near  the coexistence curve \cite{WZ-75,SH-78,Lawrie-81}, 
i.e.,  for $y\to0$,
\begin{eqnarray}
v\equiv 1+x =  c_1 y^{1-\epsilon/2} + c_2 y + 
+ d_1 y^{2-\epsilon} + d_2 y^{2-\epsilon/2} + d_3 y^2 + \ldots
\label{expcoex1}
\end{eqnarray}
where $\epsilon \equiv 4 - d$.  This expansion has been derived essentially
from an $\epsilon$-expansion analysis.  In
three dimensions it predicts an expansion of $v$ in powers of
$y^{1/2}$, or equivalently an expansion of $f(x)$ in powers of
$v$ for $v\to 0$.

The asymptotic expansion  of the $d$-dimensional  equation of state at
the coexistence curve was computed  analytically in the framework
of the large-$N$ expansion~\cite{PV-99},  using the $O(1/N)$  formulae
reported in  Ref.~\cite{BW-73}.    It turns out that
the  expansion
(\ref{expcoex1})  does not hold  for  values of the dimension
$d$ such that
\begin{equation}
2 < d = 2 + {2 m\over n} < 4, \quad {\rm for}\quad 
0< m < n,
\label{speciald}
\end{equation}
with $m,n\in\ \N$.
In particular, in three dimensions one finds~\cite{PV-99}
\begin{equation}
f(x)=v^2\left\{1 + {1\over N}\left[
  f_1(v) + f_2(v)\ln v  \right] 
    + O(N^{-2})\right\} ,
\label{fx3d}
\end{equation}
where the functions $f_1(v)$ and $f_2(v)$ have a regular
expansion in powers of $v$. Moreover,
$f_2(v)=O(v^2)$,
so that logarithms affect the expansion only at the
next-next-to-leading order.  A possible interpretation of the
large-$N$ result is that the expansion (\ref{fx3d}) holds for all
values of $N$, so that Eq.\ (\ref{expcoex1}) is not correct due to the
presence of logarithms.  The reason of their appearance is 
unclear, but it does not contradict the conjecture that the
behavior near the coexistence curve is controlled by the
zero-temperature infrared-stable Gaussian fixed point.  In this case
logarithms would not be unexpected, as they usually appear in
reduced-temperature asymptotic expansions around Gaussian fixed points
(see, e.g., Ref.~\cite{BB-85}).

\subsubsection{Parametric representations} \label{sec-1.5.6}

The analytic properties of the equation of state
can be implemented in a simple way by introducing appropriate
parametric representations \cite{Schofield-69,SLH-69,Josephson-69}.
One may parametrize $M$ and $t$ in terms of two 
new variables $R$ and $\theta$ according to\footnote{It is also possible 
to generalize the expression for $t$, writing $t = R\, k(\theta)$. The 
function $k(\theta)$ must satisfy the obvious requirements: 
$k(0) > 0$, $k(\theta_0) < 0$, $k(\theta)$ decreasing in 
$0\le \theta\le \theta_0$.}
\begin{eqnarray}
|M| &=& m_0 R^\beta m(\theta) ,\nonumber  \\
t &=& R (1-\theta^2), \nonumber \\
|H| &=& h_0 R^{\beta\delta}h(\theta),\label{parrepg}
\end{eqnarray}
where $h_0$ and $m_0$ are normalization constants.
The variable $R$ is nonnegative and measures
the distance from the critical point in the $(t,H)$ plane;
the critical behavior is obtained for $R\to 0$.
The variable $\theta$  parametrizes the displacements along the lines
of constant $R$. 
The line $\theta=0$ corresponds to the high-temperature phase $t>0$ and $H=0$;
the line $\theta=1$ to the critical isotherm $t=0$;
$\theta=\theta_0$, where $\theta_0$ is the smallest positive zero
of $h(\theta)$, to the coexistence curve $T<T_c$ and $H\to 0$.
Of course, one should have $\theta_0 > 1$, $m(\theta)>0$ for 
$0< \theta\le \theta_0$, and $h(\theta)> 0$ for 
$0< \theta< \theta_0$.
The functions $m(\theta)$ and  $h(\theta)$ must be
analytic in the physical interval $0\le\theta<\theta_0$ in order to satisfy the
requirements of regularity of the equation of state (Griffiths' analyticity).
Note that the mapping (\ref{parrepg}) is not invertible when
its Jacobian vanishes, which occurs  when
\begin{equation}
Y(\theta) \equiv (1-\theta^2)m'(\theta) + 2\beta\theta m(\theta)=0.
\label{Yfunc}
\end{equation}
Thus, the parametric representation 
is acceptable only if $\theta_0<\theta_l$,
where $\theta_l$ is the smallest positive
zero of the function $Y(\theta)$.
The functions $m(\theta)$ and $h(\theta)$ are odd\footnote{
This requirement guarantees that the equation of state has an
expansion in odd powers of $|M|$, see Eq. \reff{Fzdef}, in the 
high-temperature phase for $|M|\to 0$. In the Ising model, this 
requirement can be understood directly, since in Eq. \reff{parrepg} 
one can use $H$ and $M$ instead of their absolute values, and thus 
it follows from the $\mathbb{Z}_2$ symmetry of the theory.}
in $\theta$, and
can be  normalized so that $m(\theta)=\theta+O(\theta^3)$ and
$h(\theta)=\theta+O(\theta^3)$. Since 
$\vec{H} = a_{11} b_1^2 t^\gamma \vec{M}$
for $|M|\to 0$, $t>0$, see Eqs. \reff{eq-stato_a} and \reff{Fzdef},
these normalization conditions imply $h_0 = a_{11} b_1^2 m_0 = m_0/C^+$. 
Following Ref.~\cite{GZ-97}, we introduce a new constant $\rho$ by
writing
\be
m_0 = {\rho\over b_1}, \qquad\qquad
h_0 = {\rho b_1 a_{11}}.
\label{eq:1.132}
\ee
Using Eqs. \reff{eq-stato_a} and (\ref{parrepg}),
one can relate the functions $h(\theta)$ and $m(\theta)$ to
the scaling functions $F(z)$ and $f(x)$. We have
\begin{eqnarray}
z = \rho \,m(\theta)\,\left( 1 - \theta^2\right)^{-\beta},\qquad
F(z(\theta)) = \rho \left( 1 - \theta^2 \right)^{-\beta\delta} h(\theta),
\label{hFrel}
\end{eqnarray}
and 
\begin{eqnarray}
x = {1 - \theta^2\over \theta_0^2 - 1} \left[ {m(\theta_0)\over
m(\theta) }\right]^{1/\beta} ,\qquad
f(x) = \left[ {m(\theta)\over m(1)}\right]^{-\delta} {h(\theta)\over h(1)}.
\label{fxmt}
\end{eqnarray}
The functions $m(\theta)$ and $h(\theta)$ are largely arbitrary. In many cases,
one simply takes $m(\theta) = \theta$. 
Even so, the normalization condition $h(\theta) \approx \theta$ 
for $\theta \to 0$ does not completely fix $h(\theta)$. 
Indeed, one can rewrite 
\be
x^\gamma = h(1) f_0^\infty (1-\theta^2)^\gamma \theta^{1-\delta}, \qquad
f(x) = \theta^{-\delta} h(\theta)/h(1).
\ee
Thus, given $f(x)$, the value $h(1)$ can be chosen arbitrarily.

For the Ising model, the expansion \reff{expansion-calF-coex}
at the coexistence curve implies a regular expansion 
in powers of $(\theta-\theta_0)$, with 
\begin{eqnarray}
m(\theta) \approx m_{f,0} + m_{f,1} (\theta-\theta_0) + \ldots,\qquad
h(\theta) \approx h_{f,1} (\theta-\theta_0) + \ldots,
\label{hcoex-Neq1}
\end{eqnarray}
with $m_{f,0}\not = 0$. 
For three-dimensional models with $N\geq 2$, Eq. \reff{fxcc} implies
\begin{eqnarray}
m(\theta) \approx m_{f,0} + m_{f,1} (\theta-\theta_0) + \ldots,\qquad
h(\theta) \approx h_{f,2} (\theta-\theta_0)^2 + \ldots,
\label{hcoex-Nge2}
\end{eqnarray}
with $m_{f,0}\not = 0$.
The logarithmic corrections discussed in Sec. \ref{sec-1.5.5}
imply that $h(\theta)$ and/or $m(\theta)$ cannot be expanded 
in powers of $\theta-\theta_0$. 

From the parametric representations (\ref{parrepg})
one can recover the singular part of the free energy. Indeed
\begin{equation}
{\cal F}_{\rm sing} = h_0 m_0 R^{2-\alpha} g(\theta),
\end{equation}
where $g(\theta)$  is the solution of the first-order differential
equation
\begin{eqnarray}
(1-\theta^2) g'(\theta) + 2(2-\alpha)\theta g(\theta) =
 \left[(1-\theta^2)m'(\theta) + 2\beta\theta m(\theta)\right] h(\theta)
\label{pp1}
\end{eqnarray}
that is regular at $\theta=1$.

The parametric representations are useful because the functions 
$h(\theta)$ and $m(\theta)$ can be chosen analytic in all the 
interesting domain  $0\le \theta < \theta_0$. This is at variance 
with the functions $f(x)$ and $F(z)$ which display a nonanalytic behavior 
for $x\to\infty$ and $z\to\infty$ respectively. 
This fact is very important from a practical point of view. 
Indeed, in order to obtain approximate expressions of the 
equation of state, one can approximate $h(\theta)$ and 
$m(\theta)$ with analytic functions. The structure of the 
parametric representation automatically ensures that the analyticity 
properties of the equation of state are satisfied.

\subsubsection{Corrections to scaling} \label{sec-1.5.7}

In the preceding sections we have only considered the asymptotic
critical behavior.
Now, we discuss the corrections that are due to the nonlinear
scaling fields in Eq. \reff{Gsing-RGscaling} with $y_i < 0$. 

Using Eq. \reff{Fsing-RG} and keeping only one irrelevant field, 
the one  with the largest
$y_i$ (we identify it with $u_3$), we have
\begin{eqnarray}
{\cal F}_{\rm sing}(u_1,u_2,u_3)  
&=& |u_1|^{d/y_1} {\cal F}_{\rm sing}({\rm sign}\, u_1, u_2
|u_1|^{-y_2/ y_1} ,u_3 |u_1|^{-y_3/ y_1})  \nonumber \\
&=& |u_1|^{d\nu} \sum_{n=0}^\infty
    f_{n,\pm} (u_2 |u_1|^{-\beta - \gamma}) (u_3 |u_1|^{\Delta})^n,
\label{F-expansion}
\end{eqnarray}
where we use the standard notations $\omega \equiv - y_3$,
$\Delta \equiv \omega\nu$. 
The presence of the irrelevant operator induces nonanalytic
corrections proportional to $|u_1|^{n \Delta}$. 
The nonlinear scaling fields 
are analytic functions of $t$, $H$, and of any parameter appearing in 
the Hamilto\-nian---we indicate them collectively by $\lambda$.
Therefore, we can write
\begin{eqnarray}
u_1 &=& t + t^2 g_{11}(\lambda) + H^2 g_{21}(\lambda) +
         O(t^3,t H^2,H^4), \nonumber \\
u_2 &=& H\left[1 + t g_{12}(\lambda) + H^2 g_{22}(\lambda) +
          O(t^2, t H^2, H^4)\right], \nonumber \\
u_3  &=& g_{13} (\lambda) + t g_{23} (\lambda) +H^2 g_{33} (\lambda) + 
        O(t^2, t H^2, H^4). 
\label{mu-expansion}
\end{eqnarray}
Substituting these expressions into
Eq.\ (\ref{F-expansion}), we see that, if
$g_{13} (\lambda)\not=0$, the singular part of the  free energy 
has corrections of order
$t^{n\Delta+m}$. These nonanalytic corrections 
appear in all quantities. Additional 
corrections are due to the background term. For instance,
the susceptibility in zero magnetic field should be written
as \cite{AF-83}
\begin{eqnarray}
\chi =\, t^{-\gamma} \sum_{m,n=0}^\infty
    \chi_{1,mn}(\lambda) t^{m\Delta + n} 
     + t^{1 -\alpha} \sum_{m,n=0}^\infty
    \chi_{2,mn}(\lambda) t^{m\Delta + n} 
    + \sum_{n=0}^\infty \chi_{3,n}(\lambda) t^{n},
\end{eqnarray}
where the contribution proportional to $t^{1 -\alpha}$ stems from the
terms of order $H^2$ appearing in the expansion of $u_1$ and $u_3$,
and the last term comes from the regular part of the free
energy.  
The regular part of the free energy has often been assumed not to
depend on $H$. If this were the case, we would have
$\chi_{3,n}(\lambda) = 0$. However, for the two-dimensional Ising
model, one can prove rigorously that $\chi_{3,0}\not=0$
\cite{KAP-86,GM-88}, showing the
incorrectness of this conjecture. For a discussion, see Ref.\
\cite{SS-00}.

Analogous corrections are due to the 
other irrelevant operators present in the theory,
and therefore we expect corrections proportional to $t^\rho$ with
$\rho = n_1 + n_2 \Delta + \sum_i m_i \Delta_i$, where $\Delta_i$ are
the exponents associated with the additional irrelevant operators. 

In many interesting instances, by choosing a 
specific value $\lambda^*$ of a parameter $\lambda$ appearing in the
Hamiltonian, one can achieve the suppression of the leading
correction due to the irrelevant operators.
It suffices to choose $\lambda^*$
such that $g_{13}(\lambda^*)=0$.  In this case, $u_3 
|u_1|^\Delta \sim t^{1 + \Delta}$, so that no terms of the form
$t^{m\Delta + n}$, with $n < m$, are  present.  In particular, the
leading term proportional to $t^\Delta$ does not appear in the
expansion. 
This class of models is particularly useful in numerical works. 
We will call them {\em improved} models, and the corresponding 
Hamiltonians will be named {\em improved} Hamiltonians.

\subsubsection{Crossover behavior} \label{sec-1.5.8}

The discussion presented in Sec. \ref{sec-1.5.1} can be generalized by 
considering a theory perturbed by a generic relevant 
operator\footnote{Here, we assume ${\cal O}$ to be an eigenoperator
of the RG transformations. This is often guaranteed by the 
specific symmetry properties of ${\cal O}$. For instance, 
the magnetic field $H$ is an eigenoperator in magnets due to the 
$\mathbb{Z}_2$-symmetry of the theory.} ${\cal O}(x)$. 
Let us consider the Hamiltonian
\begin{equation}
{\cal H} = {\cal H}_0 + \sum_x h_o {\cal O}(x)
\end{equation}
and assume that the theory
is critical for $h_o = h_{o,c}$. 
The singular part of the Gibbs free energy for $t\to 0$ and 
$\Delta h_{o} \equiv h_o - h_{o,c}\to 0$ can be
written as
\begin{equation}
{\cal F}_{\rm sing}(t,h_o) \approx |t|^{d\nu}
 \widehat{\cal F}_\pm \left(\Delta h_o |t|^{-y_o \nu}\right),
\end{equation}
where $y_o$ is the RG dimension of ${\cal O}$. It is customary to
define the crossover exponent $\phi_o$ as 
$\phi_o = y_o \nu$.
Correspondingly, we have 
\begin{eqnarray}
&&\langle {\cal O} (x)\rangle_{\rm sing} \approx 
        |t|^{\beta_o} a_\pm\left(\Delta h_o |t|^{-\phi_o}\right),\\
&&\sum_x \langle {\cal O} (0) {\cal O} (x)\rangle_{\rm conn} \approx 
    |t|^{-\gamma_o} b_\pm\left(\Delta h_o |t|^{-\phi_o}\right),
\end{eqnarray}
where
\begin{eqnarray}
\beta_o = 2 - \alpha - \phi_o, \qquad
\gamma_o = 2 \phi_o + \alpha - 2. \label{expcross}
\end{eqnarray}
The functions $\widehat{\cal F}_\pm(x)$, $a_\pm(x)$, and 
$b_\pm(x)$ are universal and are usually referred to as crossover functions.

There are several interesting cases in which this formalism applies. 
For instance, one can consider the Gaussian theory and 
${\cal O} = \phi^4$. This gives the crossover behavior from the 
Gaussian to the Wilson-Fisher point that will be discussed in 
Sec.~\ref{crossover}. In $O(N)$ models, it is interesting to consider 
the case in which the operator is a linear combination of the 
components of the spin-two tensor \cite{FP-72,Wegner-72,FN-74,Aharony-76}
\begin{equation}
{\cal O}^{ab}_2 = \phi^a \phi^b - {1\over N} \delta^{ab} \phi^2.
\label{Oab}
\end{equation}
Such an operator is relevant for the description of the breaking of the 
$O(N)$ symmetry down to $O(M)\oplus O(N-M)$, $N > M$. 
Note that the crossover exponent and the crossover functions do not 
depend on the value of $M$. Higher-spin operators are also of interest. 
We report here the spin-3 and spin-4 operators:
\begin{eqnarray}
{\cal O}^{abc}_3 &=& \phi^a \phi^b \phi^c - 
         {1\over N+2} \phi^2 \left( 
         \delta^{ab} \phi^c + \delta^{ac} \phi^b + \delta^{bc} \phi^a\right),
\\
{\cal O}^{abcd}_4 &=& \phi^a \phi^b \phi^c \phi^d - 
       {1\over N+4} \phi^2 \left( 
        \delta^{ab} \phi^c \phi^d + \delta^{ac} \phi^b \phi^d + 
        \delta^{ad} \phi^b \phi^c + \delta^{bc} \phi^a \phi^d + 
        \delta^{bd} \phi^a \phi^c + \delta^{cd} \phi^a \phi^b \right) 
\nonumber \\
   && + {1\over (N+2)(N+4)} (\phi^2)^2 \left(
         \delta^{ab} \delta^{cd} + \delta^{ac} \delta^{bd} + 
         \delta^{ad} \delta^{bc} \right),
\end{eqnarray}
which are symmetric and traceless tensors.
In the following we will name $\phi_n$, $\beta_n$, and $\gamma_n$ the 
exponents associated with these spin-$n$ perturbations of the 
$O(N)$ theory.

The operators reported here are expected to be the most relevant ones for 
each spin. Other spin-$n$ operators with smaller RG dimensions can be obtained
by multiplying by powers of $\phi^2$ and adding derivatives.

\subsection{The two-point correlation function of 
the order parameter} \label{sec-1.6}

The critical behavior of the two-point correlation function $G(x)$ of
the order parameter is relevant to the description of 
scattering phenomena with light and neutron sources.
RG predicts the scaling behavior \cite{TF-75}
\be
\widetilde{G}(q) \approx |t|^{-\gamma} 
   Z(t |M|^{-1/\beta}, q |t|^{-\nu}),
\label{Gscalingform}
\ee
where $\widetilde{G}(q)$ is the 
Fourier transform of $G(x)$ and
$Z(y_1,y_2)$ is  universal apart from trivial rescalings.
Here we discuss the behavior for $H=0$. Results for 
$H\not = 0$ can be found in Refs. \cite{TF-75,BLZ-74b}.

\subsubsection{The high-temperature critical behavior} \label{sec-1.6.1}

In the high-temperature phase
we can write
\cite{FB-67,FA-73,FA-74,TF-75}
\be
\widetilde{G}(q) \approx {\chi\over g^+(y)},
\label{scaling-G-HT}
\ee
where $y \equiv q^2 \xi^2$, $\xi$
is the second-moment correlation length, 
and $g^+(y)$ is a universal function. 
In the Gaussian theory
the function $\widetilde{G}(q)$ has a very simple form, the 
so-called Ornstein-Zernike behavior,
\be 
\widetilde{G}(q) \approx {\chi\over 1 + q^2 \xi^2}.
\label{gaubeh}
\ee
Such a formula is by definition exact for $q\to 0$. However, as $q$ 
increases, there are significant deviations.

For $y\to 0$, $g^+(y)$ has a regular expansion in powers of $y$, i.e.
\be
g^+(y) = 1 + y + \sum_{n=2} c^+_n y^n,
\label{gypiu}
\ee
where $c^+_n$, $n\ge 2$, are universal constants.

For $y\to\infty$, the function $g^+(y)$ follows the Fisher-Langer law
\cite{FL-68}
\begin{equation}
g^+(y)^{-1} \approx {A^+_1\over y^{1 - \eta/2}}
  \left(1 + {A^+_2\over y^{(1-\alpha)/(2 \nu)}} +
            {A^+_3\over y^{1/(2\nu)}}\right).
\label{FL-law}
\ee
Other two interesting quantities, $\xi_{\rm gap}$ and $Z_{\rm gap}$,
characterize the 
large-distance behavior of $G(x)$. Indeed, for $t>0$
the function decays exponentially for large $x$ according to:
\be
G(x) \approx {Z_{\rm gap}\over 2 (\xi_{\rm gap})^d} \, 
    \left({2 \pi |x|\over  \xi_{\rm gap}}\right)^{-(d-1)/2}  
     e^{-|x|/\xi_{\rm gap}}.
\label{Gx-largex-generale}
\ee
Then, we can define the universal ratios 
\begin{eqnarray} 
S_M^+ \equiv \lim_{t\to 0^+} {\xi^2\over \xi^2_{\rm gap}} = 
    \left({f^+\over f^+_{\rm gap}}\right)^2, \qquad
S_Z^+ \equiv \lim_{t\to 0^+} {\chi \over \xi^2_{\rm gap} Z_{\rm
    gap}},
\label{SMdef}
\end{eqnarray}
and $Q^+_\xi \equiv (S_M^+)^{-1/2}$.
If $y_0$ is the negative zero of $g^+(y)$ that is closest to the origin,
then
\begin{eqnarray}
S_M^+ = {|y_0|}, \qquad
S_Z^+= \left.{\partial g^+(y) \over  \partial y} \right|_{y=y_0}.
\end{eqnarray}

\subsubsection{The low-temperature critical behavior} \label{sec-1.6.2}

For scalar models, the behavior in the low-temperature phase is 
analogous, and the same formulae hold. In particular, 
Eqs. \reff{gypiu} and \reff{FL-law} are valid, but of course with different 
functions and coefficients, i.e. $g^-(y)$, $c^-_i$, $A_i^-$, and so on.
The coefficients $A_n^-$ are related to the coefficients $A^+_n$. 
A short-distance expansion analysis \cite{HS-73,BLZ-74b} gives
\begin{eqnarray}
{A_1^+\over A_1^-} = U_2^{-1} U_\xi^{\gamma/\nu}, \qquad
{A_2^+\over A_2^-} = - U_0 U_\xi^{(1-\alpha)/\nu}, \qquad
{A_3^+\over A_3^-} = - U_\xi^{1/\nu},
\label{relazioni_Aipiu_Aimeno}
\end{eqnarray}
where  $U_0$, $U_2$, and $U_\xi$ have been defined in
Table~\ref{notationsur}.
 
For vector systems the behavior is more complex, since 
the correlation function at zero momentum diverges 
at the coexistence line \cite{HM-65,BWW-72,PP-73}. 
For small $H$, $t<0$, and $q\to 0$, 
the transverse two-point function behaves as \cite{PP-73,FBJ-73}
\be
\widetilde{G}_T(q) \approx {M^2\over M H + \rho_s q^2},
\label{GTfuoriH}
\ee
where $\rho_s$ is the stiffness constant.\footnote{We mention 
that the stiffness constant can also be written
in terms of the helicity modulus \cite{FBJ-73}.}
For $t\to 0$ the stiffness 
constant goes to zero as \cite{HH-69,FBJ-73}
\be
\rho_s = \rho_{s0} (-t)^s,
\ee
where the exponent $s$ is given by the hyperscaling relation 
$s = (d-2) \nu$.
From the stiffness constant one may define a correlation length on the coexistence 
curve by
\be
\xi_T = \rho_s^{-1/(d-2)} = \rho_{s0}^{-1/(d-2)} (-t)^{-\nu}.
\label{def-xiT}
\ee
For $H= 0$ the correlation function diverges for $q\to 0$ 
as $1/q^2$, implying the algebraic decay of $G_T(x)$ for large $x$, i.e.
\be
G_T(x) \approx M^2 {\Gamma(d/2) \pi^{-d/2}\over 2(d-2)} 
    \left( {\xi_T\over |x|}\right)^{d-2}.
\ee
For $|x|\to \infty$ and $|H|\to 0$, 
the longitudinal correlation function is related 
to the transverse one. 
On the coexistence curve we have \cite{PP-73}
\be
\widetilde{G}_L(q) \sim {M^2 \xi_T^d \over (q\xi_T)^{d-4}}\, ,
\ee
and, in real space,
\be
G_L(x) \sim M^2 \left( {\xi_T\over |x|}\right)^{2d - 4}.
\ee

\subsubsection{Scaling function associated with the 
correlation length} \label{sec-1.6.3}

From the two-variable scaling function (\ref{Gscalingform})
of the two-point function one may derive scaling 
functions associated with the correlation lengths $\xi_{\rm gap}$ and $\xi$ 
defined  in Eqs.~(\ref{xigap}) and (\ref{xism}). One may write
in the scaling limit
\begin{eqnarray}
\xi^2(M,t) = (B^c)^{2\delta/d} M^{-2\nu/\beta} f_\xi(x),\qquad
\xi_{\rm gap}^2(M,t) = (B^c)^{2\delta/d} M^{-2\nu/\beta} f_{\rm gap}(x),
\end{eqnarray}
where $x \equiv B^{1/\beta} t M^{-1/\beta}$ is the variable introduced in 
Sec. \ref{sec-1.5.2}. The normalization of the functions is such to
make $f_\xi(x)$ and $f_{\rm gap}(x)$ universal. This follows
from two-scale-factor universality, i.e. the assumption 
that the singular part of the free energy in a correlation volume 
is universal. Using the scaling relations \reff{scaling-calF} and
\reff{normalizzazione-calA2} for the Helmholtz free-energy 
density, we obtain in the scaling limit
\be
{\cal A}_{\rm sing}(M,t) [\xi(M,t)]^d = {1\over \delta + 1} A_2(x) f_\xi(x).
\ee
Since $A_2(x)$ is universal, it follows that also $f_\xi(x)$ is 
universal. The same argument proves that $f_{\rm gap}(x)$ is universal.
Note the following limits:
\begin{eqnarray}
&&\lim_{x \to \infty} x^{2\nu} f_\xi(x) = ( R_\chi Q_c )^{2/d},  
\\
&& f_\xi(0) = (f^c)^2 (B^c)^{2/d}  =
   \left({R_4^+\over g_4^+}\right)^{2/d} 
   \left({Q_2\over\delta}\right)^{2\nu/\gamma}
   (R_\chi)^{-4\beta/d\gamma}, \\
&&    
f_\xi(-1) = U_\xi^{-2} ( R_\chi Q_c )^{2/d}. 
\label{fxi-atxm1}
\end{eqnarray}
Similar equations hold for $f_{\rm gap}(x)$. 
Of course, Eq. \reff{fxi-atxm1} applies only to scalar systems.

One may also define a scaling function associated with the ratio
$\xi^2/\chi$, that is
\begin{equation}
\xi^2/\chi = (B^c)^{(2-d)\delta/d} M^{-\eta\nu/\beta} f_Z(x),
\label{xi2suchi-scaling}
\end{equation}
where $f_Z(x)$ is a universal function that is related to 
the scaling functions $f_\xi(x)$ and $f(x)$ by
\begin{equation}
f_Z(x) = f_\xi(x) \left[ \delta f(x) - \case{1}{\beta} x f'(x) \right].
\end{equation}
For vector systems, in Eq. \reff{xi2suchi-scaling} one should consider
the longitudinal susceptibility $\chi_L$.
Similarly we define a
scaling function in terms of the variable $z$ 
defined in Eq. \reff{def-zvariable}, i.e.
\begin{equation}  
\xi^2/\chi = (-a_{11})^{2/d} \left(g_4^+\right)^{-2/d} t^{\eta\nu} F_Z(z),
\end{equation}
where the normalization factor is chosen so that $F_Z(0)=1$, and 
$a_{11}$ is defined in Eq. \reff{def-abconstant}. 
The function $F_Z(z)$ is universal and is related 
to $f_Z(x)$ defined in \reff{xi2suchi-scaling} through
the relation
\be
z^{\eta\nu/\beta} F_Z(z) = \, 
   {1\over f_0^\infty} z_0^{\eta\nu/\beta} \,
   \left(R_\chi Q_c\right)^{-2/d} \, f_Z(x),
\ee
where the universal constants $f_0^\infty$ and $z_0$ are defined in 
Sec. \ref{sec-1.5.3}.
Note that $\xi^2/\chi$ must always be positive by 
thermodynamic-stability requirements
(see, e.g., Refs. \cite{Griffiths-66,Stanley-71}). 
Indeed, this combination is related to
the intrinsically positive free energy associated with nonvanishing
gradients $|\nabla M|$ throughout the critical region. 

One may also consider parametric representations of the correlation lengths
$\xi$ and $\xi_{\rm gap}$ supplementing those for the equation
of state, cf. Eq.~(\ref{parrepg}). We write
\begin{eqnarray}
\xi^2/\chi = R^{-\eta\nu} a(\theta) , \qquad
\xi_{\rm gap}^2/\chi = R^{-\eta\nu} a_{\rm gap}(\theta) . 
\end{eqnarray}
Given the parametric representation (\ref{parrepg}) 
of the equation of state, the
normalizations of $a(\theta)$ and $a_{\rm gap}(\theta)$ are not
arbitrary but are fixed by two-scale-factor universality. We have
\begin{eqnarray}
&&a(0)=  h_0^{1-2/d} m_0^{-1-2/d} g_4^{-2/d} 
         \left[6(\gamma + h_3 - m_3)\right]^{2/d}, \nonumber \\ 
&&a_{\rm gap}(0) = (Q^+_\xi)^2 \,a(0), 
\end{eqnarray}
where $h_3 = d^3 h/d\theta^3(\theta=0)$, 
      $m_3 = d^3 m/d\theta^3(\theta=0)$.

\subsubsection{Scaling corrections} \label{sec-1.6.4}

One may distinguish two types of scaling corrections
to the scaling limit of the correlation function:
\begin{itemize}
\item[(a)]
Corrections due to operators that are rotationally invariant.
Such corrections are always pre\-sent, both in continuum systems and in 
lattice systems, and are controlled by the exponent $\omega$.
\item[(b)] 
Corrections due to operators that have only the lattice symmetry.
Such corrections are not pre\-sent in rotationally invariant systems,
but only in models and experimental systems on a lattice. 
The operators that appear depend on the lattice type.
These corrections are controlled by another exponent $\omega_{\rm NR}$.
\end{itemize} 
In three dimensions, corrections of type (b) are weaker than corrections of 
type (a), i.e. $\omega < \omega_{\rm NR}$. 
Therefore, rotational invariance is recovered before the disappearance 
of the rotationally-invariant scaling corrections. 
In two dimensions instead and on the square lattice, corrections 
of type (a) and (b) have exactly the same exponent \cite{CCCPV-00}.

\section{Numerical determination of critical quantities}
\label{sec-2}

In this section we review the numerical methods 
that have been used in the study of statistical systems at criticality.
In two dimensions many nontrivial models can be solved exactly, 
and moreover there exists a powerful tool, conformal field theory,
that gives exact predictions for the critical exponents
and for the behavior at the critical point.
In three dimensions there is no theory providing 
exact predictions at the critical point. Therefore, one must
resort to approximate methods. 
The most precise results have been 
obtained from the analysis of high-temperature (HT) expansions, from 
Monte Carlo (MC) simulations, and using perturbative 
field-theore\-tic\-al (FT)
methods. 
For the Ising model, one can also consider low-temperature (LT)
expansions (see, e.g., Ref.~\cite{Parisibook}). 
The results of LT analyses are less precise than those obtained using 
HT or MC techniques.
None\-theless, LT series are important since they give direct
access to LT quantities. 
We will not review them here since they are conceptually similar to 
the HT expansions.\footnote{The interested reader can find  
LT expansions in Refs. \cite{SGMME-73,BE-79,BCL-92,GE-93,Vohwinkel-93,AT-95}
and references therein.}

\subsection{High-temperature expansions}
\label{sec-2.1}

The HT expansion is one of the most efficient
approaches to the study of critical phenomena. For the models that we are
considering, present-day computers and a careful use of graph techniques 
\cite{Wortis-74,LW-88,NR-90,Reisz-95-2,Campostrini-00} 
allow the  generation of quite long series. In three dimensions,
for $\chi$ and 
\begin{equation}
\mu_2 \equiv \sum_x |x|^2 G(x), 
\label{mu2}
\end{equation}
the two quantities 
that are used in the determination of the critical exponents, 
the longest published series are the following:
\begin{itemize}
\item[(a)] Ising model: 25 orders on the body-centered cubic (bcc) 
lattice \cite{Campostrini-00} and on the 
simple cubic (sc) lattice \cite{Campostrini-00,CPRV-02};
\item[(b)] spin-$S$ Ising model
for $S=1,3/2,2,5/2,3,7/2,4,5,\infty$: 25 orders on the sc and bcc lattices \cite{BC-02-2};
\item[(c)] Improved Hamiltonians for $N=1,2,3$ on the sc lattice,
see Sec. \ref{sec-2.3.2}: 25 orders for the Ising case
\cite{CPRV-99,CPRV-02} and 20 orders for $N=2,3$ 
\cite{CPRV-00-es,CHPRV-01,CHPRV-02};
\item[(d)] Klauder, double-Gaussian, and Blume-Capel mo\-del for generic 
values of the coupling: 21 orders on the bcc lattice \cite{NR-90};
\item[(e)] $N$-vector model for generic values of $N$: 
21 orders on the bcc, sc, and diamond lattices \cite{BC-97-2,CPRV-98};
\item[(f)] $N$-vector model for $N=0$ (the generation of the HT series is 
equivalent to the enumeration of self-avoiding walks):
26 orders for $\chi$ on the sc lattice \cite{MJHMJG-00}.
\end{itemize}
On the bcc and sc lattice, Campostrini {\em et al.} 
\cite{Campostrini-00,CPRV-02} generated 25th-order series 
for the most general model with nearest-neighbor interactions
in the Ising universality class; they are available on request. 
Analogously, for $N=2,3$,  20th-order series for general models
on the sc lattice may be obtained from the 
authors of Refs.~\cite{CHPRV-01,CHPRV-02}. 

Series for the zero-momentum
$2n$-point correlation functions 
can be found in Refs.
\cite{BC-98,CPRV-99,CPRV-00-es,Campostrini-00,CHPRV-01,CHPRV-02,CPRV-02,BC-02-2}. 
Other HT series can be found in Refs.~\cite{KYT-77,McKenzie-79}.
In two dimensions, the longest published series 
for the $N$-vector model for generic $N$ are the following: 
triangular lattice, 15 orders \cite{CPRV-96-XY,CPRV-96-Nge3};
square lattice, 21 orders \cite{BC-96,CPRV-96-XY,CPRV-96-Nge3};
honeycomb lattice, 30 orders \cite{CPRV-96-XY,CPRV-96-Nge3}.

The analysis of the HT series requires an extrapolation to the critical 
point. Several different methods have been developed in the years. 
They are reviewed, e.g., in Refs. \cite{Guttmann-rev-89,Baker-book}.

The nonanalytic scaling corrections 
with noninteger exponents discussed in Sec. \ref{sec-1.5.7} are 
the main obstacle for a precise determination of 
universal quantities.  Their presence 
causes a slow convergence and 
introduces a large (and dangerously undetectable)
systematic error in the results of the HT analyses.
In order to obtain precise estimates of the critical parameters, the
approximants of the HT series should properly allow for the confluent
nonanalytic corrections 
\cite{Nickel-82,Gaunt-82,Zinn-Justin-81,CFN-82,Adler-83,GR-84,FC-85}.  
Second- or higher-order integral (also called differential) approximants 
\cite{GJ-72,HB-73,FA-79,RGJ-80}
are, in principle, able to describe nonanalytic correction terms.
However, the extensive numerical work that has been done
shows that in practice, with the series of moderate length that 
are available today, no unbiased analysis is able to 
take effectively into account nonanalytic correction-to-scaling
terms \cite{Zinn-Justin-81,Nickel-82,ND-81,CFN-82,Adler-83,NR-90}.
In order to deal with them, one must use
biased methods in which the presence of a nonanalytic term 
with exponent $\Delta$ is imposed, see, e.g., Refs.\
\cite{Roskies-81,ND-81,AMP-82,Privman-83,BC-97-2,PV-gr-98,BC-98,BC-00}.  

There are several different methods that try to handle properly 
the nonanalytic corrections, at least the leading term. 
For instance, one may use the method  proposed
in Ref.\ \cite{Roskies-81} and generalized in Refs.\ \cite{AMP-82,Privman-83}.
The idea is to perform the change of variables---we will call it 
Roskies transform---
\begin{equation}
z= 1 - (1 - \beta/\beta_c)^\Delta,
\label{RTr}
\end{equation}
so that the leading nonanalytic terms in $(\beta_c - \beta)$ become analytic in
$(1-z)$. The new series has weaker nonanalytic corrections, of order 
$(1-z)^{\Delta_2/\Delta}$ and $(1-z)^{1/\Delta}$ (here $\Delta_2$ 
is the second irrelevant exponent, $\Delta_2 = \nu\omega_2$), and thus 
analyses of these new series should provide more reliable estimates. 
Note, however, that the change of variable \reff{RTr} requires
the knowledge of $\beta_c$ and $\Delta$. Therefore, there is 
an additional source of error due to the uncertainty on these two quantities.
A substantially  equivalent method
consists in using suitably biased integral approximants 
\cite{BC-97-2,BC-98,BC-00},
again fixing $\Delta$ and $\beta_c$. It is also possible to 
fit the coefficients with the expected large-order behavior, 
fixing the subleading exponents, as it was done, e.g., in 
Refs.~\cite{MJHMJG-00,CPRV-02}. All these methods work reasonably 
and appear to effectively take into account the corrections to scaling.

A significant improvement of the HT results is obtained using improved
Hamiltonians (see Sec. \ref{sec-1.5.7}),
i.e. Hamiltonians 
that do not couple with the irrelevant operator 
that gives rise to the leading scaling correction of order $t^\Delta$.
In improved models, such correction does not appear in the expansion of {\em any}
thermodynamic quantity near the critical point.
Thus, standard analysis techniques
are much more effective,
since the main source of systematic error has  been eliminated.

\begin{figure}[t]
\centerline{\psfig{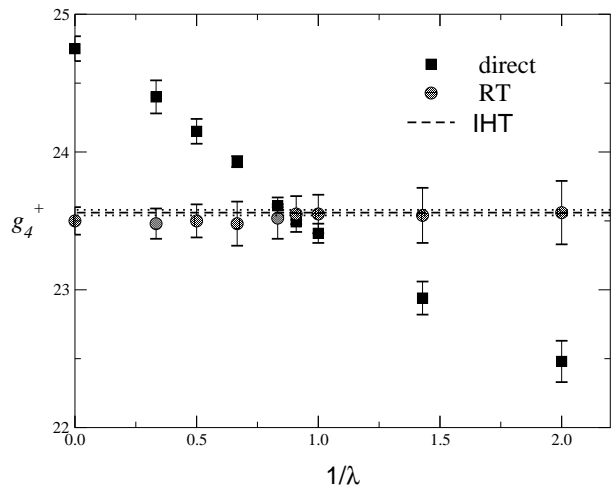}}
\caption{
Estimates of $g_4^+$ obtained from an unbiased analysis (direct) of the 
HT series and from the analysis (RT) of the series 
obtained by means of the Roskies transform \reff{RTr},
for the $\phi^4$ lattice  model. The dashed line marks the
more precise estimate (with its error) derived from the analysis of
an improved HT expansion in Ref.~\protect\cite{CPRV-02}, $g_4^+=23.56(2)$.
}
\label{figg}
\end{figure}

In order to illustrate the role played by the 
nonanalytic corrections, we consider
the zero-momentum four-point coupling $g_4$
defined in the high-tempera\-tu\-re phase by
\begin{equation}
g_4 \equiv - {3N\over N+2} {\chi_4\over \chi^2 \xi^d},
\label{grdef}
\end{equation}
which, in the critical limit, converges to the hyperuniversal constant
$g_4^+$ defined in Table~\ref{notationsur}.
In Fig.~\ref{figg} we show some results concerning 
the three-dimensional Ising universality class.
They were obtained in Ref.~\cite{CPRV-99} 
from the analysis of the HT series of $g_4$ (using $\chi_4$ to 18th order)
for the $\phi^4$ Hamiltonian \reff{latticephi4} and  several values of
$\lambda$. We report an unbiased analysis (direct)
and an analysis using the transformation \reff{RTr} with 
$\Delta = 1/2$ (RT). 
It is evident that the first type of analysis is unreliable, since one obtains
an estimate of $g_4^+$ that is not independent of $\lambda$ within the 
quoted errors, 
which are obtained as usual from the spread of the 
approximants. For instance, the analysis of the
series of the standard Ising model, corresponding to
$\lambda=\infty$, gives results that differ by more than 5\% from
the estimate obtained from the second analysis, 
while the spread of the approximants is
much smaller. The estimates obtained from the
transformed series are independent of $\lambda$ within error bars, giving
the estimate $g_4^+\approx 23.5$.
Such independence clearly indicates that the transformation \reff{RTr}
is effectively able to take into account the
nonanalytic behavior. 
Moreover, the result  is in good agreement with the more precise
estimate obtained using improved Hamiltonians, i.e. \cite{CPRV-02} 
$g_4^+=23.56(2)$ (see Sec. \ref{g4rev}).

\subsection{Monte Carlo methods}
\label{sec-2.2}

The Monte Carlo (MC) method is a powerful technique for the 
simulation of statistical systems. Its main advantage is its 
flexibility. Of course, results
will be more or less precise depending on the efficiency of the algorithm.
Systems with an $N$-vector order parameter and $O(N)$ symmetry 
are quite a special case, since there exists an efficient algorithm
with practically no critical slowing down: the Wolff algorithm 
\cite{Wolff-89-90}, 
a generalization of the Swendsen-Wang algorithm \cite{SW-87} for the 
Ising model (see Ref. \cite{CEPS-91-92} for a general discussion).  
The original algorithm was defined for the $N$-vector model,
but it can be applied to general $O(N)$ models by simply adding a Metropolis 
test \cite{BT-89}. 

In this section we describe different methods for obtaining 
critical quantities from MC simulations. After discussing
the standard infinite-volume methods, we present
two successful techniques. One is based on real-space 
RG transformations, the second one makes use of the finite-size scaling (FSS) 
theory. Finally, we discuss nonequilibrium methods that represent a 
promising numerical technique for systems with slow relaxation.

\subsubsection{Infinite-volume methods}
\label{sec-2.2.1}

Traditional MC simulations determine the critical behavior from 
infinite-volume data. In this case, the analysis of the MC data 
is done in two steps. In order to determine 
the critical behavior of a quantity $S$, 
one first computes ${S}(\beta,L)$ for 
fixed $\beta$ and several values of $L$ and determines 
\be
{S}_\infty(\beta) = \lim_{L\to\infty} {S}(\beta,L),
\ee
by performing an extrapolation in $L$. 
For $L\to \infty$, 
\be 
{S}(\beta,L) \approx {S}_\infty(\beta) + a L^p\, 
   e^{-L/\xi_{\rm gap}},
\ee
where $\xi_{\rm gap}$ is the exponential correlation length and 
$p$ an observable-dependent exponent. Such a rapid
convergence usually makes the finite-size effects negligible compared to the 
statistical errors for moderately large values of $L/\xi_{\rm gap}$. 
In the HT phase a ratio $L/\xi_{\rm gap}\approx 5$-7 is usually sufficient,
while in the LT phase larger values must be considered: 
for the three-dimensional Ising model, Ref. \cite{CH-97} 
used $L/\xi_{\rm gap}\approx 20$.
Finite-size effects introduce a severe limitation on the values of
$\xi_{\rm gap}$ that can be reached,
since $L\lesssim 100$-200 in present-day three-dimensional MC simulations.

Once the infinite-volume quantities ${S}_\infty(\beta) $ are determined,
exponents and amplitudes are obtained by fitting the numerical 
results to the corresponding expansion: 
\be
{S}_\infty (\beta) = a |\beta_c - \beta|^{-\sigma }
   + b |\beta_c - \beta|^{-\sigma+\Delta} + \ldots
\label{fit2.4}
\ee
Of course, one cannot use too many unknown parameters in the fit 
and often only the leading term in Eq. \reff{fit2.4} is kept. However, this is 
the origin of large systematic errors: 
It is  essential to keep into account 
the leading nonanalytic correction with exponent $\Delta$. 

Again, in order to show the importance of the nonanalytic scaling corrections,
we present in Fig. \ref{figg-MC} some numerical 
results \cite{KL-96,BK-96} for the four-point coupling $g_4$.
A simple extrapolation of the MC data 
to a constant gives $g_4^+= 24.5(2)$ \cite{KL-96}
which is inconsistent with the result of Ref.~\cite{CPRV-02},  
$g_4^+ = 23.56(2)$. On the other hand, 
a fit that takes into account the 
leading correction to scaling gives $g_4^+ = 23.7(2)$ \cite{PV-gr-98},
which is in agreement with the above-reported estimate.

\begin{figure}[t]
\centerline{\psfig{width=8truecm,angle=-90,file=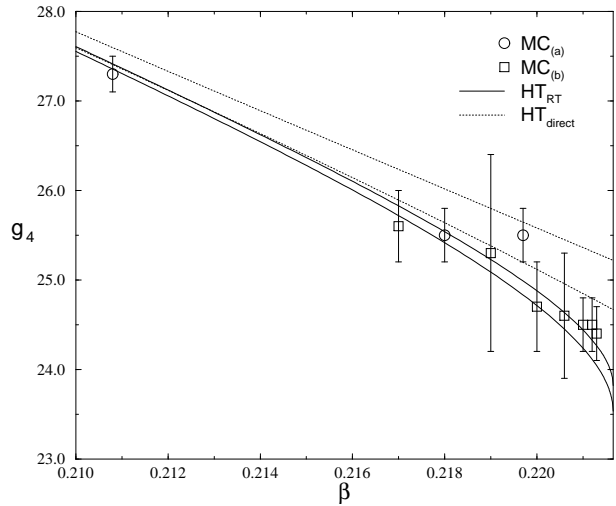}}
\caption{
MC results for the four-point coupling $g_4$ for the 
three-dimensional Ising model: 
(a) from Ref.~\protect\cite{BK-96}; (b) from Ref.~\protect\cite{KL-96}.
For comparison we also report the extrapolation of the 
18th-order HT series of Ref.~\protect\cite{CPRV-99} 
by means of a direct analysis ($\rm HT_{\rm direct}$)
and of an analysis that uses the transformation \reff{RTr} ($\rm HT_{RT}$). 
For each of these extrapolations we report two lines corresponding 
to the one-error-bar interval. 
}
\label{figg-MC}
\end{figure}

\subsubsection{Monte-Carlo renormalization group}
\label{sec-2.2.2}

Here we briefly outline the real-space RG
which has been much employed
in numerical MC RG studies.\footnote{There exist other numerical 
methods based on the real-space RG. Among others, we should mention 
the works on approximate RG transformations that followed the ideas 
of Kadanoff and Migdal \cite{Kadanoff-75-76,KHY-76,Migdal-76}. 
For a general review, see, e.g., Refs. \cite{NvL-76,BvL-82}.} 
This method has been amply reviewed in the literature, see, e.g.,
Refs.~\cite{WK-74,Swendsen-82}.

The main idea of the RG approach 
is to reduce the number of degrees of freedom of the system by
integrating out the short-range fluctuations.
In the real-space RG this is performed
by block-field transformations \cite{Kadanoff-66}.
In a block-field transformation, a block with $l^d$ sites on the original
lattice is mapped into a site of the blocked lattice.
A block field $\phi_B$ is then constructed from the field $\phi$ of the original
lattice, according to rules that should leave unchanged the 
critical modes, eliminating only the noncritical ones. 
The Hamiltonian ${\cal H}_B$ of the blocked system is defined as
\begin{equation}
\exp \left[ - {\cal H}_B(\phi_B) \right] = 
\int D \phi \, {\cal M}(\phi_B,\phi) \exp\left[ - {\cal H}(\phi) \right],
\label{HB}
\end{equation}
where ${\cal M}(\phi_B,\phi)$ denotes the kernel of the block-field
transformation.
Then, the lattice spacing of the blocked lattice is rescaled to one. 

RG transformations are defined in the infinite-dimensional space of
all possible Hamiltonians. If ${\cal H}$ is written as 
\begin{equation}
{\cal H}(K_1,K_2,\ldots;O)  = \sum_a K_a O_a,
\end{equation}
where $O_a$ are translation-invariant functions of the field $\phi$
and $K_a$ are the corresponding couplings, 
the RG transformation induces a mapping 
\begin{equation}
K \rightarrow K' = R(K).
\end{equation}
The renormalized couplings $K'$ are assumed to be analytic functions
of the original ones.\footnote{This assumption should be taken with care.
Indeed, it has been proved 
\cite{GP-78-79,Israel-81,vEFS-94,vEFK-94,BKL-98,vEF-99} 
that in many specific cases
real-space RG transformations are singular. 
These singularities reflect the mathematical fact that RG transformations
may transform a Gibbs measure into a new one that is non-Gibbsian
\cite{vEFS-94,vEFK-94,BKL-98,vEF-99}. 
In approximate RG studies, these singularities appear as discontinuities 
of the RG map, see, e.g., Ref. \cite{Salas-95}. }

The nonanalytic behavior at the critical point is obtained 
by iterating the RG transformation an infinite number of times.
Continuous phase transitions are associated with the fixed points
$K^*$ of the RG transformation. Critical exponents are determined
by the RG flow in the  neighborhood of the fixed point.   
If we define
\begin{equation}
T_{ab} = \left. \partial K'_a \over \partial K_b \right|_{K=K^*},
\label{def-T}
\end{equation}
the eigenvectors of $T$ give the linearized scaling fields.
The corresponding eigenvalues can be written as $\lambda_i=l^{y_i}$,
where $y_i$ are the RG dimensions of the scaling fields.

An exact RG transformation is defined in the space of Hamiltonians 
with an infinite number of couplings. However, in practice a numerical implementation
of the method requires a truncation of the Hamiltonians considered. 
Therefore, any method that is based on real-space RG transformations chooses
a specific basis, trying to keep those terms that are more important 
for the description of the critical modes. As a general rule, one keeps
only terms with a small number of fields and that are 
localized (see, e.g., Ref. \cite{Blote-etal-89}). 
The precision of the method depends crucially on the choice for
the truncated Hamiltonian and for the RG transformation.

In numerical MC studies, given a MC generated configuration $\{\phi\}$,
one generates a series of blocked configurations 
$\{\phi^{(i)}\}$, with $i=0$ corresponding to the original configuration,
by applying the block-field transformation. Correspondingly, one 
computes the operators $O^{(i)} \equiv O(\phi^{(i)})$. Then, 
one determines the matrices
\begin{eqnarray}
&& \hskip -0.8truecm 
A_{ab}^{(i)} = 
   \left \langle \left( O_a^{(i)} - \langle O_a^{(i)} \rangle \right)
\left( O_b^{(i)} - \langle O_b^{(i)} \rangle \right) \right\rangle,
\nonumber \\
&& \hskip -0.8truecm B_{ab}^{(i)} = 
   \left\langle \left( O_a^{(i)} - \langle O_a^{(i)} \rangle \right)
\left( O_b^{(i-1)} - \langle O_b^{(i-1)} \rangle \right) \right\rangle
\end{eqnarray}
and the matrix $T^{(i)}$ \cite{Swendsen-79}
\begin{equation}
\sum_b A_{ab}^{(i)} T_{bc}^{(i)} = B_{ac}^{(i)}.
\end{equation}
If all possible couplings were considered, the matrix $T^{(i)}$ would 
converge to $T$ defined in Eq. \reff{def-T} for $i\to\infty$.
In practice, only a finite number of couplings and 
a finite number of iterations is used. These approximations can be
partially controlled by checking
the convergence of the results with respect to the number of couplings 
and of RG iterations.

\subsubsection{Finite-size scaling}
\label{sec-2.2.3}

Finite-size effects in critical phenomena have been the object of 
theoretical studies for a long time:
see, e.g., Refs. \cite{Barber-83,Privman-FSS,Cardy-book,Cardy-review} for reviews.
Only recently, due to the progress in the preparation
of thin films, this issue has begun being investigated experimentally,
see, e.g., Refs. 
\cite{HMKW-93,LRJC-93,AC-96,AC-96-2,MG-97,BFAE-90,Fullerton-etal,%
Elmers-etal,Lipa-etal-00,KG-01,KMG-00}.
FSS techniques are particularly important in
numerical work. With respect to the infinite-volume methods, 
they do not need to satisfy the condition $\xi_{\rm gap} \ll L$. 
One can work with $\xi_{\rm gap} \sim L$ and thus is better able to 
probe the critical regime. FSS MC simulations are at present one of the 
most effective techniques for the determination of critical quantities.
Here, we will briefly review the main ideas behind FSS and report 
several relations
that have been used in numerical studies to determine the critical quantities.

The starting point of FSS is the generalization of 
Eq. \reff{Gsing-RGscaling} for the singular part of 
the Helmholtz free energy of a sample of linear size $L$ 
\cite{FB-72,Barber-83,PF-84,BLH-95}:
\begin{eqnarray}
{\cal F}_{\rm sing}(u_t,u_h,\{u_i\},L) = 
b^{-d} {\cal F}_{\rm sing}( b^{y_t} u_t, b^{y_h} u_h, \{b^{y_i} u_i\}, L/b), 
\end{eqnarray}
where $u_t\equiv u_1$, $u_h\equiv u_2$, $\{u_i\}$ with $i\geq 3$ 
are the scaling fields associated
respectively with the reduced temperature, magnetic field, and the other
irrelevant operators. Choosing $b = L$, we obtain 
\begin{eqnarray}
{\cal F}_{\rm sing} (u_t, u_h, \{u_i\},L)  = 
 L^{-d} 
{\cal F}_{\rm sing}( L^{y_t} u_t, L^{y_h} u_h, \{L^{y_i} u_i\},1),
\label{FscalL}
\end{eqnarray}
from which, by performing the appropriate derivatives with respect to
$t$ and $H$, one finds the scaling behavior of the thermodynamically
interesting quantities. We again assume that
$u_t$ and $u_h$ are the only relevant 
scaling fields, and thus,
 neglecting correction of order $L^{y_3} = L^{-\omega}$,
we can simply set $u_i = 0$ in the previous equation.
Using Eq.~(\ref{FscalL}) one may obtain the FSS behavior 
of any thermodynamic quantity $S$.
Considering $S(\beta,L)$ for $H=0$, if 
$S_\infty(\beta) \equiv S(\beta,\infty)$ 
behaves as $t^{-\sigma}$ for $t\rightarrow 0$, then we have
\begin{equation}
S(\beta,L) = L^{\sigma/\nu} 
\left[ f_S\left(\xi_\infty/L\right) + O\left(L^{-\omega},
\xi_\infty^{-\omega}\right)\right],
\label{stl1}
\end{equation}
where $\xi_\infty(\beta)$ is the correlation length in 
the infinite-volume limit.
We do not need to specify which definition we are using.
For numerical studies, it is convenient to rewrite
this relation in terms of a correlation length $\xi(\beta,L)$ defined
in a finite lattice. Then, one may rewrite the above equation as
\begin{eqnarray}
S(\beta,L) =  L^{\sigma/\nu} 
\left[ \bar{f}_S\left(\xi(\beta,L)/L\right) + O\left(L^{-\omega},
\xi^{-\omega}\right)\right].
\label{stl2}
\end{eqnarray}
FSS methods can be used to determine $\beta_c$, critical exponents, and 
critical amplitudes. Below we will review a few of them
(we assume everywhere $H=0$, but much can be generalized to 
$H\not= 0$). 

In order to determine $\beta_c$, a widely used method is the ``crossing" 
method. Choose a thermodynamic 
quantity $S(\beta,L)$ for which $\sigma = 0$ or $\sigma/\nu$ 
is known and define $R(\beta,L) \equiv  S(\beta,L) L^{-\sigma/\nu}$. 
Then consider pairs $(L_1,L_2)$ and  
determine the solution $\beta_{\rm cross}$ of the equation \cite{Binder-81}
\be 
R(\beta_{\rm cross},L_1) = R(\beta_{\rm cross},L_2).
\ee
If $L_1$ and $L_2$ diverge, $\beta_{\rm cross}$ converges to $\beta_c$ with 
corrections of order $L_1^{-\omega-1/\nu}$, $L_2^{-\omega-1/\nu}$, and 
thus it provides an estimate of $\beta_c$. 
A widely used quantity is the Binder  cumulant $Q$, 
\be
Q = \, {\langle M^4 \rangle\over \langle M^2\rangle^2},
\end{equation}
where $M$ is the magnetization. Other choices that have been considered
are $\xi/L$, 
generalizations of the Binder cumulant using higher powers of the 
magnetization, and the ratio of the partition function with 
periodic and antiperiodic boundary conditions 
\cite{HPV-99,Hasenbusch-99,CHPRV-01}.
%\footnote{ Precise estimates of $Q^*$ have been recently obtained:
%$Q^* = 0.62393 (13{+}35{+}5)$ \cite{HPV-99}
%(where the error is given as the sum of three contributions: the first
%is the statistical error,  the second and the third account for
%corrections to scaling), and $Q^*=0.62358(15)$ \cite{BST-99}.}

The determination of the critical exponents can be performed using several
different methods. 
One of the oldest approaches is the phenomenological renormalization
of Nightingale \cite{Nightingale-76-77}. One fixes a temperature 
$\beta_1$ and two sizes $L_1$ and $L_2$ and then determines $\beta_2$ 
so that 
\be
 { \xi(\beta_2, L_2)\over \xi( \beta_1,  L_1) } = {L_2\over L_1}.
\ee
Neglecting scaling corrections, in the FSS regime $\beta_1$ and 
$\beta_2$ are related by
\be
     (\beta_2 - \beta_c)  = \left( {L_{1}\over L_2} \right)^{1/\nu} 
     (\beta_{1} - \beta_c).
\label{phen-renorm}
\ee
In Ref. \cite{Nightingale-76-77} the method is implemented iteratively,
using $L_2 = L_1 + 1$. Starting from $\beta_0$, $L_0$, by
using Eq. \reff{phen-renorm}
one obtains a sequence of estimates $\beta_i$, $\nu_i$ that 
converge to $\beta_c$ and $\nu$ respectively.
It is also possible to consider a magnetic field, obtaining in this case 
also the exponent $(\beta+ \gamma)/\nu$. 

Critical exponents can also be determined by studying 
thermodynamic quantities at the critical point.
In this case, 
\begin{equation}
S(\beta_c,L)\sim L^{\sigma/\nu},
\end{equation}
neglecting scaling
corrections. Thus, one can determine $\sigma/\nu$ by simply studying the
$L$-dependence. For example, $\gamma/\nu$ and $\beta/\nu$ can be determined
from
\begin{eqnarray}
\chi(\beta_c,L) \sim L^{\gamma/\nu}, \qquad
|M|(\beta_c,L) \sim L^{\beta/\nu}.
\label{FSS-exponents-1}
\end{eqnarray}
The exponent $\nu$ can be determined by studying the $L$-dependence
of derivatives with respect to $\beta$. Indeed,
\be
\left. {\partial S(\beta,L)\over \partial\beta}\right|_{\beta=\beta_c} \sim
   L^{(\sigma + 1)/\nu},
\label{FSS-exponents-2}
\ee
which can be obtained from Eq. \reff{stl1} 
using $\xi_\infty \sim |t|^{-\nu}$.
This method has the drawback that an estimate of $\beta_c$ is needed.
Moreover, since $\beta_c$ is usually determined only at the end of the
runs, one must take into account the fact that the available
numerical results correspond to $\beta\not = \beta_c$.
There are then two possibilities: one may compute $S(\beta_c,L)$
 using the reweighting method \cite{FMPPT-82,FS-88},
or include correction terms proportional to
$(\beta - \beta_c) L^{1/\nu}$
in the fit Ansatz \cite{BLH-95,BST-99}.
In both cases, the method requires
$(\beta - \beta_c) L^{1/\nu}$ to be small.
One may also consider 
$\beta_c$ as a free parameter and determine 
it by fitting $S(\beta,L)$ near the critical
point \cite{BLH-95,BST-99}.

It is possible to avoid using $\beta_c$.
In Refs. \cite{BFMM-96,BFMM-96a,BFMM-97,BFMM-98,BFMMPR-99}
one fixes $L_1$ and $L_2$ and then determines $\beta$, for instance 
by reweighting the data,  so that 
\begin{equation}
{\xi(\beta,L_1)\over \xi(\beta,L_2)} = {L_1\over L_2}.
\label{Spagnoli-FSS-1}
\end{equation}
Then, the exponent $\sigma$ is obtained from
\begin{equation}
{S(\beta,L_1)\over S(\beta,L_2) }= \left( {L_1\over L_2} \right)^{\sigma/\nu} ,
\label{Spagnoli-FSS-2}
\end{equation}
neglecting scaling corrections.
When $L_1$ and $L_2$ go to infinity, this estimate converges towards 
the exact value. 
Due to the presence of cross-correlations, this method gives 
results that are more precise than those obtained by studying the theory at 
the critical point.

A somewhat different approach is proposed in 
Ref. \cite{Hasenbusch-99}. 
One introduces an additional
quantity $R(\beta,L)$ such that $R(\beta_c,L)\to R^*$ for $L\to\infty$.
Then, one fixes a value $\overline{R}$ ---for practical purposes it 
is convenient to choose $\overline{R}\approx R^*$--- and, for each $L$,
determines $\beta_f(L)$ from 
\be 
R(\beta_f(L),L) = \overline{R}.
\ee
Finally, one considers $S(\beta_f(L),L)$ which still
behaves as $L^{\sigma/\nu}$ for large $L$. 
Due to the presence of cross-correlations, the error 
on $S$ at fixed $R$ turns out to be  smaller than the error on 
$S$ at fixed $\beta$.
With respect to the approach of 
Refs. \cite{BFMM-96,BFMM-96a,BFMM-97,BFMM-98,BFMMPR-99},
this method  has the advantage of avoiding a tuning on two different lattices. 

The FSS methods that we have described are effective in determining the 
critical exponents. However, they cannot be used to compute 
amplitude ratios or other infinite-volume quantities. 
For this purpose, however, one can still use FSS methods
\cite{LWW-91,Kim-93-94,CEFPS-95,CEPS-95,MPS-96a,MPS-96b,BBGW-98,PC-99,FS-99}. 
The idea consists in rewriting Eq. \reff{stl2} as 
\be 
{S(\beta,sL)\over S(\beta,L)} = 
   \hat{f}\left(\xi(\beta,L)/L\right) + O(L^{-\omega},\xi^{-\omega}),
\label{stl3}
\ee
where $s$ is an arbitrary (rational) number.
In the absence of scaling corrections, one may  proceed as follows. 
First, one performs several runs for fixed $L$, determining 
$S(\beta,sL)$, $S(\beta,L)$, $\xi(\beta,sL)$, and $\xi(\beta,L)$. 
By means of a suitable interpolation, this provides the 
function $\hat{f}(\xi(\beta,L)/L)$ for $S$ and $\xi$.  Then, 
$S_\infty(\beta)$ and $\xi_\infty(\beta)$ are obtained 
from $S(\beta,L)$ and $\xi(\beta,L)$ 
by iterating Eq. \reff{stl3} and the corresponding equation for 
$\xi(\beta,L)$. Of course, in practice one must be very careful about scaling
corrections, increasing systematically $L$ till the results 
become independent of $L$ within error bars.

Finally, we note that
scaling corrections 
represent the main source of error in all these methods. 
They should not be neglected
in order to get reliable estimates of the critical exponents.
The leading scaling correction, which is of order $O(L^{-\omega})$, 
is often important, and should be taken into account
by performing fits with Ansatz
\be
   a L^{\sigma/\nu} + b L^{\sigma/\nu-\omega}.
\ee
As we have already stressed previously, these difficulties can be 
partially overcome if one uses an improved Hamiltonian. 
In this case the leading scaling corrections 
are absent and a naive fit to the leading behavior gives reliable 
results at the level of present-day statistical accuracy. 
However, in practice improved Hamiltonians are known 
only approximately, so that one may worry of the systematic 
error due to the residual correction terms. In Sec.~\ref{sec-2.3.1} we will
discuss how to keep this systematic error into account.

\subsubsection{Dynamic methods} \label{sec-2.2.4}

Nonequilibrium dynamic methods are numerical techniques that
compute static and dynamic critical exponents by studying
the relaxation process towards equilibrium. They are especially
convenient in systems with slow dynamics,
since they allow the determination of the critical exponents
without ever reaching thermal equilibrium. Two slighly different
techniques have been developed: the nonequilibrium-relaxation (NER)
method, see Refs. \cite{Ito-93,IO-99,OO-00,IHOO-00,OI-00} and 
references therein, and the short-time critical dynamics (STCD) 
method, see Refs. \cite{LSZ-95,LSZ-96,LSZ-98,SZ-00} and 
references therein. 

In the NER method one studies the long-time relaxation towards 
equilibrium. In this limit, the nonequilibrium free energy
scales as \cite{Suzuki-76-77}
\begin{equation}
{\cal F} (t,H,L,\tau) = 
 L^{-d} \widehat{F}(t L^{y_t}, H L^{y_h}, \tau L^{z}),
\end{equation}
where $\tau$ is the dynamics time, $z$ is a new critical exponent, 
and subleading corrections have been omitted. 
The method bears some similarities with the FSS methods described before. 
One first determines the critical point, and then studies the 
dynamics at criticality determining the exponents from the 
large-time (instead of large-$L$) behavior of correlation functions. 
Since one does not need to reach 
equilibrium, large volumes can be considered. Moreover, since 
correlations increase with increasing $\tau$, 
one can avoid finite-size effects 
by stopping the dynamics when the correlation length is some fraction of 
the size. In order to determine the critical point, one may monitor
the behavior of the time-dependent magnetization $m(t,\tau)$. 
For $t\not=0$, $m(t,\tau)$ converges to its asymptotic value exponentially,
while at the critical point $m(0,\tau) \sim \tau^{-\beta/(z\nu)}$. 
It suffices to consider 
\begin{equation}
f(t,\tau) = {d \ln m(t,\tau)\over d \ln \tau}
\end{equation} 
that diverges for $t > 0$, goes to zero for $t < 0$, and converges to 
a constant for $t=0$. Once $T_c$ is determined, the critical exponents 
can be obtained from the behavior of powers of $m(t,\tau)$
and of their derivatives with respect to $t$ at the critical point. 

The STDC method is similar and uses again dynamic scaling. Here,
one assumes that, besides the universal behavior in the long-time regime,
there exists another universal stage of the relaxation at early
(macroscopic) times.
The scaling behavior of this regime has been shown for the dynamics of model
A in Ref.~\cite{JSS-89}, but it is  believed to be a general characteristic
of dynamic critical phenomena.

\subsection{Improved Hamiltonians}
\label{sec-2.3}

As we already stressed, one of the main sources of systematic errors 
is the presence of nonanalytic 
corrections controlled by the RG eigenvalue $y_3 = - \omega$. 
A way out exploits improved models, i.e. models for which 
there are no corrections with exponent $\omega$: No terms 
of order $|t|^{\omega\nu}=|t|^\Delta$ appear in infinite-volume quantities 
and no terms of order $L^{-\omega}$ in FSS variables. 
Such Hamiltonians cannot be determined analytically and one must use 
numerical methods. Some of them will be presented below. 

\subsubsection{Determinations of the improved Hamiltonians}
\label{sec-2.3.1}

In order to determine an improved Hamiltonian, one may consider 
a one-parameter family of models, parame\-tri\-zed, say, by $\lambda$,
that belong to the given universality class. Then, one may consider a specific 
quantity and find numerically a value $\lambda^*$ 
for which the leading correction to scaling is absent.  
According to RG theory, at $\lambda^*$ the leading 
scaling correction gets suppressed
in {\em any} quantity. Note that, within a given one-parameter 
family of models, nothing guarantees that 
such a value of $\lambda$ can be found. For instance, 
in the large-$N$ limit of 
the lattice $\phi^4$ theory \reff{latticephi4} 
no positive value of $\lambda$ exists that achieves the suppression of
the leading scaling corrections \cite{CPRV-99}.
For a discussion in the continuum, see, e.g., 
Refs.~\cite{BB-90,Zinn-Justin-book}.

The first attempt to exploit improved Hamiltonians is due to 
Chen, Fisher, and Nickel \cite{CFN-82}. They
studied two classes of two-parameter models, the bcc scalar
double-Gaussian and Klauder models,
that are expected to belong to the Ising universality class and   
that interpolate between
the spin-$1/2$ Ising model and the Gaussian model. 
They  showed that improved models with suppressed
leading corrections to scaling can be obtained by tuning the
parameters (see also Refs.\ \cite{FC-85,NR-90}).  
The main difficulty of the method is the 
precise determination of $\lambda^*$.
In Refs.\ \cite{CFN-82,GR-84} the partial differential approximant
technique was used; however, the error on $\lambda^*$
was relatively large, and the final results represented only a
modest improvement with respect to standard (and much simpler) analyses
using biased approximants.

One may determine the improved Hamiltonian 
by comparing the results of a ``good" and of a ``bad" analysis of the 
HT series \cite{FC-85,CPRV-99}. Considering again the zero-momentum 
four-point coupling 
$g_4$, we can for instance determine $\lambda^*$ from the results 
of the analyses presented in Sec. \ref{sec-2.1}. The improved model 
corresponds to the value of $\lambda$
for which the unbiased analysis gives results that 
are consistent with the analysis that takes into account 
the subleading correction. 
From Fig. \ref{figg}, we see that in the interval $1.0< \lambda < 1.2$ 
the two analyses coincide and thus, we can estimate 
$\lambda^* = 1.1(1)$. This estimate is consistent with  
the result of Ref. \cite{Hasenbusch-99}, i.e. $\lambda^*=1.10(2)$ 
obtained using the MC method based on FSS, but is much less precise.

In the last few years many numerical studies
\cite{BLH-95,HPV-99,BFMM-98,BFMMPR-99,Hasenbusch-99,%
Hasenbusch-99-h,Hasenbusch-00,CHPRV-01,CHPRV-02}
have shown that improved Hamiltonians
can be accurately determined by means of MC simulations,
using FSS methods.

There are several methods that can be used. A first class of methods
is very similar in spirit to the crossing technique employed for 
the determination of $\beta_c$. In its simplest implementation
\cite{CPRV-99} one considers 
a quantity $R(\beta,\lambda,L)$ such that, for $L\to\infty$,
$R(\beta_c(\lambda),\lambda,L)$
converges to a universal constant $R^*$, which is 
supposed to be known.
Standard scaling arguments predict for $L\to \infty$
\begin{eqnarray}
R (\beta_c(\lambda),\lambda,L) \approx 
R^* + 
   a_1 (\lambda) L^{-\omega} + 
   a_2 (\lambda) L^{-2 \omega} + \ldots 
+   b_1 (\lambda) L^{-\omega_2} \ldots
\end{eqnarray}
where $\omega_2 \equiv - y_4$ is the next-to-leading 
correction-to-scaling exponent. 
In order to evaluate $\lambda^*$, which is the value for 
which $a_1(\lambda) = 0$,  one can determine $\lambda^{\rm eff}(L)$ from the 
equation
\begin{equation}
{R} (\beta_c(\lambda^{\rm eff} (L)),
          \lambda^{\rm eff} (L),L) = {R}^*,
\label{equation-lambdastar}
\end{equation}
where we assume $R^*$ and $\beta_c(\lambda)$ to be known.
For $L\to \infty$, $\lambda^{\rm eff}(L)$ converges to $\lambda^*$
with corrections of order $L^{\omega - \omega_2}$.
In practice, neither $R^*$ nor $\beta_c(\lambda)$ are known exactly. 
It is possible to avoid these problems by considering 
two different quantities $R_1$ and $R_2$ that have a universal 
limit for $L\to \infty$ \cite{Hasenbusch-99,CHPRV-01}.
First, we define $\beta_f(\lambda,L)$ by
\begin{equation}
\label{betafix}
R_1(\beta_f,\lambda,L) = \bar{R}_{1}  ,
\label{betaf-def}
\end{equation}
where $\bar{R}_{1}$ is a fixed value taken from the range of $R_1$.
Approximate estimates of $\lambda^*$  are then obtained by solving 
the equation
\begin{equation}
{R}_2(\beta_f(\lambda,L),\lambda, L) = 
   {R}_2(\beta_f(\lambda, b L),\lambda, b L) .
\end{equation}
for some value of $b$. 

Alternatively \cite{BFMM-98,BFMMPR-99,Hasenbusch-99,HT-99,CHPRV-01},
one may determine the size 
of the corrections to scaling for two values of $\lambda$ 
which are near to $\lambda^*$, but not too near in order to have a good signal,
and perform a linear interpolation. 
In the implementation of Refs. \cite{BFMM-98,BFMMPR-99} one considers 
the corrections to Eq. \reff{Spagnoli-FSS-2}, while 
Refs. \cite{Hasenbusch-99,HT-99,CHPRV-01} consider the 
corrections to a RG-invariant quantity at fixed $\beta_f$, see 
Eq. \reff{betaf-def}.

We should note that all these numerical methods provide only 
approximately improved Hamiltonians. Therefore, leading 
corrections with exponent $\omega$ are small but
not completely absent.
However, it is possible to evaluate the residual systematic error 
due to these terms. The idea is the following \cite{Hasenbusch-99}.
First, one considers a specific quantity that behaves as 
\be 
    S(L) = a L^{\sigma/\nu} (1 + b(\lambda) L^{-\omega} + \ldots).
\ee
Then, one studies numerically $S(L)$ in the approximately improved model, 
i.e. for $\lambda = \lambda^*_{\rm est}$ ($\lambda^*_{\rm est}$ is the 
estimated value of $\lambda^*$), and in a model in which the 
corrections to scaling are large: for instance in the $N$-vector model
corresponding to
$\lambda = \infty$. Finally, one determines numerically an upper bound 
on $b(\lambda^*_{\rm est})/b(\infty)$. RG theory guarantees that this 
ratio is identical for any quantity. Therefore, 
one can obtain an upper bound on the residual irrelevant corrections, 
by computing the correction term in the $N$-vector model---this is easy since 
corrections are large---and multiplying the result by the factor 
determined above.

\subsubsection{List of improved Hamiltonians} 
\label{sec-2.3.2}

We list here the improved models that have been 
determined so far. We write 
the partition function in the form 
\be 
  Z = \int \prod_i\,  d\mu(\phi_i)  
    \exp\left[\beta \sum_{<ij>} \phi_i\cdot\phi_j \right],
\ee
where $\phi_i$ is an $N$-dimensional vector and the sum is 
extended over all nearest-neighbor pairs $<ij>$.
For $N=1$ the following improved models have been determined:
\begin{enumerate}
\item Double-Gaussian model: 
\begin{eqnarray}
d\mu(\phi) = d\phi\, \left\{ 
        \exp\left[ - {(\phi + \sqrt{y})^2\over 2 (1-y)}\right]  
+       \exp\left[ - {(\phi - \sqrt{y})^2\over 2 (1-y)}\right] 
        \right\},
\label{double-Gaussian}
\end{eqnarray}
with $0< y < 1$. The improved model has been determined on the bcc 
lattice from the study of HT series.  The improved model 
corresponds to 
$y^* = 0.87(4)$ \cite{CFN-82}, $y^* = 0.87(1)$ \cite{GR-84}, 
$y^* = 0.90(3)$ \cite{FC-85}, $y^*\approx 0.85$ \cite{NR-90}.
\item Klauder model:
\be
d\mu(\phi) =\, d\phi\, |\phi|^{y/(1-y)} 
        \exp\left( - {\phi^2\over 2(1-y)}\right),
\label{Klauder}
\ee
with $0< y < 1$. The improved model has been determined on the bcc 
lattice from the study of HT series. 
The improved model corresponds to 
$y^* = 0.81(6)$ \cite{CFN-82}, $y^* = 0.815(35)$ \cite{FC-85}.
\item $\phi^4$-$\phi^6$ model:
\begin{eqnarray}
d\mu(\phi) =\, d\phi\, 
\exp\left[ - \phi^2 - \lambda (\phi^2 - 1)^2 - 
          \lambda_6 (\phi^2 - 1)^3\right].
\label{phi4-phi6}
\end{eqnarray}
The couplings corresponding to improved models have been determined 
by means of MC simulations.
On the sc lattice, the Hamiltonian is improved for these values of 
the couplings: (a)\footnote{
This is the estimate used in Ref.~\cite{CPRV-99}, which was 
derived from the MC results  of
Ref.~\cite{Hasenbusch-99}. There, the result
$\lambda^*=1.095(12)$ was obtained by fitting the data for lattices 
of size $ L \ge 16$.
Since fits using also data for smaller lattices, i.e.
with  $L \ge 12$ and  $L \ge 14$, gave consistent results,
one might expect that the systematic error is at most as large as 
the statistical one \cite{Hasenbusch-pc}.}
$\lambda^* = 1.10(2)$, $\lambda_6^* = 0$ 
\cite{Hasenbusch-99}; 
(b) $\lambda^* = 1.90(4)$, $\lambda_6^* = 1$ \cite{CPRV-99}.
\item Blume-Capel model:
\begin{eqnarray}
d\mu(\phi) =\, d\phi\, \left[\delta(\phi - 1) + \delta(\phi + 1) 
    + e^D \delta(\phi)\right].
\label{Blume-Capel}
\end{eqnarray}
On the sc lattice the Hamiltonian is improved for 
$D^* \approx 0.7$ \cite{BLH-95},
$D^* = 0.641(8)$ \cite{Hasenbusch-99-h}.
\end{enumerate} 
For $N=2,3,4$ and for  the $\phi^4$ theory \reff{latticephi4} 
on a simple cubic lattice, $\lambda^*$ 
has been determined 
by means of FSS MC simulations \cite{HT-99,CHPRV-01,CHPRV-02,Hasenbusch-00}. 
The estimates of 
$\lambda^*$ are the following \cite{CHPRV-01,CHPRV-02,Hasenbusch-00}:
\begin{eqnarray}
\hskip -1truecm 
\lambda^* &=& 2.07(5)\hphantom{2.} \qquad\qquad \hbox{\rm for $N = 2$} , \\
\hskip -1truecm 
\lambda^* &=& 4.6(4)\hphantom{22.} \qquad\qquad \hbox{\rm for $N = 3$} , \\
\hskip -1truecm 
\lambda^* &=& 12.5(4.0)            \qquad\qquad \hbox{\rm for $N = 4$} .
\end{eqnarray}

For $N=2$ the dynamically dilute XY model has also been considered:
\be
d\mu(\phi) =\, d^2\phi\,\left[\delta(\phi_1)\delta(\phi_2) + 
   {e^D\over 2\pi} \delta(1 - |\phi|)\right],
\label{ddXY}
\ee
where $\phi = (\phi_1,\phi_2)$. The Hamiltonian is improved for 
$D^* = 1.02(3)$ \cite{CHPRV-01}. Again, the improved theory 
has been determined by means of MC simulations.

Also models with extended interactions have been considered,
such as \cite{BLH-95,HPV-99}
\be
{\cal H} = \beta\Bigl[ \sum_{<ij>} \sigma_i\sigma_j + 
                    y \sum_{[ij]} \sigma_i\sigma_j \Bigr],
\label{Ising-n-3n}
\ee
where $\sigma_i=\pm1$, the first sum is extended over all nearest-neighbor
pairs $<ij>$, and the second one over all third nearest-neighbor pairs 
$[ij]$. In Ref. \cite{BLH-95} a significant reduction of the 
subleading corrections was observed for  $y\approx 0.4$. 
However, the subsequent analysis of Ref. \cite{HPV-99} found $y^*\ltapprox 0.25$. 

\subsection{Field-theoretical methods}
\label{sec-2.4}

Field-theoretical methods 
can be divided into two classes: 
(a) perturbative approaches based on the $\phi^4$ continuum Hamiltonian
\begin{eqnarray}
{\cal H} = \int d^d x 
\left[ {1\over 2} \partial_\mu \phi(x) \partial_\mu \phi(x) +  
{r\over2} \phi(x)^2 +  {u\over 4!} \phi(x)^4 \right];
\label{Hphi4}
\end{eqnarray}
(b) nonpertubative approaches based on approximate solutions of 
Wilson's RG equations.

The oldest perturbative method is the $\epsilon$ expansion in which 
the expansion parameter is $\epsilon = 4 - d$ \cite{WF-72}. 
Subsequently, 
Parisi \cite{Parisi-80} pointed out the possibility of using 
pertubation theory directly at the physical dimensions $d=3$ and $d=2$.
In the original works \cite{BNGM-77,Parisi-80} 
the theory was renormalized at zero momentum. Later,
a four-dimensional minimal subtraction scheme 
without $\epsilon$ expansion was also proposed \cite{Dohm-85,SD-89,SD-90}.
With a slight abuse of language, we will call the first ``traditional" method
as the fixed-dimension expansion approach, although also in the second case 
the dimension is fixed. The second approach will be named 
the minimal subtraction scheme without $\epsilon$ expansion.

The nonperturbative approach has a very long history 
\cite{WH-73,WK-74,GR-76,NR-82,NR-84} and it has been 
the subject of extensive work even recently, see, e.g.,
Refs. \cite{Morris-98,BTW-99,BB-00} and references therein. 
A brief discussion will be presented here. 
We only mention that for $O(N)$ vector models 
the estimates of the critical parameters
are less precise  than those obtained in studies using perturbative approaches.

\subsubsection{The fixed-dimension expansion}
\label{sec-2.4.1}

In the fixed-dimension expansion one works directly in $d=3$ or 
$d=2$. In this case the theory is super-renormalizable since the 
number of primitively divergent diagrams is finite. 
%the one-loop tadpole for $d=2,3$ and the two-loop ``setting-sun" diagram in $d=3$. 
One may regularize the corresponding  integrals by keeping $d$ arbitrary 
and performing an expansion in $\epsilon=3-d$ or $\epsilon=2-d$. 
Poles in $\epsilon$ appear in divergent diagrams. Such divergences
are related to the necessity of performing an infinite renormalization of the 
parameter $r$ appearing in the bare Hamiltonian, see, e.g., the discussion
in Ref.~\cite{BB-85}. This problem
can be avoided by replacing $r$ with the mass $m$ 
defined by 
\be
m^{-2} = \, {1\over \Gamma^{(2)}(0)} \, 
     \left. {\partial \Gamma^{(2)}(p^2) \over \partial p^2}\right|_{p^2=0},
\ee
where the function $ \Gamma^{(2)}(p^2)$ is related 
to the one-particle irreducible two-point function  by
\be
\Gamma^{(2)}_{ab} (p) = \delta_{ab} \Gamma^{(2)}(p^2).
\ee
Perturbation theory in terms of $m$ and $u$ is finite. The critical limit is obtained for 
$m\to 0$. To handle it, one considers appropriate 
RG functions. Specifically, one defines
the zero-momentum four-point coupling $g$ and the
field-renormalization constant $Z_\phi$ by
\bea
&&\Gamma^{(2)}_{ab}(p) = \delta_{ab} Z_\phi^{-1} \left[ m^2+p^2+O(p^4)\right],
\label{ren1} \\
&&\Gamma^{(4)}_{abcd}(0) = 
Z_\phi^{-2} m^{4-d}
{g\over 3}\left(\delta_{ab}\delta_{cd} + \delta_{ac}\delta_{bd} +
                \delta_{ad}\delta_{bc} \right),
\label{ren2}
\eea
where $\Gamma^{(n)}_{a_1,\ldots,a_n}$ are one-particle irreducible 
correlation functions. 
Then, one defines a coupling-renormalization constant $Z_u$  and a
mass-renormalization constant $Z_t$ by
\begin{eqnarray}
u = m^{4-d} g Z_u Z_\phi^{-2}, \qquad
\Gamma^{(1,2)}_{ab}(0)=\, \delta_{ab} Z_t^{-1}, \label{ren3}
\end{eqnarray}
where $\Gamma^{(1,2)}_{ab}(p)$ is the one-particle irreducible
two-point function with an insertion of ${1\over2}\phi^2$.
The renormalization constants are determined as perturbative expansions
in powers of $g$. The fixed point of the model is determined by
the nontrivial zero $g^*$ of the $\beta$-function
\begin{eqnarray}
   \beta(g) = m \left. {\partial g\over \partial m}\right|_{u} 
=   (d-4) g\left[ 1 + g {d\over dg}\log (Z_u Z_\phi^{-2})\right]^{-1}.
\end{eqnarray}
Note that the fixed-point value $g^*$ coincides with the critical value 
$g_4^+$ of $g_4$, cf. Eq.~(\ref{grdef}).
Then, one defines
\begin{eqnarray}
\hskip -1truecm 
\eta_\phi(g) &=& \left. {\partial \ln Z_\phi \over \partial \ln m}
         \right|_{u}
= \beta(g) {\partial \ln Z_\phi \over \partial g},
\\
\hskip -1truecm 
\eta_t(g) &=& \left. {\partial \ln Z_t \over \partial \ln m}
         \right|_{u}
= \beta(g) {\partial \ln Z_t \over \partial g}.
\end{eqnarray}
Finally, the critical exponents are given by
\begin{eqnarray}
\eta &=& \eta_\phi(g^*),
\label{eta_fromtheseries} \\
\nu &=& \left[ 2 - \eta_\phi(g^*) + \eta_t(g^*)\right] ^{-1},
\label{nu_fromtheseries} \\
\omega &=& \beta'(g^*),
\end{eqnarray}
where $\omega$ is the exponent associated with the
leading irrelevant operator.\footnote{This is not always 
correct \cite{CCCPV-00}.
Indeed, $\beta'(g^*)$ is always equal to the  exponent of the first
nonanalytic correction in $g(m)$. Usually, the first correction is due to
the leading irrelevant operator, but this is not necessarily the case.
In the two-dimensional Ising model, the first correction in $g(m)$ is
related to the presence of an analytic background in the free energy, and
$\beta'(g^*) = \gamma/\nu=7/4$, while $\omega=2$. See the discussion in 
Sec. \ref{sec-2.4.3}.}
All other exponents can be obtained using the scaling 
and hyperscaling relations. 

Since this method is based on zero-momentum renormalization 
conditions, it is not well suited for the study of vector models 
in the LT phase. In this case, 
the minimal-subtraction scheme without $\epsilon$ expansion can be used. 
Since it is strictly related to the 
$\epsilon$ expansion, it will be presented in the next section. 

The longest available series for the critical exponents can be found in 
Refs. \cite{BNGM-77,MN-91,AS-95} for $d=3$ and in 
Ref. \cite{OS-00} for $d=2$.  
More precisely, in three dimensions the critical exponents and the 
$\beta$-function are known to six loops for generic values of $N$ 
\cite{AS-95}. For $N=0,1,2,3$ seven-loop series for 
$\eta_\phi$ and $\eta_t$ were computed in 
Ref. \cite{MN-91}. They are reported in the Appendix of Ref. \cite{GZ-98}. 
In two dimensions, five-loop series are available for all values of $N$
\cite{OS-00}.
Perturbative expansions of some
universal amplitude ratios involving HT quantities
and of the $2n$-point renormalized coupling 
constants $g_{2n}^+$ can be found in Refs.~\cite{BB-85} and
\cite{SOUK-99} respectively. For the scalar theory an extension 
of the method \cite{BBMN-87} allowed to obtain the free energy 
below the critical temperature and therefore all universal amplitude ratios 
defined from zero-momentum quantities. 
For the study of the LT phase of the Ising model, a 
slightly different approach was developed in Refs. 
\cite{MH-94,GKM-96}, which also allowed the computation of ratios involving the 
correlation length. 

\subsubsection{The $\epsilon$ expansion} 
\label{sec-2.4.2}

The $\epsilon$ expansion 
\cite{WF-72} is based on the observation that, for $d=4$, 
the theory is essentially Gaussian. One  considers 
the standard perturbative expansion, and then transforms it
into an expansion in powers of $\epsilon\equiv 4 - d$. 
In practice, the method works as in the fixed-dimension expansion. 
One first determines the expansion of the
renormalization constants 
$Z_u$, $Z_\phi$, and $Z_t$ in powers of the coupling $g$. 
Initially, they were obtained by 
requiring the normalization conditions
\reff{ren1}, \reff{ren2}, and \reff{ren3}. However, in this framework it 
is simpler to use the minimal-subtraction scheme \cite{tHV-72}. 
Once the renormalization constants are determined, one computes 
the RG functions $\beta(g)$, $\eta_\phi(g)$, and $\eta_t(g)$ as in
Sec.~\ref{sec-2.4.1}. 
The fixed-point value $g^*$ is obtained by solving the equation $\beta(g^*) = 0$
perturbatively in $\epsilon$.
Once the expansion of $g^*$ is available,
one obtains the expansion of the exponents,  by expressing
$\eta_\phi(g^*)$ and $\eta_t(g^*)$ in powers of $\epsilon$. 
Notice that, in the minimal-subtraction scheme, $g$ is not related to
$g_4$.

In this scheme, five-loop series for the exponents were computed in 
Refs.~\cite{CGLT-83,KNSCL-93}.  The equation of state
and several amplitude ratios were determined in Refs. 
\cite{BWW-72,WZ-74,NA-85,Bervillier-86,PV-00}. 

The minimal-subtraction scheme without $\epsilon$ ex\-pan\-sion 
\cite{Dohm-85,SD-89,SD-90} is strictly
related. The functions $\beta(g)$, $\eta_\phi(g)$, and $\eta_t(g)$ are 
the minimal-subtraction functions. However, here 
$\epsilon$ is no longer considered
as a small quantity but it is set to its physical value,\footnote{
Note that the dependence on $\epsilon$ of the above-defined RG 
functions is trivial. The exponents $\eta_\phi(g)$ and 
$\eta_t(g)$ are independent of $\epsilon$, 
$\beta(g) = - \epsilon g + b(g)$ and $b(g)$ is independent of $\epsilon$.}
i.e. in three dimensions one simply sets 
$\epsilon = 1$. Then,  the procedure is identical to that presented
for the fixed-dimension expansion. A nontrivial zero $g^*$ 
of the $\beta$-function is determined and the exponent series are computed for 
this value of $g^*$. 
The method is well suited for the study of universal LT
properties of vector systems, and indeed, perturbative series 
for several amplitude ratios
have been computed \cite{BSD-97,LMSD-98,SLD-99,SMD-00}.

\subsubsection{Resummation of the perturbative series}
\label{sec-2.4.3}

FT perturbative expansions are divergent. Thus,
in order to obtain accurate results, an appropriate resummation 
is required. This can be done by exploiting their Borel summability, 
that has been proved for the fixed-dimension expansion of the 
$O(N)$ $\phi^4$ theory in $d<4$
\cite{EMS-75,MS-77,FO-76,MR-85} and has been conjectured  
for the $\epsilon$ expansion.
If we consider a quantity $S(g)$ that has
a perturbative expansion 
\be 
S(g) \approx \sum s_k g^k,
\ee
the large-order behavior of the coefficients is given by
\be
s_k \sim k! \,(-a)^{k}\, k^b \,\left[ 1 + O(k^{-1})\right], 
\label{lobh}
\end{equation}
with $a>0$. 
Here, the perturbative coupling is the renormalized 
coupling constant of the fixed-dimension expansion, but the 
same discussion applies to the $\epsilon$ expansion,
replacing $g$ by $\epsilon$.
Note that the value of the constant $a$ 
is independent of the particular quantity considered,
unlike the constant $b$.
The constants $a$ and $b$ can be determined by
means of a steepest-descent calculation in which
the relevant saddle point is a finite-energy solution (instanton)
of the classical field equations with negative 
coupling~\cite{Lipatov-77,BLZ-77},
see also Refs.~\cite{Zinn-Justin-book,Parisibook}.

In order to resum the perturbative series, 
we introduce the Borel-Leroy transform $B(t)$ of $S(g)$, 
\begin{equation}
S(g) = \int_0^\infty t^c e^{-t} B(t),
\label{defSg-Bt}
\end{equation}
where $c$ is an arbitrary number. 
Its series expansion is given by
\be 
B_{\rm exp}(t) = \sum_k {s_k\over \Gamma(k + c + 1)} t^k.
\label{Bexpansion}
\ee
The constant $a$ that characterizes the large-order behavior 
of the original series is related to the
singularity $t_s$ of the Borel transform $B(t)$ that is nearest 
to the origin: $t_s=-1/a$.  
The series $B_{\rm exp} (t)$ is convergent 
in the disk $|t| < |t_s| = 1/a$ of the complex plane, and also on the 
boundary if $c >  b$. In this domain, one can  compute 
$B(t)$ using $B_{\rm exp} (t)$. However, in order to compute the 
integral \reff{defSg-Bt},
one needs $B(t)$ for all positive values of $t$. 
It is thus necessary to perform an analytic continuation of $B_{\rm exp} (t)$. 

The analytic continuation may be achieved using \cite{BNGM-77}
Pad\'e approximants to the series (\ref{Bexpansion}).
A more refined procedure exploits the knowledge of the large-order
behavior of the expansion, and in particular of the constant $a$.
One performs an Euler transformation \cite{LZ-77}
\begin{equation}
y(t) = {\sqrt{1 + a t} - 1\over \sqrt{1 + a t} + 1 },
\label{cfmap}
\end{equation}
that allows to rewrite $B(t)$ in the form
\be
B(t) = \sum_k f_k \, [y(t)]^k.
\label{Borel-2}
\ee
If all the singularities belong to the real interval $[-\infty,-t_c]$,
the expansion \reff{Borel-2} converges everywhere in the complex 
$t$-plane except on the negative axis for $t < - t_c$. 
After these transformations, one obtains a new expansion for the 
function $S(g)$:
\be
S(g) \approx  \int^\infty_0 dt\, e^{-t} t^c\, \sum_k f_k\,  [y(tg)]^k.
\label{Borel-final}
\ee
This sequence of operations has transformed the original divergent series
into an integral of a convergent one, which can then be studied 
numerically. Notice that 
the convergence of the integral \reff{Borel-final}, that is 
controlled by the analytic properties of $S(g)$,
is not guaranteed. For instance,
if $S(g)$ has a cut for $g\ge g^*$---we will show below that this  occurs
in the fixed-dimension expansion---then the integral does not 
converge for $g>g^*$.

We mention that one may also use Eq. \reff{Borel-2} when $a$ is 
not known, considering $a$ as a free parameter that can be optimized
in the resummation procedure \cite{MV-98-d}.

A different resummation method is used by Kleinert
\cite{Kleinert-93,Kleinert-95,JK-00}. Instead of using the perturbative 
series in terms of the renormalized coupling constant $g$, he considers
the expansion in terms of the bare coupling $u$. Since perturbative 
series in three dimensions 
are expressed in terms of $\overline{u} = u/m$, the critical
results are obtained by evaluating the perturbative expressions
in the limit $\overline{u} \to \infty$. Similar extrapolations are
used in the context of polymers, i.e. in the $\phi^4$ theory in the 
limit $N\to 0$; see, e.g., Ref. \cite{dCCJ-85}. Of course, the extrapolation 
is here a tricky point. Refs. \cite{Kleinert-93,Kleinert-95} 
use a variational method. Essentially, one introduces a new parameter 
such that the exact expressions are independent of it. 
In  the truncated series  the new parameter is a nontrivial 
variable that is fixed by requiring the results to be stationary with 
respect to its variation. The variational method transforms an 
initially divergent series in a convergent sequence of approximations. 

The convergence of all the resummation methods depends on the analytic behavior 
at $g=g^*$. In particular, the convergence may be rather slow 
if the resummed function is nonanalytic at $g^*$.

Singularities---predicted 
long ago in Refs. \cite{Parisi-80,Nickel-82,Nickel-91}---appear
in the fixed-dimension expansion renormalized at zero momentum. 
To understand the problem, following Nickel \cite{Nickel-82}, 
let us consider the zero-momentum four-point
coupling $g_4$---as we already remarked,
it coincides with the perturbative coupling $g$ defined in 
Eq.~\reff{ren2}---as a function of the reduced temperature $t$. For
$t\to 0$ we can write down an expansion of the form
\begin{eqnarray}
g_4 = g^{+}_4 \,
         \Bigl[ 
&&
\hskip -0.5truecm
1 + a_1 t + a_2 t^2 + \ldots +
                b_1 t^{\Delta} + b_2 t^{2 \Delta} 
    + \ldots  \nonumber \\
&&    
\hskip -0.5truecm
+c_1 t^{\Delta+1} + \ldots +
    d_1 t^{\Delta_2} + \ldots 
    + e_1 t^\gamma + \ldots \Bigr]\; ,
\label{grintermsTmTc}
\end{eqnarray}
where $\Delta$, $\Delta_2$, $\ldots$ are subleading exponents.
The correction proportional to $t^\gamma$ is due to the presence
of an analytic background in the free energy.

Starting from
Eq. (\ref{grintermsTmTc}), one may easily compute the $\beta$-function.
Since the mass gap $m$ scales analogously, one obtains the following 
expansion:\footnote{This is the generic behavior when $\Delta < 1$. 
In some models, for instance in the two-dimensional nearest-neighbor
Ising model, $\Delta>\gamma$ and $a_1 = 0$.
In this case, Eq. (\ref{betafunction}) is still correct \cite{CCCPV-00}
if $\gamma$ and $\Delta$ are interchanged. Moreover, $\alpha_1 = - \gamma/\nu$
in this case. If $a_1\not=0$ and $\Delta > 1$, we have $\alpha_1 = - 1/\nu$
and corrections $(\Delta g)^\Delta$, $(\Delta g)^{\Delta_2}$, etc.}
\begin{eqnarray}
 \beta(g_4) \equiv  m {\partial g_4\over \partial m} = && 
\hskip -0.5truecm
   \alpha_1 \Delta g + \alpha_2 (\Delta g)^2 +
               \ldots + 
        \beta_1 (\Delta g)^{1\over \Delta} +
        \beta_2 (\Delta g)^{2\over \Delta} + \ldots \nonumber \\
&&     
\hskip -0.5truecm
+\gamma_1 (\Delta g)^{1+{1\over \Delta}} + \ldots +
        \delta_1 (\Delta g)^{\Delta_2\over \Delta} + \ldots
        +\zeta_1 (\Delta g)^{\gamma\over \Delta} + \ldots,
\label{betafunction}
\end{eqnarray}
where $\Delta g = g^+_4 - g_4$.
Eq. (\ref{betafunction}) clearly shows the presence of several
nonanalytic terms with exponents depending on $1/\Delta$, $\Delta_i/\Delta$,
and $\gamma/\Delta$.

As pointed out by Sokal \cite{Sokal-94,LMS-95} 
(see also Ref. \cite{BB-97} for a discussion in Wilson's RG setting),
the nonanalyticity of the RG functions can also be understood within
Wilson's RG approach. We repeat his argument here. Consider the
Gaussian fixed point which, for $3\le d < 4$, has a two-dimensional
unstable manifold ${\cal M}_u$: The two unstable directions correspond to the
interactions $\phi^2$ and $\phi^4$. The continuum
field theories are in a one-to-one correspondence with Hamiltonians
on ${\cal M}_u$. Thus, the FT RG is nothing but
Wilson's RG restricted to ${\cal M}_u$. The point is that there is no
reason why it should approach the fixed point along a direction orthogonal 
to all other subleading irrelevant operators.
Barring miracles, the approach should have
nonzero components along any of the irrelevant directions. But,
if this happens, nonanalytic terms are present in any RG  function.

This issue has been investigated in the framework of the $1/N$
expansion \cite{PV-gr-98}, computing the asymptotic behavior
of $\beta(g)$ for $g\to g^*$ to next-to-leading order in $1/N$
and for $2<d<4$. The result shows that nonanalytic terms are present 
consistently with Eq.~(\ref{betafunction}). Indeed,
corrections of order 
$(\Delta g)^{1 + 1/\Delta}$ and/or $(\Delta g)^{\Delta_2/\Delta}$
appear. No term proportional to 
$(\Delta g)^{1/\Delta}$ is found, which implies $a_1 = 0$ 
in Eq. \reff{grintermsTmTc}.
In Ref.~\cite{CCCPV-00} the computation is extended to two dimensions,
finding again nonanalytic terms.

These singularities may cause a slow convergence 
in the resummations of the perturbative series. 
In Refs. \cite{CCCPV-00,CaPeVi-00}
some simple test functions were considered and it was shown
that large discrepancies should be expected if the first nonanalytic 
exponent is small. For instance, if a function $f(g)$ behaves as the 
$\beta$ function, i.e. 
\begin{equation}
f(g) \approx  a (g-g_0) + b (g - g_0)^{1+p}
\end{equation}
for $g\to g_0$, 
a relatively precise 
estimate of $a$ is obtained if $p \gtapprox 1$, while for small values
of $p$ 
the estimate is largely incorrect and, even worse, 
the errors, which are obtained as usual by stability criteria, are far too 
small. It is important to note that these discrepancies 
are not related to the fact that the series are divergent. They would 
be present even if the perturbative expansions were convergent.\footnote{
The reader may consider $f(g) = (1 - g)^p$ and try to compute $f(1)$ 
from its Taylor expansion around $g=0$. Since for $p\to 0^+$, $f(g)\to 1$ 
pointwise for all $g<1$, for small values of $p$ any extrapolation 
provides an estimate $f(1)\approx 1$, clearly different from the 
exact value $f(1) = 0$.} 

Of course, the interesting question is whether these nonanalyticities 
are relevant in the $\phi^4$ perturbative series.
In three dimensions and for $N=0,1,2,3$,
$\Delta\approx 0.5$ and $\Delta_2/\Delta$
is approximately 2 \cite{GR-76,NR-84}. Thus, the leading nonanalytic term
has exponent $\Delta_2/\Delta$ and is rather close to an analytic one.
Therefore, we expect small effects in three dimensions,
and indeed the FT results
are in substantial agreement with the estimates obtained in
MC and HT studies.
The situation worsens in the two-dimensional case. In 
the Ising model Ref. \cite{CCCPV-00}
predicted a nonanalytic term $(\Delta g)^{8/7}$, 
while for $N\ge 3$ logarithmic corrections, such as
$(\Delta g)/\log\, \Delta g$, are predicted on general grounds 
and found explicitly in a large-$N$ calculation \cite{CCCPV-00}. Since the 
first nonanalytic term has a small exponent, 
large deviations are expected \cite{CCCPV-00}. Indeed, two-dimensional 
estimates differ significantly from the theoretically expected results 
\cite{LZ-77,OS-00}.

On the other hand, 
for the $\epsilon$ expansion it has been argued that no
nonanalyticities are present in universal quantities. 
This has also been argued for the fixed-dimension
expansion in the minimal subtraction scheme \cite{Schafer-94,BB-94},
but it is difficult to verify in the absence of a 
nonperturbative definition of the scheme. 

Finally, we mention that it is possible to improve the 
$\epsilon$-expansion results for a quantity $R$ 
if its value is known for some 
values of the dimensions.
The method was originally proposed in Refs. 
\cite{LZ-87b,desCloizeaux-81,desCloizeaux-Jannink_book} where 
exact values in two 
or one dimension were used in the analysis of the $\epsilon$ expansion. 
The method of Ref.~\cite{LZ-87b} works as follows.
One considers a quantity $R$ such that 
for $\epsilon = \epsilon_1$ the exact value $R_{\rm ex}(\epsilon_1)$ is
known. Then, one defines
\begin{equation}
\overline{R}(\epsilon) = \left[
  {R(\epsilon) - R_{\rm ex}(\epsilon_1) \over (\epsilon - \epsilon_1)}\right],
\end{equation}
and a new quantity
\begin{equation}
   R_{\rm imp}(\epsilon) = R_{\rm ex}(\epsilon_1) +
         (\epsilon - \epsilon_1) \overline{R}(\epsilon) .
\end{equation}
New estimates of $R$ at $\epsilon = 1$ can then be obtained by
resumming the $\epsilon$ expansion of 
$\overline{R}(\epsilon)$  and then computing $R_{\rm imp}(1)$.
The idea behind this method is very simple. If, for instance, the value
of $R$ for $\epsilon = 2$ is known, one uses as zeroth-order
approximation at $\epsilon = 1$ the value of the linear interpolation
between $\epsilon = 0$ and $\epsilon = 2$ and then uses the series
in $\epsilon$ to compute the deviations. If the interpolation is a
good approximation, one should find that the series that gives the
deviations has smaller coefficients than the original one.
Consequently, also the errors in the resummation are reduced. 

In Ref. \cite{PV-gr-98} 
this strategy was generalized to the case in which 
one knows the exact value of $R$ for more than one value of $\epsilon$.
If exact values
$R_{\rm ex}(\epsilon_1)$, $\ldots$, $R_{\rm ex}(\epsilon_k)$ are known
for a set of dimensions $\epsilon_1$, $\ldots$, $\epsilon_k$,
$k \ge 2$, then one defines
\begin{equation}
Q(\epsilon) = \sum_{i=1}^k \left[
    {R_{\rm ex}(\epsilon_i) \over (\epsilon - \epsilon_i)}
    \prod_{j=1,j\not=i}^k (\epsilon_i - \epsilon_j)^{-1} \right],
\end{equation}
and
\begin{equation}
\overline{R}(\epsilon) =
   {R(\epsilon) \over \prod_{i=1}^k (\epsilon - \epsilon_i)} -
   Q(\epsilon) ,
\end{equation}
and finally
\begin{equation}
   R_{\rm imp}(\epsilon) = \left[ Q(\epsilon) +
      \overline{R}(\epsilon) \right]
      \prod_{i=1}^k (\epsilon - \epsilon_i)  .
\label{serieconstrained}
\end{equation}
One can easily verify that the expression
\begin{equation}
   \left[ Q(\epsilon) +
      \overline{R}(0) \right]
      \prod_{i=1}^k (\epsilon - \epsilon_i)
\label{interpolation}
\end{equation}
represents the $k$th-order polynomial interpolation
a\-mong the points $\epsilon=0,\epsilon_1,...,\epsilon_k$.
Again the resummation procedure is applied to the 
$\epsilon$ expansion of $\overline{R}(\epsilon)$
and the final estimate is obtained by computing $R_{\rm imp}(1)$.
Such a technique was successfully used in many different cases, 
see Refs. \cite{PV-gr-98,PV-ef-98,PV-99,PV-00}.

\subsubsection{Nonperturbative methods}
\label{CRGth}

Critical exponents and several universal properties have been 
obtained using nonperturbative FT methods based on 
approximate solutions of continuous RG equations (CRG).
The starting point of this approach  is
an exact functional differential RG equation.
Various proposals of RG equations
have been considered in the literature,
see, e.g., 
Refs.~\cite{WK-74,WH-73,Polchinski-84,NC-77,Wetterich-93,Morris-94,BDM-93,Ellwanger-94}.
For instance, one may write down a RG equation for the
average action $\Gamma_k[\phi]$,
which is a functional of the fields 
and depends on a coarse-graining scale $k$. 
The dependence of the average action $\Gamma_k$ on the 
scale $k$ is described by the flow equation \cite{Wetterich-93,Morris-94,BDM-93,Ellwanger-94}  
\begin{equation}
\partial_t \Gamma_k =
{1\over 2} {\rm Tr} \; \left( \Gamma^{(2)}_k + R_k\right)^{-1}
\partial_t R_k,
\end{equation}
where $\Gamma^{(2)}_k$ is the second functional derivative
of the average action, $t\equiv\ln k$, and $R_k(q^2)$
is an infrared regulator at the momentum scale $k$.
In the infrared limit  $k\rightarrow 0$, the functional
$\Gamma_k$ yields the Helmholtz free energy 
in the presence of a position-dependent 
magnetic field, which is usually called effective action in
this context.
Except for a few trivial cases, 
this functional equation cannot be solved exactly, so that one 
must perform approximations and/or truncations and 
use numerical methods.
A systematic scheme of truncations is provided by 
the derivative expansion (DE), which is a 
functional expansion of the average action in powers of momenta and
requires a sufficiently small anomalous dimension of the field,
i.e. $\eta\ll 1$ \cite{Golner-86,ILF-90}.
In particular, for a scalar theory one may write 
\begin{equation}
\Gamma_k[\phi] = \int d^d x 
\left[ U_k(\phi^2) + {1\over 2} Z_k(\phi^2) (\partial_\mu \phi)^2 
+ O(\partial^4)\right].
\end{equation}
The lowest order of the DE is the so-called local potential approximation
(LPA), see, e.g., Ref.~\cite{BB-00}. It 
includes the potential $U_k(\phi^2)$ and a standard
kinetic term, i.e. it assumes $Z_k$ to be a constant. 
This implies $\eta=0$,
and thus it is expected to provide a good
starting point only  when $\eta\ll 1$. For example, in the
two-dimensional Ising case, where $\eta=1/4$ is not
particularly small, the LPA is unable to
display the expected fixed-point structure \cite{FB-91,Morris-95}.
A variant of the LPA \cite{TW-94} assumes $Z_k$ dependent on $k$.
In this improved approximation (ILPA) $\eta$ is not equal to zero and 
it is determined
from the behavior of the propagator, still assuming $\eta\ll 1$.
The first correction in the DE 
(1st DE) takes into account the dependence
on $\phi^2$ in $Z_k(\phi^2)$. The next order involves terms with
four derivatives, and so on.
The convergence properties of the DE are still not
clear. In particular, it seems rather sensitive to the
choice of the infrared regulator $R_k(q^2)$.
See, e.g., Refs.~\cite{Morris-94-2,MT-99,LPS-00,Litim-01,Litim-02} for discussions of 
this point.  It is therefore difficult to estimate the uncertainty of 
the results obtained by this approach. 
Moreover, the technical difficulties increase very rapidly
with the order of the DE, so that
only the first order has been effectively implemented for the
$N$-vector model. 
Within a level of the DE, one may also consider an expansion of the 
coefficient functions $U_k(\phi^2)$, 
$Z_k(\phi^2)$, $\ldots$, in powers of the field $\phi$. 
Usually, the results of this expansion show a relatively fast convergence, 
while the dependence on the DE order seems to be more important,
see, e.g., the list of results reported in Ref.~\cite{BTW-99}.
We refer to Refs.~\cite{BTW-99,BB-00} for a more 
detailed discussion about the various key
ingredients of the method, which are essentially the choice
of the continuous RG equation, of the infrared regulator,
and of the approximation scheme.

A similar approach is the scaling-field method
\cite{GR-76,NR-82,NR-84}.
The starting point is again a continuous
RG equation, but the peculiarity is the use of 
an expansion in scaling fields to
transform the original equation into an infinite
hierarchy of nonlinear differential equations for the scaling fields.

\section{The Ising universality class}
\label{Ising} 

\subsection{Physical relevance}

The Ising model is one of the most studied models in the 
theory of phase transitions,
not only because it is considered as the prototype of statistical
systems showing a nontrivial power-law critical behavior,
but also because it describes several physical systems.  
Indeed, many 
systems characterized by short-range interactions and a scalar order
parameter undergo a critical transition belonging 
to the Ising universality class. We mention 
the liquid-vapor transition in simple fluids, the 
transitions in multicomponent fluid mixtures, in uniaxial
antiferromagnetic materials, and in micellar systems 
(see Sec. \ref{sec3.1.1}). Experiments in this area are still very
numerous, the most part focusing on the critical behavior of 
simple and complex fluids, which have a large variety of industrial and
technological applications. Many experiments on the static and dynamic 
critical behavior of these systems have been performed in 
microgravity environment, on the Space Shuttle, on the Mir space station,
and using specially-designed rockets; a new generation is 
currently developed for the International Space Station
\cite{Beysens-01,LAI-00,BG-01,LI-01}. In particular, a new experiment
(MISTE) \cite{BHZ-00} will be flown in 2005 and is supposed to provide 
high-precision data for the critical behavior of $^3$He. 
We should mention that Ising criticality is also observed in several
models that are relevant for high-energy physics, see Sec. \ref{sec3.1.2}.

\subsubsection{Experimental systems}
\label{sec3.1.1}

The most important physical systems  belonging to the Ising universality class
may be divided into different classes:

\begin{itemize}
\item[(i)] 
{\bf Liquid-vapor transitions.}
The order parameter is $\rho-\rho_c$, where $\rho$ is the density 
and $\rho_c$ its value at the critical point. 
The Ising-like continuous transition occurs at the end of  
the first-order liquid-gas transition line in the pressure-temperature
plane; see Fig. \ref{figHT}. The liquid-vapor transition does not 
have the $\mathbb{Z}_2$ symmetry which is present in magnetic
systems.  Therefore, in fluids one observes 
$\mathbb{Z}_2$-noninvariant corrections to scaling, which are absent in magnets.
For a general review see, e.g., Ref.~\cite{PR-review}.
For a discussion of the mapping of the fluid Hamiltonian onto 
a magnetic one, see also Ref.~\cite{Brilliantov-98}.

\item[(ii)] {\bf Binary mixtures.}
One considers here two fluids. The order parameter 
is the concentration and the transition corresponds to the mixing 
of the two liquids (or gases): on the one side of the transition the two fluids 
are separated, on the other side they are mixed. 
Binary mixtures also undergo a liquid-vapor transition as described 
in (a). Similar transitions also occur in solids, for instance in  
$\beta$-brass, and in several complex fluids, such as polymer solutions 
and polymer blends (see, e.g., 
Ref. \cite{desCloizeaux-Jannink_book}), 
colloidal suspensions \cite{CPR-00-01}, and solutions 
of biological proteins (see, e.g., Refs. \cite{BBPOB-91,APLOHSB-98}). 

\item[(iii)] {\bf Coulombic systems.}
The critical behavior of ionic fluids has been rather controversial.
Originally, on the basis of the experimental results,
see the discussion in Refs. \cite{Fisher-94-96,Stell-95-96}, 
electrolytes were divided in solvophobic---in this case criticality was driven
by short-range forces---and Coulombic---phase separation was driven by 
the long-range Coulomb force. Solvophobic electrolytes were supposed 
to have Ising behavior, while Coulombic systems were expected to be 
mean-field like. At present, there is a general consensus that all
ionic systems show Ising criticality, although the Ising window may be 
extremely small, so that one observes a mean-field--to--Ising crossover,
see Refs. \cite{Anisimov-etal_95,Jacob-etal_98,GAS-01,AS-review} and 
references therein. Recent experiments 
\cite{BBB-97,WBLSKS-98,KWSW-99,HMK-00,BJ-01} and numerical simulations
\cite{LFP-01,CLW-02} confirm this scenario.

\item[(iv)] {\bf Micellar systems.} 
Micellization is the process of aggregation of certain surfactant 
mole\-cu\-les in
dilute aqueous solutions. The onset of micellization, i.e. the concentration
at which the aggregation process begins, can be regarded as a second-order 
phase-transition point \cite{Hall-72,Anisimov-book}.

\item[(v)] {\bf Uniaxial magnetic systems.}
These systems are those that inspired the Ising Hamiltonian. They are magnetic 
systems in which the crystalline structure favors the alignment along 
a specific direction. Experimental systems often display 
antiferromagnetism, but, as we discussed in Sec. \ref{sec-1.3}, 
on bipartite lattices ferromagnetic and antiferromagnetic criticality are 
closely related. Because of the crystalline structure, 
these systems are not rotationally invariant. Thus, there are 
corrections to scaling that are not present in fluids.

\end{itemize}

\subsubsection{Ising systems in high-energy physics}
\label{sec3.1.2}

Continuous transitions belonging to the three-di\-men\-sio\-nal Ising 
universality
class are expected in some theories relevant for high-energy physics. 
We mention:
\begin{itemize}
\item[(i)]
The  finite-temperature transition in the electroweak theory, 
which is  relevant for the initial evolution of the universe.
RG arguments and lattice simulations \cite{KLRS-96,RTKLS-98} show that 
in the plane of the temperature
and of the Higgs mass there is a line of first-order transitions, 
which eventually ends at a second-order transition point.
Such a transition is argued to belong to the Ising universality class.
\item[(ii)]
An Ising-like continuous transition is predicted at 
finite temperature and finite barion-number 
chemical potential in the theory of strong interactions 
(QCD) \cite{HJSSV-98,BR-99}.
\item[(iii)]
For large values of the quark mass, the finite-temperature 
transition of QCD is of first order.
With decreasing the quark mass, 
the first-order transition should persist up to a critical value,
where the transition becomes continuous and 
is expected to be Ising-like \cite{PW-84}.
\item[(iv)]
The chiral phase transition with three massless flavored  quarks is 
expected to be of first order \cite{PW-84}.
The first-order phase transition should persist for $m_{\rm quark}>0$
up to a critical value of the quark mass. 
At this critical point, the transition is continuous
and it has been conjectured \cite{GPS-87} and verified numerically
\cite{KLS-01} that it belongs to the Ising universality class.
\item[(v)]
The finite-temperature transition of the 
four-dimensional $SU(2)$ gauge theory \cite{PW-84},
which has been much studied as a prototype of nonabelian gauge theories, 
belongs to the Ising universality class.
\end{itemize}

\subsection{The critical exponents}
\label{critexp}

\subsubsection{Theoretical results}
\label{thexpising}

The Ising universality class has been studied 
using several theoretical approaches.
In Tables \ref{expIsingthHT}, \ref{expIsingthother}, \ref{expIsingthMC}, 
and \ref{expIsingthFT} we present several estimates 
of the critical exponents obtained by various methods, such as
HT expansions, LT expansions, MC simulations, FT methods, etc.

\begin{table*}
\caption{
Theoretical estimates of the critical exponents obtained from HT expansions 
for the three-dimensional Ising universality class.  
See text for explanation of the symbols in the column ``info.''
We indicate with an asterisk (${}^*$) the estimates that have been
obtained using the scaling relations $\gamma =(2 - \eta)\nu $,
$2 - \alpha = 3 \nu $, $\beta=\nu(1+\eta)/2$
(when the error was not reported by the authors, we used 
the independent-error formula to estimate it).
}
\label{expIsingthHT}
\footnotesize
%\tiny
\hspace*{-1cm}    % Move table leftwards, so it doesn't run off the right
\tabcolsep 4pt        % Less than the usual 6pt
\begin{center}
\begin{tabular}{rlcllllll}
\hline
\multicolumn{1}{c}{Ref.}& 
\multicolumn{1}{c}{info}& 
\multicolumn{1}{c}{order}& 
\multicolumn{1}{c}{$\gamma$}& 
\multicolumn{1}{c}{$\nu$}& 
\multicolumn{1}{c}{$\eta$}&
\multicolumn{1}{c}{$\alpha$}& 
\multicolumn{1}{c}{$\beta$}&
\multicolumn{1}{c}{$\Delta=\omega\nu$} \\   
\hline  
\cite{CPRV-02} $_{2002}$  & IHT $\phi^4,\phi^6,{\rm s}$-1 & 25th
& 1.2373(2)   &  0.63012(16) & 0.03639(15) & 0.110(2)$_a$ 0.1096(5)$^*$ 
  & 0.32653(10)$^*$ & 0.52(3)  \\

\cite{CPRV-99} $_{1999}$ & MC+IHT $\phi^4$ & 20th & 1.2372(3)   &  0.6301(2) & 0.0364(4) &
0.1097(6)$^*$ & 0.32652(15)$^*$ & 
\\
\cite{CPRV-99} $_{1999}$  & IHT $\phi^4,\phi^6,{\rm s}$-1 & 20th
& 1.2371(4)   &  0.63002(23) & 0.0364(4) & 0.1099(7)$^*$ & 0.32648(18)$^*$ & 
\\
\cite{BC-02} $_{2002}$ & s-$\case{n}{2}$ bcc & 25th  
& 1.2371(1)   &  0.6299(2) & 0.0360(8)$^*$ & 0.1103(6)$^*$ & 0.3263(3)$^*$ & \\
\cite{BC-00}  $_{2000}$  &  s-$\case{1}{2}$ bcc & 25th 
& 1.2375(6)  &  0.6302(4)    & 0.036(2)$^*$  &  0.1094(12)$^*$  & 0.3265(7)$^*$  & \\
\cite{BC-97-2} $_{1997}$ &  s-$\case{1}{2}$ bcc & 21st 
& 1.2384(6)  &  0.6308(5)    & 0.037(2)$^*$  &  0.1076(15)$^*$  & 0.3270(8)$^*$  & \\
\cite{BC-02} $_{2002}$ & s-$\case{n}{2}$ sc & 25th  
& 1.2368(10)   &  0.6285(20) & 0.032(6)$^*$ & 0.114(6)$^*$ & 0.3243(30)$^*$ & 
\\
\cite{BC-00}  $_{2000}$  &  s-$\case{1}{2}$ sc  & 23rd 
& 1.2378(10) &  0.6306(8)    & 0.037(3)$^*$  &  0.1082(24)$^*$  & 0.3270(13)$^*$  & \\
\cite{BC-97-2} $_{1997}$ &  s-$\case{1}{2}$ sc  & 21st 
& 1.2388(10) &  0.6315(8)    & 0.038(3)$^*$  &  0.1055(24)$^*$  & 0.3278(13)$^*$  & \\
\cite{BC-99}  $_{1999}$    &   s-$\case{1}{2}$ sc,bcc & 21st  
&    &  0.631(2)$^*$    &   & 0.106(6)   &   & \\
\cite{SA-98} $_{1998}$ & s-$\case{1}{2}$ sc & 21st & 
1.239(2) & 0.630(9) & 0.033(28)$^*$ & 0.11(3)$^*$ & 0.326(13)$^*$ & \\ 
\cite{GE-94}  $_{1994}$    &  s-$\case{1}{2}$ sc & 25th   &    &  0.6330(13)$^*$  &   & 0.101(4)   &   & \\
\cite{BCGS-94}  $_{1994}$    &  s-$\case{1}{2}$ sc & 23rd &    &  0.6320(13)$^*$   &   & 0.104(4)   & & \\
%\cite{GE-93}  $_{1993}$    &     &    &  0.6300(17)$^*$   &   & 0.110(5)   &   & \\
\cite{NR-90}  $_{1990}$ &  DG bcc & 21st & 1.237(2)     &  0.6300(15) &
0.0359(7) & 0.11(2)$_a$ 0.110(5)$^*$ & 0.3263(8)$^*$ & 0.52(3)  \\
\cite{Guttmann-87}  $_{1987}$ &  s-$\case{1}{2}$,1,2, bcc & 21st & 1.239(3) &
0.632$^{+2}_{-3}$ & 0.040(9) & 0.104$^{+9}_{-6}$$\,^*$ &  0.328(4)$^*$ &  \\
\cite{FC-85}   $_{1985}$  &  DG,K bcc & 21st & 1.2395(4)    &  0.632(1)
& 0.039(4)$^*$ & 0.105(7)$_a$ & 0.3283(15)$^*$ & 0.54(5) \\
\cite{GR-84}   $_{1984}$  &  DG bcc & 21st  &  1.2378(6)    &0.63115(30)  & 0.0375(5) & 
0.1066(9)$^*$ &  0.3278(6)$^*$   & 0.52(3) \\
\cite{FV-83}  $_{1983}$   &  s-$S$ bcc & 21st & 1.242$^{+3}_{-5}$    &
0.634$^{+3}_{-4}$   &  & 0.098$^{+12}_{-9}$$\,^*$ &   & \\
\cite{CFN-82}  $_{1982}$   &  DG,K bcc & 21st  & 1.2385(15)    &     &  &  &   & \\
\cite{Zinn-Justin-81}  $_{1981}$ &   s-$S$ bcc & 21st & 1.2385(25)
&0.6305(15) & 0.0357(6)$^*$ &0.1085(45)$^*$  &  0.3265(26)$^*$ &   \\
\cite{ND-81}  $_{1981}$ &   DG bcc & 21st & 1.237(3)   &  0.630(3) &
0.036(2)$^*$ & 0.110(9)$^*$ &  0.327(5)$^*$ &   \\
\cite{Roskies-81}  $_{1981}$ &   s-$S$ bcc & 21st & 1.240(2)   &
0.628(2) & 0.025(7)$^*$ &0.116(6)$^*$ &  0.322(3)$^*$ &   \\
\hline
\end{tabular}
\end{center}
\end{table*}

\begin{table*}

\caption{
Other theoretical estimates of the critical exponents for the 
three-dimensional Ising universality class.
See text for explanation of the symbols in the column ``info.''
We indicate with an asterisk (${}^*$) the estimates that have been
obtained using the scaling relations $\gamma =(2 - \eta)\nu $,
 $2 - \alpha = 3 \nu $, $\beta=\nu(1+\eta)/2$.
}
\label{expIsingthother}
\footnotesize
%\tiny
%\hspace*{-2cm}    % Move table leftwards, so it doesn't run off the right
%\tabcolsep 4pt        % Less than the usual 6pt
%\doublerulesep 1.5pt  % Less than the usual 2pt
\begin{center}
\begin{tabular}{rllllll}
\hline
\multicolumn{1}{c}{Ref.}& 
\multicolumn{1}{c}{info}& 
\multicolumn{1}{c}{$\gamma$}& 
\multicolumn{1}{c}{$\nu$}& 
\multicolumn{1}{c}{$\eta$}&
\multicolumn{1}{c}{$\alpha$}& 
\multicolumn{1}{c}{$\beta$}\\
\hline  
\cite{SA-98} $_{1998}$ & LT s-$\case{1}{2}$ & 
1.24(1) & 0.629(4)$^*$ & 0.030(7)$^*$ & 0.112(11)$^*$ & 0.324(2)  \\ 
\cite{AT-95}  $_{1995}$    & LT s-$\case{1}{2}$   &    & 0.624(10) &   & 0.13(3)$^*$   &    \\
\cite{GE-93}  $_{1993}$    & LT s-$\case{1}{2}$   & 1.251(28) & 0.625(2) & 0.05(3)$^*$  & 0.125(6)$^*$   & 0.329(9)  \\
\cite{BCL-92}  $_{1992}$    & LT s-$\case{1}{2}$  & 
1.177(11)$^*$ &  0.598(1)$^*$  & 0.031(17)$^*$ & 0.207(4)   & 0.308(5)   \\
\cite{Pelizzola-95}  $_{1995}$ & CVM s-$\case{1}{2}$ & 
1.239(3) & 0.630(3)$^*$ & 0.038(9)$^*$ & 0.109(9)$^*$ & 0.325(4)  \\
\cite{CAM-95}  $_{1995}$ & CAM s-$\case{1}{2}$ & 1.237(4) & 0.631(2)$^*$ & 0.039(8)$^*$ & 0.108(5) & 0.327(4)  \\

\cite{PHO-93}  $_{1993}$ & FSS HA s-$\case{1}{2}$ & 
1.23(1)   & 0.627(2) & 0.038(17)$^*$ & 0.12(2) & 0.324(3)  \\
\cite{OHZ-91}  $_{1991}$ & LT HA s-$\case{1}{2}$ & 
1.255(10) & 0.64(1)  & 0.04(3)$^*$  & 0.096(8) & 0.320(3)  \\
\cite{HHO-90}  $_{1990}$ & HT HA s-$\case{1}{2}$ & 
1.241(3)  & 0.636(4) & 0.049(13)$^*$  & 0.10(2) &   \\
\cite{HJ-86}  $_{1986}$ & FSS HA s-$\case{1}{2}$ & 
1.236(8) & 0.627(4)  & 0.029(18)$^*$ & 0.119(12)$^*$ & 0.332(6)  \\
\cite{Henkel-84}  $_{1984}$ & FSS HA s-$\case{1}{2}$ & 
         & 0.629(2)  & & 0.11(1) & 0.324(9)  \\
\hline
\end{tabular}
\end{center}
\end{table*}

\begin{table*}
\caption{
Estimates of the critical exponents from MC simulations for the 
three-dimensional Ising universality class.  
See text for explanation 
of the symbols in the column ``info.''
We indicate with an asterisk (${}^*$) the estimates that have been
obtained  using the scaling relations $\gamma =(2 - \eta)\nu $,
 $2 - \alpha = 3 \nu $, $\beta=\nu(1+\eta)/2$
(when the error was not reported by the authors, we used 
the independent-error formula to estimate it).
}
\label{expIsingthMC}
\footnotesize
%\tiny
\hspace*{-2cm}    % Move table leftwards, so it doesn't run off the right
\tabcolsep 4pt        % Less than the usual 6pt
%\doublerulesep 1.5pt  % Less than the usual 2pt
\begin{center}
\begin{tabular}{rlllllll}
\hline
\multicolumn{1}{c}{Ref.}& 
\multicolumn{1}{c}{info}& 
\multicolumn{1}{c}{$\gamma$}& 
\multicolumn{1}{c}{$\nu$}& 
\multicolumn{1}{c}{$\eta$}&
\multicolumn{1}{c}{$\alpha$}& 
\multicolumn{1}{c}{$\beta$}&
\multicolumn{1}{c}{$\omega$} \\   
\hline  
\cite{Hasenbusch-99-h}  $_{1999}$ & FSS $\phi^4$
& 1.2366(15)$^*$ &  0.6297(5) & 0.0362(8) & 0.1109(15)$^*$ &0.3262(4)$^*$ & 0.845(10) \\
\cite{BST-99}  $_{1999}$ &  FSS s-$\case{1}{2}$$_{nn,3n}$ & 1.2372(13)$^*$ & 0.63032(56)  &
0.0372(10) & 0.1090(17)$^*$ & 0.3269(5)$^*$ & 0.82(3) \\
\cite{Hasenbusch-99}  $_{1999}$ &  FSS $\phi^4$
&1.2367(11)$^*$ &  0.6296(7) & 0.0358(9) & 0.1112(21)$^*$ & 0.3261(5)$^*$ & 0.845(10) \\
\cite{BFMMPR-99}  $_{1999}$ &    FSS s-$\case{1}{2}$ & 
1.2353(25)$^*$ &  0.6294(10) & 0.0374(12) & 0.1118(30)$^*$& 0.3265(4)$^*$ &0.87(9) \\
\cite{HPV-99}  $_{1999}$ &  FSS $\phi^4$ & 1.2366(11)$^*$ & 0.6298(5)
& 0.0366(8) &0.1106(15)$^*$&  0.3264(4)$^*$ &\\
\cite{CMTC-99} $_{1999}$ &  FSS s-$\case{1}{2}$& &  & 0.036(2) & & & \\
\cite{HP-98}  $_{1998}$ &  FSS s-$\case{1}{2}$&  &  0.6308(10)$^*$ && 0.1076(30)  &  &\\
\cite{BLH-95}  $_{1995}$  &  FSS s-$\case{1}{2}$$_{nn,3n}$, s-1 & 1.237(2)$^*$ & 0.6301(8)  & 0.037(3)
&0.110(2)$^*$ & 0.3267(10)$^*$ & 0.82(6) \\
%\cite{BK-93}  $_{1993}$  &  FSS & & &  &&  & 0.85(4) \\
%\cite{Hasenbusch-93}  $_{1993}$  &  FSS & & 0.630(2)  &  &&  &  \\
\cite{FL-91}  $_{1991}$  &  FSS s-$\case{1}{2}$ & 
1.239(7)$^*$  & 0.6289(8)&0.030(11) & 0.1133(24)$^*$& 0.3258(44)$^*$ &  \\

\cite{Itakura-99}  $_{1999}$  &  MCRG $\phi^4$ 
&   & 0.653(10)  & & 0.04(3)$^*$ &  & 0.7(2) \\
\cite{BHHMS-96}  $_{1996}$ &  MCRG s-$\case{1}{2}$$_{nn,2n,3n}$ &
1.2378(27)$^*$ & 0.6309(12) & 0.038(2)  &  0.1073(36)$^*$&  0.3274(9) &  \\
\cite{GT-96}  $_{1996}$ &  MCRG s-$\case{1}{2}$ & 1.234(4)$^*$ & 0.625(1) & 0.025(6) &0.125(3)$^*$ &0.320(2) & 0.7  \\
\cite{BGHP-92}  $_{1992}$  & MCRG s-$\case{1}{2}$& 1.232(4)$^*$ & 0.624(2)  &
0.026(3) &0.128(6)$^*$&  0.3201(13)$^*$ & 0.80--0.85 \\
\cite{Blote-etal-89}  $_{1989}$  &  
MCRG s-$\case{1}{2}$ & 1.242(10)$^*$  & 0.6285(40)& 0.024(8) & 0.114(12)$^*$& 0.3218(32)$^*$ &  \\
\cite{PSWW-84} $_{1984}$ & MCRG s-$\case{1}{2}$ & 1.238(11)$^*$ &
0.629(4) &0.031(5)&  0.113(12)$^*$& 0.324(3)$^*$& \\

\cite{TB-96}  $_{1996}$ &  S s-$\case{1}{2}$ & &  & &  & 0.3269(6) & \\
\cite{IS-91}  $_{1991}$ &  S s-$\case{1}{2}$ & &  & &  & 0.324(4) &  \\

\cite{ADH-00}  $_{2000}$  &  FSS DS s-$\case{1}{2}$ &   & 0.6280(15)  & & 0.1160(45)$^*$&  & 0.745(74) \\
\cite{ABV-90}  $_{1990}$  &  FSS DS s-$\case{1}{2}$ &   & 0.6285(19) & & 0.1145(57)$^*$&  &  \\
\cite{BSBCT-87}  $_{1987}$  &  FSS DS s-$\case{1}{2}$ &   & 0.6295(10)  & & 0.1115(30)$^*$&  &  \\  
\cite{Marinari-84}  $_{1984}$  &  FSS DS s-$\case{1}{2}$ &   & 0.62(1)  & & 0.14(3)$^*$&  &  \\

\cite{IHOO-00}  $_{2000}$  &  NER s-$\case{1}{2}$ & 
1.255(18)$^*$  & 0.635(5)  & 0.024(18)$^*$ & 0.14(2)$^*$ & 0.325(5)  & \\
\cite{Ito-93}  $_{1993}$  &  NER s-$\case{1}{2}$ &  
 & 0.6250(25)  & & 0.125(8)$^*$ &  & \\

\cite{JMSZ-99}  $_{1999}$ & STCD s-$\case{1}{2}$ & 
1.244(7)$^*$ & 0.6327(20) & 0.035(6)$^*$ & 0.102(6)$^*$ & 0.3273(17) & \\

\cite{JV-00}  $_{2000}$ & FSS s-$\case{1}{2}$ PRL & 1.2331(13)$^*$
& 0.6299(5) & 0.0424(13) & 0.1103(15)$^*$  & 0.3249(6)$^*$ & \\

\cite{PV-02}  $_{2002}$ & FSS SU(2) GT & & 0.6298(28) & & 0.111(8)$^*$  & & \\

\cite{OFP-01} $_{2001}$ & fluid &
1.245(25)  & 0.63(3) & & 0.11(9)$^*$ & 0.322(18) & \\

\hline
\end{tabular}
\end{center}
\end{table*}

\begin{table*}
\caption{
FT  estimates of the critical exponents for the 
three-dimensional Ising universality class.  See text for explanation 
of the symbols in the column ``info.''
}
\label{expIsingthFT}
\footnotesize
%\tiny
%\hspace*{-2cm}    % Move table leftwards, so it doesn't run off the right
\tabcolsep 4pt        % Less than the usual 6pt
%\doublerulesep 1.5pt  % Less than the usual 2pt
\begin{center}
\begin{tabular}{rlllllll}
\hline
\multicolumn{1}{c}{Ref.}& 
\multicolumn{1}{c}{info}& 
\multicolumn{1}{c}{$\gamma$}& 
\multicolumn{1}{c}{$\nu$}& 
\multicolumn{1}{c}{$\eta$}&
\multicolumn{1}{c}{$\alpha$}& 
\multicolumn{1}{c}{$\beta$}&
\multicolumn{1}{c}{$\omega$} \\   
\hline  
\cite{JK-00}   $_{2001}$ & $d=3$ exp & 1.2403(8)  &  0.6303(8)  & 0.0335(6)  & 0.1091(24) & 0.3257(5) & 0.792(3) \\
\cite{GZ-98}   $_{1998}$& $d=3$ exp & 1.2396(13)  &  0.6304(13)  & 0.0335(25)  & 0.109(4) & 0.3258(14) & 0.799(11) \\
\cite{YG-98} $_{1998}$ &$d=3$ exp  &1.243 & 0.632 &  0.034 & 0.103 & 0.327 & \\
\cite{AS-95}   $_{1995}$& $d=3$ exp & 1.239  &  0.631  & 0.038  & 0.107 & 0.327 & \\
\cite{MN-91}   $_{1991}$   &  $d=3$ exp & 1.2378(6)\{18\}&  0.6301(5)\{11\}& 0.0355(9)\{6\} &0.1097(15)\{33\}&   &\\
\cite{LZ-77}  $_{1977}$ & $d=3$ exp & 1.241(2)  &  0.6300(15)  & 0.031(4)  & 0.1100(45)  & 0.3250(15) &  0.79(3) \\
\cite{GZ-98} $_{1998}$ &$\epsilon$ exp  &1.2355(50)  &
0.6290(25) & 0.0360(50) & 0.113(7) & 0.3257(25)& 0.814(18) \\
\cite{GZ-98}  $_{1998}$ & $\epsilon$ exp bc & 1.2380(50)  &  0.6305(25) & 0.0365(50) &0.108(7)&  0.3265(15) & \\
\cite{YG-98} $_{1998}$ &$\epsilon$ exp  &1.242 & 0.632 & 0.035 & 0.104& 0.327 & 0.788 \\
\cite{NR-84}       $_{1984}$      & SFM  & 1.23(2)   &  0.626(9)   &
0.040(7) & 0.122(27)  &0.326(5) & 0.85(7) \\

\cite{Litim-02}       $_{2002}$      & CRG (LPA) &   &  0.6495   &
& 0.0515 & & 0.6557 \\
\cite{SW-99}       $_{1999}$      & CRG (1st DE) &   1.2322 &  0.6307   &
0.0467 &0.1079 & 0.3300 & \\
\cite{Comellas-98}       $_{1998}$      & CRG (1st DE)  &  1.218 &  0.622 &  0.042  & 0.134 & 0.324&  \\
\cite{Morris-97}   $_{1997}$     &  CRG (1st DE) &  1.203 &  0.618(14)
& 0.054 & 0.146(42)& 0.326 & 0.90(9) \\
\cite{BTW-96}       $_{1994}$      & CRG ILPA &   1.258 &  0.643   &  0.044 & 0.071 & 0.336 & \\
\cite{TW-94}       $_{1994}$      & CRG ILPA &   1.247 &  0.638   &  0.045 & 0.086 & 0.333 & \\
\cite{Morris-94-2}       $_{1994}$      & CRG LPA &   1.32 &  0.66 & 0 & 0.02 & 0.33 & 0.63  \\
\cite{Golner-86}       $_{1986}$      & CRG (1st DE)  &   &  0.617(8) &  0.024(7)  & 0.149(24) & &  \\
\hline
\end{tabular}
\end{center}
\end{table*}

We begin by reviewing the results obtained by 
employing HT expansion techniques, which appear to 
be the most precise ones.
In Table~\ref{expIsingthHT} we report those obtained
in the last two decades.
Older estimates are reviewed in Ref.~\cite{Adler-83}.

Refs.\ \cite{CPRV-99,CPRV-02} consider three specific improved Hamiltonians
on the simple cubic (sc) lattice,
see Sec. \ref{sec-2.3.2}:
the $\phi^4$-$\phi^6$ lattice model \reff{phi4-phi6} for 
$\lambda^* = 1.10(2)$, $\lambda_6^* = 0$
\cite{Hasenbusch-99}, and 
$\lambda^* = 1.90(4)$, $\lambda_6^* = 1$ \cite{CPRV-99};
the Blume-Capel model \reff{Blume-Capel} for 
$D^* = 0.641(8)$ \cite{Hasenbusch-99-h}.
For each improved model, the 25th-order HT expansions
of $\chi$ and $\mu_2$ were analyzed  using integral 
approximants of various orders and ratio methods. 
The comparison of the results obtained using these 
three improved Hamiltonians 
provides a strong check of the expected
reduction of systematic errors in the HT results
and an estimate of the residual errors due to the subleading
confluent corrections to scaling.
The estimates of the critical exponents obtained in such a way
are denoted by IHT in Table~\ref{expIsingthHT}.
The comparison of the results obtained from the analyses of the 
20th- and 25th-order series, cf. Refs.~\cite{CPRV-99}
and \cite{CPRV-02} respectively, shows that
the results are stable---within the quoted errors---with respect to the
number of terms of the series.
We also report (MC+IHT) a biased analysis of the
20th-series of the improved $\phi^4$ model
using the MC estimate of $\beta_c$, i.e.
$\beta_c=0.3750966(4)$ for $\lambda=1.10$ 
\cite{Hasenbusch-99}.\footnote{The analysis of the 25th-order series provides
the estimate $\beta_c=0.3750975(5)$, in reasonable agreement
with the MC result.}

Ref.~\cite{BC-02} reports results obtained by
analyzing 25th-order series for generic spin-$S$ (s-${n\over2}$) 
models on the simple cubic
(sc) and on the body-centered cubic (bcc) 
lattice, using a ratio method and fixing $\Delta$ (in most of the analyses
$\Delta = 0.504$ was used). 
The final estimates of the critical exponents were essentially
obtained from the results of the models with $S=1,{3\over2},2$ on the
bcc lattice, and in particular from the spin-$\case{3}{2}$ model, 
which, according
to the authors, provides the most stable results
with respect to the value of $\Delta$ chosen in the
analysis.\footnote{
We note that the spin-$\case{3}{2}$ model on the bcc
lattice is an almost improved model. Indeed,
let us consider the lattice Hamiltonian
\[
{\cal H} = {\cal H}_{3/2} + D \sum_i s_i^2 
\]
where ${\cal H}_{3/2}$ is the spin-$\case{3}{2}$ Hamiltonian,
$s_i=\case{3}{2},\case{1}{2},-\case{1}{2},-\case{3}{2}$
and $D$ is an irrelevant parameter.
We estimated  the value $D^*$ corresponding to 
an improved Hamiltonian using MC simulations and
FSS techniques. We found $D^*=-0.015(20)$,
showing that the spin-$\case{3}{2}$ model
is almost improved.
This explains the approximate independence on the choice of $\Delta$
observed in Ref.~\cite{BC-02}.}
They are consistent with the IHT results  of 
Refs. \cite{CPRV-99,CPRV-02}.
Refs. \cite{BC-00,BC-99,BC-97-2} present results obtained by
analyzing series for the spin-$\case{1}{2}$ (s-$\case{1}{2}$) 
model on the sc
and bcc lattices. They were essentially obtained 
using biased approximants, fixing $\beta_c$ and $\Delta$.
With increasing the order of the series,
the estimates show a trend towards the results obtained using
improved Hamiltonians.

The estimates of Refs.\ 
\cite{ND-81,CFN-82,GR-84,FC-85,NR-90} were obtained from the
analysis of 21st-order expansions
for two families of models, the Klauder (K) and the
double-Gaussian (DG) models on the bcc lattice, see Eqs. \reff{Klauder} 
and \reff{double-Gaussian}, which depend on an
irrelevant parameter $y$ and interpolate between the Gaussian model
and the  spin-1/2 Ising model. 
In Refs.~\cite{CFN-82,FC-85} the double expansion of $\chi$ in the inverse
temperature $\beta$ and the irrelevant parameter $y$ was analyzed employing
two-variable partial differential approximants, devised to reproduce the
expected scaling behavior in a neighborhood of $(y^*,\beta_c(y^*))$ in 
the $(y,\beta)$ plane.
The estimate of $\gamma$ of Ref.~\cite{FC-85} is significantly 
higher than the most recent estimates of Refs.~\cite{CPRV-02,CPRV-99,BC-02,BC-00}.
The same series were analyzed using a different 
method in Ref.~\cite{NR-90}: the estimate of $\gamma$ was lower and 
in agreement with the IHT result.
As pointed out in Ref.~\cite{NR-90}, the discrepancy is
strictly correlated with the estimate of $y^*$: the estimates of $\gamma$ 
increase with $y^*$, and thus the larger value of $\gamma$ of 
Ref.~\cite{FC-85} is due to the fact that a larger value of $y^*$ is used. 
The results of Refs.~\cite{Roskies-81,Zinn-Justin-81,FV-83} were obtained
by analyzing 21st-order expansions  for  spin-$S$ models on the bcc lattice,
computed by Nickel \cite{Nickel-82}.

The HT-expansion analyses usually focus on $\chi$ and $\mu_2$, or 
equivalently on $\xi^2=\mu_2/(2d\chi)$, 
and thus they provide direct estimates of $\gamma$ 
and $\nu$. The other exponents can be obtained using scaling 
relations. The specific-heat exponent $\alpha$ can be estimated
independently, although the results are not so precise as those 
obtained using the hyperscaling relation
$\alpha=2-3\nu$. One can obtain $\alpha$ from the analysis 
of the HT expansion of the specific heat \cite{BC-99,GE-94,BCGS-94}, 
and, on bipartite lattices, from the analysis 
of the magnetic susceptibility $\chi$ at the antiferromagnetic
singularity $\beta=-\beta_c$ \cite{FC-85,NR-90}.
In Table \ref{expIsingthHT}
we added a subscript $a$ to these latter estimates of $\alpha$. 
In particular, the precise estimate $\alpha=0.110(2)$ obtained
in Ref.~\cite{CPRV-02} provides  a stringent check of the
hyperscaling relation $\alpha+3\nu=2$. Indeed, 
using the estimate $\nu=0.63012(16)$, we obtain 
\begin{equation}
\alpha + 3\nu=2.000(2).
\end{equation}

Results obtained by analyzing the LT expansions
of the Ising model (see, e.g., Refs.\cite{SA-98,AT-95,GE-93,BCL-92}) are
consistent (with the exception of the results of Ref.~\cite{BCL-92}), 
although much less precise than, the HT results.
They are reported in Table \ref{expIsingthother}. There, we also 
report results obtained using 
the so-called cluster variation method (CVM) \cite{Kikuchi-51,Pelizzola-95},
and a generalization of the mean-field approach, the so-called
coherent-anomaly method (CAM) \cite{CAM-95}. 
Moreover, we show results obtained exploiting a Hamiltonian
approach (HA) \cite{Henkel-84,HJ-86,HHO-90,OHZ-91,PHO-93},
supplemented with finite-size-scaling (FSS) techniques,
<HT and LT expansions.

There are several MC determinations of the critical exponents. 
The most precise results have been obtained using the FSS methods 
described in Sec. \ref{sec-2.2.3},
see, e.g., Refs.
\cite{Hasenbusch-99-h,BST-99,Hasenbusch-99,BFMMPR-99,HPV-99,CMTC-99,%
HP-98,BLH-95,BK-93,Hasenbusch-93,FL-91,PV-02} and the 
results denoted by FSS in Table \ref{expIsingthMC}.
The results of Refs.~\cite{Hasenbusch-99,HPV-99} were obtained by
simulating an improved $\phi^4$ lattice Hamiltonian and an 
improved Blume-Capel model, see Sec. \ref{sec-2.3.2}. 
Ref.~\cite{Hasenbusch-99-h} reports the estimates
resulting from the combination of these results.
In Refs.~\cite{BST-99,BHHMS-96,BLH-95} the Ising model 
\reff{Ising-n-3n} with nearest-neighbor
($nn$) and third-neighbor ($3n$) interactions was
considered, using values of $y$ that reduce the scaling corrections.
The critical exponents have also been computed using 
the MC RG method presented in Sec. \ref{sec-2.2.2} (MCRG) 
\cite{Itakura-99,BHHMS-96,GT-96,BGHP-92,Blote-etal-89,PSWW-84},
by fitting the infinite-volume data 
to the expected scaling behavior (S) \cite{TB-96},  
from the FSS of the partition-function zeros in the complex-temperature 
plane \cite{IPZ-83}
determined from the density of states (FSS DS)
\cite{ADH-00,ABV-90,BSBCT-87,Marinari-84},
by studying the nonequilibrium relaxation (NER) \cite{IHOO-00,Ito-93} and
the short-time  critical dynamics (STCD) \cite{JSS-89,JMSZ-99}.
The MC results for Ising systems have been recently
reviewed in Ref.~\cite{BL-01}. The authors summarize
the available MC results for spin models 
proposing the following estimates
for the RG dimensions $y_t$, $y_h$, and $\omega$: 
$y_t = 1.588(2)$, $y_h = 2.482(2)$, and $\omega=0.83(4)$. Correspondingly, 
$\gamma = 1.2368(30)$, $\nu=0.6297(8)$, and $\beta=0.3262(13)$.

In Ref.~\cite{JV-00} a MC study of the Ising model on
three-dimensional lattices with connectivity disorder 
is reported: The results provide
evidence that the critical behavior on quenched Poissonian
random lattices (PRL), see, e.g., Ref.~\cite{Itzykson-83},
is identical to that on regular lattices.
Ref.~\cite{PV-02} presents results for the four-dimensional
SU(2) gauge theory at the deconfinement transition
that is expected to belong to the Ising 
universality class  \cite{PW-84}.

Numerical methods have also been applied to the study of the critical 
behavior of fluids, see 
Refs.~\cite{BV-98,Caillol-98,Panagiotopoulos-00,Litniewski-01,OFP-01} and 
references therein. Results obtained by MC and molecular-dynamics 
simulations are much less precise than those obtained in spin models,
because of the absence of efficient algorithms and of the lack of 
$\mathbb{Z}_2$-symmetry. Note however that, unlike spin models, 
fluid simulations allow 
to study the additional singularities that are present in 
systems without $\mathbb{Z}_2$-symmetry, e.g., the singularity of the 
diameter of the coexistence curve or the Yang-Yang anomaly, see, e.g.,
Ref. \cite{OFP-01}. We should also mention MC results for the 
restricted primitive model of electrolytes, where charged hard spheres
interact through Coulomb potential. Extensive FSS analyses
confirm that this system belongs to the Ising universality class
and give: $\gamma = 1.24(3)$, $\nu = 0.63(3)$ \cite{LFP-01};
$\nu = 0.66(3)$, $\beta/\nu \approx 0.52$ \cite{CLW-02}.

Finally, we should mention a numerical study of Ising ferrofluids 
\cite{NPR-98}. Such systems are expected to show Ising behavior 
with Fisher-renormalized critical exponents \cite{Fisher-68}
because of the presence of configurational annealed disorder.
Ref.~\cite{NPR-98} quotes $\beta/\nu = 0.54(2)$, $0.51(2)$, 
$\gamma/\nu = 1.931(8)$, 1.92(2) and $1/\nu_{\rm ren} = 1.47(4)$,
1.54(3) (different results correspond to different analyses), 
to be compared with $\beta/\nu = 0.5182(1)$, $\gamma/\nu = 1.9636(2)$, 
$1/\nu_{\rm ren} = (\alpha-1)/\nu = 1.4130(4)$ \cite{CPRV-02}. 
Some discrepances are observed, especially for $\nu_{\rm ren}$. 
This is not unexpected, since Fisher renormalization can usually be observed 
only very near to the critical point, see, e.g., Ref. \cite{FS-70}, 
mainly due to the presence of corrections of order $t^\alpha$
\cite{MF-01}. 

Let us turn to the results obtained in the FT approaches that 
are presented in Table \ref{expIsingthFT}.
In the fixed-dimension approach, the perturbative series of 
Refs. \cite{BNGM-77,MN-91}
were reanalyzed in Ref.~\cite{GZ-98},
using the resummation method of Ref.~\cite{LZ-77}
(see also Ref.~\cite{Zinn-Justin-00}).
Comparing with the HT and MC results, 
we note that there are small discrepancies for $\gamma$, $\eta$, and $\omega$.
These deviations are probably 
due to the nonanalyticity of the RG functions for $g=g^*$ 
that we discussed in Sec. \ref{sec-2.4.3}. 
Similar results were obtained in Refs.~\cite{JK-00,Kleinert-98}, 
using different methods of analysis, but still neglecting the 
confluent singularities at the infrared-stable fixed point. 
The errors reported there seem to be rather  optimistic,
especially if compared with those obtained in Ref.~\cite{GZ-98}.
The analysis of Ref.~\cite{MN-91} allowed for a more general
nonanalytic behavior of the $\beta$-function. In Table
\ref{expIsingthFT}, we quote two errors for the results of Ref.\ 
\cite{MN-91}: the first one (in parentheses) 
is the resummation error, and the second one (in braces) 
takes into account the uncertainty of $g^*$, which is estimated to
be approximately 1\%. To estimate the second error we used the results of
Ref.~\cite{GZ-98} where the dependence of the exponents on $g^*$ is 
given. Concerning the $\epsilon$ expansion,
we report  estimates obtained by
standard analyses and constrained
analyses \cite{LZ-87} (denoted by ``bc'') that make use of  
the exact two-dimensional values, employing the method discussed in 
Sec. \ref{sec-2.4.3}.  Ref.~\cite{YG-98} analyzes
the $\epsilon$ series using a method based on self-similar 
exponential approximants.

Other estimates of the critical exponents have been 
obtained by nonperturbative FT methods based on 
approximate solutions of continuous RG equations (CRG),
see Sec.~\ref{CRGth}.
They are less precise than the above-presented methods, 
although much work has been
dedicated to their improvement (see, e.g., the recent
reviews \cite{BTW-99,BB-00} and references therein).  
Table~\ref{expIsingthFT} reports some results obtained by CRG methods.
This is not a complete list, but it should give an overview
of the state of art of this approach. 
Additional CRG results are reported and compared in Refs.~\cite{BTW-99,BB-00}.
There, one can also find a detailed discussion of the key
ingredients of the method, which are essentially the choice
of the continuous RG equation, of the infrared regulator,
and of the approximation scheme such as derivative expansion,
field expansion, etc.
The CRG estimates apparently improve (in the sense that 
they get closer to the more precise estimates obtained by other methods) 
when better truncations are considered, see, e.g., 
Ref.\cite{BTW-99,BB-00},
and in particular passing from the lowest to the first order of the DE.

The results of Ref.~\cite{NR-84} were obtained using a similar
approach, the so-called 
scaling-field method (SFM). 
The results for the critical
exponents $\gamma$, $\nu$, and $\eta$ 
(see Table \ref{expIsingthFT}) are considerably less
precise than those based on perturbative approaches. 
But it is interesting to note that 
the authors were able to estimate additional subleading 
exponents, such as the  next-to-leading irrelevant exponent, obtaining
$\omega_2=1.67(11)$.

Many systems undergoing phase transitions in the Ising universality
class do not have the $\mathbb{Z}_2$-symmetry that is present in the standard
Ising model. In these cases the $\mathbb{Z}_2$-symmetry is 
effectively realized only at the
critical point. Asymmetry gives rise to scaling corrections only:
Some of them are due to the mixing of the thermodynamic variables, while 
other are due to a new class of $\mathbb{Z}_2$-odd operators.
The leading one is characterized by a new critical exponent $\omega_A$
\cite{NZ-81,Nicoll-81,ZZ-82,NR-84}. The exponent
$\omega_A$ has been computed to $O(\epsilon^3)$ in the framework of
the $\epsilon$ expansion \cite{NZ-81,Nicoll-81,ZZ-82},
using the scaling-field method \cite{NR-84}, and the LPA
in the framework of the Wegner-Houghton equation \cite{Tsypin-01}. 
These calculations suggest
a rather large value for $\omega_A$, i.e. $\omega_A\gtapprox 1.5$. 
For example, Ref.~\cite{NR-84} reports $\Delta_A\equiv
\omega_A\nu=1.5(3)$, and Ref.~\cite{Tsypin-01}
gives $\omega_A = 1.691$.
In many experimental papers, a value $\Delta_A\approx 1.3$ is often
assumed, see, e.g., Refs. \cite{GAS-01,KS-99}.
These results show that contributions due to the antisymmetric operators
are strongly suppressed, even with respect to 
the leading $\mathbb{Z}_2$-symmetric scaling corrections, 
that scale with $\Delta\approx 0.5$.

Finally, we mention  the results obtained for the universal critical exponent 
$\omega_{\rm NR}$ describing how the
spatial anisotropy, which is present in physical systems with cubic 
symmetry such as uniaxial magnets, vanishes when approaching the
rotationally-invariant fixed point \cite{CPRV-98}, see Sec. \ref{sec-1.6}.  
The most accurate estimate of $\omega_{\rm NR}$ has been  obtained by analyzing
the IHT expansions of  the first non-spherical moments of the two-point 
function of the order parameter \cite{CPRV-99},
obtaining\footnote{We signal the presence of a misprint in Ref.~\cite{CPRV-99}
concerning the estimate of $\omega_{\rm NR}$, which is there called $\rho$.}  
$\omega_{\rm NR} = 2.0208(12)$,
which is very close to the Gaussian value $\omega_{\rm NR}=2$.
FT results \cite{CPRV-99,CPRV-98} are consistent,
although considerably less precise.

In conclusion, taking into account the sources of systematic errors 
of the various methods, we believe that all the results 
presented in this section, and especially those obtained from HT and MC methods,
can be summarized by the following estimates:
\begin{eqnarray}
\gamma &=& 1.2372(5) , \nonumber \\
\nu &=& 0.6301(4) ,\nonumber \\
\eta &=& 0.0364(5) ,\nonumber \\
\alpha &=& 0.110(1) ,\nonumber \\
\beta &=& 0.3265(3) ,\nonumber \\
\delta &=& 4.789(2) ,\nonumber \\
\omega &=& 0.84(4). \label{ourest} 
\end{eqnarray}
In our opinion, 
these numbers and their errors
should represent quite safe estimates of the critical exponents.

\subsubsection{Experimental results}
\label{experimentsIsing}

Many experimental results can be found in the literature. 
For recent reviews, see, e.g., 
Refs.~\cite{Anisimov-book,PHA-91,BLH-95,Belanger-00}.
In Table \ref{expIsingsper} we report some experimental results
for the critical exponents, most of them published after 1990. It
is not a complete list of the published results, but it
may be useful to get an overview of the experimental state of the art.
The results for the various systems 
substantially agree, although, looking in more detail,
one may find small discrepancies.
The agreement with the theoretical results supports 
the RG theory of critical phenomena,
although experimental results are substantially less accurate than
the theoretical ones. 

In Refs. \cite{KDYNK-97,KKD-97,KKD-98} polydisperse
polymeric solutions were studied. While monodisperse solutions behave 
as an ordinary binary mixture, polydispersion causes a Fisher 
renormalization \cite{Fisher-68} of the exponents. The results 
reported in Table \ref{expIsingsper} have been obtained using 
$\alpha = 0.1096(5)$ \cite{CPRV-02}. Fisher renormalization is also
observed in the results of Ref. \cite{CPR-00-01} that studied the 
phase separation of a colloidal dispersion in the presence of soluble 
polymers, in the results for dilute polymer blends\footnote{In the 
absence of dilution, standard critical exponents are expected, see,
e.g., Ref. \cite{SKK-96}. However, in a recent experiment 
\cite{SFW-02}, Fisher renormalized exponents were observed also in this 
case. The reason is unclear.} of Ref. \cite{YHNDH-93}, and in the 
results of Ref. \cite{MW-99} for ternary mixtures. We should also mention
the results of Ref. \cite{MCARO-93} that observed the expected doubling of the 
exponents at a double critical point in a liquid mixture with upper and 
lower consolute critical point, and the result of Ref. \cite{RBAMB-01}
$\gamma/\beta = 3.83(11)$. Finally, Ref. \cite{AKBQWW-92} 
measured the surface-tension exponent $\mu$, finding $\mu = 1.27(1)$,
in good agreement with the hyperscaling prediction $\mu = 2\nu$.

\begin{table*}
\caption{
Experimental estimates of the critical exponents for the 
three-dimensional Ising universality class. lv\ denotes the liquid-vapor 
transition in simple fluids and mx\ the mixing transition in multicomponent 
fluid mixtures and in complex fluids;
ms\ refers to a uniaxial magnetic system, mi\  to a micellar system, and Cb\ to 
the mixing transition in Coulombic systems. The results indicated by 
$({}^\dag)$ have been
obtained from Fisher-renormalized exponents.
}
\footnotesize
%\hspace*{-2.7cm}    % Move table leftwards, so it doesn't run off the right
%\tabcolsep 4pt        % Less than the usual 6pt
%\doublerulesep 1.5pt  % Less than the usual 2pt
\begin{center}
\begin{tabular}{crlllll}
\hline
\multicolumn{1}{c}{}& 
\multicolumn{1}{c}{Ref.}& 
\multicolumn{1}{c}{$\gamma$}& 
\multicolumn{1}{c}{$\nu$}& 
\multicolumn{1}{c}{$\eta$}&
\multicolumn{1}{c}{$\alpha$}& 
\multicolumn{1}{c}{$\beta$} \\
\hline  
lv 
& \cite{SNFR-00} $_{2000}$ & 1.14(5) & 0.62(3) & & & \\
& \cite{HS-99} $_{1999}$  &  & & & 0.1105$^{+0.0250}_{-0.0270}$ &  \\
&\cite{DLMFL-98}  $_{1998}$ & & & 0.042(6) & & \\ 
&\cite{KASW-95}  $_{1995}$  &  & & & & 0.341(2)  \\
&\cite{APB-94}   $_{1994}$  &  & & & 0.111(1) & 0.324(2)  \\
&\cite{SN-93}   $_{1993}$    &  & & & 0.1075(54)  & \\
%%%%%%  &\cite{Edwards-84}  $_{1984}$  &  & & & 0.1084(23) & \\
& \cite{PC-84}   $_{1984}$  &  1.233(10) & & & &0.327(2) \\
\hline
mx
& \cite{RODR-02}  $_{2002}$   & & & & 0.12(1) & \\
& \cite{NGMJ-01}  $_{2001}$   &  & & & 0.111(2) &  \\
& \cite{OJ-01}  $_{2001}$   &  & & & 0.106(26) &  \\
& \cite{CPR-00-01} $_{2000}$ & 1.236(9)$^\dag$ & 0.631(9)$^\dag$ &&& 0.330(23)$^\dag$ \\
& \cite{MW-99}  $_{1999}$   & 1.32(6)$^\dag$ & 0.70(4)$^\dag$ & 0.058(16) & & \\
& \cite{MW-99}  $_{1999}$   & 1.244(42) & 0.636(31) & 0.045(11) & & \\
& \cite{RJ-98}  $_{1998}$   &  & & & 0.104(11) &  \\
& \cite{KKD-98} $_{1998}$   & 1.22(3)$^\dag$ & 0.62(2)$^\dag$ &&& \\
& \cite{KKD-97} $_{1997}$   & 1.23(4)$^\dag$ & 0.64(2)$^\dag$ &&& \\
& \cite{KDYNK-97} $_{1997}$ & & & & & 0.335(5),0.323(4)$^\dag$ \\
& \cite{SKK-96} $_{1996}$   & 1.25(2) & 0.63(2) & 0.038(3) & & 0.327(3) \\
& \cite{Flewelling-etal-96}  $_{1996}$   &  & & & 0.107(6) &  \\
& \cite{HBT-96}  $_{1996}$ &  & & & 0.111(2) & \\
& \cite{HGBT-95} $_{1995}$   &  & & & 0.103(3), 0.113(3) & \\
& \cite{Wiegand-etal-94}  $_{1994}$  &  & 0.621(3) &&& \\
& \cite{KMTO-94} $_{1994}$ &  1.09(3) & & & & \\
&\cite{AnSh-94}  $_{1994}$      &  & & & & 0.324(5)  \\
&\cite{AnSh-94}  $_{1994}$      &  & & & & 0.329(2)  \\
&\cite{ASWZ-94}  $_{1994}$  &  & & & & 0.329(4)  \\
&\cite{ASWZ-94}  $_{1994}$  &  & & & & 0.333(2)  \\
&\cite{YHNDH-93} $_{1993}$   &  & 0.60(2)$^\dag$  & & & \\
&\cite{SWW-93}   $_{1993}$   &  & 0.610(6) & & & \\
&\cite{AKBQWW-92} $_{1992}$  & 1.23(3) & 0.631(1) & & & \\
&\cite{WGW-92}   $_{1992}$   &  & & &  0.105(8) &  \\
&\cite{BWW-92}   $_{1992}$   &  & &  &  &  0.336(30) \\  
&\cite{DLC-89}   $_{1989}$   &  1.228(39) & 0.628(8) & 0.0300(15) & & \\
& \cite{Jacobs-86}  $_{1986}$  &  1.26(5) & 0.64(2) & & & \\
&\cite{HKKK-85}  $_{1985}$  &  1.24(1) & 0.606(18) & & 0.077(44) & 0.319(14) \\
\hline
Cb
& \cite{BJ-01}     $_{2001}$ & & & & & 0.328(10) \\
& \cite{WBLSKS-98} $_{1998}$ & & 0.641(3) & & & 0.34(1) \\
\hline
mi
&\cite{Shimofure-etal-99}  $_{1999}$    &   1.26(5) & 0.63(2) & & & 0.329(3) \\
&\cite{LBW-97}  $_{1997}$    & 1.242(4) & 0.642(10) & & & \\
&\cite{SBW-94}  $_{1994}$    &   1.216(13) & 0.623(13) & 0.039(4) & & \\
&\cite{ZBW-94}  $_{1994}$    &   1.237(7) & 0.630(12) & & & \\
& \cite{AB-93} $_{1993}$   &   & & & & 0.34(8) \\
  & \cite{AB-93-2} $_{1993}$  &   1.18(3) & 0.60(2) & & & \\
&\cite{SW-92} $_{1992}$      &   1.17(11) & 0.65(4)   & & & \\
&\cite{HKMKK-91} $_{1991}$  &   1.25(2) & 0.63(1)    & & & \\
\hline
ms 
&\cite{Kats-etal-01}  $_{2001}$ & 1.14(7) & & & & 0.34(2) \\
&\cite{MMB-95b} $_{1995}$    &  & & & 0.11(3) & \\
&\cite{MMB-95}  $_{1995}$    &  & & & 0.11(3) &  \\
&\cite{MMZPSD-94} $_{1994}$    &  & & & 0.10(2) & \\
&\cite{MDN-94}  $_{1994}$    &  & & & & 0.325(2) \\
&\cite{SAASCE-93} $_{1993}$ & & & & 0.11(3) &  \\
&\cite{SPKT-93}  $_{1993}$  &   1.25(2) & & & & 0.315(15)  \\
&\cite{BY-87}  $_{1987}$      &   1.25(2) & 0.64(1)  & & & \\
&\cite{BNKJLB-83} $_{1983}$ && & & 0.110(5) & 0.331(6)  \\\hline
\end{tabular}
\end{center}
\label{expIsingsper}
\end{table*}

\subsection{The zero-momentum four-point coupling constant}
\label{g4rev}

The zero-momentum four-point coupling constant $g_4$ defined in 
Eq.~\reff{grdef} plays an important role in the
FT perturbative expansion at fixed dimension,
see Sec.~\ref{sec-2.4.1}.
In this approach, any
universal quantity is obtained from a perturbative expansion  
in powers of $g\equiv g_4$ computed at $g=g^*\equiv g_4^+$.

In Table~\ref{g4ising} we review the estimates of $g_4^+$
obtained by exploiting various approaches. 
The most precise HT estimates have been determined following 
essentially two strategies to handle the problem of confluent
corrections. One, used in Refs. \cite{CPRV-02,CPRV-99},
is based on the analysis of HT expansions for improved models.
The other one, used in Refs.~\cite{BC-02,BC-98,PV-gr-98,BC-97-1}, 
employs appropriate biased approximants (fixing $\beta_c$ and $\Delta$)
to reduce the effect of the confluent
singularities.  Refs.~\cite{CPRV-02,CPRV-99}  analyzed the 
HT expansion of $g_4$ (with $\chi_4$ computed to  21st order)
for three improved models: 
the $\phi^4$-$\phi^6$ model \reff{phi4-phi6} and the Blume-Capel model
\reff{Blume-Capel}, see Sec.~\ref{sec-2.3.2}.
The small difference between the results of Refs.~\cite{CPRV-02} and
\cite{CPRV-99} was mainly due to the different analyses employed,
and much less to the fact 
that the series used in Ref.~\cite{CPRV-99} were shorter.
Indeed, the more robust analysis of Ref.~\cite{CPRV-02} applied
to the 18th-order series of Ref.~\cite{CPRV-99} gives $g_4^+=23.54(4)$.
Ref.~\cite{BC-02} analyzed the series of $g_4$ (using $\chi_4$ to 23rd
order) for the spin-$S$ models on the sc and bcc lattices;
its final estimate was essentially given by the results of 
the spin-$\case{3}{2}$ model on the bcc lattice.
All other HT results that take
into account the leading scaling corrections are in substantial agreement.
The authors of Ref.~\cite{BB-92} performed a dimensional expansion of the
Green functions around $d=0$ ($d$-exp.). The analysis of these series allowed
them to obtain a quite precise estimate of $g_4^+$ in three dimensions.  

Ref.~\cite{Kim-99} reports the results of a MC simulation, in which
a FSS technique was used to obtain
$g_4$ for large correlation lengths. An estimate of $g^+_4$---in  
agreement with the IHT result---was  
obtained by properly taking into account the 
leading scaling correction.
Ref.\ \cite{BFMM-98} reports MC results obtained from simulations 
of the $\phi^4$ lattice Hamiltonian \reff{latticephi4} with $\lambda=1$, 
which is close to the optimal value
$\lambda^*\approx 1.10$. In Ref.\ \cite{BFMM-98} no final estimate is 
reported.  
The value we report in Table \ref{g4ising} is the result quoted in 
Ref.~\cite{CPRV-99}, obtained by fitting  
their data.
The result of Ref.\ 
\cite{Tsypin-94} was obtained by studying the probability
distribution of the average magnetization (a similar approach 
was also used in Ref.~\cite{RLJ-98}). The other
estimates were obtained from fits to data in the neighborhood 
of $\beta_c$. The MC estimates of Refs.\ \cite{BK-96,KL-96} were
larger because scaling corrections were neglected, as
shown in Ref.\ \cite{PV-gr-98}.  

FT estimates are substantially consistent.
In the $d$=3 fixed-dimension approach, $g_4^+$ is determined from the zero 
of the corresponding Callan-Symanzik $\beta$-function, 
obtained by resumming its perturbative six-loop series \cite{BNGM-77}.  
The results of Refs.\cite{GZ-98,MN-91,LZ-77}
are in substantial agreement with the HT estimates.
The $\epsilon$-expansion result of Refs.\cite{PV-00,PV-gr-98} was obtained
from a constrained analysis---see Sec. \ref{sec-2.4.3}---of 
the $O(\epsilon^4)$ series using
the known values of $g_4^+$ for $d=0,1,2$.
In Table \ref{g4ising} we also report estimates obtained
using the nonperturbative continuous RG (CRG) approach
\cite{TW-94,Morris-97,SW-99}.
Other estimates of $g_4^+$, which do not appear in Table~\ref{g4ising},
can be found in Refs. 
\cite{Baker-76,NS-79,BK-79,BCGRS-81,FSW-82,FB-82,%
Wheater-84,Weston-89,KP-93,BK-95}.

\begin{table*}
\caption{
Estimates of $g_4^+$ for the three-dimensional Ising universality class.
}
\label{g4ising}
\footnotesize
%\hspace*{-2.7cm}    % Move table leftwards, so it doesn't run off the right
%\tabcolsep 4pt        % Less than the usual 6pt
%\doublerulesep 1.5pt  % Less than the usual 2pt
\begin{center}
\begin{tabular}{llllll}
\hline
\multicolumn{1}{c}{HT}& 
\multicolumn{1}{c}{MC}& 
\multicolumn{1}{c}{$\epsilon$ exp}& 
\multicolumn{1}{c}{$d=3$ exp}&
\multicolumn{1}{c}{$d$ exp}& 
\multicolumn{1}{c}{CRG} \\
\hline  
23.56(2) \cite{CPRV-02} & 23.6(2) \cite{Kim-99} & 23.6(2) \cite{PV-00} & 23.64(7) \cite{GZ-98} 
   & 23.66(24) \cite{BB-92}   & 24.3 \cite{SW-99} \\
  23.52(5) \cite{BC-02} & 23.4(2) \cite{PV-gr-98,BFMM-98} & 23.33
 \cite{GZ-98} & 23.46(23) \cite{MN-91}  &  & 21(4) \cite{Morris-97}\\
  23.49(4) \cite{CPRV-99} & 23.3(5) \cite{Tsypin-94} &  & 23.71 \cite{SOUK-99}   & & 28.9 \cite{TW-94}  \\
  23.57(10)  \cite{BC-98} & 25.0(5) \cite{BK-96} & &  23.72(8) \cite{LZ-77}  & &   \\
  23.55(15) \cite{PV-gr-98}  & 24.5(2) \cite{KL-96} & &   &  &   \\
  23.69(10) \cite{BC-97-1}  & & &                       &  &   \\
  24.45(15) \cite{ZLF-96} & &  & & &  \\
  23.7(1.5) \cite{Reisz-95} & &  & & &  \\
\hline
\end{tabular}
\end{center}
\end{table*}

\subsection{The critical equation of state}
\label{eqstIsing}

The equation of state relates the magnetic field $H$, the magnetization $M$,
and the reduced temperature $t\equiv (T-T_c)/T_c$.
In the lattice gas, the explicit mapping shows that 
the variables playing the role of $H$ and $M$ are $\Delta\mu\equiv\mu-\mu_c$ 
and $\Delta\rho\equiv\rho-\rho_c$ respectively, 
where $\mu$ is the chemical potential and $\rho$ the density, and the 
subscript $c$ indicates the values at the critical point. 
However, the lattice-gas model has an additional $\mathbb{Z}_2$ symmetry
that is not present in real fluids. In this case, $H$ and $M$ are 
usually assumed \cite{RM-73,LKG-81} to be combinations of 
$\Delta\mu$ and $\Delta\rho$,
i.e. 
\be
M = \alpha_M \Delta\mu + \alpha_\rho \Delta\rho, \qquad\qquad
H = \beta_M \Delta\mu + \beta_\rho \Delta\rho, 
\label{scalingfields-fluids}
\ee
where $\alpha_M$, $\alpha_\rho$, $\beta_M$, and $\beta_\rho$ are 
nonuniversal constants. Such an Ansatz has been recently challenged in Ref. 
\cite{FO-00}. It was suggested that also a pressure term proportional to
$\Delta p\equiv p-p_c$ should be added in
Eq. \reff{scalingfields-fluids} 
and some evidence
was presented for this additional mixing \cite{OFU-00}
(see also the critique of Ref. \cite{KWASK-02}). Similar mixings are expected
in mixtures.

\subsubsection{Small-magnetization expansion of the Helmholtz 
free energy in the HT phase}
\label{HTEFising}

As discussed in Sec.~\ref{sec-1.5.2}, for small values of $M$ and $t>0$, 
the scaling function $A_1(z)$, which corresponds to the Helmholtz 
free energy, and the equation-of-state scaling function $F(z)$ can be parametrized 
in terms of the universal constants $r_{2n}$, see Eqs. \reff{Fzdef} 
and \reff{r2jgreen}.

Table \ref{EFising} reports the available estimates of $r_6$,
$r_8$, and $r_{10}$.  
The results of Ref.\cite{CPRV-02} were obtained by
analyzing the IHT expansions of $r_6$, $r_8$, and $r_{10}$ 
to order 19, 17, and 15 respectively for three improved lattice models, 
the $\phi^4$-$\phi^6$ model \reff{phi4-phi6} and the 
improved Blume-Capel model \reff{Blume-Capel}. 
Additional results were obtained from HT expansions
\cite{BC-97-1,ZLF-96,Reisz-95} and MC simulations
\cite{Tsypin-94,KL-96,Kim-99} of the Ising model. 
The results of Ref. \cite{LF-96} were obtained from the analysis
of 14th-order virial expansions for a binary fluid model consisting 
of Gaussian molecules.
 The MC results
do not agree with the results of other approaches, especially those of
Refs.\ \cite{Kim-99,KL-96}, where FSS
techniques were employed.  But one should consider the difficulty of such
calculations due to the subtractions that must be performed 
in order to compute the irreducible correlation functions.  
In the framework of the $\epsilon$ expansion, 
the $O(\epsilon^3)$ series of $r_{2n}$ were derived from the 
$O(\epsilon^3)$ expansion of the equation of state \cite{WZ-74,NA-85,BWW-72}.
Ref.~\cite{PV-ef-98} performed
a constrained analysis---the method is described in Sec. 
\ref{sec-2.4.3}---exploiting the known values of $r_{2n}$ for $d=0,1,2$. 
In the framework of the fixed-dimension expansion, 
Refs.\cite{GZ-98,GZ-97} analyzed the five-loop series computed
in Refs. \cite{BBMN-87,HD-92}.
Rather good  estimates of $r_{2n}$ were also
obtained in Ref.\ \cite{Morris-97} (see also Ref.\ \cite{TW-94})  using 
the CRG method, although the estimate of $g_4^+$ by the same method is not equally
good. CRG methods seem to be  quite effective for the determination of 
zero-momentum quantities such as $r_{2n}$, but are imprecise 
for quantities that involve derivatives of correlation functions,
as is the case for $g_4^+$.
This is not unexpected since the Ansatz used to
solve the RG equation is based on a derivative expansion.

\begin{table*}
\caption{
Estimates of $r_6$, $r_8$, and $r_{10}$ for the three-dimensional 
Ising universality class. 
We also mention the estimate $r_{10}=-10.6(1.8)$ 
obtained in Refs.~\protect\cite{CPRV-02,CPRV-99} 
by studying the equation of state.
}
\label{EFising}
\footnotesize
\begin{center}
\begin{tabular}{llllll}
\hline
\multicolumn{1}{c}{}& 
\multicolumn{1}{c}{HT}& 
\multicolumn{1}{c}{$\epsilon$ exp}& 
\multicolumn{1}{c}{$d=3$ exp}&
\multicolumn{1}{c}{CRG} &
\multicolumn{1}{c}{MC} \\
\hline  
$r_6$ &  2.056(5)  \cite{CPRV-02} & 2.058(11) \cite{PV-ef-98} & 2.053(8) \cite{GZ-98} 
&  2.064(36) \cite{Morris-97} & 2.72(23) \cite{Tsypin-94} \\
&   1.99(6) \cite{BC-97-1} & 2.12(12) \cite{GZ-98} & 2.060
\cite{SOUK-99} &   1.92 \cite{TW-94} & 3.37(11) \cite{Kim-99} \\
&   2.157(18) \cite{ZLF-96} &                         &                 &        & 3.26(26)~\cite{KL-96}  \\
&   2.25(9) \cite{LF-96} & & & &    \\
&   2.5(5) \cite{Reisz-95} & & &  &   \\
$r_8$ & 2.3(1) \cite{CPRV-02}  & 2.48(28) \cite{PV-ef-98} & 2.47(25) \cite{GZ-98} 
& 2.47(5) \cite{Morris-97} & \\
&  2.7(4) \cite{BC-97-1} & 2.42(30) \cite{GZ-98} & &   2.18
\cite{TW-94} & \\
$r_{10}$ & 
$-$13(4) \cite{CPRV-99}& $-$20(15) \cite{PV-ef-98} & $-$25(18) \cite{GZ-98} 
&   $-$18(4) \cite{Morris-97} & \\
&  $-$4(2) \cite{BC-97-1} &  $-$12.0(1.1) \cite{GZ-98} & & &  \\
\hline
\end{tabular}
\end{center}
\end{table*}

\subsubsection{Approximate parametric representa\-tions of the 
equation of state: The general formalism}
\label{eqstising}

In order to obtain
approximate representations of the equation of state, it is convenient to use 
the parametric model described in Sec. \ref{sec-1.5.6}, i.e. to rewrite 
$H$, $t$, and $M$ in terms of the two variables $\theta$ and $R$,
see Eq. \reff{parrepg}.
The advantage in using parametric representations is that all the 
analytic properties of the equation of state are automatically 
satisfied if $h(\theta)$ and $m(\theta)$ are analytic and satisfy 
a few simple constraints: 
(a) $h(\theta)>0$, $m(\theta)>0$, $Y(\theta)\not = 0$
for $0<\theta<\theta_0$; 
(b) $m(\theta_0) > 0$; 
(c) $\theta_0 > 1$.
Here $\theta_0$ is the positive zero of 
$h(\theta)$ that is nearest to the origin and $Y(\theta)$ is defined in 
Eq. \reff{Yfunc}. 

In general, in order to obtain an approximation of the equation of state, 
one can proceed as follows (see, e.g., Refs. 
\cite{BHK-75,HAHS-76,SLS-78,FZU-99}):
\begin{enumerate}
\item[(a)] 
One chooses some parametrization of $h(\theta)$ and $m(\theta)$ 
depending on $k$ parameters such that $h(\theta)$ and $m(\theta)$ are 
odd and $h(\theta) = \theta + O(\theta^3)$ and 
$m(\theta) = \theta + O(\theta^3)$ for $\theta\to 0$.
\item[(b)] One chooses $\tilde{k}\ge k$ universal quantities that 
can be derived from
the equation of state and that are known independently, for instance 
from a MC simulation, from the analysis of HT and/or LT series, 
or from experiments. Then, one uses them to determine the 
$k$ parameters defined in (a).
\item[(c)] The scale factors $m_0$ and $h_0$ are determined 
by requiring the equation of state 
to reproduce two nonuniversal amplitudes.
\end{enumerate}
For the functions $h(\theta)$ and $m(\theta)$, polynomials 
are often used. There are many reasons for this choice. First, 
this choice makes the expressions simple and 
analytic calculations easy. 
Moreover, the simplest representation, the so-called 
``linear" model, is already a good approximation 
\cite{SLH-69}. Such a model is defined by 
\begin{eqnarray} 
     m(\theta) = \theta, \qquad
     h(\theta) = \theta + h_3 \theta^3.
\label{linmod}
\end{eqnarray} 
The value of $h_3$ can be computed by considering a universal 
amplitude ratio. Ref. \cite{SLH-69} considered $U_2 \equiv C^+/C^-$, 
and observed that, for all acceptable values of $h_3$, the linear model gave
values of  $U_2$ that were larger than the HT/LT estimates. Therefore,
the  best approximation corresponds to setting $h_3=\bar{h}_3$, 
where $\bar{h}_3$ is the value of $h_3$ that
minimizes $U_2$, i.e.
\be
    \bar{h}_3 = {\gamma (1 - 2 \beta)\over \gamma - 2 \beta},
\label{h3-linearmodel}
\ee
which is the solution of the equation
\be 
  \left. {d U_2\over d h_3}\right|_{h_3 = \bar{h}_3} = 0.
\label{stazionarieta-linearmodel}
\ee
Later, Ref. \cite{WZ-74} showed that 
the stationarity condition ${d R/d h_3}|_{h_3 = \bar{h}_3}
= 0$ is satisfied for any invariant ratio $R$, where in the equation 
one uses the linear-model expression for $R$ as a function of $h_3$.
Numerically, using the results for the critical 
exponents reported in Sec. \ref{critexp}, the choice $h_3 = \bar{h}_3$
gives $U_2 \approx  4.83$, 
which is in relatively good agreement with the most accurate MC estimate 
$U_2 =  4.75(3)$ \cite{CH-97}. Thus, the linear parametric model with 
$h_3 = \bar{h}_3$ (sometimes called ``restricted" linear model)
is already a good zeroth-order approximation. Then, one may think 
that higher-order polynomials provide better approximations. 

A second argument in favor of polynomial representations 
is provided by the $\epsilon$ expansion. Setting
\begin{eqnarray} 
     m(\theta) = \theta, \qquad
     h(\theta) = \theta + \sum_{n=1}^k h_{2n+1} \theta^{2n+1},
\label{parametric-2}
\end{eqnarray} 
one can prove \cite{WZ-74,CPRV-99} that, for each 
$k$, one can fix the coefficients of the polynomial $h(\theta)$ 
so that the represention is exact up to order $\epsilon^{k+2}$. 

We mention that
alternative nonpolynomial representations have been introduced in the 
literature. 
Motivated by the desire of extending the Helmholtz free energy 
into the unstable two-phase region below the critical temperature, 
Refs. \cite{FU-90,FZ-98,FZU-99} considered trigonometric representations. 
They will be discussed in Sec. \ref{trir}.

Let us now discuss how to determine the parameters appearing in 
$m(\theta)$ and $h(\theta)$. If $\tilde{k} = k$, the number of unknowns 
is equal to the number of conditions and thus we can fix the 
$k$ parameters by requiring the approximate equation of state 
to reproduce these values. Since the equations are nonlinear, 
this is not always possible,
as in the case of the linear model (\ref{linmod}).
In these
cases, one may determine the values of the parameters that give 
the least discrepancy. 
One may also consider the case
$\tilde{k}> k$. This may be convenient if the input data 
have large errors. The $k$ parameters may be fixed by means 
of a standard fitting procedure. In
Refs. \cite{GZ-97,CPRV-99,CHPV-00-2,CPRV-02} it was proposed 
to consider $k = \tilde{k} + 1$. 
In this case, one must specify an additional condition to completely
determine the parametric functions.  In Ref. \cite{GZ-97} it was proposed
to use the parametric representation \reff{parametric-2}, 
fixing the $k$ parameter $h_3$,..., $h_{2k+1}$ so that 
$|h_{2k+1}|$ is as small as possible. In Refs. \cite{CPRV-99,CHPV-00-2,CPRV-02}
a variational approach was used; it will be described below.

In order to fix the parametric functions one must choose several 
zero-momentum universal ratios. 
One possibility \cite{GZ-97} consists in matching
the small-magnetization expansion of the free energy in the HT phase,
i.e. the coefficients $r_{2n}$, cf. Eq.~(\ref{Fzdef}),
which can be determined
either by FT or HT methods, cf. Sec. \ref{HTEFising}. 
Starting from Eq. \reff{parametric-2} and 
requiring the approximate parametric representation 
to give the correct $(k-1)$ universal ratios $r_6$, $r_8$, 
$\ldots$, $r_{2k+2}$, one finds the relations
\begin{equation}
   h_{2n+1} = \sum_{m=0}^n c_{nm} 6^m (h_3 + \gamma)^m {r_{2m+2}\over (2m+1)!},
\label{hcoeff}
\end{equation}
where
\begin{equation}
   c_{nm} = {1\over (n-m)!} \prod_{k=1}^{n-m} 
   (2\beta m - \gamma + k - 1),
\end{equation}
and $r_2 = r_4 = 1$. 
Moreover, by requiring that $F(z)=z + {1\over 6} z^3 + ...$,
one obtains 
\begin{equation}
   \rho^2 = 6 (h_3 + \gamma)
\label{h3intermsofrho}
\end{equation}
for the parameter $\rho$ defined in Eq. \reff{eq:1.132}.
In Ref. \cite{GZ-97} the parameter $h_3$, or equivalently $\rho$, 
which is left undetermined, was fixed
by minimizing $|h_{2k+1}|$. 

The same polynomial approximation scheme was considered in
Refs. \cite{CPRV-99,CPRV-02}, but, at variance with Ref. \cite{GZ-97}, 
a variational approach was used to fix $h_3$. 
As discussed in Sec.~\ref{sec-1.5.6},
in the parametrization (\ref{parametric-2}) of $h(\theta)$ 
one can choose one parameter at will. 
Of course, this is true only in the exact case.
In an approximate parametrization the results depend on all parameters
introduced. However, one may still
require that they have some approximate independence from 
one of the parameters appearing in $h(\theta)$. This procedure 
is exact for $k\to\infty$. In practice, one fixes $h_3$
by requiring the approximate function
$f_{\rm approx}^{(k)}(x,h_3)$ to have the smallest possible dependence 
on $h_3$.
Thus, one sets $h_3 = h_{3,k}$, where $h_{3,k}$ is 
a solution of the global stationarity condition
\begin{equation}
  \left. 
  {\partial f_{\rm approx}^{(k)}(x,h_3) \over 
   \partial h_3}\right|_{h_3 = h_{3,k}} 
  = 0
\label{globalstationarity}
\end{equation}
for all $x$. Equivalently one may require that, 
for {\em any} universal ratio $R$ that 
can be obtained from the equation of state, its approximate 
expression $R_{\rm approx}^{(k)}$ obtained using the 
parametric representation satisfies 
\begin{equation}
  \left.
  {d R_{\rm approx}^{(k)}(h_3) \over d h_3}\right|_{h_3 = h_{3,k}}
  = 0.
\label{globalstationarity2}
\end{equation}
The existence of such a value of $h_3$ is a nontrivial mathematical fact.
The stationary value $h_{3,k}$ is the solution of the 
algebraic equation \cite{CPRV-99}
\begin{equation}
\left[ 2 (2\beta - 1) (h_3 + \gamma) {\partial\over \partial h_3} - 2 \gamma 
   + 2 k\right] h_{2k + 1} = 0,
\label{stationaritycondition}
\end{equation}
where $h_{2k+1}$ is given in Eq. \reff{hcoeff}.
Note that the restricted linear model (\ref{linmod}),\reff{h3-linearmodel}
represents the lowest order ($k=1$) of this systematic approximation scheme.

The same method was used in Ref. \cite{CHPV-00-2}, where, beside  
the coefficients $r_{2n}$, the universal constant 
$F_0^\infty$ that parametrizes the large-$z$ behavior of the 
function $F(z)$, see Eq. \reff{asyFz}, was used.
The parameters $h_5$, $\ldots$, $h_{2k-1}$ 
are fixed by matching  the first 
$(k-2)$ universal parameters $r_{2n}$, $n=3,...,k$. 
They are thus given by Eq. \reff{hcoeff}. Then, one sets
\be
h_{2k+1} = \rho^{\delta - 1} F_0^\infty - 1 
   - \sum_{n=1}^{k-1} h_{2n+1},
\ee
so that the parametric representation is exact for large values of 
$z$, i.e. it gives $F(z) \approx F_0^\infty z^\delta$ with 
the correct amplitude. The coefficient $h_3$ can be still determined
using the global stationarity condition \reff{globalstationarity},
which is again a nontrivial property.

Note that it is not possible to employ the variational method  using 
other generic amplitude ratios as input parameters. 
Indeed, the proof that Eq. \reff{globalstationarity} holds 
independently of $x$ requires identities that are valid only for 
very specific choices of amplitude ratios. At present, the procedure 
is known to work only for the two sets of amplitude ratios we mentioned above: 
(a) $r_6$, ..., $r_{2k+2}$; (b) $F_0^\infty$, $r_6$, ..., $r_{2k}$. 
The first set of amplitude ratios was used in Refs. \cite{CPRV-99,CPRV-02} 
in three dimensions, where
no sufficiently precise estimate of $F_0^\infty$ exists.
The second set was used in Ref. \cite{CHPV-00-2} in two dimensions, 
since in that case $F_0^\infty$ is known to high precision.

\subsubsection{Approximate critical equation of state}
\label{sec-3.3.3}

\begin{table*}
\caption{
Polynomial approximations of $h(\theta)$ using the variational approach
for several values of the parameter $k$, cf. Eq.~(\ref{parametric-2}).
Results from Ref. \cite{CPRV-02}.
}
\label{trht}
\vspace{0cm}
\footnotesize
\hspace*{0cm}    % Move table leftwards, so it doesn't run off the right
\tabcolsep 4pt        % Less than the usual 6pt
\doublerulesep 1.5pt  % Less than the usual 2pt
\begin{center}
\begin{tabular}{clcl}
\hline
\multicolumn{1}{c}{$k$}& 
\multicolumn{1}{c}{$h(\theta)/\theta$}& 
\multicolumn{1}{c}{$\theta_0^2$}& 
\multicolumn{1}{c}{$h(\theta)/[\theta(1 - \theta^2/\theta^2_0)]$}\\
\hline  
1  &  $ 1 - 0.734732 \theta^2 $ & $\quad$ 1.36104 $\quad$   & 1 \\
2  &  $ 1 - 0.731630 \theta^2 + 0.009090 \theta^4 $ &  $\quad$ 1.39085
$\quad$ &  
$1 - 0.0126429 \theta^2$ \\
3  &  
  $ 1 - 0.736743 \theta^2 + 0.008904 \theta^4 - 0.000472 \theta^6 $ & 
$\quad$  1.37861 $\quad$  &    $1-0.0113775 \theta^2 + 0.0006511 \theta^4$ \\
\hline
\end{tabular}
\end{center}
\end{table*}

The variational method outlined in the preceding section was
applied in Refs.~\cite{CPRV-99,CPRV-02}. In these works  the IHT results
for $\gamma$, $\nu$, $r_6$, and $r_8$ were used as input
parameters, obtaining polynomial approximations (\ref{parametric-2})
with $k=1,2,3$.
The corresponding polynomials $h(\theta)$ are reported in Table~\ref{trht}.
Note that the absolute values of the high-order coefficients rapidly
decrease and their signs alternate,
supporting the effectiveness of the approximation scheme. 

\begin{figure}[tb]
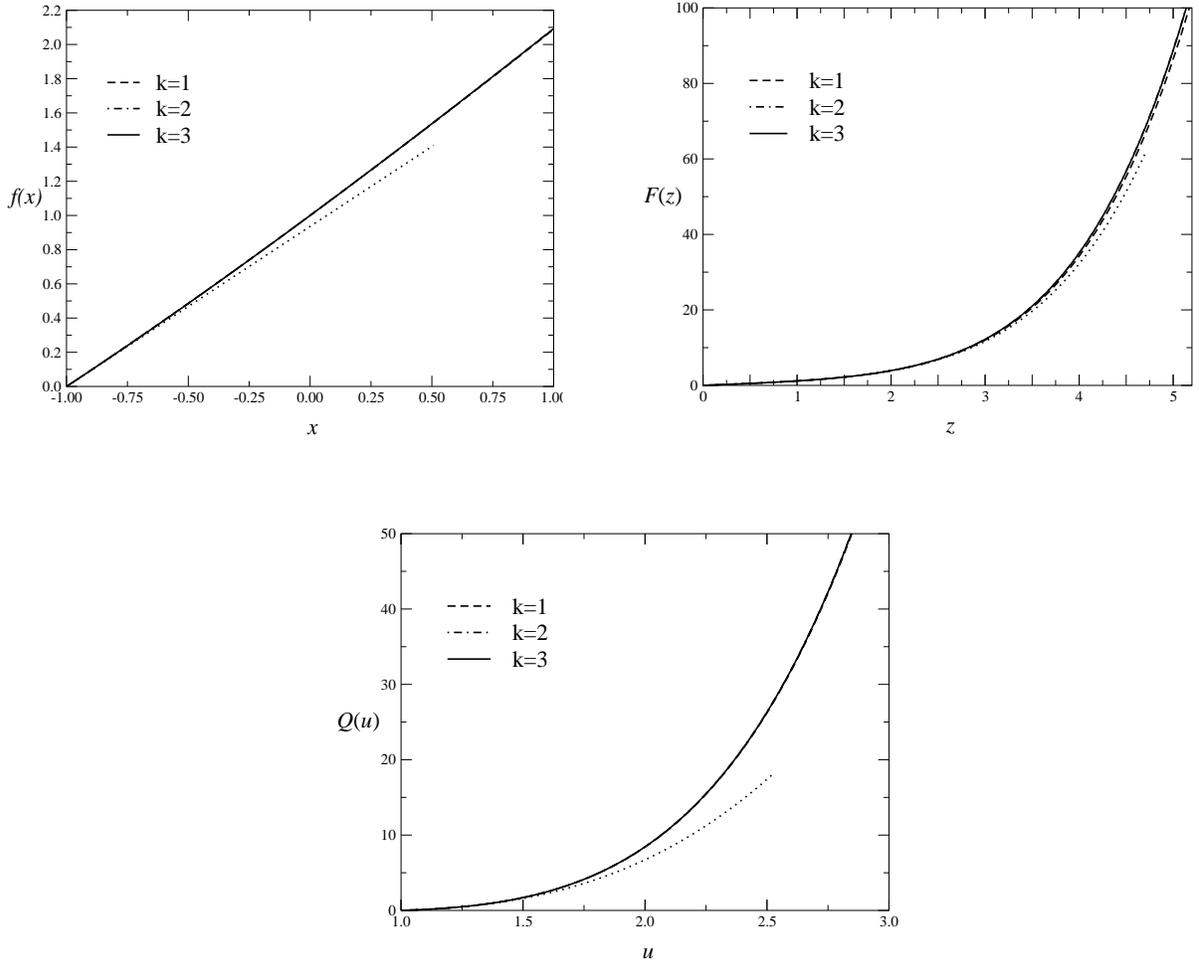

\hspace{0cm}
\begin{tabular}{cc}
\hskip -0.5truecm
\psfig{width=7.5truecm,angle=0,file=fxI.eps} &
\hskip 0.5truecm
\psfig{width=7.4truecm,angle=0,file=FzI.eps} \\[20mm]
\multicolumn{2}{c}{\psfig{width=7.5truecm,angle=0,file=quI.eps}}
\end{tabular}
\vspace{0cm}
\caption{
The scaling functions $f(x)$, $F(z)$, and $Q(u)$. 
We also plot the following asymptotic behaviors (dotted lines): 
$f(x)$ at the coexistence
curve, i.e., $f(x)\approx f_1^{\rm coex} (1+x)$
for $x\rightarrow -1$;
$F(z)$ at the HT line, i.e., 
$F(z)\approx z + \frac{1}{6} z^3 + 
\frac{1}{120} r_6 z^5$ for $z\rightarrow 0$;
$Q(u)$ at the coexistence
curve, i.e., $Q(u) \approx 
(u-1) + \frac{1}{2} v_3 (u-1)^2 + \frac{1}{6} v_4 (u-1)^3$
for $u\rightarrow 1$.
Results from Ref. \cite{CPRV-02}.
}
\label{figfxI}
\end{figure}

\begin{figure}[tb]
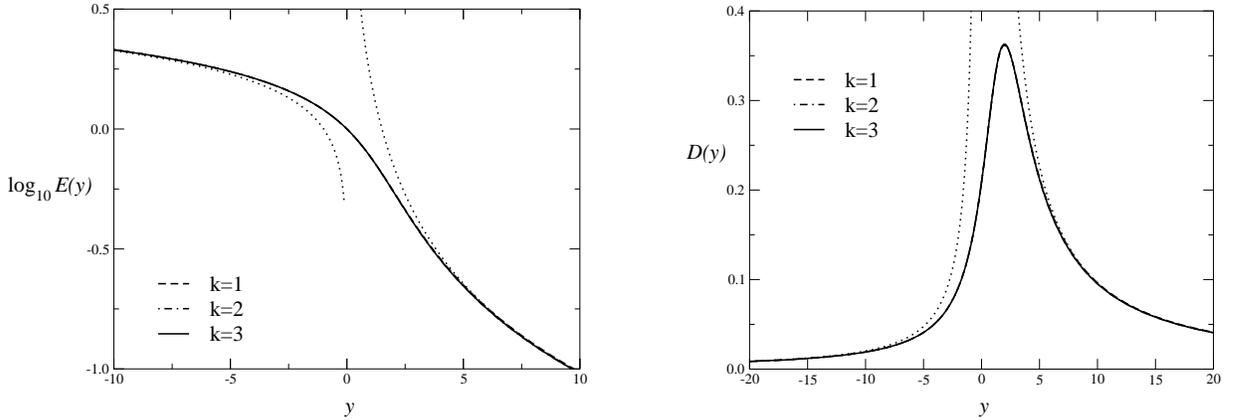

\vspace{0cm}
\hspace{0cm}
\begin{tabular}{cc}
\hskip -0.5truecm
\psfig{width=7.8truecm,angle=0,file=ewI.eps} &
\hskip 0.8truecm
\psfig{width=7.2truecm,angle=0,file=dd.eps} \\
\end{tabular}
\vspace{0cm}
\caption{
The scaling functions $E(y)$ and $D(y)$.
We also plot their asymptotic behaviors
(dotted lines):
$E(y)\approx R_\chi y^{-\gamma}$ for $y\rightarrow +\infty$, and
$E(y)\approx (- y)^{\beta}$ for $y\rightarrow -\infty$;
$D(y)\approx R_\chi y^{-\gamma}$ for $y\rightarrow +\infty$, and
$D(y)\approx \beta(- y)^{-\gamma}/f_1^{\rm coex}$ for $y\rightarrow -\infty$;
Results from Ref. \cite{CPRV-02}.
}
\label{figEwI}
\end{figure}

Figures~\ref{figfxI}
%%\ref{figFzI}, \ref{figPhiuI}, \ref{figCuI} 
and \ref{figEwI}  show 
the scaling functions $f(x)$, $F(z)$, $Q(u)$,
$E(y)$, and $D(y)$,
as obtained from $h(\theta)$ for $k=1,2,3$.
The curves for $k=1,2,3$ show a good convergence with increasing $k$:
Differences are hardly visible in the figures.  
The results for $k=2,3$ are consistent within the errors induced by
the uncertainty on the input parameters, indicating that 
the error due to the truncation is at most of the same order 
of the error induced by the input data. 
The asymptotic behaviors of the scaling functions are  
also shown in the figures. 

%% \begin{figure}[tb]
%% \vspace{0cm}
%% \hspace{0cm}
%% \centerline{\psfig{width=7.5truecm,angle=0,file=FzI.eps}}
%% \vspace{0cm}
%% \caption{
%% The scaling function $F(z)$. 
%% We also plot the small-$z$ expansion (dotted line), i.e., 
%% $F(z)\approx z + \frac{1}{6} z^3 + 
%% \frac{1}{120} r_6 z^5$ for $z\rightarrow 0$.
%% Results from Ref. \cite{CPRV-02}.
%% }
%% \label{figFzI}
%% \end{figure}
%% 
%% \begin{figure}[tb]
%% \vspace{0cm}
%% \hspace{0cm}
%% \centerline{\psfig{width=7.5truecm,angle=0,file=quI.eps}}
%% \vspace{0cm}
%% \caption{
%% The scaling function $Q(u)$. 
%% We also plot the asymptotic behavior of $Q(u)$ at the coexistence
%% curve (dotted line), i.e., $Q(u) \approx 
%% (u-1) + \frac{1}{2} v_3 (u-1)^2 + \frac{1}{6} v_4 (u-1)^3$
%% for $u\rightarrow 1$.
%% Results from Ref. \cite{CPRV-02}.
%% }
%% \label{figPhiuI}
%% \end{figure}

Some results concerning the scaling functions are \cite{CPRV-02}:
$f_0^\infty = R_\chi^{-1}= 0.6024(15)$,
which is related to the large-$x$ behavior of $f(x)$,
cf. Eq.~(\ref{largexfx});
$f_1^0=1.0527(7)$, $f_2^0=0.0446(4)$, $f_3^0=-0.0254(7)$,
which are related to the expansion at $x=0$ of $f(x)$,
cf. Eq.~(\ref{expansionfx-xeq0}); 
$f_1^{\rm coex} = 0.9357(11)$, $f_2^{\rm coex} = 0.080(7)$,
which are related to the behavior of $f(x)$ at the coexistence curve,
cf. Eq.~(\ref{coexcfx});
$F_0^\infty = 0.03382(15)$,
which is related to the large-$z$ behavior of $F(z)$,
cf. Eq.~(\ref{asyFz});
$v_3=6.050(13)$, $v_4=16.17(10)$, that are related
to the expansion of $Q(u)$ around $u=1$, cf.
Eq.~(\ref{sviluppoBu});
the scaling function $D(y)$ has a maximum for $y_{\rm max} = 1.980(4)$, 
corresponding
to the crossover or pseudocritical line, the value at the maximum 
is $D(y_{\rm max}) = 0.36268(14)$.

Also Refs.~\cite{GZ-97,GZ-98,Zinn-Justin-00} determined 
parametric representations of the equation of state 
starting from the small-magnetization expansion in the HT phase.
Instead of the variational approach, they fixed the additional
coefficient by minimizing
the absolute value of the highest-order term  of $h(\theta)$.
The results are consistent with
those obtained using the variational approach.

Other approximate representations of the equation of state are reported 
in Refs.~\cite{FZ-98,FZU-99}, see also Sec.~\ref{trir}. The parametric 
functions were determined  using several HT,LT universal amplitude ratios.

Finally, we mention that the equation of state has been computed
to $O(\epsilon^3)$ in the framework of the FT
$\epsilon$ expansion \cite{WZ-74,NA-85,BWW-72}.
It has also been studied using CRG methods, up to first order
in the derivative expansion. 
Results can be found in Refs.~\cite{BTW-96,SW-99,BTW-99}.

%% \begin{figure}[tb]
%% \vspace{0cm}
%% \hspace{0cm}
%% \centerline{\psfig{width=7.5truecm,angle=0,file=dr.eps}}
%% \vspace{0cm}
%% \caption{
%% The scaling function $D_R(y_R)\equiv D(y)/D(y_{\rm  max})$ versus
%% $y_R\equiv y/y_{\rm max}$, cf. Eq.~(\ref{defDw}). 
%% We also plot its asymptotic behaviors (dotted lines):
%% $D_R(y_R)\approx R_\chi y_{\rm max}^{-\gamma} D(y_{\rm max})^{-1}
%% y_R^{-\gamma}\approx 1.97 y_R^{-\gamma}$ for $y_R\rightarrow +\infty$, and
%% $D(y_R)\approx 
%% \beta (f_1^{\rm coex})^{-1} y_{\rm max}^{-\gamma} D(y_{\rm max})^{-1}
%% (-y_R)^{-\gamma}\approx 0.413 (-y_R)^{-\gamma}$ for $y_R\rightarrow -\infty$.
%% Results from Ref. \cite{CPRV-02}.
%% }
%% \label{figCuI}
%% \end{figure}

\subsubsection{Trigonometric parametric representa\-tions}
\label{trir} 

Refs.~\cite{FU-90,FZ-98,FZU-99} considered 
the possibility of determining a parametric representation
of the equation of state that also describes the 
two-phase region below the critical temperature.  
In the classical mean-field equation of state  
that describes a first-order transition,
one finds a characteristic van der Waals (vdW)
loop that represents an isothermal analytic continuation of 
the equation of state
through the coexistence curve. For $t\to 0^-$, it has the simple
cubic form $H\propto M(M^2 - M_0^2)$,
which shows the classical critical exponents.
The properties of the vdW loop are relevant for classical theories
of surface tension, interfaces, spinodal decomposition, etc. 
(see, e.g., Refs.~\cite{RW-82,CH-58}). In Refs.~\cite{FU-90,FZ-98,FZU-99} 
the authors search for representations of the equation of state
that, on the one hand, describe the 
nonclassical critical behavior of the system,
and, on the other hand, have a good analytic continuation
in the two-phase region. As also mentioned by the authors,
the existence of a full vdW loop is not guaranteed in nonclassical theories, 
because of the presence of essential singularities at the coexistence
curve \cite{FF-70,Fisher-67,Andreev-64,Isakov-84}, which preclude
the possibility of performing the required analytic continuation.
Nevertheless, in Refs.~\cite{FU-90,FZ-98,FZU-99} the classical thermodynamic 
picture was assumed and an
analytic continuation of the critical equation of state was performed.
They looked for parametric representations
that give a reasonable realization of the vdW loop
from their analytic continuation to the two-phase region.
Polynomial representations do not offer a natural description of vdW
loops. This is because the analytic continuation of these
equations of state fails to generate a closed, continuous
vdW loop inside the two-phase region.
To overcome this problem, Ref.~\cite{FU-90} proposed 
an interpolation scheme that ensured the 
expected vdW loop: The traditional parametric representations
are retained, but a new angular variable $\phi$  
is introduced to describe the two-phase region only, 
the values $\phi=\pm 1$ being assigned to the phase boundaries, while
$\phi=0$ corresponds to the coexistence-curve diameter $M=0$, $H=0$.
Thus, new angular functions describing the two-phase region
are required, satisfying a matching condition
that ensures the smoothness of the equation of state
across the phase boundaries $H=0$, $t<0$. 
However, according to the authors, this procedure is not the optimal one.
They noted that a natural description of the vdW loop 
requires analytic periodicity
with period $2 \theta_p$ with $\theta_p>\theta_0$.
Therefore, they proposed an alternative approach based on 
trigonometric parametric representations: 
\begin{eqnarray}
&M = m_0\, R^\beta \,m(\theta), \qquad 
&t = R  \,k(\theta), \nonumber \\
&{\cal A}_{\rm sing} = n_0 \, R^{2-\alpha} \,n(\theta), \qquad
&{\xi^2\over 2\chi} = R^{-\eta\nu} \,a(\theta) , 
\end{eqnarray}
where the traditional expression for $H$ is replaced by a direct 
parametrization of the singular part of the free energy ${\cal A}_{\rm sing}$. 
The parametric functions $m(\theta)$, $k(\theta)$, $n(\theta)$, and 
$a(\theta)$ are chosen to guarantee a closed analytic vdW loop:
\begin{eqnarray}
&&m(\theta) = {\sin (q\theta)\over q}, \qquad
k(\theta) =\left[ 1 - {2b^2\over q^2}  
   \left(1 - \cos (q\theta) \right)\right], \qquad
n(\theta) =  1 + \sum_{i=1}^j c_i k(\theta)^i,\nonumber \\
&&a(\theta) = a_0 \left[ 1 + {2 a_2\over q^2} \left( 1 - \cos (q\theta) \right)
+{a_4\over q^4} \left( 1 - \cos (q\theta) \right)^2 \right],
\end{eqnarray}
where $q,b,c_i,a_i$ are parameters that are determined by
using known universal amplitude ratios.
As shown in Ref.~\cite{FZU-99}, 
this parametrization is able to provide a good fit to
the universal amplitude ratios and  reasonable vdW loops.
They found that the shape of the vdW loop for three-dimensional Ising
systems near criticality differs significantly from the classical form. 
In particular, the spinodal should be closer to the coexistence curve
and the size of the vdW loop is smaller by approximately a factor of two.

\subsubsection{Universal amplitude ratios}
\label{sec-3.3.3b}

In this section we report the estimates of several
universal amplitude ratios, see Table \ref{notationsur} for definitions.
Those involving only zero-momentum
quantities, such as the specific heat and the magnetic susceptibility,
can be derived from the equation of state.
Estimates of universal ratios
involving correlation-length amplitudes, such as 
$Q^+$, $R_\xi^+$, and $Q_c$, can be obtained  using the
estimate of $g_4^+$. For instance,
$Q^+ = R_4^+ R_c^+/g_4^+$.
Other universal ratios can be derived  by
supplementing the above-reported results with the estimates of 
$w^2$ and  $Q^-_\xi$ (which may be estimated by analyzing 
the corresponding LT expansions
\cite{CPRV-99,PV-gr-98,AT-95,Vohwinkel-93})
and the estimate of $Q^+_\xi$ (see Sec.~\ref{twopointf} and 
Table \ref{ci3d}).  
Moreover, in Ref.~\cite{CPRV-99}
estimate of $Q_\xi^c$ and $Q_2$ were obtained from
approximate parametric representations of the correlation lengths $\xi$ and
$\xi_{\rm gap}$, such as
\begin{eqnarray}
a(\theta) = a(0) \left(1 + c \theta^2\right), 
\qquad
  a_{\rm gap}(\theta) = a_{\rm gap}(0) 
      \left(1 + c_{\rm gap} \theta^2\right), 
\label{atheta-approx}
\end{eqnarray}
where $c$ and $c_{\rm gap}$ were obtained using 
the IHT estimates of $U_\xi$ and $U_{\xi_{\rm gap}}$. 

In Table \ref{summaryeqstisingLaTe} we  report the results obtained using
the parametric representations reported in Sec.~\ref{sec-3.3.3} (IHT-PR),
from the analysis of 
HT and LT expansions (HT,LT), and from MC simulations.
The results for $U_2$ and $U_\xi$ of Ref.~\cite{BC-00} were
obtained by taking $\beta_c$ and $\Delta$ as external inputs; the 
reported errors do not take into account
the uncertainties on $\beta_c$ and $\Delta$, which should not be negligible.
The IHT-PR estimates agree nicely with the most recent MC results,
especially with those reported in
Ref.\ \cite{CH-97}, which are quite precise. There is a discrepancy  
only for $U_0$: The
estimates reported in Ref.\ \cite{HP-97} are slightly larger.
On the other hand, there is good agreement with the rather precise
experimental result of Ref.~\cite{NGMJ-01}. 
It is worth mentioning that the result of
Ref.\ \cite{ES-99} for $U_2$ was obtained by 
simulating a four-dimensional $SU(2)$
lattice gauge model at finite temperature.

Table \ref{summaryeqstisingLaTe} also reports some experimental results for 
binary mixtures, liquid-vapor
transitions, and uniaxial antiferromagnetic systems. 
They should give an overview of the level of
precision reached by experiments.
Some of the experimental data are taken from Ref. \cite{PHA-91}.
Sometimes, we report a range of values without a corresponding reference: 
this roughly summarizes the
results reported in the corresponding table of Ref. \cite{PHA-91}
and should give an idea of the range of the experimental results.

Table \ref{summaryeqstisingFT} shows the results 
obtained by FT methods.
FT estimates are consistent, although in general less precise.
We mention that the results denoted by ``$d=3$
exp'' were obtained using different schemes, see 
Sec. \ref{sec-2.4}: the traditional zero-momentum scheme
\cite{BBMN-87,BB-85}, the minimal subtraction
without $\epsilon$ expansion \cite{LMSD-98,SD-89}, and the expansion in
the LT coupling $u\equiv 3 w^2$ 
\cite{GKM-96}.  Refs.~\cite{GZ-97,GZ-98} 
used the fixed-dimension expansion and the $\epsilon$ expansion 
to determine the universal coefficients $r_6$, $r_8$, and $r_{10}$,
which were then used to obtain an approximate parametric 
representation of the critical equation of state.  
The corresponding amplitude ratios are denoted by FT-PR.

\begin{table*}
\caption{
Estimates of universal quantities, see Table \ref{notationsur} for 
definitions.
The results are obtained by combining HT results and the parametric 
representation of the equation of state (IHT--PR),
from the analysis of 
high- and low-temperature expansions (HT,LT), and from 
Monte Carlo simulations (MC).
For the experimental results:
ms\ denotes a
magnetic system; bm\ a binary mixture; lv\ a liquid-vapor transition;
mi\ a micellar system. 
Experimental estimates without reference are taken from 
Ref. \cite{PHA-91}.
}
\label{summaryeqstisingLaTe}
\footnotesize
\hspace*{-2cm}    % Move table leftwards, so it doesn't run off the right
\tabcolsep 4pt        % Less than the usual 6pt
\doublerulesep 1.5pt  % Less than the usual 2pt
\begin{center}
\begin{tabular}{lllll}
\hline
\multicolumn{1}{c}{}& 
\multicolumn{1}{c}{IHT--PR \cite{CPRV-02,CPRV-99}}& 
\multicolumn{1}{c}{HT,LT}& 
\multicolumn{1}{c}{MC}& 
\multicolumn{1}{c}{experiments}\\
\hline  
$U_0$& 0.532(3)& 0.523(9) \cite{LF-89}& 0.560(10) \cite{HP-98} &  0.536(5) bm \cite{NGMJ-01} \\
     &         & 0.51 \cite{BHK-75}   & 0.550(12) \cite{HP-98} & 0.538(17) lv \cite{SN-93} \\
     &         &                      & 0.567(16) \cite{HP-98} & 0.54(2) ms \cite{BNKJLB-83} \\
     &         &                      & 0.45(7) \cite{Marinari-84} & 
                                             0.55(6) ms \cite{MMB-95b} \\
     & &&& 0.47--0.53 lv \\
     &         &                      &                        &  0.54--0.58 bm \\   
     &         &                      &                        &  0.52--0.56 ms \\
$U_2$ &  4.76(2) &  4.762(8) \cite{BC-00}&4.75(3) \cite{CH-97}& 4.3(3) bm \cite{ZBB-83}\\
      &          & 4.95(15) \cite{LF-89}     &  4.72(11) \cite{ES-99}  & 4.5--5.3 lv \\
      &          & 5.01 \cite{TF-75}     &                         & 4.9(5) ms \cite{CC-80}\\
      &          &                       &                         & 4.6(2) ms \cite{BY-87}\\
$U_4$ &  $-$9.0(2)& $-$9.0(3) \cite{ZLF-96} & &  \\
$R_c^+$ & 0.0567(3) & 0.0581(10) \cite{ZLF-96}& &  0.050(15) bm \cite{ZBB-83} \\
        &           &                         & &  0.04--0.06 lv \\
$R_c^-$ & 0.02242(12) & & &  \\
$R_4^+$ & 7.81(2)   & 7.94(12) \cite{FZ-98}   & & \\
$R_4^-$ & 93.6(6) & 107(13) \cite{ZLF-96,FZ-98} &  &  \\
$R_\chi$ &  1.660(4) & 1.57(23) \cite{ZF-96,FZ-98}& &  1.75(30) bm \cite{ZBB-83} \\
&  & & &  1.69(14) lv \cite{NB-89} \\
$w^2$ &   & 4.75(4) \cite{PV-gr-98}         & 4.77(3) \cite{CH-97} &  \\
      &   & 4.71(5) \cite{ZLF-96,Fisher-pv} & & \\
$U_\xi$ & 1.956(7)& 1.963(8) \cite{BC-00} & 1.95(2) \cite{CH-97} & 2.0(4) bm \cite{HTKK-86}  \\
        &          &  1.96(1) \cite{LF-89} & 2.06(1) \cite{RZW-94}&  1.9(2) bm \cite{ZBB-83} \\
        &          &  1.96 \cite{TF-75} & & 1.89(4) ms \cite{BY-87} \\
        &          &  & & 1.93(10) ms \cite{CC-80} \\
$U_{\xi_{\rm gap}}$ & 1.896(10)& & & \\
$Q^+$ & 0.01880(8) & 0.01899(11) \cite{BC-02} & 0.0193(10) \cite{HP-98} & 
          0.023(4) lv \cite{HS-99} \\ 
&  & 0.0202(9) \cite{BC-99} &   &  0.0187(13) bm \cite{NGMJ-01} \\
&  & 0.01880(15) \cite{LF-89} &  &  0.016(4) mi \cite{LBW-97} \\
&  & &  &  0.018--0.022 bm \\
$Q^-$ & 0.00472(5) & 0.00477(20) \cite{FZ-98} & 0.0463(17) \cite{HP-98} &    \\
$Q_c$ & 0.3315(10) & 0.324(6) \cite{FZ-98} & 0.328(5) \cite{CH-97} & 0.3--0.4 bm  \\
  & & & & 0.34(19) bm \cite{WBLSKS-98}  \\
  & & & & 0.36(3) bm \cite{Jacobs-86}  \\
  & & & & 0.29(4) bm \cite{AKJ-86}  \\
  & & & & 0.3--0.4 lv  \\
$Q_\xi^+$ & 1.000200(3) & 1.0001 \cite{FZ-98} &  & \\
$Q_\xi^c$ & 1.024(4) & 1.007(3) \cite{FZ-98}  & &  \\
$Q_\xi^-$ &  & 1.032(4) \cite{CPRV-99} & 1.031(6) \cite{CHP-99,ACCH-97} & \\
& &  1.037(3) \cite{FZ-98} & &  \\
$Q_2$ & 1.195(10) & 1.17(2) \cite{ZF-96,FZ-98} & & 1.1(3) bm \cite{ZBB-83} \\
$v_3$ & 6.050(13) & 6.44(30) \cite{FZ-98,ZLF-96} & & \\
$v_4$ & 16.17(10) & & & \\
$g_3^-$ & 13.19(6) & 13.9(4) \cite{ZLF-96} & 13.6(5) \cite{Tsypin-97} &   \\
$g_4^-$ & 76.8(8) & 85 \cite{ZLF-96} & 108(7) \cite{Tsypin-97} & \\
$P_m$ & 1.2498(6) & & & \\
%$P_c$ & 0.3933(7) & & & \\
$R_p$ & 1.9665(10) & & & \\
$R_\sigma$ & & & 0.1040(8) \cite{HP-97}     & \\
           & & & 0.1056(19) \cite{ACCH-97}  &   \\
           & & &  0.098(2) \cite{ZF-96}   &   \\
$R_\sigma^+$ & & & 0.40(1) \cite{Hasenbusch-99-h}     & 0.38(3) \\
             & & & 0.377(11) \cite{ZF-96,FZ-98}   & 0.41(4) bm \cite{MW-96} \\
             & & & & 0.33(6) mi \cite{LBW-97} \\
\hline
\end{tabular}
\end{center}
\end{table*}

In Tables 
\ref{summaryeqstisingLaTe} and \ref{summaryeqstisingFT}
we also report the universal amplitude ratios 
$R_\sigma$ and $R_\sigma^+$ involving the surface-tension
amplitude, see Table \ref{notationsur}
for definitions.

\begin{table*}
\caption{
FT estimates of universal quantities, see Table \ref{notationsur} for 
definitions.
We report results obtained in the $\epsilon$ expansion 
($\epsilon$ exp), in the fixed-dimension expansion in $d=3$ in 
different schemes (see text)
($d=3$ exp),  using a parametric equation of state 
and $d=3$ and $\epsilon$-expansion results 
($d=3$ and $\epsilon$ FT--PR) \protect\cite{GZ-97,GZ-98},
and in the continuous RG approach (CRG).
}
\label{summaryeqstisingFT}
\footnotesize
\tabcolsep 3.2pt        % Less than the usual 6pt
\doublerulesep 1.5pt  % Less than the usual 2pt
\begin{center}
\begin{tabular}{cccccc}
\hline
\multicolumn{1}{c}{}& 
\multicolumn{1}{c}{$\epsilon$ exp}& 
\multicolumn{1}{c}{$d=3$ exp}& 
\multicolumn{1}{c}{$d=3$ FT--PR \cite{GZ-98}}&
\multicolumn{1}{c}{$\epsilon$ FT--PR \cite{GZ-98}}&
\multicolumn{1}{c}{CRG} \\
\hline  
$U_0$& 0.524(10) \cite{NA-85,Bervillier-86} & 0.540(11)\cite{LMSD-98} & 0.537(19) & 0.527(37) & \\
     & & 0.541(14) \cite{BBMN-87} & & & \\

$U_2$ &  4.9 \cite{NA-85} & 4.77(30) \cite{BBMN-87} & 4.79(10) & 4.73(16) & 4.966 \cite{SW-99}\\
      &  4.8 \cite{BLZ-74,AH-76} &   4.72(17) \cite{GKM-96} &   &  & 4.29 \cite{BTW-96} \\

$U_4$ &  &  &  $-$9.1(6) & $-$8.6(1.5) &  \\
$R_c^+$ & &  & 0.0574(20) & 0.0569(35)  & \\

$R_4^+$ & &  & 7.84  & 8.24(34) &  \\

$R_\chi$ &1.67 \cite{NA-85,Bervillier-86}  & 1.7 \cite{BBMN-87} & 1.669(18) & 1.648(36) &  1.647 \cite{SW-99} \\
&  &  &  & & 1.61 \cite{BTW-96} \\

$w^2$ &   & 4.73 \cite{GKM-96} & & &  \\

$U_\xi$ & 1.91 \cite{BLZ-74} & 2.013(28) \cite{GKM-96} && & 2.027 \cite{SW-99}\\
        &          & 2.04(4) \cite{MH-94}  &&&  1.86 \cite{BTW-96} \\

$Q^+$ & 0.0197 \cite{BG-80,Bervillier-86} & 0.01968(15) \cite{BB-85} &&&  \\
$Q_c$ & & 0.331(9) \cite{BBMN-87}  &&& \\
$Q_\xi^+$ & 1.00016(2) \cite{CPRV-99} & 1.00021(3) \cite{CPRV-98}& && \\
$Q_2$ &  1.13 \cite{BLZ-74} && && \\
$v_3$ &  5.99(5) \cite{PV-99}  & & 6.08(6) & 6.07(19) & \\
$v_4$ & 15.8(1.4) \cite{PV-99} & & &&\\
$g_3^-$ & 13.06(12)\cite{PV-99} &  & &&  \\
$g_4^-$ & 75(7)\cite{PV-99} && & & \\
$R_\sigma$ & 0.055 \cite{BF-84}& 0.1065(9) \cite{Munster-98} & && \\
\hline
\end{tabular}
\end{center}
\end{table*}

\subsection{The two-dimensional Ising universality class}
\label{eqstising2d}

\subsubsection{General results}
\label{sec-3.4.1}

In two dimensions a wealth of exact results exists.
Many exact results have been obtained for the 
simplest model belonging to this universality class, the 
spin-$1/2$ Ising model. 
For the square-lattice Ising model we mention:
the exact expression of the free energy along the $H=0$ axis \cite{Onsager_44}, 
the two-point correlation function for $H=0$ \cite{WMTB-76}, 
and the spontaneous magnetization on the coexistence curve \cite{Yang-52}.
For a review, see, e.g., Ref. \cite{McCoy-95}.
Moreover, in the critical limit several 
amplitudes are known to high precision, see, e.g., Refs. 
\cite{Nickel-99,ONGP-00}. 
Besides, at the critical point one can use conformal field theory. 
This provides the exact spectrum of the theory, i.e. all the 
dimensions of the operators present in the model. 
In particular, one finds \cite{CCCPV-00,CHPV-01}
that the first rotationally-invariant 
correction-to-scaling operator has dimension 
$y_3 = - 2$, i.e. $\omega = 2$.\footnote{It is interesting to note that 
such correction does not appear in the nearest-neighbor 
lattice Ising model, which is 
thus an exactly improved model. There is no mathematical proof, but in 
the years a lot of evidence has been collected \cite{CHPV-01}. 
In particular, no such 
correction is found in the susceptibility for $H=0$ and $t>0$
\cite{Nickel-99,ONGP-00}, in the free energy along the critical 
isotherm \cite{CH-00}, in the mass gap \cite{CHPV-01,CCCPV-00} for $H=0$,
and in some finite-size quantities \cite{CHPV-01,Salas-02}. 
We should also notice that it has been claimed 
sometimes that $\omega= 4/3$. Such a statement is partially incorrect.
Indeed, such exponent only appears in the Wegner expansion of 
some quantities and correlations that 
provide a nonunitary extension of the
Ising universality class, but not in the expansion 
of standard thermodynamic variables.
For a detailed discussion, see Ref.\cite{CCCPV-00}.} 
Moreover, the exponent $\omega_{NR}$ that 
gives the corrections related to the breaking of the rotational invariance
can be exactly predicted \cite{CPRV-98,CH-00,CCCPV-00,CHPV-01}: 
$\omega_{NR} = 2$ on the square lattice and $\omega_{NR} = 4$ on the 
triangular lattice.

\begin{table}
\caption{
Critical exponents and universal amplitude ratios for the 
two-dimensional Ising universality class, 
taken from Refs.\protect\cite{WMTB-76,Delfino-98,CPRV-99,CHPV-00,CCCPV-00}.
Since the specific heat diverges logarithmically,
the specific-heat amplitudes $A^{\pm}$ are defined by
$C_H\approx - A^\pm \log t$. See Sec. \ref{sec-1.3} for the 
definitions of the other amplitudes. 
The definition of $R^\pm_c$ and $Q^\pm$ differ from those given in Table 
\ref{notationsur} because of the absence of $\alpha$, which is 
zero in this case. The value of $\omega_{NR}$ depends on the lattice that is 
considered: The reported values refer to the square (sq) and triangular (tr)
lattices respectively.  }
\label{Ising2dex}
\footnotesize
\begin{center}
\begin{tabular}{lc}
\hline
$\gamma$ &  7/4 \\
$\nu$ &  1 \\
$\eta$ &  1/4 \\
$\beta$ &  1/8 \\
$\delta$ &  15 \\
$\omega$ &  2 \\
$\omega_{NR}$ &  2 (sq), \, 4 (tr) \\ \hline
$U_0\equiv A^+/A^-$ & 1  \\
$U_2\equiv C^+/C^-$ & 37.69365201    \\
$R_c^+\equiv A^+C^+/B^2 $ & 0.31856939 \\
$R_c^-\equiv A^- C^-/B^2 $ & 0.00845154 \\
$R_\chi\equiv Q_1^{-\delta}\equiv C^+ B^{\delta-1}/(B^c)^\delta$ 
& 6.77828502  \\
$w^2\equiv C^- /[ B^2 (f^-)^2]$ &  0.53152607 \\ 
$U_\xi\equiv f^+/f^- $ &  3.16249504 \\
$U_{\xi_{\rm gap}}\equiv f^+_{\rm gap}/f^-_{\rm gap}$ & 2 \\
$Q^+ \equiv A^+ (f^+)^2$  &  0.15902704  \\
$Q^- \equiv A^- (f^-)^2 $  &  0.015900517 \\
$Q^+_\xi\equiv f^+_{\rm gap}/f^+$ & 1.000402074 \\
$Q^c_\xi\equiv f^c_{\rm gap}/f^c$&  1.0786828  \\
$Q^-_\xi\equiv f^-_{\rm gap}/f^-$&  1.581883299 \\
$Q_2\equiv (f^c/f^+)^{2-\eta} C^+/C^c$ &  2.8355305\\
\hline
\end{tabular}
\end{center}
\end{table}

Additional results have been obtained  using the $S$-matrix 
approach to two-dimensional integrable theories and in particular 
the thermodynamic Bethe Ansatz 
(for a review, see, e.g., Ref.~\cite{McCoy-95}).
Indeed, 
the quantum field theories that describe the critical regime 
for $H=0$ and $t=0$ are integrable and one can compute 
the corresponding $S$-matrices. While for $H=0$ the $S$-matrix is trivial, 
for two-particle scattering $S=-1$, on the critical isotherm 
the $S$-matrix solution is complex  with a nontrivial mass spectrum
\cite{zam}. 
A related method is the form-factor approach, 
which uses the knowledge of the $S$-matrix to set up a system of
recursive functional equations for the form factors.
By solving this system, one can in principle compute exactly
all the form factors, thus performing an analytic continuation of
the $S$-matrix off mass shell \cite{KW-78,Smirnov-92}.
Once the form factors are known, one can compute the correlation
functions of the fundamental field, as well as of other
composite operators, by inserting complete sets of scattering states
between them. This gives the correlation functions as infinite
series of convolution products of form factors.

In Table~\ref{Ising2dex} we report some exact results and some 
high-precision estimates of the amplitude ratios that have 
been obtained using the approaches that we mentioned above. 
In Table~\ref{sumgr} we report estimates of 
the zero-momentum four-point coupling $g_4^+$, for which
very precise estimates have been recently obtained
by various methods.

\begin{table*}[tb]
\caption{ 
Estimates of $g_4^+$ for the two-dimensional Ising universality class. 
We report the
existing results obtained using 
transfer-matrix techniques combined with RG scaling
(TM+RG), the form-factor approach (FF),
high-temperature expansions (HT), Monte Carlo simulations (MC),
field theory (FT) based on the $\epsilon$ expansion
and the fixed-dimension $d=2$ expansion,
and a method based on a dimensional expansion around $d=0$
($d$ exp).
}
\label{sumgr}
\footnotesize
\begin{center}
\begin{tabular}{rll}
\hline
\multicolumn{1}{c}{Ref.}& 
\multicolumn{1}{c}{Method}& 
\multicolumn{1}{c}{$g_4^+$}\\
\hline
\cite{CHPV-00-2,CHPV-00} $_{2000}$ &         TM+RG &  14.697323(20)   \\
\cite{BNNPSW-00}    $_{2000}$ &       FF &  14.6975(1)   \\
\cite{PV-gr-98}   $_{1998}$ &        HT &  14.694(2)   \\
\cite{BC-96}     $_{1996}$ &      HT &  14.693(4)   \\
\cite{ZLF-96}    $_{1996}$ &       HT &  14.700(17)\\
\cite{BNNPSW-00}   $_{2000}$ &        MC &  14.69(2)   \\
\cite{Kim-99}   $_{2000}$ &        MC &  14.7(2)   \\
\cite{PV-00}   $_{2000}$ &        FT $\epsilon$ exp &  14.7(4) \\
\cite{OS-00}   $_{2000}$ &        FT $d=2$ exp &  15.4(3) \\
\cite{LZ-77}   $_{1977}$ &        FT $d=2$ exp &  15.5(8) \\
\cite{BB-92}   $_{1992}$ &       $d$ exp &  14.88(17) \\
\hline
\end{tabular}
\end{center}
\end{table*}

The two-dimensional Ising universality class is 
also of experimental interest.
Indeed, there exist several uniaxial antiferromagnets which present a strongly 
enhanced in-plane coupling and an easy-axis anisotropy
(see, e.g., Refs.~\cite{PHA-91,HI-80,WP-98} for some experimental results), 
and have therefore a two-dimensional Ising critical behavior. 
Ising behavior has also been observed in several order-disorder and 
structural transitions: in monolayers of carbon monoxide and 
C$_2$F$_6$ physisorbed on graphite \cite{FC-93,WA-93,AFEK-98},
in adsorbed hydrogen on Ni \cite{BSVP-95}, and in GaAs(001) surfaces
\cite{LBADEBT-00}. We also mention an 
experimental study of the Yang-Lee edge singularities in 
FeCl$_2$ \cite{BKAK-01}.

\subsubsection{The critical equation of state: exact results}
\label{sec-3.4.2}

The behavior of the free energy for the two-dimensional Ising model 
is somewhat different from that described in Sec. \ref{sec-1.5}. 
The reason is that in this case there are resonances among the 
RG eigenvalues with the subsequent appearance of logarithmic 
terms. Because of the resonance between the identity and the
thermal operator, the singular part of the Gibbs free energy 
becomes \cite{Wegner-76}
\begin{eqnarray}
{\cal F}_{\rm sing}(H,t) = t^2 \widehat{\cal F}_1(Ht^{-15/8}) + 
t^2 \log |t|\, \widehat{\cal F}_{1,\rm log}(Ht^{-15/8}),
\end{eqnarray}
where irrelevant terms have been discarded. Note that additional 
resonances involving subleading operators are expected, and thus 
additional logarithmic terms should be present: such terms, involving 
higher powers of $\log |t|$, have been found in a high-precision 
analysis of the susceptibility for $H=0$ in the HT phase \cite{ONGP-00}.
The exact results for the free energy at $H=0$ and the numerical 
results for the higher-order correlation functions at zero momentum 
show that $\widehat{\cal F}_{1,\rm log}(x)$ 
is constant \cite{AF-83}.\footnote{There is evidence that such property
holds even if we consider the contributions of the irrelevant scaling 
fields \cite{CHPV-01,Salas-02}.} 
Indeed, if this function were nontrivial, then one would obtain
$\chi_n \sim |t|^{-\gamma_n} \log |t|$ for $|t|\to 0$, a behavior that has 
not been observed. The constant is easily related to the amplitudes
of the specific heat for $H\to 0$ defined in Eq. \reff{CH-log-def}. 
The analyticity for $t=0$, $H\not=0$  implies
\be
    A^+ = A^- \equiv A,
\ee
so that 
\be
{\cal F}_{\rm sing}(H,t) = t^2 \widehat{\cal F}_1(Ht^{-15/8}) 
    + {A\over 2} t^2 \log |t|.
\ee
For the Helmholtz free energy similar formulae holds. Using 
the notations of Sec. \ref{sec-1.5.2} we write
\begin{eqnarray}
{\cal A}_{\rm sing}(M,t) = a_{11} t^2 A_1(z) + {A\over2} t^2 \log |t|
=  a_{20} M^{16} A_2(x) + {A\over2} t^2 \log |t|,
\end{eqnarray}
where $a_{11}$ and $a_{20}$ are defined in Eqs. \reff{def-abconstant} and 
\reff{def-a20constant}, the 
variables $z$ and $x$ in Eqs. \reff{def-zvariable} and 
\reff{def-xvariable}, and the functions 
$A_1(z)$ and $A_2(x)$ are normalized as in Sec. \ref{sec-1.5.2}.
The presence of the logarithmic term gives rise to logarithms in the 
expansions of $A_1(z)$ for $z\to\infty$ and $A_2(x)$ for $x\to 0$. 
Indeed, the analyticity of ${\cal A}_{\rm sing}(M,t)$ for $t=0$, 
$|M|\not=0$ implies 
\begin{eqnarray}
A_1(z) &=& z^{16} \sum_{n=0} a_{1,n} z^{-8n} + a_{1,\rm log} \log z ,
\label{espansioneA1z-Ising2d}\\
A_2(x) &=& \sum_{n=0} a_{2,n} x^{n} + a_{2,\rm log} x^2 \log |x|.
\label{espansioneA2x-Ising2d}
\end{eqnarray}
The constant $a_{1,\rm log}$ and $a_{2,\rm log}$ are easily expressed in terms 
of invariant amplitude ratios: 
\begin{eqnarray} 
    a_{1,\rm log} = {4 A\over a_{11}} = 4 Q^+ g_4^+, \qquad
    a_{2,\rm log} = - {A\over 2 a_{20} B^{16}} = 
              - {8 R^+_c\over R_\chi}.
\end{eqnarray}
For the equation of state we have 
\be
{H} = {\partial{\cal A}\over \partial {M}} =
          a_{11} b_1 t^{15/8} F(z) =
          (B^c)^{-15} M^{15} f(x),
\ee
where $F(z)$ and $f(x)$ are defined in Eq. \reff{eq-stato}. 
The properties of these two functions are described in Sec. \ref{sec-1.5.3}. 
Using Eqs. \reff{espansioneA1z-Ising2d} and \reff{espansioneA2x-Ising2d} 
we can compute the coefficients $F_2^\infty$ and $f_2^0$ appearing in the 
expansions of $F(z)$ and of $f(x)$ for $z\to\infty$ and $x\to0$ 
respectively, cf. Eqs. \reff{asyFz} and \reff{expansionfx-xeq0}.
We have
$F_2^\infty = a_{1,\rm log}$ and
$f^0_2 = -\case{1}{2} a_{2,\rm log}$.

A detailed study of the analytic properties of the
critical equation of state can be found in Ref.~\cite{FZ-01}.

\subsubsection{Approximate representations of the equation of state}
\label{sec-3.4.3}

The equation of state in the whole $(t,H)$ plane is not known
exactly, only approximate results are available.
Approximate parametric representations have been determined in Ref.~\cite{CHPV-00-2},
using the variational approach presented in Sec. \ref{eqstising}. 
Specifically, the parametrization \reff{parametric-2} was used,
and the $k$ parameters $h_3,\ldots,h_{2k+1}$ were determined by requiring 
the approximate representation to reproduce the $(k-2)$ invariant 
ratios $r_{2n}$, $n:3,\ldots,k$, and the large-$z$ behavior of the 
function $F(z)$, $F(z)\approx F_0^\infty z^\delta$, and to satisfy 
the global stationarity condition \reff{globalstationarity}; see 
Sec. \ref{eqstising} for details of the method. 

In order to apply the method, good estimates of the coefficients 
$r_{2n}$, which parametrize the small-magnetization expansion
of the Helmholtz free energy, and of $F_0^\infty$ are needed. 
The latter constant can be obtained from the results of Table 
\ref{Ising2dex} and the precise estimate of $g_4^+$ 
of Ref. \cite{CHPV-00-2}
reported in Table~\ref{sumgr}.
Indeed,
\begin{eqnarray}
R_4^+\equiv {g_4^+ Q^+\over R^+_c}=7.336774(10),\qquad
F_0^\infty = R_\chi (R^+_4)^{(1-\delta)/2}=
5.92357(6) \times 10^{-5}.  
\end{eqnarray}
Accurate estimates of the first coefficients $r_{2n}$, see
Table~\ref{eqstd2app}, have been recently
determined in Refs.~\cite{CHPV-00-2,CPRV-00}, using transfer-matrix 
techniques and general RG properties.
Another approach is presented in Ref.~\cite{FZ-01},
where the authors exploit the analytic properties
of the free energy to write down a dispersion relation.
Approximate expressions for the corresponding kernel
are obtained using the knowledge of the behavior of the 
free energy at the Yang-Lee edge singularity \cite{Fisher-78,Cardy-85}.
The estimates of first few $r_{2n}$
obtained in this approach, see Table~\ref{eqstd2app}, 
are in good agreement with the results of Ref. \cite{CHPV-00-2}.
The comparison worsens for the higher-order coefficients, showing the 
limitations of the approximation employed.
The coefficients appearing in the expansion of the
scaling function $Q(u)$ around $u=1$, 
cf. Eq. \reff{sviluppoBu}, have also been determined. 
We report the results \cite{PV-99}
$v_3=33.011(6)$, $v_4=48.6(1.2)$ from LT expansions 
and  \cite{FZ-01} $v_3=33.0502$, $v_4=48.0762$ from an 
appropriate dispersion relation. Estimates of $v_n$ for $n>4$ 
are reported in Ref. \cite{PV-99} and can also be derived from the 
results of Ref. \cite{FZ-01}.

\begin{table*}
\caption{
Estimates of the coefficients $r_{2n}$. 
They have been determined 
by transfer-matrix techniques supplemented with RG 
results (TM+RG) \protect\cite{CHPV-00-2}, using 
approximate parametric representations of the equation of state 
(TM+RG+PR) \protect\cite{CHPV-00-2}, and by means of a dispersive
approach (Disp) \protect\cite{FZ-01}.
}
\label{eqstd2app}
\footnotesize
\begin{center}
\begin{tabular}{lccc}
\hline
    &  TM+RG \cite{CHPV-00-2} & TM+RG+PR \cite{CHPV-00-2} & Disp \cite{FZ-01} \\
\hline
$r_6$ & 3.67867(7) &         &  3.67797    \\
$r_8$ & 26.041(11)&          &  26.0332 \\
$r_{10}$ & 284.5(2.4)  &    & 286.12 \\
$r_{12}$ & $4.2(7) \times 10^3$ & $4.44(6) \times 10^3$ & 4215 \\
$r_{14}$ & &  $8.43(3) \times 10^4$ & $-$7356 \\
%$F_1^\infty$ & 0.0211(2) \\
\hline
\end{tabular}
\end{center}
\end{table*}

\begin{figure}[tb]
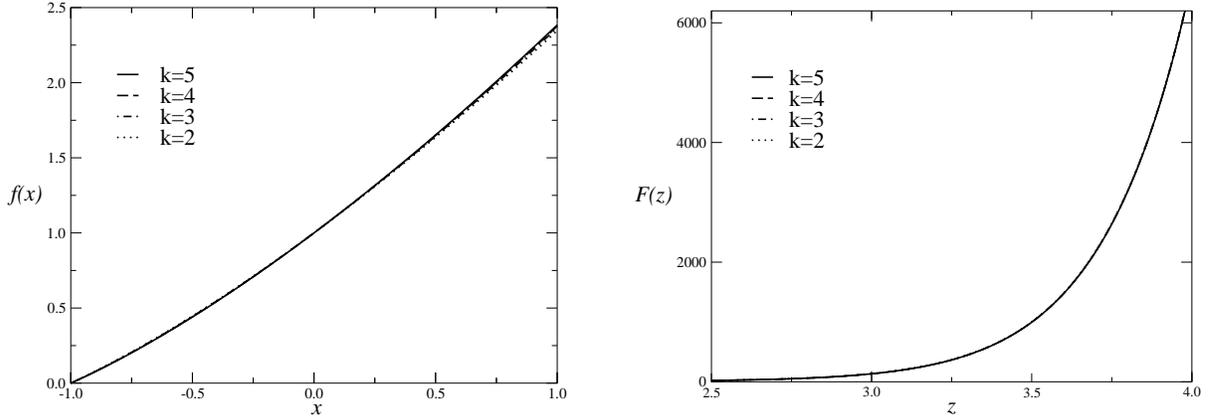

\hspace{0cm}
\vspace{0cm}
\begin{tabular}{cc} 
\hskip -0.5truecm
\psfig{width=7.5truecm,angle=0,file=fxI2d.eps} &
\hskip  0.5truecm
\psfig{width=7.5truecm,angle=0,file=FzI2d.eps} \\
\end{tabular}
\vspace{0cm}
\caption{
The scaling functions $f(x)$ and $F(z)$ as obtained
from the polynomial
approximations (\ref{parametric-2}) for $k=2,3,4,5$.
Results from Ref. \cite{CHPV-00-2}.
}
\label{figfxI2d}
\end{figure}

Finally,
we present the results of Ref. \cite{CHPV-00-2} for the equation of state.
In Table~\ref{trht2d}, for $k=2,3,4,5$, we report the polynomials $h(\theta)$
obtained  using the global stationarity condition 
\reff{globalstationarity} and the central values of the input parameters
$F^\infty_0$, $r_6$, $r_8$, $r_{10}$.
In  Fig.~\ref{figfxI2d} we show the scaling functions
$f(x)$ and $F(z)$, as obtained from ${h}(\theta)$ for $k=2,3,4,5$.
The convergence is satisfactory. The scaling function $F(z)$ is determined
with a relative uncertainty of at most a few per thousand 
in the whole region $z\geq 0$. 
The convergence is slower at the
coexistence curve, so that the error on the 
fuction $f(x)$ is of  a few per cent.

\begin{table*}
\caption{
Polynomial approximations of $h(\theta)$ 
obtained using the variational approach
for several values of the parameter $k$,
cf. Eq.~(\ref{parametric-2}).
The reported expressions correspond to the central values
of the input parameters. Results from Ref. \cite{CHPV-00-2}.
}
\label{trht2d}
\footnotesize
\begin{center}
\begin{tabular}{ccl}
\hline
\multicolumn{1}{c}{$k$}& 
\multicolumn{1}{c}{$\theta_0^2$}& 
\multicolumn{1}{c}{${h}(\theta)/[\theta (1 - \theta^2/\theta_0^2)]$ }\\ 
\hline  
2  &   1.15278 & $ 1 - 0.208408 \theta^2 $\\
3  &   1.15940 & $ 1 - 0.215675 \theta^2 - 0.039403 \theta^4 $\\
4  &   1.16441 & 
   $ 1 - 0.219388 \theta^2 - 0.041791 \theta^4  - 0.013488 \theta^6 $\\
5  &  1.16951 & 
   $ 1 - 0.222389 \theta^2 - 0.043547 \theta^4  - 0.014809 \theta^6  - 0.007168 \theta^8 $\\
\hline
\end{tabular}
\end{center}
\end{table*}

%% \begin{figure}[tb]
%% \hspace{0cm}
%% \vspace{0cm}
%% \centerline{\psfig{width=7.5truecm,angle=0,file=FzI2d.eps}}
%% \vspace{0cm}
%% \caption{
%% The scaling function $F(z)$ obtained
%% from the polynomial approximations (\ref{hexpn2}) with $k=2,3,4,5$.
%% Results from Ref. \cite{CHPV-00-2}.
%% }
%% \label{figFzI2d}
%% \end{figure}

\subsection{The two-point function of the order parameter}
\label{twopointf}

We shall discuss here the two-point function of the order parameter,
that is relevant in the description of scattering phenomena,
see Sec. \ref{sec3.5.3}. We mention that also the energy-energy
correlation function has been computed \cite{NA-97,CPV-02}.
It is relevant in the description of elastic deformations in fluids
and it can be measured via sound-attenuation techniques \cite{NA-97}.

We shall concentrate on the experimentally relevant case $H=0$. 
Results on the whole $(t,H)$ plane and on the critical isotherm 
can be found in Refs.~\cite{TF-75,CDK-74}.
For the two-dimensional case we mention that the large-distance
expansion of the two-point function on the critical isotherm,
i.e. for $t=0$ and $H\neq 0$, has been determined 
using the form-factor approach in Refs.~\cite{DM-95,DMS-96}.

\subsubsection{High-temperature phase} 
\label{sec-3.5.1}

As discussed in Sec.~\ref{sec-1.6.1}, the two-point correlation function
$\widetilde{G}(q)$ has the scaling form \reff{scaling-G-HT}. 
For $y\equiv q^2 \xi^2\to 0$, the function $g^+(y)$ has the 
expansion \reff{gypiu}.
In Table \ref{ci3d} we report the estimates of the
first few coefficients $c_n^+$,  obtained 
from the analysis of HT expansions, from the FT fixed-dimension expansion,
and from the $\epsilon$ expansion. There, we also report the invariant 
ratios $S_M^+$ and $S_Z^+$, see Eq. \reff{SMdef}, 
that parametrize the large-distance behavior of $G(x)$.

\begin{table*}
\caption{
Estimates of $c_{i}^+$, $S_M^+$, and $S_Z^+$  for the three-dimensional Ising 
universality class. Note $Q^+_\xi = \left(S_M^+\right)^{-1/2}$.
}
\label{ci3d}
\footnotesize
\begin{center}
\begin{tabular}{llll}
\hline
\multicolumn{1}{c}{}& 
\multicolumn{1}{c}{HT}& 
\multicolumn{1}{c}{$\epsilon$ exp}& 
\multicolumn{1}{c}{$d=3$ exp}\\ 
\hline  
$c_2^+$  
& $-$3.90(6)$\times 10^{-4}$ \cite{CPRV-02} & $-$3.3(2) $\times
10^{-4}$ \cite{CPRV-99} & 
$-$4.0(5) $\times 10^{-4}$ \cite{CPRV-98}  \\
& $-$3.0(2)$\times 10^{-4}$ \cite{CPRV-98}& &  \\
& $-$5.5(1.5)$\times 10^{-4}$, $-$7.1(1.5)$\times 10^{-4}$ \cite{TF-75} & & \\
$c_3^+$ 
& 0.882(6)$\times 10^{-5}$ \cite{CPRV-02} &  0.7(1) $\times 10^{-5}$ \cite{CPRV-99} & 1.3(3) $\times 10^{-5}$ \cite{CPRV-98} \\ 
& 1.0(1)$\times 10^{-5}$ \cite{CPRV-98} & &  \\ 
& 0.5(2)$\times 10^{-5}$, 0.9(3)$\times 10^{-5}$ \cite{TF-75} & & \\
$c_4^+$ 
& $-0.4(1) \times 10^{-6}$ \cite{CPRV-02} & 
$-$0.3(1)$\times 10^{-6}$ \cite{CPRV-99} & $-$0.6(2)$\times 10^{-6}$\cite{CPRV-98} \\ 
$S_M^+$ & 0.999601(6) \cite{CPRV-02} & 0.99968(4) \cite{CPRV-99} & 0.99959(6) \cite{CPRV-98}  \\ 
& 0.99975(10) \cite{CPRV-98} & &    \\ 
$S_Z^+$ & 1.000810(13)\cite{CPRV-02} & & \\
\hline
\end{tabular}
\end{center}
\end{table*}

The coefficients $c_n^+$ show the pattern
\begin{equation}
|c_n^+|\ll |c_{n-1}^+|\ll...\ll |c_2^+| \ll 1
\label{patternci}
\end{equation}
for $n\geq 3$.
This is in agreement with the
theoretical expectation that the singularity of $g^+(y)$ nearest to the
origin is the three-particle cut \cite{FS-75,Bray-76}.
If this is the case, the convergence radius $r_g$ of the Taylor
expansion of $g^+(y)$ is $r_g=9S_M^+$.  Since $S_M^+\approx 1$, 
at least asymptotically we should have
\begin{equation}
c_{n+1}^+\approx -{1\over 9}c_n^+.
\label{pattern-cip1-ci}
\end{equation}
This behavior was checked explicitly in the large-$N$ limit of the 
$N$-vector model \cite{CPRV-98}.
In two dimensions, the critical two-point function can be written
in terms of the solutions of a Painlev\'e differential equation
\cite{WMTB-76} and it can be verified explicitly that $r_g=9S_M^+$.  
In Table \ref{ci2d} we report the
values of $c_i^+$ for the two-dimensional Ising model. They are
taken from Refs. \cite{TMC-75,CPRV-99}. 

For large $y$ the function $g^+(y)$ follows the Fisher-Langer law 
\reff{FL-law}. The coefficients $A^+_n$ have been computed to three 
loops in Ref. \cite{Bray-76}. In three dimensions one obtains the 
estimates
$A_1^+ \approx  0.92$, $A_2^+ \approx  1.8$, and
$A_3^+ \approx  -2.7$.
In two dimensions the Fisher-Langer law must be modified since $\alpha = 0$. 
In this case, for large values of $y$, $g^+(y)$ behaves as
\begin{equation}
g^+(y)^{-1} \approx {A^+_1\over y^{7/8}}
  \left(1 + {A^+_2\over y^{1/2}}\, \log y +
            {A^+_3\over y^{1/2}}\right),
\label{FL-law-Ising2d}
\end{equation}
where the coefficients are \cite{TMC-75}  $A_1^+ \approx 0.413840$,
$A_2^+ \approx 0.802998$, and  $A_3^+ \approx 0.395345$. 

In the years, several parametrizations of the scaling function $g^+(y)$ 
have been proposed, see, e.g., Refs. \cite{FB-67,TF-75,TMC-75,Bray-76}. 
The most successful approximation is the one  proposed by Bray 
\cite{Bray-76}. It is based on a dispersive approach
\cite{FS-75,FB-79} and, by definition, it has the correct large-$y$ 
behavior \reff{FL-law} and has the pattern
\reff{pattern-cip1-ci} built in. In this approach one fixes the 
values of the exponents and of the sum $A_2^++A_3^+$ and determines 
an approximation of $g^+(y)$. The accuracy of the results can be 
evaluated by comparing the predictions for $c^+_n$ and for 
the coefficients $A_i^+$ with those obtained above.
Using Bray's parametrization one obtains
$A_1^+ \approx 0.918$, $A_2^+ \approx 2.55$,  
$A_3^+ \approx - 3.45$,
$c_2^+ \approx - 4.2\times 10^{-4}$, and
$c_3^+ \approx 1.0 \times 10^{-5}$.
These estimates  are in reasonable agreement with those 
reported in Table \ref{ci3d}
and with the $\epsilon$-expansion results for $A^+_n$.

Bray's approach was also applied in two dimensions. A slightly different
approximation that makes use of the high-precision results for 
$A_1^+$, $A_2^+$, and  
$A_3^+$ reproduces the results of Ref. \cite{TMC-75} with a 
maximum error of 0.03\%. 

\begin{table}[tb]
\caption{
Values of $c_i^\pm$ and $S_M^\pm$ for the two-dimensional Ising universality
class.  Results from Refs.~\protect\cite{TMC-75,CPRV-99}.
}
\label{ci2d}
\footnotesize
\begin{center}
\begin{tabular}{ll}
\hline
\multicolumn{1}{c}{HT phase}& 
\multicolumn{1}{c}{LT phase}\\ 
\hline
$S_M^+ =    0.999196337056$               &  $S_M^- = 0.399623590999$  \\
$c_2^+ =-0.7936796064\times 10^{-3}$ &  $c_2^- =-0.42989191603$ \\
$c_3^+ =    0.109599108\times 10^{-4}$  &  $c_3^- = 0.5256121845$    \\
$c_4^+ = -0.3127446\times 10^{-6}$    &  $c_4^- =-0.8154613925$  \\
$c_5^+ =    0.126670 \times 10^{-7}$    &  $c_5^- = 1.422603449$     \\
$c_6^+ = -0.62997\times 10^{-9}$      &  $c_6^- =-2.663354573$   \\
\hline
\end{tabular}
\end{center}
\end{table}

The three-dimensional correlation function was studied in
Ref. \cite{MPV-02} by means of a MC simulation. The function
$g^+(y)$ was determined with 0.5\% (resp. 1\%) precision 
up to $q\xi\approx 5$ (resp. 30). The numerical results were used to determine
an interpolation that reproduces the MC results for $y$ small and has the 
Fisher-Langer behavior \reff{FL-law} behavior for $y\to\infty$.

\subsubsection{Low-temperature phase} 
\label{sec-3.5.2}

In the LT phase, one introduces a scaling function $g^-(y)$ that is 
defined as $g^+(y)$ in Eq. \reff{scaling-G-HT}.
For $y\to 0$, also $g^-(y)$
admits a regular expansion of the form (\ref{gypiu}) with 
different coefficients $c_n^-$. With respect to the HT case, 
for $y$ small the deviations from the Gaussian (Ornstein-Zernike) behavior 
are larger. 
From the $\epsilon$ expansion at two loops, Ref. \cite{CDK-74} obtains
$c_2^- \approx -2.4 \times 10^{-2}$ and
$c_3^- \approx 3.9 \times 10^{-3}$, 
in reasonable agreement with the series estimates of Ref. \cite{TF-75}:
$c_2^- \approx -1.2(6) \times 10^{-2}$ and
$c_3^- \approx 7(3) \times 10^{-3}$.
The larger deviations from the Gaussian behavior are confirmed by the
estimates of $S_M^-$:
$S_M^-=0.938(8)$ \cite{CPRV-99} and
$S_M^-=0.930(6)$ \cite{FZ-98} from the analysis of the LT expansion,
and $S_M^-=0.941(11)$ \cite{CHP-99,ACCH-97}
from MC simulations.
Such a different behavior is probably related to the different 
analytic structure of the two-point function in the LT phase. Indeed,
perturbative arguments indicate the presence of a two-particle cut
in the LT phase \cite{CDK-74,FS-75,Bray-76}.
Thus, the convergence radius of the small-$y$ expansion is 
expected to be at most $4S_M^-$, and asymptotically
$c_{n+1}^-\approx - 0.27\, c_n^-$.
For large values of $y$, $g^-(y)$ follows the Fisher-Langer law \reff{FL-law}
with different coefficients $A_n^-$. They can be derived from $A_n^+$ 
using Eq. \reff{relazioni_Aipiu_Aimeno}. These relations have been
checked in Ref. \cite{CDK-74} to two-loop order in the 
$\epsilon$ expansion. 

The mass spectrum of the model in the LT phase was investigated in 
Refs.~\cite{CHP-99,ACCH-97,Provero-98,CHPZ-00}
using numerical techniques. 
In particular, Ref.~\cite{CHP-99} reports MC results obtained from
simulations of the standard Ising model and of 
the improved $\phi^4$ lattice model \reff{latticephi4}
at $\lambda=1.10$ (see Sec. \ref{sec-2.3.2}), 
and provides evidence for a state with $M_2< 2M$,
where $M$ is the mass of the fundamental state, 
i.e. ${M_2/M} = 1.83(3)$,
that is below the pair-production threshold.
This second state should appear 
as a pole in the Fourier transform of the two-point function.

The two-dimensional Ising model shows even larger deviations from Eq.\ 
(\ref{gaubeh}), as one can see from the estimates of $S_M^-$ and $c_i^-$
reported in Table \ref{ci2d}.  Note that in the LT
phase of the two-dimensional Ising model the singularity at
$k^2=-1/\xi^2_{\rm gap}$ of $\widetilde{G}(k)$ is not a simple pole, but a
branch point\footnote{In the particle interpretation of Ref. \cite{WMTB-76},
this is due to the fact that the lowest propagating state in the LT
phase is a two-particle state.} \cite{WMTB-76}.  
As a consequence, the convergence radius of the
expansion around $y=0$ is $S_M^-$.  

For large values of $y$, $g^{-}(y)$ behaves according to Eq. 
\reff{FL-law-Ising2d}, with different coefficients $A_n^-$. 
They are given by \cite{TMC-75}: $A_1^- \approx 2.07993$, 
$A_2^- \approx -0.253913$, and $A_3^- \approx -0.709701$. 
Bray's approximation has been  also applied to the LT phase, see Ref. \cite{MPV-02}. 

\subsubsection{Experimental results} \label{sec3.5.3}

In scattering experiments one measures the scattering cross-section
\begin{equation}
{d^2\sigma\over dq d\omega}
\end{equation}
 where $q$ is the exchanged momentum vector and 
$\omega$ the corresponding frequency (energy).
This cross-section is proportional to the dynamic structure factor 
$S(q,\omega)$. In the critical limit, $S(q,\omega)$ is dominated by the 
elastic Rayleigh peak, whose width goes to zero as $t\to 0$. 
Thus, in this limit only elastic scattering is relevant and 
\be
{d\sigma\over dq} \propto \int d\omega S(q,\omega) = \widetilde{G}(q).
\ee
The momentum-transfer vector $q$ is related to the scattering angle 
$\theta$ by 
\begin{equation}
q = {4 \pi\over \lambda} \sin {\theta\over2},
\label{q-vs-theta}
\end{equation}
where $\lambda$ is the wavelength of the radiation (neutrons) in the
scattering medium. Note that scattering data can be directly related to 
$\widetilde{G}(q)$ only if multiple scattering can be neglected. 
See Ref.~\cite{Anisimov-book} for a discussion. 

Several experiments determined the scaling functions $g^\pm(y)$ in 
magnetic systems \cite{Belanger-00} and in fluids. 
In the HT phase, because of the smallness of the coefficients 
$c^{+}_n$, the Ornstein-Zernike approximation $g^+(y) \approx 1 + y$ 
can be used up to $y\approx 30$. For larger values of $y$ 
it is necessary to take into account the anomalous behavior
\cite{BBC-82,CBS-79,SBSWC-80,DLC-89,Izumi-89,JSS_92,SKK-96,LMBWH-97,%
BBB-97,BC-97-98,DLMFL-98,BRCB-00,BCB-01}. 
The large-momentum  behavior of $g^+(y)$ has been extensively studied.
In particular, the exponent $\eta$ and the constant $A_1^+$
have been determined:  
$\eta = 0.017(15)$, $A_1^+= 0.96(4)$ and
$\eta \approx 0.030(25)$, $A_1^+\approx 0.95(4)$ (two different
parametrizations of the structure factor are used)
\cite{CBS-79};
$\eta = 0.0300(15)$, $A_1^+\approx 0.92(1)$ \cite{DLC-89};
$\eta = 0.042(6)$, $A_1^+\approx 0.915(21)$ \cite{DLMFL-98}.
No unbiased determination of $A_2^+$ and $A_3^+$ is available.
Fixing $A_2^+ + A_3^+ = -0.9$ (the $\epsilon$-expansion result
of Ref. \cite{Bray-76}), Ref. \cite{DLMFL-98}
obtains $A_2^+ = 2.05(80)$ and $A_3^+ = - 2.95(80)$, in reasonable
agreement with the $\epsilon$-expansion predictions. 

A very precise determination of $g^+(y)$ was obtained in Ref. \cite{DLMFL-98}.
From the analysis of scattering data for CO$_2$, the function $g^+(y)$
was determined up to $y = 1600$. These experimental results are 
in good agreement with the theoretical determinations 
(within an accuracy of approximately 1\%).

\subsubsection{Turbidity} \label{sec3.5.4}
The turbidity $\tau$ is defined as the attenuation of the transmitted
light intensity per unit optical path length due to the scattering
with the sample. Explicitly, it is given by
\begin{equation}
\tau \sim \int d\Omega\, \widetilde{G}(q) \sin^2 \Phi,
\end{equation} 
where $\Omega$ is the scattering solid angle, $q$ is given by 
Eq. \reff{q-vs-theta}, and $\Phi$ is the angle 
between the polarization of the incoming radiation (or neutrons) 
and the scattering wave vector.

If $k_0 = 2 \pi n/\lambda$ is the
momentum of the incoming radiation in the medium, $\lambda$ the
corresponding wavelength in vacuum, $n$ the refractive index, for small
$k_0 \xi$ the Puglielli-Ford expression \cite{PF-70} can be used:
\begin{equation}
\tau_{\rm PF} = \tau_0 t^{-\gamma}
  \left[ {2 a^2 + 2 a + 1\over a^3} \log(2 a + 1) -
         {2 (a + 1)\over a^2}\right],
\end{equation}
where $a = 2 k_0^2 \xi^2$ and $\tau_0$ is a temperature-independent 
constant. Deviations are less than 1\% (resp. 3\%) for 
$k_0 \xi \ltapprox 15$ (resp. 90). An extensive discussion of the deviations 
from the Puglielli-Ford expression is given in Ref. \cite{MPV-02}. 
In particular, in the experimentally relevant interval $k_0 \xi \ltapprox 100$,
the turbidity can be computed using the expression \cite{MPV-02}
\begin{eqnarray}
\tau =
   \tau_{\rm PF} \left[
   0.666421 
    + 0.242399 \left(1 + 0.0087936 Q_0^2\right)^{0.018195} +
             0.0911801 \left(1 + 0.09 Q_0^4\right)^{0.0090975}\right],
\label{phen-turb}
\end{eqnarray}
where $Q_0 \equiv k_0\xi$.
Other results for the turbidity can be found in 
Refs.~\cite{Ferrell-91,CLL-72}. However, as discussed in Ref. \cite{MPV-02}, 
they predict a turbidity that is larger than Eq. \reff{phen-turb}, which is 
based on the 
most accurate approximations of the structure factor available today.

The turbidity $\tau$ is larger than $\tau_{\rm PF}$
since $g^+(y)$ increases slower for $y\to \infty$ than the
Ornstein-Zernike approximation. However, this is apparently in contrast
with the experimental results for the binary fluid mixture
methanol-cyclohexane presented in Ref. \cite{JLMW-99}.

\section{The three-dimensional $XY$ universality class}
\label{XY}

\subsection{Physical relevance}
\label{XYd3}

The three-dimensional $XY$ universality class is characterized by 
a two-component order parameter and 
effective short-range interactions with U(1) symmetry. 
The most interesting representative
of this universality class is the
superfluid transition of $^4$He along the $\lambda$-line $T_\lambda(P)$.
It provides an exceptional opportunity for a very accurate experimental
test of the RG predictions, 
because of the weakness of the singularity in the compressibility of 
the fluid, of the purity of the samples, and of the possibility
of performing experiments 
in a microgravity environment, for instance on the Space Shuttle as the 
experiment reported in Ref. \cite{LSNCI-96}, thereby achieving a 
significant reduction of the
gravity-induced broadening of the transition.  
Exploiting these favorable conditions, the specific heat of liquid helium
was measured 
to within a few nK from the $\lambda$-transition \cite{LSNCI-96},
i.e. very deep in the critical region, where the scaling corrections 
are small. 
Ref. \cite{LSNCI-96} obtained\footnote{
Ref. \cite{LSNCI-96} reported $\alpha=-0.01285(38)$ and $A^+/A^-=1.054(1)$.
But, as mentioned in footnote [15] of Ref.~\cite{Lipa-etal-00}, 
the original analysis
was slightly in error. Ref.~\cite{Lipa-etal-00} reports the new estimates
$\alpha=-0.01056$ and $A^+/A^-=1.0442$. The error reported here is 
a private communication of J. A. Lipa, quoted in Ref. \cite{CHPRV-01}.}
the very precise estimate $\alpha = - 0.01056(38)$.
This result represents a challenge for theorists. Only recently have
the theoretical estimates reached a comparable  accuracy.

Beside $^4$He, there are many other systems that undergo an $XY$ transition. 
First of all, one should mention ferromagnets or antiferromagnets with 
easy-plane anisotropy, which is the original characterization of the 
$XY$ universality class. 

An $XY$ behavior is observed in systems that exhibit phase transitions
characterized by the establishement of a density wave. Indeed, 
the order parameter of density waves in a uniaxial system
is the complex amplitude $\phi_1$, associated with the
contribution 
${\rm Re} \, \phi_1 e^{iq_0 z}$
to the density modulation, where $q_0$ is the wavelength of the
modulation. Interesting
examples in solids are  charge-density wave (see,
e.g., Refs.~\cite{FMAB-84,Girault-etal-89}) and spin-density wave
systems  (see, e.g., Ref.  \cite{Fawcett-88}). 
Similar phenomena occur in liquid crystals,
in which several transitions are expected 
to belong to the $XY$ universality class
\cite{DeGennes_73,NH-80,BN-81,LB-82,Anisimov-book,Singh-00}.
We should mention the 
nematic--smectic-A phase transition, that
corresponds to the establishement of a one-dimensional
mass-density wave along the direction of the orientational order,
although experiments have found a wide range of effective exponents 
that are often quite different from the $XY$ predictions, see
Refs.~\cite{Garland-etal-93,GN-94} and references therein. 
The smectic-A--hexatic-B transition should by either $XY$ 
or first-order \cite{ABBL-86}; again experimental results are 
contradictory, see Ref. \cite{KG-98} and references therein.
The same behavior is expected for the smectic-A--smectic-C and the 
smectic-A--chiral-smectic-C transitions \cite{DeGennes_73}; in this case
it is found experimentally 
that the $XY$ window is very small and one usually observes 
a crossover from mean-field to $XY$ critical behavior,
see, e.g., Refs. \cite{EWTY-95,EY-98} and references therein. 
Finally, we should mention the nematic-to-lamellar phase transition,
which is similar to the nematic--smectic-A transition
\cite{SBSLK-98}.

$XY$ criticality is expected in materials that undergo a 
phase transition from a normal (disordered) HT phase to a LT incommensurate
modulated phase in one direction \cite{CB-78}. Such a transition is expected in
some rare-earth metals like Er and Tm that are longitudinally modulated.
The experimental evidence is however quite controversial, see
Refs. \cite{HCFBZ-92,LCH-93,HHTG-95}. A similar transition 
is observed in some insulating crystals of type 
A$_2$BX$_4$ \cite{Cummins-90}, 
where $A^+$ is a monovalent cation like K$^+$ or Rb$^+$, and 
BX$_4^{--}$ is a divalent tetrahedral anion like 
ZnCl$_4^{--}$ or ZnBr$_4^{--}$.

The $XY$ model is relevant for superconductors, as long 
as one is able to neglect the fluctuations of the magnetic potential.
We mention that an inverted $XY$-scaling scenario is invoked 
in the description of superconductors in the extreme type-II region,
where the transition is expected to be of second order,
see, e.g., Ref.~\cite{MHS-01}. 
The idea is to use duality arguments to map the Ginzburg-Landau model 
with a U(1) gauge field with temperature parameter $\tau$
into an XY model with inverted temperature $-\tau$ \cite{DH-81,KKS-94}.
High-temperature superconductors for small magnetic fields are also
found to show $XY$ behavior \cite{Schneider-02} 
(for a different point of view, see
Refs. \cite{RJM-95,JRRE-00,CPGSB-01}), both for the statics and the dynamics, 
see, e.g., Refs.
\cite{OHL-94,BC-00a,Krylov-00,KYCLPP-00,NINSK-00,HER-00,RRPP-01,
RRCPP-01} and references therein.

The Peierls transition in CuGeO$_3$ and in some organic materials
has been identified with an $XY$ transition. Indeed, the latest 
intensity measurements give estimates of $\beta$ that are in good agreement
with the theoretical predictions, see Refs. \cite{LGD-98,LG-99} and 
references therein. On the other hand, scattering experiments 
either do not observe critical scattering or observe anomalous line shapes 
with exponents $\gamma$ and $\nu$ much larger than expected, see, e.g., 
Refs. \cite{HFBHSHU-95,LG-99} and references therein.

\subsection{The critical exponents}
\label{expXY}

\subsubsection{Theoretical results}

In Table \ref{XYexponents} we report the 
theoretical estimates of the critical exponents. 

\begin{table*}[t]
\caption{Estimates of the critical exponents for the three-dimensional $XY$ universality class.
We indicate with an asterisk (${}^*$) the estimates we
obtained using the hyperscaling relation $2 - \alpha = 3 \nu $
or the scaling relation $\gamma =(2 - \eta)\nu $.
When the error was not reported by the authors, we used 
the independent-error formula to estimate it.
}
\label{XYexponents}
\footnotesize
%\hspace*{-2.7cm}    % Move table leftwards, so it doesn't run off the right
\tabcolsep 2.5pt        % Less than the usual 6pt
%\doublerulesep 1.5pt  % Less than the usual 2pt
\begin{center}
\begin{tabular}{rlllllll}
\hline
\multicolumn{2}{c}{Ref.}& 
\multicolumn{1}{c}{info}& 
\multicolumn{1}{c}{$\gamma$}& 
\multicolumn{1}{c}{$\nu$}& 
\multicolumn{1}{c}{$\eta$}&
\multicolumn{1}{c}{$\alpha$}& 
\multicolumn{1}{c}{$\omega$} \\   
\hline  
\cite{CHPRV-01} & $_{2001}$ & MC+IHT, $\phi^4$, dd$XY$ 
&1.3177(5)   & 0.67155(27)     & 0.0380(4) & $-$0.0146(8)$^*$ & \\ 

\cite{CPRV-00} & $_{2000}$  &  IHT, $\phi^4$ &1.3179(11)    & 0.67166(55)     & 0.0381(3) & $-$0.0150(17)$^*$ & \\ 
\cite{BC-99} & $_{1999}$ & HT, $XY$ sc     & 
   & 0.671(3)$^*$    & & $-$0.014(9) & \\
\cite{BC-99} & $_{1999}$ & HT, $XY$ bcc   & 
   & 0.674(2)$^*$    & & $-$0.022(6) & \\
\cite{BC-97-2} & $_{1997}$ & HT, $XY$ sc     & 
1.325(3)    & 0.675(2)    & 0.037(7)$^*$ & $-$0.025(6)$^*$ & \\
\cite{BC-97-2} & $_{1997}$ & HT, $XY$ bcc   & 
1.322(3)    & 0.674(2)        & 0.039(7)$^*$ & $-$0.022(6)$^*$ & \\
\cite{BCG-93} & $_{1993}$ & HT, $XY$ sc   & 1.315(9)    & 0.68(1)        & 0.07(3)$^*$ & $-$0.04(3)$^*$ & \\
\cite{FMW-73} & $_{1973}$ & HT $XY$, easy-plane & 1.318(10)    & 0.670(6)  & 0.04(1)$^*$  & $-$0.02(3)$^*$& \\

\cite{DMC-01} & $_{2002}$ & MC FSS $XY$ & & & 0.037(2)   & & \\

\cite{CHPRV-01} & $_{2001}$ & MC FSS, $\phi^4$, dd$XY$ & 1.3177(10)$^*$&
0.6716(5)   & 0.0380(5)   & $-$0.0148(15)$^*$ & 0.795(9) \\

\cite{HT-99} & $_{1999}$    & MC FSS, $\phi^4$ & 1.3190(24)$^*$& 0.6723(11)  & 0.0381(4)   & $-$0.0169(33)$^*$ & 0.79(2) \\
\cite{KL-99} &  $_{1999}$   & MC FSS, easy-plane& 1.315(12)$^*$ & 0.6693(58)  & 0.035(5)    &  $-$0.008(17)$^*$ & \\
\cite{NM-99} &  $_{1999}$   & MC FSS, easy-plane& 1.320(14)$^*$ & 0.670(7)  & 0.0304(37)    &  $-$0.010(21)$^*$ & \\
\cite{BFMM-96} & $_{1996}$  & MC FSS, $XY$ & 1.316(3)$^*$  & 0.6721(13)  & 0.0424(25)    &   $-$0.0163(39)$^*$ & \\
\cite{SM-95} & $_{1995}$    & MC FSS, $XY$ &               & 0.6724(17)  & &  $-$0.017(5)$^*$ & \\ 
\cite{GH-94,GH-94-2} & $_{1994}$    & MC FSS, AF Potts & 1.310(10)$^*$  & 0.664(4) & 0.027(9) &  +0.008(12)$^*$ & \\
\cite{GH-93} & $_{1993}$    & MC FSS, $XY$ & 1.307(14)$^*$ & 0.662(7)    & 0.026(6)    &  +0.014(21)$^*$ & \\
\cite{Janke-90} & $_{1990}$ & MC FSS, S, $XY$& 1.316(5)      & 0.670(2)    & 0.036(14)$^*$  & $-$0.010(6)$^*$ & \\

\cite{JK-00} & $_{2001}$ & FT $d=3$ exp & 1.3164(8) & 0.6704(7) & 0.0349(8) & $-$0.0112(21)  & 0.784(3) \\
\cite{GZ-98} & $_{1998}$ & FT $d=3$ exp & 1.3169(20)& 0.6703(15) & 0.0354(25) & $-$0.011(4) & 0.789(11) \\
\cite{MN-91} & $_{1991}$ & FT $d=3$ exp & 1.3178(10)\{28\} & 
0.6715(7)\{17\} & 0.0377(6)\{7\} & $-$0.0145(21)\{51\}  &  \\
\cite{LZ-77} & $_{1977}$ & FT $d=3$ exp & 1.3160(25) & 0.669(2) & 0.033(4) & $-$0.007(6) & 0.780(27) \\
\cite{GZ-98} & $_{1998}$ & FT $\epsilon$ exp & 1.3110(70) & 0.6680(35) & 0.0380(50) & $-$0.004(11) & 0.802(18) \\

\cite{NR-84} & $_{1984}$ & SFM  & 1.31(2)   &  0.672(15)   & 0.043(7) &$-$0.016(45)$^*$ & 0.85(7) \\

\cite{GW-01,BTW-99} &  $_{2001}$      & CRG (1st DE)&   1.299 &  0.666  & 0.049 & +0.002 &  \\
\cite{MT-98}  &  $_{1998}$      & CRG (1st DE) &   1.27 &  0.65   & 0.044 & +0.05  &  \\
\cite{TW-94}  &  $_{1994}$      & CRG ILPA &   1.371 &  0.700   & 0.042 & $-$0.100 &  \\
\hline
\end{tabular}
\end{center}
\end{table*}

Accurate results for the critical exponents
have been obtained by combining 
MC simulations based on FSS techniques  and 
HT expansions for improved Hamiltonians \cite{CHPRV-01,CPRV-00,HT-99}.
On the one hand,  
one exploits the effectiveness of FSS MC simulations to determine the
critical temperature and the parameters of 
the improved Hamiltonians \cite{CHPRV-01,HT-99}.
On the other hand, using this information,
one exploits the effectiveness of IHT to
determine the critical exponents \cite{CHPRV-01,CPRV-00}, 
especially  when a precise estimate of $\beta_c$ is available. 
Two improved Hamiltonians were considered in Ref.~\cite{CHPRV-01}, the 
lattice $\phi^4$ model \reff{latticephi4}  for $\lambda^* = 2.07$,  
and the dynamically dilute $XY$ model \reff{ddXY} (dd$XY$) for 
$D^* = 1.02$, cf. Sec.~\ref{sec-2.3.2}.
An accurate MC study \cite{CHPRV-01} employing FSS techniques
provided estimates of $\lambda^*$ and $D^*$, of 
the inverse critical temperature $\beta_c$ for 
several values of $\lambda$ and $D$,
and estimates of the critical exponents (see Table~\ref{XYexponents}).
Using the linked-cluster expansion technique, the HT expansions of 
$\chi$ and $\mu_2 = \sum_x |x|^2 G(x)$
were computed to 20th order for these two Hamiltonians. 
The analyses were performed  using the estimates of $\lambda^*$,
$D^*$, and $\beta_c$ obtained from the MC simulations. The 
results are denoted by MC+IHT in Table~\ref{XYexponents}. 
The critical exponent $\alpha$ was derived using 
the hyperscaling relation $\alpha=2 - 3\nu$, obtaining 
$\alpha = - 0.0146(8)$ \cite{CHPRV-01}. 

The HT results of Refs.~\cite{BC-97-2,BC-99} were obtained by analyzing
21st-order HT expansions for the standard $XY$ model on 
the simple (sc) and body-centered (bcc) cubic lattices. 
To take into account the subleading corrections, they employed
approximants biased with the MC estimate of  $\beta_c$ and 
with the FT result for $\Delta$.

Most MC results reported in Table~\ref{XYexponents} have been
obtained using FSS techniques. Only Ref. \cite{Janke-90} determines 
the critical exponents from the behavior of infinite-volume quantities 
near the critical point (``S" in the column info in Table \ref{XYexponents}).
Refs.~\cite{BFMM-96,SM-95,GH-93,Janke-90} present results for 
the standard $XY$ model, Refs.~\cite{KL-99,NM-99} for a 
classical ferromagnetic XXZ model with no coupling for $s_z^2$ (this 
is the model that in the old literature was called $XY$ model; in 
Table~\ref{XYexponents} we refer to it as ``easy-plane"), 
and Ref.~\cite{GH-94} for the
three-state antiferromagnetic Potts model on a simple cubic lattice
(AF Potts) that has been conjectured 
\cite{BGJ-80} to be
in the $XY$ universality class.\footnote{Actually, the 
authors of Ref. \cite{BGJ-80} argued,
using RG arguments, that 
the effective Hamiltonian for the HT transition of the three-state
antiferromagnetic Potts model on a simple cubic lattice 
is in the same universality class of the two-component 
$\phi^4$ theory with cubic anisotropy.
As we shall discuss in Sec.~\ref{lsec-cubic}, 
in the two-component case, the stable fixed point of the cubic Hamiltonian
is the $O(2)$ symmetric one. Therefore, the HT continuous transition of
the three-state antiferromagnetic Potts model 
belongs to the $XY$ universality class.
We mention that other transitions are expected
for lower values of the temperature (see, e.g., Ref.\cite{RRS-98}
and references therein).
}

Refs.~\cite{GZ-98,JK-00,MN-91,LZ-77} report FT results obtained by
analyzing the fixed-dimension expansion. 
The perturbative series of the $\beta$-function and of the exponents 
are known to six-loop 
\cite{BNGM-77} and seven-loop order \cite{MN-91} respectively.
In Refs.~\cite{GZ-98,LZ-77} the resummation is  performed by
using the method presented in Sec. \ref{sec-2.4.3}, based on a Borel 
transform and a conformal mapping that makes use of the large-order 
behavior of the series.
Ref.~\cite{JK-00} (see also Refs.~\cite{Kleinert-98,KS-01}) 
employs a resummation method based on a variational technique:
as in the Ising case, the errors seem to be rather optimistic,
especially for $\omega$. Using the same method, 
Ref.~\cite{KS-01} reports the estimate $\alpha=-0.01126(10)$.
The analysis of Ref.\cite{MN-91} allows for a more general
nonanalytic behavior of the $\beta$-function.
In Table \ref{XYexponents}, we quote two errors for the results of Ref.\ 
\cite{MN-91}: the first one (in parentheses) 
is the resummation error, the second one (in braces) 
takes into account the uncertainty of $g^*$, which is estimated to
be approximately $1$\%. To estimate the second error we used the results of
Ref.~\cite{GZ-98} where the dependence of the exponents on $g^*$ is given.
Consistent results are also obtained from the analysis of Ref. 
\cite{GZ-98} of the 
$O(\epsilon^5)$ series computed in the framework
of the $\epsilon$ expansion~\cite{CGLT-83,KNSCL-93}.
In Table~\ref{XYexponents} we also report results
obtained by approximately solving
continuous RG (CRG) equations, to the lowest (ILPA) 
and first order (1st DE) of the derivative expansion \cite{TW-94,GW-01,MT-98}.
The agreement among the theoretical calculations is overall good.

There also exist estimates of the crossover exponents associated with 
the spin-$n$ operators, see Sec.~\ref{sec-1.5.8}.
The crossover exponent $\phi_2$ associated with 
the spin-two tensor field
describes the instability of the O(2)-symmetric 
theory against anisotropy \cite{FP-72,Wegner-72,FN-74,Aharony-76}. 
It is thus relevant for the description of 
multicritical phenomena, for instance the critical behavior near 
a bicritical point where two critical Ising lines meet, 
giving rise to a critical theory with enlarged
$O(2)$ symmetry, see, e.g., Refs. \cite{PJF-74,Fisher-75,KNF-76}. 
The exponent $\phi_2$ has been determined 
using various approaches, obtaining
$\phi_2 = 1.184(12)$ by the analysis of the six-loop expansion
in the framework of the FT fixed-dimension expansion
\cite{CPV-02}; 
$\phi_2\approx 1.15$ 
by setting $\epsilon=1$ in the corresponding $O(\epsilon^3)$ 
series \cite{Yamazaki-74};
$\phi_2=1.175(15)$ by HT expansion
techniques \cite{PJF-74}.
Correspondingly, $\beta_2 = 2 - \alpha - \phi_2 = 0.831(12)$, 0.86, 
0.840(15). 
The exponent $\phi_4$ can be computed from the theoretical results 
for $O(N)$ models with a cubic-symmetric perturbation, see
Sec. \ref{lsec-cubic}. Using the results  of Ref. \cite{CPV-00}, 
we obtain $\phi_4 = -0.069(5)$, $\beta_4 = 2.084(5)$. 
For generic values of $n$, Ref. \cite{ABBL-86} found
\be
\beta_n \approx \beta n + {1\over 2} \nu x_n n(n-1),
\label{betan-theor}
\ee
where $x_n \approx 0.3 - 0.008 n$, using
a two-loop calculation in the fixed-dimension expansion.
For comparison, note that 
Eq. \reff{betan-theor} gives
$\beta_2 \approx 0.89$, $\beta_4 \approx 2.5$ (we use the estimates of 
$\beta$ and $\nu$ of Ref. \cite{CHPRV-01})
to be compared with the above-reported results.
For $n=3$ it gives $\beta_3 \approx 1.60$, so that $\phi_3 \approx 0.41$.
Note that only the spin-2 and the spin-3 operators are relevant perturbations. 
Higher-spin perturbations do not change the critical theory.

We also mention Refs. \cite{NA-97,CPV-02}, 
where the two-point correlation function of the spin-two 
operator was computed.

\subsubsection{Experimental results}

\begin{table*}[tbp]
\caption{Experimental estimates of the critical exponents for 
the three-dimensional $XY$ universality class. 
Here BCPS stands for bis(4-chlorophenyl)sulfone, 7APCBB for 
4'-$n$-heptyloxycarbonylphenyl-4'-(4''-cyanobenzoyloxy)-benzoate.
The exponent $\zeta$ is given by $\zeta \equiv 2 \gamma - 3 \nu$.
We indicate with an asterisk (${}^*$) the estimates we
obtained using the hyperscaling relation $2 - \alpha = 3 \nu $.
}
\label{XYexponents-expt}
\footnotesize
%\hspace*{-2.7cm}    % Move table leftwards, so it doesn't run off the right
\tabcolsep 2.5pt        % Less than the usual 6pt
%\doublerulesep 1.5pt  % Less than the usual 2pt
\begin{center}
\begin{tabular}{rrllllll}
\hline
\multicolumn{2}{c}{Ref.}& 
\multicolumn{1}{c}{Material}& 
\multicolumn{1}{c}{$\gamma$}& 
\multicolumn{1}{c}{$\nu$}& 
\multicolumn{1}{c}{$\alpha$}& 
\multicolumn{1}{c}{$\beta$}& 
\multicolumn{1}{c}{$\zeta$}\\
\hline
\cite{AL-00} &  $_{2000}$  & ${}^4$He &  
     & 0.66758(6)   & $-$0.00274(18)$^*$ & & \\
\cite{LSNCI-96,Lipa-etal-00} &  $_{1996}$  & ${}^4$He &  
     & 0.67019(13)$^*$  & $-$0.01056(38) & & \\
\cite{GMA-93} &  $_{1993}$   & ${}^4$He  & 
     & 0.6705(6)       &    $-$0.0115(18)$^*$ & &\\
\cite{Swanson-etal-92} &  $_{1992}$ & ${}^4$He     & 
     & 0.6708(4)       &    $-$0.0124(12)$^*$ & &\\
\cite{SA-84} &  $_{1984}$ & ${}^4$He     &   
     & 0.6717(4)       &    $-$0.0151(12) & & \\
\cite{LC-83} &  $_{1983}$ & ${}^4$He     &  
     & 0.6709(9)$^*$   & $-$0.0127(26) & & \\

\cite{RKS-95b} & $_{1995}$ & Gd$_2$IFe$_2$ &
  1.320(65) & & & 0.347(17) & \\
\cite{RKS-95b} & $_{1995}$ & Gd$_2$ICo$_2$ &
  1.315(65) & & & 0.345(17) & \\
\cite{RKS-95b} & $_{1995}$ & Gd$_2$BrFe$_2$ &
  1.316(65) & & & 0.345(17) & \\

\cite{Wu-etal-94} & $_{1994}$ & 7APCBB &
  1.34(14) & & & & \\

\cite{DPIM-02} & $_{2002}$ &  Rb$_2$ZnBr$_4$ &
  1.317(30) & & & & 0.64(1) \\
\cite{SII-00} & $_{2000}$ &  Cs$_2$HgCl$_4$ &
  & & & &  0.615(25) \\
\cite{DP-00}  & $_{2000}$ & BCPS &
  & & & &  0.69(2) \\
\cite{SII-99} & $_{1999}$ & Cs$_2$CdBr$_4$ &
  & & & &  0.62(2) \\
\cite{SII-99} & $_{1999}$ & Cs$_2$HgBr$_4$ &
  & & & &  0.50(2) \\
\cite{KKY-98} & $_{1998}$ & Rb$_2$ZnCl$_4$ & 
  & & & 0.36(1) & \\
\cite{KKY-98} & $_{1998}$ & K$_2$ZnCl$_4$ & 
  & & & 0.375(10) & \\
\cite{KKY-98} & $_{1998}$ & (NH$_4$)$_2$ZnCl$_4$ & 
  & & & 0.365(10) & \\
\cite{KKY-98} & $_{1998}$ & [N(CH$_3$)$_4$]$_2$ZnCl$_4$ & 
  & & & 0.365(10) & \\
\cite{Zinkin-etal-96} & $_{1996}$ & Rb$_2$ZnCl$_4$ & 
     1.28(9) & 0.66(2) & +0.02(6)$^*$ & & \\
\cite{AM-83} &  $_{1983}$ & Rb$_2$ZnCl$_4$ &
1.26$^{+0.04}_{-0.02}$ & 0.683(15) & $-$0.049(45)$^*$ & & \\
\hline
\end{tabular}
\end{center}
\end{table*}

In Table \ref{XYexponents-expt} we report some experimental results for 
systems that are supposed to belong to the $XY$ universality class. 
They should be compared with the theoretical results of 
Table \ref{XYexponents}. Note that, using the theoretical results of 
Ref. \cite{CHPRV-01}, one obtains $\beta = 0.3485(2)$ and 
$\zeta\equiv 2\gamma-3\nu = 0.6205(6)$.  Table \ref{XYexponents-expt} 
is not a complete list, but it should give an idea of the quality
of the results. 
The most accurate results have been obtained from 
the $\lambda$-transition of $^4$He.
In particular, the  estimate of $\alpha$ reported 
in Refs. \cite{LSNCI-96,Lipa-etal-00} appears very precise.  
It was obtained by 
measuring the specific heat in the LT phase up to a few nK from 
the $\lambda$-transition,
and by fitting the data to the RG behavior 
\begin{equation}
C_H(t) = A |t|^{-\alpha}\left( 1 + C |t|^\Delta + D t \right) + B,
\label{Chscaling}
\end{equation} 
where $t\equiv (T-T_c)/T_c\to 0^-$ and $\Delta$ was fixed equal to $1/2$.
In this respect,
it should be noticed that, due to the small value of $\alpha$, a fit
to Eq.~(\ref{Chscaling}) 
requires very accurate data for very small $t$; otherwise, it is 
very difficult to distinguish the 
nonanalytic term from the analytic background. 
Note that the  estimate of $\alpha$ reported in 
Refs.~\cite{LSNCI-96,Lipa-etal-00}
does not agree with the comparably precise
theoretical estimates of Ref. \cite{CHPRV-01}. 
It is not clear whether this disagreement is significant,
or it is due to an underestimate of the experimental and/or 
theoretical errors.
The recent experimental estimate  of $\nu$ reported in
Ref.~\cite{AL-00}, determined from a measurement of the second sound, 
does not help to clarify this issue, because the quoted error does not
include the systematic effects due to satellite modes and the uncertainty in the 
temperature scale calibration, which are expected to be much larger.
Therefore, the situation  calls for further theoretical and experimental
investigations. 
A new generation of experiments in microgravity environment
that is currently in preparation
\cite{Nissen_etal-99} should clarify the issue from the experimental side.

Estimates of the critical exponents in other systems are not very precise. 
It is difficult to measure $\nu$ in liquid crystals.
Indeed, the intrinsic anisotropy of these systems gives rise 
to strong anisotropic scaling 
corrections. As a consequence, the effective exponents $\nu$, 
that are obtained by fitting the
correlation length in different directions,  are apparently different.
Structural transitions 
give apparently better estimates. In particular, in these systems 
the exponent $\zeta \equiv 2 \gamma - 3 \nu$ is directly determined 
in NMR experiments.

Experiments have also measured higher-harmonic exponents.
Analysis of the experimental data near the smectic-C--tilted-hexatic-I
transition gives $\phi_2 = 1.16(7)$, $\phi_3 = 0.40(17)$, 
$\beta_n = \beta[n + 0.295 n (n-1)]$ for $2\le n\le 9$ 
\cite{ABBL-86,Brock-etal-86}.  
The exponent $\phi_2$ was also measured for the bicritical point
in GdAlO$_3$ \cite{RG-77} $\phi_2=1.17(2)$. 
In Ref. \cite{Zinkin-etal-96} the estimates 
$\beta_2 = 0.87(1)$, $\beta_3 = 1.50(4)$ were obtained for 
Rb$_2$ZnCl$_4$.
Older experimental estimates are reported in Ref.~\cite{PHA-91,Zinkin-etal-96}.
In a liquid crystal at the smectic-C--tilted-hexatic-I
transition, the structure factor 
$G_2(x-y) \equiv \langle O_{2}^{ab}(x) O_{2}^{ab}(y) \rangle$ 
was measured using x-ray scattering techniques 
\cite{Wu-etal-94}. The results, reanalyzed in Ref. \cite{Aharony-etal-95}, 
are in good agreement with the theory \cite{NA-97,CPV-02}.

\subsection{The critical equation of state}
\label{eqstXY}

The critical equation of state of the
three-dimensional $XY$ universality class is of direct
experimental interest for magnetic systems, but it cannot be 
observed in the $\lambda$-transition in ${}^4$He.
Indeed, in this case  
the order parameter is related to the
complex quantum amplitude of helium atoms. Therefore,
the ``magnetic'' field $H$ does not correspond to an 
experimentally accessible external field. 
Only universal amplitude ratios of 
quantities formally defined at zero external momentum, such
as $U_0\equiv A^+/A^-$, are here of physical relevance.

\subsubsection{Small-magnetization expansion of the free energy 
in the HT phase}
\label{r2jXY}

In Table~\ref{EFXY} we report a summary of the available results for
the zero-momentum four-point coupling $g_4^+$, cf. Eq. (\ref{grdef}), 
and for the coefficients $r_{6}$, $r_8$, and $r_{10}$ that  
parametrize the small-magnetization expansion of the 
Helmholtz free energy, cf. Eq. (\ref{Fzdef}).

The results of Refs.~\cite{CHPRV-01,CPRV-00-es} were obtained by analyzing
HT series for two improved Hamiltonians.
The small difference in the results for $g_4^+$ of 
Refs.~\cite{CHPRV-01} and \cite{CPRV-00-es} is 
essentially due to a different method of analysis. 
The result of Ref.~\cite{CHPRV-01} should be  more reliable.
Refs. \cite{BC-98,PV-gr-98,Reisz-95} considered the 
HT expansion  of the standard $XY$ model.
In the fixed-dimension FT approach, $g_4^+$
is obtained from the zero of the corresponding Callan-Symanzik 
$\beta$-function.
Note the good agreement between the HT and the FT estimates.
In the same framework $g_6^+=r_6 (g_4^+)^2$ and 
                      $g_8^+=r_8 (g_4^+)^3$ 
were estimated from the analysis
of the corresponding four- and three-loop series respectively~\cite{SOUK-99}. 
The authors of Ref.~\cite{SOUK-99} argued
that the uncertainty on their estimate of $g_6^+$ is approximately 0.3\%,
while they considered their value for $g_8^+$ much less accurate.
The $\epsilon$-expansion estimates were obtained from constrained
analyses of the $O(\epsilon^4)$ series of $g_4^+$  
and of the $O(\epsilon^3)$ series
of $r_{2j}$~\cite{PV-00,PV-ef-98,PV-gr-98}.

\begin{table}
\caption{
Estimates of $g_4^+$, $r_{6}$, $r_8$, and $r_{10}$.
We also mention the estimate $r_{10}=-10(1)$ obtained 
by studying the equation of state \protect\cite{CHPRV-01},
see Sec.~\ref{pareqstXY}. 
}
\label{EFXY}
\footnotesize
%\hspace*{-2.7cm}    % Move table leftwards, so it doesn't run off the rig3ht
\tabcolsep 4pt        % Less than the usual 6pt
%\doublerulesep 1.5pt  % Less than the usual 2pt
\begin{center}
\begin{tabular}{clll}
\hline
\multicolumn{1}{c}{}& 
\multicolumn{1}{c}{HT}& 
\multicolumn{1}{c}{$d=3$ exp}&
\multicolumn{1}{c}{$\epsilon$ exp} \\
\hline  
$g_4^+$   & 21.14(6) \cite{CHPRV-01} & 21.16(5)  \cite{GZ-98} & 
               21.5(4) \cite{PV-00,PV-gr-98}\\
            & 21.05(6) \cite{CPRV-00-es} &  21.11  \cite{MN-91} & \\
            & 21.28(9) \cite{BC-98}   & 21.20(6) \cite{LZ-77} & \\
            & 21.34(17) \cite{PV-gr-98} & & \\

$r_6$ & 1.950(15) \cite{CHPRV-01}  &   1.967 \cite{SOUK-99} & 1.969(12) \cite{PV-00,PV-ef-98} \\
            & 1.951(14) \cite{CPRV-00-es}& &   \\
            & 2.2(6) \cite{Reisz-95}  & &   \\
 
$r_8$       & 1.44(10) \cite{CHPRV-01}   &  1.641 \cite{SOUK-99} & 2.1(9) \cite{PV-00,PV-ef-98} \\
            & 1.36(9)\cite{CPRV-00-es}      & &  \\ 

$r_{10}$ & $-$13(7) \cite{CHPRV-01}   & & \\
\hline
\end{tabular}
\end{center}
\end{table}

\subsubsection{Approximate representations of the equation of state}
\label{pareqstXY}

The results of Sec.~\ref{r2jXY} can be used to determine approximate
 parametric representations of the critical equation of state.
In Refs.~\cite{CPRV-00-es,CHPRV-01} the parametric representation 
\reff{parrepg} was considered, approximating the functions $m(\theta)$ 
and $h(\theta)$ by polynomials, and requiring 
$h(\theta) \sim (\theta - \theta_0)^2$ for $\theta\rightarrow \theta_0$,
to reproduce the
correct leading singular behavior at the coexistence curve.

Two polynomial schemes were considered: 
\begin{eqnarray}
\hbox{\rm scheme A}: && \qquad
m(\theta) = \theta \Bigl( 1 + \sum_{i=1}^n c_{i}\theta^{2i}\Bigr) , \qquad
h(\theta) = \theta \left( 1 - \theta^2/\theta_0^2 \right)^2, \label{scheme1}
\\
\hbox{\rm scheme B}: && \qquad
m(\theta) = \theta, \qquad
h(\theta) = \theta \left( 1 - \theta^2/\theta_0^2 \right)^2\Bigl( 1
+ \sum_{i=1}^n c_{i}\theta^{2i}\Bigr). \label{scheme2}
\end{eqnarray}
In both schemes
$\theta_0$ and the $n$ coefficients $c_{i}$ are determined 
by matching the small-$z$ expansion of the scaling function $F(z)$,
i.e.  using the $(n+1)$ estimates of $r_6,...r_{6+2n}$. 
In this case, a variational approach analogous to that 
presented in Sec. \ref{eqstising} cannot be employed. Indeed, 
for the class of functions that are considered here---with a double 
zero at $\theta_0$---there is no globally valid stationary solution.

\begin{figure}[tb]
\hspace{0cm}
\begin{tabular}{cc}
\psfig{width=6truecm,angle=-90,file=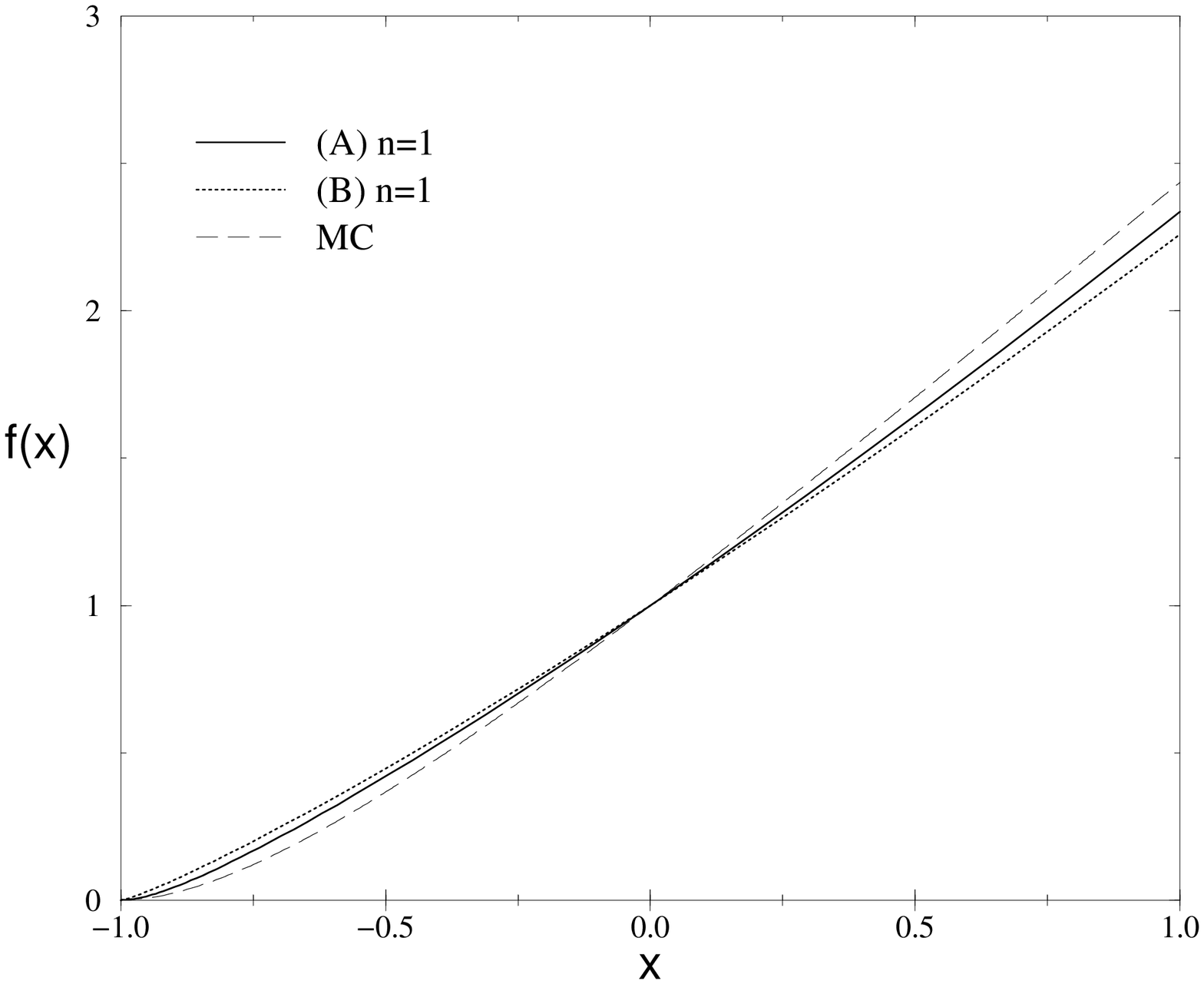} & 
\hskip 1truecm
\psfig{width=6truecm,angle=-90,file=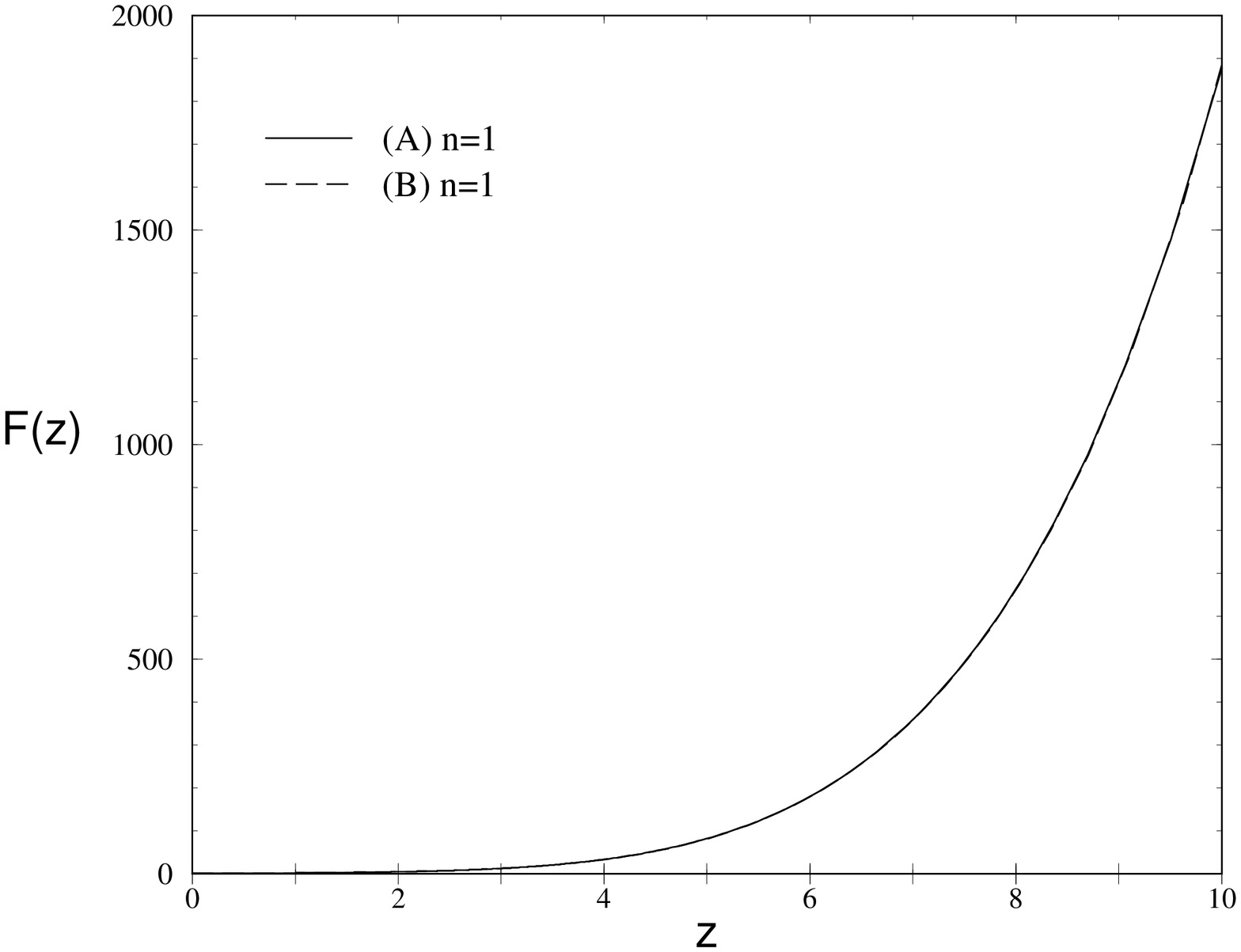} \\
\end{tabular}
\vspace{0cm}
\caption{
The scaling functions $f(x)$ and $F(z)$ for the $XY$ universality class. 
We report the results of Ref.~\cite{CHPRV-01} for schemes A and B, 
and the MC results of Ref. \cite{EHMS-00}.
}
\label{figfxXY}
\end{figure}

%% \begin{figure}[tb]
%% \hspace{0cm}
%% \centerline{\psfig{width=5.5truecm,angle=-90,file=FzXY.eps}}
%% \vspace{0cm}
%% \caption{
%% The scaling function $F(z)$ for the $XY$ universality class. Results from 
%% Ref. \cite{CHPRV-01}.
%% }
%% \label{figFzXY}
%% \end{figure}

Figure \ref{figfxXY} shows the scaling functions $f(x)$ and $F(z)$,
as obtained from schemes A and B with $n=1$, using
the MC+IHT estimates for $\gamma$, $\nu$, $r_6$, and $r_8$.
The two approximations of $F(z)$ are practically indistinguishable 
in Fig.~\ref{figfxXY}.
One obtains a rather precise estimate of the constant $F_0^\infty$
that parametrizes the large-$z$ behavior of $F(z)$, cf. Eq.~(\ref{asyFz}),
$F_0^\infty=0.0302(3)$.
The approximate parametric representations are
less precise at the coexistence curve,
as one may observe by comparing the corresponding curves of $f(x)$. 
At the coexistence curve, where $f(x)\approx c_f(1+x)^2$, 
one obtains only a rough estimate of $c_f$, i.e. $c_f=4(2)$.
A more precise determination 
of the equation of state near the coexistence curve was 
achieved by means of a MC 
simulation of the standard $XY$ model \cite{EHMS-00}. 
In particular, they obtained the precise estimate $c_f=2.85(7)$.
The MC data are well interpolated
in a relatively large region of values of $x$ around $x=-1$
by a power-law behavior of the type (\ref{expcoex1}), 
including the first three terms of the expansion (up to $y^{3/2}$).
This fact does not necessarily rule out the presence of the logarithms 
found in the $1/N$ expansion, cf. Eq.~(\ref{fx3d}). Since they are 
of order $(1+x)^2$ with respect to the leading term,
they are hardly distinguishable  from simple power terms
in numerical works.
In Fig.~\ref{figfxXY} we
also plot the interpolation of Ref. \cite{EHMS-00} of their MC data.

We finally mention that the critical equation of state is known 
to $O(\epsilon^2)$ in the framework of the $\epsilon$ expansion \cite{BWW-72}.

\subsubsection{Universal amplitude ratios}
\label{unra}

The most interesting universal amplitude ratio
is related to the  specific heat, i.e. 
$U_0\equiv A^+/A^-$, because its estimate can be
compared with the accurate experimental results for
the superfluid transition in $^4$He.
Table~\ref{summaryeqstXY} reports estimates of $U_0$ obtained
by various approaches.
The results of
Refs.~\cite{CHPRV-01,CPRV-00-es,EHMS-00} have been obtained from the
equation of state.
We note that most of the theoretical and experimental
estimates of $U_0$ reported  in Table \ref{summaryeqstXY}
are strongly correlated with the value of $\alpha$ considered.
In particular,
the difference between the 
experimental estimate $U_0=1.0442$ of Refs.\cite{LSNCI-96,Lipa-etal-00}
and the theoretical result $U_0=1.062(4)$ of Ref.~\cite{CHPRV-01} 
is a direct consequence of the difference in the values of $\alpha$ used 
in the 
analyses, i.e. $\alpha=-0.01056(38)$ in the analysis of the experimental data
of Refs.\cite{LSNCI-96,Lipa-etal-00},
and $\alpha=-0.0146(8)$ in the theoretical study of the equation of state
of Ref.~\cite{CHPRV-01}. We also mention that
the IHT--PR result of Ref.~\cite{CPRV-00-es} and the FT result of Ref. \cite{LMSD-98} 
were obtained using $\alpha=-0.01285(38)$, 
while the FT analysis of Ref.~\cite{KV-00} used the value $\alpha=-0.01056$.
In all cases the correlation between the estimates of $U_0$ and $\alpha$
is well described by the phenomenological relation 
$U_0\approx 1 - 4\alpha$ \cite{HAHS-76}, which was derived in the framework of
the $\epsilon$ expansion.  As suggested in Ref. \cite{BHK-75},
one may consider the quantity
\be
R_\alpha = {1 - U_0\over \alpha},
\ee
which is expected to be less sensitive to the value of $\alpha$.
For this quantity one finds $R_\alpha = 4.3(2)$ from the parametric
representation \cite{CHPRV-01} and $R_\alpha = 4.39(26)$ from the 
FT method employing  minimal subtraction without 
$\epsilon$ expansion \cite{SMD-00}. 
These results are consistent with the 
experimental estimate $R_\alpha \approx 4.19$ of Refs.
\cite{LSNCI-96,Lipa-etal-00}.
Accurate results for the specific heat of the $XY$ model,
obtained by  high-statistics MC simulations, 
have been recently reported in Ref.~\cite{CEHMS-02}.
The authors stress the difficulty to extract a
satisfactory  estimate of $\alpha$ by measuring the specific heat. 
A fit to the data with the expected
RG behavior \reff{Chscaling} does not even
allow to exclude a logarithmic behavior, i.e. $\alpha=0$.
This is not unexpected: 
The small value of $\alpha$ makes difficult---both numerically and 
experimentally---distinguishing 
the $O(t^{-\alpha})$ term from the constant background.
According to the authors of Ref.~\cite{CEHMS-02}, 
the best one can do is to determine the ratio $U_0$ 
as a function of $\alpha$. 
They report the expression
\begin{equation}
U_0 = 1 - 4.20(5) \alpha + O(\alpha^2),
\end{equation}
and therefore, $R_\alpha=4.20(5)$.

Table~\ref{summaryeqstXY} also reports estimates of other
universal ratios, such as $R_\xi^+$, $R_c$, $R_\chi$, $R_4$,
and $R_\xi^{\rm T}$.
In addition, we mention the results reported in Ref.~\cite{CEHMS-02}
as functions of $\alpha$:
\begin{eqnarray}
&&R_\xi^+ = 0.3382(14) - 0.72(10) \alpha + 0.9(1.1) \alpha^2 ,
\nonumber \\
&&R_\xi^{\rm T} = 1.158(36) - 0.696 \alpha + 0.97 \alpha^2 .
\end{eqnarray}
Using the estimate $\alpha = -0.0146(8)$, they  give respectively
$R_\xi^+=0.349(3)$ and $R_\xi^{\rm T}\approx 1.17$.

\begin{table*}
\caption{
Estimates of universal amplitude ratios obtained using different approaches.
The $\epsilon$-expansion estimates of $R_c$ and $R_\chi$ have been obtained by setting $\epsilon=1$
in the $O(\epsilon^2)$ series calculated in Refs.\protect\cite{AH-76,AH-77,AM-78}.
}
\label{summaryeqstXY}
\footnotesize
%\hspace*{-2.7cm}    % Move table leftwards, so it doesn't run off the right
\tabcolsep 4pt        % Less than the usual 6pt
%\doublerulesep 1.5pt  % Less than the usual 2pt
\begin{center}
\begin{tabular}{ccccccc}
\hline
 & IHT--PR & HT & MC & $d$=3 exp & $\epsilon$ exp & experiments \\
\hline
$U_0$
     &1.062(4) \cite{CHPRV-01} &  & 1.12(5) \cite{EHMS-00,CEHMS-02} & 
1.056(4)\cite{LMSD-98}&1.029(13)\cite{Bervillier-86} & 
1.0442 \cite{LSNCI-96,Lipa-etal-00} \\
&1.055(3) \cite{CPRV-00-es} &  & & 1.045 \cite{KV-00}&   & 1.067(3) \cite{SA-84}\\
     &        &  & & &  & 1.058(4) \cite{LC-83} \\
     &        & & &  &   & 1.088(7) \cite{TW-82} \\

$R_\alpha$
     &4.3(2) \cite{CHPRV-01} &  & 4.20(5) \cite{CEHMS-02} & 4.39(26)
     \cite{SMD-00} & &  4.19 \cite{LSNCI-96,Lipa-etal-00} \\

$R_\xi^+$ 
& 0.355(3) \cite{CHPRV-01} & 0.361(4) \cite{BC-99}&  &
     0.3606(20)\cite{BB-85,BG-80}& 0.36 \cite{Bervillier-76}  & \\

$R_c$    
&  0.127(6) \cite{CHPRV-01}&                  &    &  0.123(3) \cite{SLD-99}      &  0.106  & \\ 
&  &                  &    &  0.130 \cite{KV-00}      &  & \\

$R_\chi$ &  1.35(7) \cite{CHPRV-01}  &    & 1.356(4) \cite{EHMS-00}    &  &  1.407   & \\

$R_4$ &  7.5(2) \cite{CHPRV-01}  &                 &     &  &  & \\

$R_\xi^{\rm T}$  &  &   &  &  0.815(10) \cite{BSD-97,SMD-00} & 1.0(2)
\cite{PHA-91,Bervillier-76,HAHS-76}  &0.85(2) \cite{SA-84} \\
\hline
\end{tabular}
\end{center}
\end{table*}

\subsection{The two-point function in the high-temperature phase}
\label{twopXY}

The two-point function of the order parameter in the HT
phase has been studied in Refs.~\cite{FA-73,FA-74,Bray-76,CHPRV-01,CPRV-98} 
by means of HT expansions and FT calculations. 
Its small-momentum scaling behavior is qualitatively similar to the Ising case,
see Sec. \ref{sec-3.5.1}. Indeed, 
the coefficients $c_i^+$ of the small-momentum expansion of the 
scaling function
$g^+(y)$, see Eq. \reff{gypiu}, satisfy the relations (\ref{patternci}). 
Their best estimates are \cite{CHPRV-01}
$c_2^+ = -3.99(4) \times 10^{-4}$,
$c_3^+ = 0.09(1) \times 10^{-4}$, and
$|c_4^+| < 10^{-6}$.
Moreover, $S_M^+=0.999592(6)$ and
$S_Z^+=1.000825(15)$.
Other results can be found in Ref.~\cite{CPRV-98}. They are obtained  
using HT methods in the standard $XY$ model and 
FT methods, such
as the $\epsilon$ and $d=3$ fixed-dimension expansions.

For large values of $y$, the function $g^+(y)$ follows the Fisher-Langer
law reported in Eq.~(\ref{FL-law}).
The coefficients 
$A_1^+$, $A_2^+$ and $A_3^+$ have been computed in the $\epsilon$ expansion to
three loops \cite{Bray-76}, obtaining
$A_1^+ \approx 0.92$, $A_2^+ \approx 1.8$, and $A_3^+ \approx - 2.7$.

One can determine approximations of $g^+(y)$ using the phenomenological 
approach of Bray \cite{Bray-76}. Such an approximation is quite accurate 
for large and small values of $y$. Indeed,
Bray's phenomenological function provides 
the estimates \cite{CHPRV-01}
$A_1^+ \approx 0.915$, 
$c_2^+ \approx -4.4\cdot 10^{-4}$, 
$c_3^+ \approx 1.1\cdot 10^{-5}$,  
$c_4^+ \approx - 5\cdot 10^{-7}$,
in good agreement
with the above-reported estimates. The results for $A_2^+$ and $A_3^+$,
$A_2^+ \approx - 24.7$, $A_3^+ \approx 23.8$, differ
significantly from the $\epsilon$-expansion results.
Note, however, that, since $|\alpha|$ is very small, the relevant
quantity in the Fisher-Langer formula is the sum $A_2^+ + A_3^+$. 
In other words, the function does not change
significantly if one uses the $\epsilon$-expansion results or the 
approximations determined using Bray's method.

\section{The three-dimensional Heisenberg universality class}
\label{O3d3}

The three-dimensional Heisenberg universality class is
characterized by a three-component order parameter, ${\rm O}(3)$
symmetry, and short-range interactions.
It describes the critical 
behavior of isotropic
magnets, for instance the Curie transition in isotropic ferromagnets 
such as Ni  and EuO, and of antiferromagnets such as 
RbMnF$_3$ at the N\'eel transition point. 
Moreover, it describes isotropic 
magnets with quenched disorder, see also Sec.~\ref{lsec-random}.
Indeed, since $\alpha < 0$, the Harris
criterion \cite{Harris-74} states that disorder is an irrelevant 
perturbation. The only effect is to introduce 
an additional correction-to-scaling term $|t|^{\Delta_{\rm dis}}$ with 
$\Delta_{\rm dis} = - \alpha$. 

Note that
the isotropic Heisenberg Hamiltonian is a simplified model for magnets.
It neglects several interactions that are present in real materials.
Among them, we should mention the presence of interactions with
cubic anisotropy due to the lattice 
structure and the dipolar interactions. 
Even if, in the RG language, 
these effects are relevant perturbations
of the Heisenberg fixed point 
\cite{Fisher-74,Aharony-76,AF-73,CPV-00},
the new critical exponents are so close to those of the 
Heisenberg universality class that the difference is experimentally
very difficult to observe, see, e.g., 
Refs.~\cite{BA-74,CPV-00,SKS-00,CPV-02-3} and reference therein.
See also Sec.~\ref{lsec-cubic} for a discussion of the 
cubic anisotropy.

\subsection{The critical exponents}
\label{expO3}

\subsubsection{Theoretical results}
\label{expO3th}

In Table \ref{O3exponents} we report the 
theoretical estimates of the critical exponents obtained
by various approaches.

\begin{table*}[tb]
\caption{Estimates of the critical exponents for the Heisenberg universality
class.
We indicate with an asterisk (${}^*$) the estimates that have been
obtained using the scaling relations $\gamma =(2 - \eta)\nu $,
$2 - \alpha = 3 \nu $, $\beta = \nu(1 + \eta)/2$, and 
$\beta\delta= \beta+ \gamma$. 
}
\label{O3exponents}
\footnotesize
\begin{center}
\begin{tabular}{rllllll}
\hline
\multicolumn{1}{c}{Ref.}& 
\multicolumn{1}{c}{info}& 
\multicolumn{1}{c}{$\gamma$}& 
\multicolumn{1}{c}{$\nu$}& 
\multicolumn{1}{c}{$\eta$}&
\multicolumn{1}{c}{$\beta$}&
\multicolumn{1}{c}{$\delta$} \\   
\hline  
\cite{CHPRV-02} $_{2002}$ & MC+IHT $\phi^4$ & 1.3960(9)
&0.7112(5)   & 0.0375(5)  &  0.3689(3)$^*$ &4.783(3)$^*$ \\

\cite{BC-97-2} $_{1997}$ & HT sc & 1.406(3)    & 0.716(2)     &
0.036(7)$^*$ & 0.3710(13)$^*$ &4.79(4)$^*$    \\
\cite{BC-97-2} $_{1997}$ & HT bcc & 1.402(3)    & 0.714(2)     &  
0.036(7)$^*$& 0.3700(13)$^*$ & 4.79(4)$^*$ \\

\cite{AHJ-93} $_{1993}$ & HT sc & 1.40(1)    & 0.712(10)     &  
0.03(3)$^*$&  0.368(6)$^*$   &  4.80(17)$^*$\\

\cite{FH-86} $_{1986}$ & HT fcc   & 1.40(3)    & 0.72(1)    &  
  0.06(5)$^*$ & 0.38(2)$^*$  & 4.68(27)$^*$ \\
\cite{MDH-82} $_{1982}$ & HT sc   & 1.395(5)    &     &  &  &  \\
\cite{RF-72} $_{1972}$ & HT sc, bcc, fcc   & 1.375$^{+0.02}_{-0.01}$
& 0.7025$^{+0.010}_{-0.005}$  &  0.043(14) & 0.366(14)$^*$  & 4.75(16)$^*$ \\

\cite{CHPRV-02} $_{2002}$ & MC FSS $\phi^4$ &1.3957(22)$^*$ &0.7113(11)
& 0.0378(6)  &  0.3691(6)$^*$ &4.781(3)$^*$ \\

\cite{Hasenbusch-00} $_{2000}$  &  MC FSS $\phi^4$ &  1.393(4)$^*$  &
0.710(2)&  0.0380(10)  &  0.3685(11)$^*$ & 4.780(6)$^*$\\

\cite{CBL-00} $_{2000}$  &  MC FSS double-exchange  & 1.3909(30) &
0.6949(38)  & & 0.3535(30) & \\

\cite{BFMM-96} $_{1996}$  & MC FSS & 1.396(3)$^*$ & 0.7128(14)  &
0.0413(16)  &  0.3711(9)$^*$ & 4.762(9)$^*$\\

\cite{BrCi-96} $_{1996}$ & MC FSS & 1.270(1)$^*$  & 0.642(2)  &
0.020(1)  &  &  \\
\cite{HJ-94} $_{1994}$  &  MC FSS  &   &  0.706(8)$^*$  &  &  &  \\
\cite{HJ-93} $_{1993}$  &  MC FSS  & 1.389(14)$^*$  &  0.704(6)  &
0.027(2)  & 0.362(3)$^*$  & 4.842(11)$^*$   \\
\cite{CFL-93} $_{1993}$  &  MC FSS sc, bcc  & 1.3812(6)$^*$  &
0.7048(30)  &  0.0250(35)   & 0.361(2)$^*$ & 4.85(20)$^*$\\
\cite{PFL-91} $_{1991}$  &  MC FSS  & 1.390(18)$^*$  &  0.706(9)  &
0.031(7)   & 0.364(5)$^*$ & 4.82(4)$^*$ \\
\cite{DHNN-91} $_{1991}$  &  MC FSS  & &  0.73(4)  & &  &  \\
\cite{NB-88}  $_{1988}$  & MC FSS &  & 0.716(40)  & &  & \\

\cite{JK-00} $_{2001}$ & FT $d=3$ exp & 1.3882(10)& 0.7062(7)  & 
0.0350(8) & 0.3655(5)$^*$  & 4.797(5)$^*$ \\

\cite{GZ-98} $_{1998}$ & FT $d=3$ exp & 1.3895(50)& 0.7073(35) &
0.0355(25) & 0.3662(25) & 4.794(14) \\

\cite{MN-91}   $_{1991}$ & 
FT $d=3$ exp & 1.3926(13)\{39\} & 0.7096(8)\{22\} &0.0374(4) & & \\

\cite{LZ-77}   $_{1977}$  & 
FT $d=3$ exp & 1.386(4)& 0.705(3) &0.033(4) & 0.3645(25) & 4.808(22) \\

\cite{GZ-98} $_{1998}$ & FT $\epsilon$ exp & 1.382(9) & 0.7045(55) &
0.0375(45) & 0.3655(5)$^*$  & 4.797(5)$^*$ \\

\cite{YG-98} $_{1998}$ & FT $\epsilon$ exp & 1.39$^*$ & 0.708 & 0.037 & 0.367$^*$ & 4.786$^*$ \\

\cite{Kleinert-00} $_{2000}$ & FT $(d-2)$ exp & & 0.695(10) & & & \\

\cite{NR-84}  $_{1984}$ & 
SFM  & 1.40(3)   &  0.715(20)   & 0.044(7) & 0.373(11) & 4.75(4)$^*$\\
\cite{BSW-01} $_{2001}$ & CRG & 1.45 & 0.74 & 0.038 & 0.37 & 4.78 \\
\cite{GW-01} $_{2001}$  & CRG (1st DE) & 1.374 & 0.704  & 0.049  &  0.369 & 4.720 \\
\cite{MT-98} $_{1998}$  & CRG (1st DE) & 1.464 & 0.745  & 0.035  &  0.386 & 4.797 \\
\cite{BTW-96} $_{1996}$  & CRG ILPA & 1.465 & 0.747  & 0.038  &  0.388 & 4.78 \\
\hline
\end{tabular}
\end{center}
\end{table*}

Accurate results for the critical exponents
have been obtained by combining 
MC simulations and HT expansions for the  improved $\phi^4$ Hamiltonian 
(\ref{latticephi4}) with $\lambda^*=4.6(4)$
\cite{CHPRV-02,Hasenbusch-00}, cf. Sec.~\ref{sec-2.3.2}.
Using the linked-cluster expansion technique, the HT expansions of 
$\chi$ and $\mu_2 \equiv \sum_x |x|^2 G(x)$
were computed to 20th order. 
The analyses were performed  using the estimates of $\lambda^*$
and $\beta_c$ obtained from the MC simulations. The 
results 
are denoted by MC+IHT in Table~\ref{O3exponents}. 
The other results reported in the table were obtained 
from the analysis of the HT series
for the standard Heisenberg model (HT),  by MC simulations (MC),
or by  FT methods (FT).  
The HT results of Ref.\ \cite{BC-97-2}
were obtained by analyzing 21st-order HT expansions for the standard
O(3)-vector model on the simple cubic (sc) and on the 
body-centered cubic (bcc) lattice.
Most MC results concern the standard Heisenberg model
and were obtained using FSS techniques 
\cite{BFMM-96,BrCi-96,HJ-94,HJ-93,CFL-93,PFL-91,DHNN-91,NB-88}.
The results of Refs.~\cite{CHPRV-02,Hasenbusch-00}  were obtained by simulating
the improved $\phi^4$ model. Ref.~\cite{CBL-00} considers 
an isotropic ferromagnet with double-exchange interactions,\footnote{
Recently, a model with competing superexchange and 
double-exchange interactions has been studied 
\cite{TL-01}. A 
preliminary analysis for the paramagnetic-ferromagnetic
transition gives $\nu = 0.720(2)$ and $\gamma = 1.438(8)$.
While $\nu$ is in reasonable agreement with the Heisenberg
value, $\gamma$ is significantly higher, so that the 
identification of this transition as a Heisenberg one is in doubt.}
whose Hamiltonian is given by \cite{AH-55}
\begin{equation}
{\cal H} = - \beta \sum_{\langle ij \rangle} \sqrt{ 1 + s_i\cdot
s_j}.
\end{equation}
The FT results of Refs.~\cite{JK-00,GZ-98,MN-91,LZ-77,YG-98,Kleinert-00} 
were derived 
by analyzing perturbative expansions in different frameworks:
fixed-dimension expansion (6th- and 7th-order series, see 
Refs. \cite{BNGM-77,MN-91}),
$\epsilon$ expansion (to $O(\epsilon^5)$, see Refs. \cite{CGLT-83,KNSCL-93}),
and $(d-2)$-expansion (to $O[(d-2)^4]$, 
see Refs. \cite{HB-78,Hikami-83,BW-86}). 
We quote two errors for the results of Ref.~\cite{MN-91}: 
the first one (in parentheses) 
is the resummation error, the second one (in braces) 
takes into account the uncertainty of the fixed-point value $g^*$,
which was estimated to be approximately 1\% in Ref.~\cite{MN-91}. 
%To estimate the second error we use the results of
%Ref.~\cite{GZ-98} where the dependence of the exponents on $g^*$ is given.
The results of Ref.~\cite{NR-84} were obtained  using the so-called 
scaling-field method (SFM). 
Refs.~\cite{BSW-01,GW-01,BTW-96,BTW-99} present results
obtained by approximately solving continuous renormalization-group
(CRG) equations for the average action.
We also mention the HT results of Ref. \cite{BC-99}:
they performed a direct determination of the exponent $\alpha$ obtaining 
$\alpha = -0.11(2)$, $-0.13(2)$ on the sc and bcc
lattice. Ref. \cite{MOF-01} computes the critical exponents for a Heisenberg
fluid by a canonical-ensemble simulation. Depending on the analysis method, 
they find $1/\nu = 1.40(1)$, $1.31(1)$, $\beta/\nu= 0.54(2)$, $0.52(1)$, 
and  $\gamma/\nu =1.90(3)$, $1.87(3)$.
Overall, all estimates are in substantial agreement. 
We only note the quite anomalous
result of Ref.~\cite{BrCi-96}, which is further discussed in
Refs. \cite{HJ-97,BrCi-97}, and 
the apparent discrepancies of the MC+IHT results with   
the MC estimates of $\eta$ of Refs.~\cite{BFMM-96,HJ-93},
and with the FT results of Ref. \cite{JK-00}. 
%However, the reliability of the
%error bars reported in Ref.~\cite{JK-00} is unclear: indeed, 
%Ref. \cite{GZ-98} analyzes the 
%same perturbative series and reports much more cautious error estimates.

Concerning the leading scaling-correction exponent $\omega$,
we mention the estimates $\omega=0.782(13)$ obtained from
the analysis of the six-loop fixed-dimension expansion \cite{GZ-98},
$\omega=0.794(18)$ from the five-loop $\epsilon$ expansion
\cite{GZ-98}, $\omega\approx 0.773$ from MC simulations \cite{Hasenbusch-00}.
Correspondingly, using \cite{CHPRV-02} $\nu = 0.7112(5)$, we have 
$\Delta = \omega\nu = 0.556(9)$, 0.565(13), 0.550.

We finally report some results for the 
crossover exponent $\phi_2$ associated with 
the spin-2 traceless tensor field 
$O^{ab}(x) = \phi^a(x) \phi^b(x) - \case{1}{3}\delta^{ab} \phi(x)^2$,
see Sec.~\ref{sec-1.5.8}, which
describes the instability of the O(3)-symmetric 
theory against anisotropy \cite{FP-72,Wegner-72,FN-74,Aharony-76}. 
The crossover exponent $\phi_2$ has been determined 
using various approaches, obtaining
$\phi_2 = 1.271(21)$ by the analysis of the six-loop expansion
in the framework of the fixed-dimension FT expansion
\cite{CPV-02}; 
$\phi_2\approx 1.22$ 
by setting $\epsilon=1$ in the corresponding $O(\epsilon^3)$ 
series \cite{Yamazaki-74};
$\phi_2=1.250(15)$ by HT expansion
techniques \cite{PJF-74}.
The exponent $\phi_4$ can be derived from the results of Ref. \cite{CPV-00}
for the $O(N)$ model with a cubic-symmetric perturbation, see 
Sec. \ref{lsec-cubic}. One finds \cite{CPV-02-4} 
$\phi_4 = 0.009(4)$.  Since $\phi_2>0$ and $\phi_4 > 0$, 
the spin-2 and the spin-4 (we also expect the 
spin-3) operators are relevant perturbations. Higher-order spin operators
are expected to be RG irrelevant.

\subsubsection{Experimental results}
\label{expO3ex}

In Table \ref{table_exponents_exp_O3} we report 
some recent experimental results for ferromagnets and
antiferromagnets. It is not 
a complete review of published results, but it is
useful to get an overview of the experimental state of the art.  
In the table we have also included results for the well-studied 
doped manganese perovskites La$_{1-x}$A$_{x}$MnO$_3$, although the 
nature of the ferromagnetic transition in these compounds is still 
unclear.\footnote{
For some dopings and some divalent cation A a first-order
transition has been observed. Moreover, in systems in which the transition 
appears to be of second order, mean-field critical exponents have been
measured. For instance, for La$_{1-x}$Sr$_x$MnO$_3$, a mean-field 
value for $\beta$ was observed in 
Refs.~\cite{LRKBMT-97,MSKMM-98,SSA-00},
while an estimate compatible with the Heisenberg value was found in
Refs. \cite{MSEHMT-96,HLHNLKNUMC-96,LBGGKSAM-97,GLGKSAM-98}.
For $x=1/3$ there also exists \cite{Ramos-etal-01} an estimate of the exponent
$\alpha$, $\alpha = - 0.14\pm 0.10$, in agreement with the 
Heisenberg value. See also the recent review \cite{SJ-01}.}

The Heisenberg universality class also describes isotropic 
magnets with quenched disorder. 
The experimental results confirm this theoretical prediction,\footnote{
In order to observe the correct exponents, it is essential to consider 
corrections to scaling in the analysis of the experimental data \cite{Kaul-85}.
All results reported in Table \ref{table_exponents_exp_O3_2}, except 
those of Ref. \cite{SHAM-99,Tsurkan-etal-99}, have been obtained by assuming 
scaling corrections of the form $(1 + a |t|^{\Delta_1} + b 
|t|^{\Delta_2})$, with $\Delta_1 = 0.11$ and 
$\Delta_2 = 0.55$. Note that the value of $\Delta_1$ is slightly lower
than the precise theoretical estimate of Ref. \cite{CHPRV-02}, 
$\Delta_1 = 0.1336(15)$, and that RG
predicts additional corrections 
of order $|t|^{2\Delta_1}$, $|t|^{3\Delta_1}$, etc., which are more 
relevant than the term $|t|^{\Delta_2}$ and should therefore be taken into 
account in the analysis of the data.}
as it can be seen 
from Table \ref{table_exponents_exp_O3_2} (older experimental results with 
a critical discussion are reported in Ref. \cite{Kaul-85}). 
Finally, we mention the experiment 
reported in Ref.~\cite{CH-88}
on Fe$_{1-x}$V$_x$ in the presence of annealed
disorder; as predicted by theory, they obtain $\beta=0.362(8)$, in agreement
with the corresponding Heisenberg exponent.

Beside the exponents $\gamma$, $\beta$, and $\delta$ there are
also a few estimates of the specific-heat exponent $\alpha$, in most 
of the cases obtained from resistivity measurements: 
$\alpha \approx -0.10$ in Fe and Ni \cite{KHM-81};
$\alpha = -0.12(2)$ in EuO \cite{SWSLPK-92}; 
$\alpha = -0.11(1)$ in Fe$_x$Ni$_{80-x}$B$_{19}$Si \cite{KR-94};
$\alpha = -0.11(1)$ in RbMnF$_3$ \cite{MMFB-96};
$\alpha \approx -0.12$ in Sr$_2$FeMoO$_6$ \cite{YCSXM-02}.	

Some experimental estimates of crossover exponent
$\phi_2$ are
reported in Ref.~\cite{PHA-91}. We mention the experimental
result $\phi_2=1.279(31)$ for the bicritical point in MnF$_2$ \cite{KR-79}.

\begin{table*}
\caption{
Experimental estimates of the critical exponents for Heisenberg systems.}
\label{table_exponents_exp_O3}
\footnotesize
\begin{center}
\begin{tabular}{rllll}
\hline
\multicolumn{1}{c}{Ref.}& 
\multicolumn{1}{c}{Material}& 
\multicolumn{1}{c}{$\gamma$}& 
\multicolumn{1}{c}{$\beta$}&
\multicolumn{1}{c}{$\delta$} \\   
\hline  
\cite{SBEW-80} $_{1980}$        & Ni     &           & 0.354(14) &       \\
\cite{Kobeissi-81} $_{1981}$    & Fe     &           & 0.367(5)  &       \\
\cite{Seeger-etal-95} $_{1995}$ & Ni     & 1.345(10) & 0.395(10) & 4.35(6) \\
\cite{RKS-95} $_{1995}$     & Gd$_2$BrC  & 1.392(8)  & 0.365(5)  &4.80(25) \\
\cite{RKS-95} $_{1995}$     & Gd$_2$IC   & 1.370(8)  & 0.375(8)  &4.68(25) \\
\cite{Zhao-etal-99} $_{1999}$ & Tl$_2$Mn$_2$O$_7$ 
                                           & 1.31(5)   & 0.44(6)   &4.65(15) \\
\cite{Barsov-etal-00} $_{2000}$ & La$_{0.82}$Ca$_{0.18}$MnO$_3$ 
                                           &           & 0.383(9)  &         \\
\cite{Zhao-etal-00} $_{2000}$ &  La$_{0.95}$Ca$_{0.05}$MnO$_3$  
                                           & 1.39(5)   & 0.36(7)   &4.75(15) \\
\cite{ArPa-00} $_{2000}$    & Gd(0001)   &           & 0.376(15) &         \\
\cite{MRBSGV-00} $_{2000}$  & Gd$_2$CuO$_4$ & 1.32(2)& 0.34(1)   &         \\
\cite{Buhrer-etal-00} $_{2000}$ & C$_{80}$Pd$_{20}$ (liq) & 1.42(5) & &    \\
\cite{Buhrer-etal-00} $_{2000}$ & C$_{80}$Pd$_{20}$ (sol) & 1.40(8) & &    \\
\cite{Brueckel-etal-01} $_{2001}$ & GdS  &           & 0.38(2)   &         \\
\cite{Yang-etal-01} $_{2001}$ & CrO$_2$  & 1.43(1)   & 0.371(5)  &         \\
\cite{HKH-01} $_{2001}$  &  La$_{0.8}$Ca$_{0.2}$MnO$_3$   
                    & 1.45    & 0.36     & \\
\cite{YCSXM-02} $_{2002}$ & Sr$_2$FeMoO$_6$ & 1.30 & 0.388 & 4.35 \\
\hline
\end{tabular}
\end{center}
\end{table*}

\begin{table*}
\caption{
Experimental estimates of the critical exponents for Heisenberg
systems with quenched disorder.}
\label{table_exponents_exp_O3_2}
\footnotesize
\begin{center}
\begin{tabular}{rllll}
\hline
\multicolumn{1}{c}{Ref.}& 
\multicolumn{1}{c}{Material}& 
\multicolumn{1}{c}{$\gamma$}& 
\multicolumn{1}{c}{$\beta$}&
\multicolumn{1}{c}{$\delta$} \\   
\hline  
\cite{KR-94} $_{1994}$ & Fe$_{10}$Ni$_{70}$Bi$_{19}$Si 
                  &    1.387(12)   & 0.378(15)   & 4.50(5)   \\
\cite{KR-94} $_{1994}$ & Fe$_{13}$Ni$_{67}$Bi$_{19}$Si 
                  &    1.386(12)   & 0.367(15)   & 4.50(5)   \\
\cite{KR-94} $_{1994}$ & Fe$_{16}$Ni$_{64}$Bi$_{19}$Si 
                  &    1.386(14)   & 0.360(15)   & 4.86(4)   \\
\cite{RK-95,RK-95b} $_{1995}$ & Fe$_{20}$Ni$_{60}$P$_{14}$B$_6$
                  &    1.386(10)   & 0.367(10)   & 4.77(5)   \\
\cite{RK-95,RK-95b} $_{1995}$ & Fe$_{40}$Ni$_{40}$P$_{14}$B$_6$
                  &    1.385(10)   & 0.364(5)    & 4.79(5)   \\
\cite{BK-97} $_{1997}$ & Fe$_{91}$Zr$_9$ 
                  &    1.383(4)    & 0.366(4)    & 4.75(5)   \\ 
\cite{BK-97} $_{1997}$ & Fe$_{89}$CoZr$_{10}$ 
                  &    1.385(5)    & 0.368(6)    & 4.80(4)   \\
\cite{BK-97} $_{1997}$ & Fe$_{88}$Co$_2$Zr$_{10}$ 
                  &    1.389(6)    & 0.363(5)    & 4.81(5)   \\
\cite{BK-97} $_{1997}$ & Fe$_{84}$Co$_6$Zr$_{10}$ 
                  &    1.386(6)    & 0.370(5)    & 4.84(5)   \\
\cite{SHAM-99} $_{1999}$ & Fe$_{1.85}$Mn$_{1.15}$Si 
                  &    1.543(20)   & 0.408(60)   & 4.74(7)   \\
\cite{SHAM-99} $_{1999}$ & Fe$_{1.50}$Mn$_{1.50}$Si 
                  & 1.274(60)      & 0.383(10)   & 4.45(19)  \\

\cite{Tsurkan-etal-99} $_{1999}$ & MnCr$_{1.9}$In$_{0.1}$S$_4$ 
                  &    1.39(1)   & 0.36(1)   & 4.814(14)   \\

\cite{Tsurkan-etal-99} $_{1999}$ & MnCr$_{1.8}$In$_{0.2}$S$_4$ 
                  &    1.39(1)   & 0.36(1)  & 4.795(10)   \\

\cite{Perumal-etal-00} $_{2000}$ & Fe$_{86}$Mn$_4$Zr$_{10}$
                  & 1.381          & 0.361       & \\
\cite{Perumal-etal-00} $_{2000}$ & Fe$_{82}$Mn$_8$Zr$_{10}$
                  & 1.367          & 0.363       & \\
\cite{PSKYRD-01} $_{2001}$ & Fe$_{84}$Mn$_6$Zr$_{10}$
                  & 1.37(3)        & 0.359       & 4.81(4)    \\
\cite{PSKYRD-01} $_{2001}$ & Fe$_{74}$Mn$_{16}$Zr$_{10}$
                  & 1.39(5)        & 0.361       & 4.86(3)    \\
\hline
\end{tabular}
\end{center}
\end{table*}

\subsection{The critical equation of state}
\label{eqstO3}

\subsubsection{Approximate representations}

The critical equation of state can be determined  
using the method described in Sec.~\ref{pareqstXY}
in the context of the $XY$ universality class, 
i.e. using the small-magnetization expansion of the free energy
to construct approximate parametric
representations following the schemes A and B,
cf. Eqs.~(\ref{scheme1}) and (\ref{scheme2}).

\begin{table*}
\caption{
Estimates of $g_4^+$, $r_{6}$, and $r_8$ for the Heisenberg universality
class. We also mention the estimate 
$r_{10}=-6(3)$ obtained 
by studying the equation of state \protect\cite{CHPRV-02}.
}
\label{summarygjO3}
\footnotesize
\tabcolsep 4pt        % Less than the usual 6pt
\begin{center}
\begin{tabular}{cllll}
\hline
\multicolumn{1}{c}{}& 
\multicolumn{1}{c}{HT}& 
\multicolumn{1}{c}{$d=3$ exp}&
\multicolumn{1}{c}{$\epsilon$ exp} &
\multicolumn{1}{c}{CRG} \\
\hline  
$g_4^+$   & 19.13(10) \cite{CHPRV-02} & 19.06(5) \cite{GZ-98} &
          19.55(12) \cite{PV-00,PV-gr-98}  & 22.35 \cite{BTW-96,BTW-99}\\
 
        & 19.31(14),$\;$ 19.27(11) \cite{BC-98}  &  19.06 \cite{MN-91} & & \\ 
        & 19.34(16) \cite{PV-gr-98}  & &  &  \\ 

$r_6$   & 1.86(4) \cite{CHPRV-02}  & 1.880 \cite{SOUK-99} &
          1.867(9) \cite{PV-00,PV-ef-98} & 1.74 \cite{TW-94} \\
        & 2.1(6) \cite{Reisz-95} & 1.884(32) \cite{PV-00} & & \\ 

$r_8$   & 0.6(2)  & 0.975 \cite{SOUK-99} &
          1.0(6) \cite{PV-00,PV-ef-98} & 0.84 \cite{TW-94}\\

$r_{10}$   & $-$15(10) & & & \\
\hline
\end{tabular}
\end{center}
\end{table*}

In Table~\ref{summarygjO3} we report a summary of the available results for
the zero-momentum four-point coupling $g_4^+$, cf. Eq. (\ref{grdef}), 
and for the coefficients $r_{6}$, $r_8$, and $r_{10}$ that  
parametrize the small-magnetization expansion of the 
Helmholtz free energy, cf. Eq. (\ref{Fzdef}).

\begin{figure}[tb]
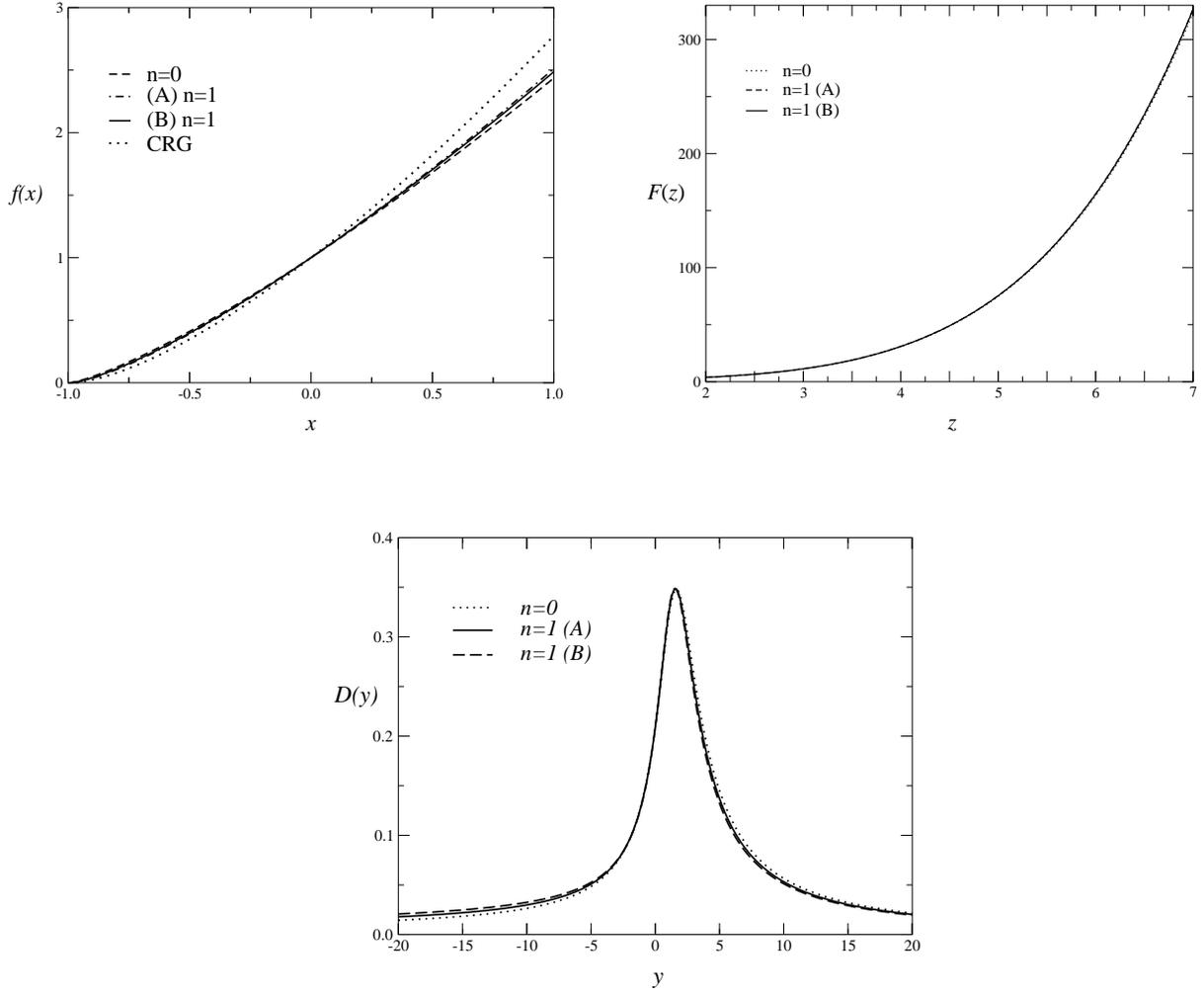

\hspace{0cm}
\begin{tabular}{cc}
\hskip -0.6truecm
\psfig{width=7.5truecm,angle=0,file=fxO3.eps} &
\hskip 0.6truecm
\psfig{width=7.5truecm,angle=-0,file=FzO3.eps} \\[12mm]
\multicolumn{2}{c}{\psfig{width=8truecm,angle=-0,file=Dy.eps}}
\end{tabular}
\vspace{0cm}
\caption{
The scaling functions $f(x)$, $F(z)$, and 
$D(y)$ 
for the Heisenberg universality class. 
All results have been obtained in  Ref. \cite{CHPRV-02},
except those labelled by CRG (Ref. \cite{BTW-96}).
}
\label{figFzO3}
\end{figure}

Figure \ref{figFzO3}
shows the scaling functions $F(z)$, $f(x)$, and $D(y)$, as obtained
in Ref. \cite{CHPRV-02}. They used 
schemes $A$ and $B$ with $n=0,1$, and
the (MC+IHT) estimates of $\gamma$, $\nu$, $r_6$, and $r_8$.
The three approximations of $F(z)$ are practically
indistinguishable, and differ at most by approximately 2\%
(the difference between the two $n=1$ curves is much smaller).
The large-$z$ behavior of $F(z)$ is well determined, indeed
$F_0^\infty = 0.0266(5)$.
The three curves for $f(x)$ 
are in substantial agreement, especially 
those with $n=1$. Indeed, the difference between them 
is within the uncertainty due to the errors on the input parameters.
These approximate parametric representations are not 
precise at the coexistence curve, providing only a rough estimate of 
the coefficient $c_f$, cf. Eq. (\ref{fxcc}),
i.e. $c_f=5(3)$.
We also report
the estimates of  the coefficients $f_n^0$, 
cf. Eq.~(\ref{expansionfx-xeq0}) obtained in Ref.~\cite{CHPRV-02}:
$f_1^0 = 1.34(5)$, $f_2^0 = 0.20(2)$, $f_3^0 = -0.10(1)$.

The scaling function $f(x)$ was also determined in Ref.~\cite{BTW-96} 
by CRG methods using the lowest order of the derivative expansion.
In Fig.~\ref{figFzO3}, together with the results of Ref.~\cite{CHPRV-02} for 
$f(x)$, we also show the approximate scaling function $f(x)$
obtained in Ref.~\cite{BTW-96}.
We note sizeable differences between the results of the two approaches.

%% \begin{figure}[tb]
%% \hspace{0cm}
%% \centerline{\psfig{width=7.5truecm,angle=0,file=fxO3.eps}}
%% \vspace{0cm}
%% \caption{
%% The scaling function $f(x)$ for the Heisenberg universality class. 
%% Results from  Refs. \cite{CHPRV-02,BTW-96}. 
%% }
%% \label{figfxO3}
%% \end{figure}
%% 
%% 
%% \begin{figure}[tb]
%% \hspace{0cm}
%% \centerline{\psfig{width=7.5truecm,angle=-0,file=duO3.eps}}
%% \vspace{0cm}
%% \caption{
%% The scaling function $D_R(y_R)\equiv D(y)/D(y_{\rm  max})$ versus
%% $y_R\equiv y/y_{\rm max}$, cf. Eq.~(\ref{defDw}). 
%% Results from  Ref. \cite{CHPRV-02}.
%% }
%% \label{figdyO3}
%% \end{figure}

\subsubsection{Universal amplitude ratios}
\label{unraO3}

In Table~\ref{univratiosO3} we report the estimates of  several
universal amplitude ratios. 
The results denoted by IHT--PR were obtained in
Ref.~\cite{CHPRV-02}  using approximate parametric representations
of the equation of state.
The FT estimates of $U_0$ were obtained from the analysis
of the fixed-dimension expansion in the 
minimal-renormalization scheme without $\epsilon$ expansion 
\cite{LMSD-98,KV-00}
and from the standard $\epsilon$ expansion to $O(\epsilon^2)$ 
\cite{Bervillier-86}.
The CRG estimate of $U_0$ and $R_\alpha$ were  obtained in
Ref. \cite{CHPRV-02}
 using the expression for $f(x)$ reported in
Refs.~\cite{BTW-96,BTW-99}; they significantly differ from the 
estimates obtained using other methods.
See, e.g., Ref.~\cite{PHA-91} for a more complete review of 
theoretical and experimental estimates.

\begin{table*}
\caption{
Estimates of universal amplitude ratios for the Heisenberg universality class.
The numbers marked by an asterisk have been obtained in Ref. \cite{CHPRV-02}
 using the results reported in the quoted references.
}
\label{univratiosO3}
\footnotesize
\begin{center}
\begin{tabular}{ccccccc}
\hline
 & IHT--PR \cite{CHPRV-02} & $d$=3 exp & $\epsilon$ exp & CRG & HT & experiments \\
\hline
$U_0$ & 1.56(4) & 1.51(4) \cite{LMSD-98} & 1.521(22) \cite{Bervillier-86} & 
            $^*$1.823 \cite{BTW-96,BTW-99} & & 1.50(5) \cite{KR-94}\\
      &         & 1.544 \cite{KV-00} & & & &   1.27(9) \cite{MMFB-96} \\
      & & & & & & 1.4(4) \cite{Ramos-etal-01} \\ 
$R_\alpha$ & 4.3(3)  &  $^*$4.4(4) \cite{LMSD-98}  & 4.56(9)
           \cite{Bervillier-86} & $^*$3.41 \cite{BTW-96,BTW-99} &&  \\
           &         &  $^*$4.46 \cite{KV-00}  &   &&&  \\

$R_\chi$ &  1.31(7) &  & 1.33 \cite{AM-78} & 1.11 \cite{BTW-96,BTW-99} & & \\

$R_C$    & 0.185(10) & 0.189(9) \cite{SLD-99} & 0.17 \cite{AH-76} & & & \\
         & & 0.194 \cite{KV-00} & & & & \\

$R_4$ & 7.8(3) & &&& & \\

$R_\xi^+$ & 0.424(3) & 0.4347(20) \cite{BB-85} & 0.42 \cite{Bervillier-76} &&
0.431(5) \cite{BC-99} & \\
& & 0.4319(17) \cite{BG-80} & & & 0.433(5) \cite{BC-99}& \\ 

$P_m$ & 1.18(2) & &&& & \\

%$P_c$ & 0.357(5) & &&& & \\

$R_p$ & 2.020(6) & &&& & \\
\hline
\end{tabular}
\end{center}
\end{table*}

\subsubsection{Comparison with the experiments} \label{CES.E}

In spite of the large number of experiments, at present there is no 
accurate quantitative study of the equation of state in the critical 
regime. We shall discuss here three different representations that 
are widely used in experimental work.

\begin{figure}[tb]
\hspace{0cm}
\centerline{\psfig{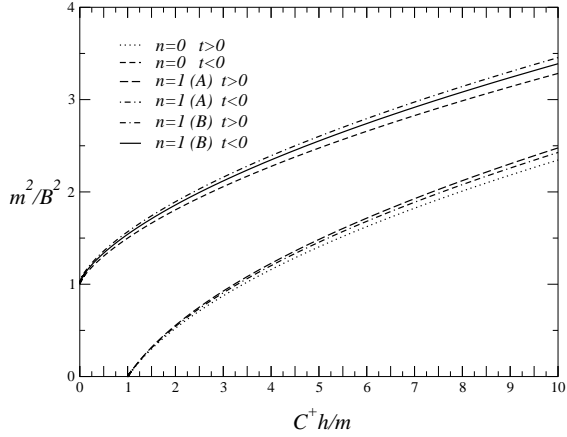}}
\vspace{0cm}
\caption{
Plot of $m^2/B^2$ versus $C^+ h/m$. From Ref. \cite{CHPRV-02}.
}
\label{figm2ohmO3}
\end{figure}

A first possibility \cite{KR-67} consists in studying the behavior of 
$h/m\equiv H |t|^{-\gamma}/M$ versus $m^2 \equiv M^2 |t|^{-2\beta}$. 
Such a function 
can be easily obtained from approximations of $f(x)$, since 
$m^2 = B^2 |x|^{-2\beta}$ and 
\begin{equation} 
   {h\over m} = k |x|^{-\gamma} f(x)
\end{equation}
where $k = \left(B_c \right)^{-\delta} B^{\gamma/\beta} = {R_\chi/C^+}$.
A plot of $m^2/B^2$ versus $C^+ h/m$ is reported in Fig. \ref{figm2ohmO3}.
It agrees qualitatively with the analogous experimental ones 
reported, e.g., in Refs. \cite{KR-94,BK-97,FKK-01}.

Often, for small $h/m$ one approximates the equation of state by
writing 
\begin{equation}
   {h\over m} = a_{\pm} + b_{\pm} m^2,
\label{Kouvell-Rodbell}
\end{equation}
where $a_{\pm}$ and $b_\pm$ are numerical coefficients depending on the phase.
Such an approximation has a limited range of validity. In the HT
phase, one obtains for $m^2 \to 0$ \cite{CHPRV-02}
\begin{eqnarray}
{h\over m} &=& {1\over C^+} \left[ 1 + {R_4\over 6} {m^2\over B^2} + 
   \sum_{n=2}^\infty {R^n_4 \ r_{2n+2}\over (2n+1)!} 
   \left({m^2\over B^2}\right)^n \right]
\nonumber \\
&\approx& {1\over C^+} \Bigl[ 1 + 1.30(5)\ {m^2\over B^2} + 
              0.94(8)\ \left({m^2\over B^2}\right)^2        
+ 0.06(2)\ \left({m^2\over B^2}\right)^3 + \cdots\Bigr].
\label{hsum-expansion}
\end{eqnarray}
From Eq. (\ref{hsum-expansion}), one sees that the 
approximation (\ref{Kouvell-Rodbell}) is valid only for very small $m^2$, 
i.e. at the 1\% level only for $m^2 \lesssim 0.01 B^2$. 
The quadratic approximation---i.e. the approximation with an 
additional $(m^2)^2$ term---has a much wider range of validity 
because of the smallness of the coefficient of $m^6$.

In the low-temperature phase, Eq. (\ref{Kouvell-Rodbell}) is theoretically
incorrect, since it does not take into account the presence of 
Goldstone modes. Indeed, for $m^2/B^2\to 1$, we have 
\begin{equation}
{h\over m} \approx {k c_f\over 4 \beta^2} \left(1 - {m^2\over B^2}\right)^2,
\label{hsum-coex}
\end{equation}
where $c_f$ is defined in Eq. (\ref{fxcc}).
Eq. (\ref{hsum-coex}) is inconsistent with the approximation
(\ref{Kouvell-Rodbell}) near the coexistence curve.

Finally, note that for $m^2$ large 
we have
\begin{equation}
{h\over m} \approx k \left({m\over B}\right)^{\delta-1}.
\end{equation}
A second form that is widely used to analyze the experimental data 
is the Arrott-Noakes \cite{AN-67} scaling equation
\begin{equation}
\left({H\over M}\right)^{1/\gamma} = a t + b M^{1/\beta},
\end{equation}
where $a$ and $b$ are numerical constants. This approximation 
is good in a neighborhood of the critical isotherm $t=0$. 
Since
\begin{equation}
\left({H\over M}\right)^{1/\gamma} k^{-1/\gamma} = 
\left({M\over B}\right)^{1/\beta} f(x)^{1/\gamma},
\end{equation}
one obtains \cite{CHPRV-02}
\begin{eqnarray}
\left({H\over M}\right)^{1/\gamma} k^{-1/\gamma} = 
\left({M\over B}\right)^{1/\beta} + 
 0.96(4)\ t 
- 0.04(2)\, t^2 \left({M\over B}\right)^{-1/\beta}
            - 0.02(2)\, t^3 \left({M\over B}\right)^{-2/\beta}
        \cdots 
\end{eqnarray}
Thus, at a 1\% level of precision the Arrott-Noakes formula is valid approximately 
for $t (M B^{-1})^{-1/\beta} \lesssim  25$ which is quite a large interval.
  
Finally, Ref.~\cite{Zhao-etal-99} reports an experimental study of
the behavior of Tl$_2$Mn$_2$O$_7$ along the crossover line,
and determines the scaling function $D(y)$, although with
different normalizations. The comparison of their results
with the theoretical curve $D(y)$ obtained 
in Ref. \cite{CHPRV-02}, see Sec.~\ref{eqstO3},
shows a very nice  quantitative agreement.

\section{Critical behavior of $N$-vector models with $N\geq 4$}
\label{ngeq4}

Among the three-dimensional $N$-vector models  with $N\geq 4$, 
the physically most relevant
ones are those with $N=4$ and $N=5$. 
The $N=4$ case is relevant for high-energy physics because it
describes the finite-temperature transition in the 
theory of strong interactions, i.e. quantum chromodynamics (QCD),  
with two light degenerate flavored quarks. 
The case $N=5$ might be relevant for superconductivity: 
indeed, an SO(5) theory has been proposed to explain the 
critical properties of high-$T_c$ 
superconductors \cite{Zhang-97}.
In this section we mainly review these two models.

For larger values of $N$, estimates of the critical exponents
can be found in Refs. \cite{BC-97-2,AS-95}.
They are obtained by analyzing 
21st-order expansions for the $N$-vector model, 
and the FT six-loop series at fixed dimension $d=3$.

In the large-$N$ limit
one can obtain analytic results based on a $1/N$ expansion. 
These results are very useful to obtain  a qualitative understanding 
of the critical behavior, but, from a quantitative point of view,
they become predictive only for rather large values of $N$, $N\gtapprox 10$ say.
We do not further discuss
this approach, but we signal the recent review
\cite{Zinn-Justin-98}, where many results and
references can be found.

We mention that the critical equation of state for the $N$-vector
model has been computed to $O(\epsilon^2)$ in the framework of the
$\epsilon$ expansion \cite{BWW-72}, 
and to $O(1/N)$ in the framework of the $1/N$ expansion \cite{BW-73}.

\subsection{The O(4) universality class}
\label{N4case}

The three-dimensional O(4) model is relevant for QCD
with two light-quark flavors at finite temperature. 
This theory shows a finite-temperature transition, in which
the quark condensate $\langle \bar{\psi}\psi\rangle$ is the order 
parameter and the quark
mass plays the role of external field.
Using symmetry arguments, it has been argued that,
if the finite-temperature transition is continuous, it should belong
to the same universality class of the three-dimensional O(4)
model \cite{PW-84,Wilczek-92,RW-93}. 
Finite-temperature simulations of lattice QCD \cite{IKKY-97,CP-PACS-99}
support the existence of a continuous phase transition. 

In Table~\ref{N4exponents} we report the theoretical results for the
critical exponents of the O(4) symmetric model.
They have been obtained by FSS MC simulations for the 
$N$-vector model \cite{KK-95,BFMM-96} and for 
an improved $\phi^4$ theory \cite{Hasenbusch-00},
from the analysis of 21st-order HT expansions \cite{BC-97-2},
using perturbative FT methods \cite{AS-95,GZ-98} and 
the nonperturbative CRG approach \cite{MT-98,GW-01}.
Numerical studies of the critical equation of state 
can be found in  Refs.~\cite{Toussaint-97,EM-00}.

\begin{table*}[t]
\caption{Estimates of the critical exponents for the O(4)-vector model.
We indicate with an asterisk (${}^*$) the estimates that have been
obtained  using the scaling relations $\gamma =(2 - \eta)\nu $,
$2 - \alpha = 3 \nu $.
}
\footnotesize
%\hspace*{-2.7cm}    % Move table leftwards, so it doesn't run off the right
%\tabcolsep 4pt        % Less than the usual 6pt
%\doublerulesep 1.5pt  % Less than the usual 2pt
\begin{center}
\begin{tabular}{rlllll}
\hline
\multicolumn{1}{c}{Ref.}& 
\multicolumn{1}{c}{info}& 
\multicolumn{1}{c}{$\gamma$}& 
\multicolumn{1}{c}{$\nu$}& 
\multicolumn{1}{c}{$\eta$}&
\multicolumn{1}{c}{$\omega$} \\   
\hline  
\cite{Hasenbusch-00} $_{2001}$  &  MC FSS $\phi^4$  &  
          1.471(4)$^*$ & 0.749(2)&  0.0365(10)  &  0.765  \\
\cite{BFMM-96} $_{1996}$  & MC FSS & 1.476(2)$^*$ & 0.7525(10)  & 0.0384(12)  & \\
\cite{KK-95} $_{1995}$  & MC FSS & 1.477(18)$^*$ & 0.748(9)  & 0.0254(38) &   \\
\cite{BC-97-2} $_{1997}$ & HT sc & 1.491(4)   & 0.759(3)  & 0.035(9)$^*$  &   \\
\cite{BC-97-2} $_{1997}$ & HT bcc & 1.484(4)   & 0.756(3) & 0.037(9)$^*$  &   \\
\cite{GZ-98} $_{1998}$ & FT $d=3$ exp & 1.456(10)      & 0.741(6) & 0.0350(45) & 0.774(20) \\
\cite{AS-95} $_{1995}$ & FT $d=3$ exp & 1.449      & 0.738 & 0.036 & \\
\cite{GZ-98} $_{1998}$ & FT $\epsilon$ exp & 1.448(15) & 0.737(8) & 0.036(4) & 0.795(30) \\
\cite{GW-01} $_{2001}$ & CRG (1st DE) & 1.443 & 0.739  & 0.047  &\\
\cite{MT-98} $_{1998}$ & CRG (1st DE) & 1.614 & 0.816  & 0.022  &\\
\hline
\end{tabular}
\end{center}
\label{N4exponents}
\end{table*}

\subsection{The O(5) universality class and the SO(5) theory of
high-$T_c$ superconductivity}
\label{N5case}

The O(5) universality class has not been much studied and, at present,
there are only a few estimates of the critical parameters. 
The critical exponents have been determined using FT methods 
\cite{AS-95}, obtaining $\gamma=1.506$, $\nu=0.766$, and $\eta=0.034$,
and by MC simulations \cite{Hu-01}, finding $\nu=0.728(18)$. 
Concerning the  spin-2 and 
spin-4 perturbations, see Sec. \ref{sec-1.5.8},
we report the following results:
$\phi_2 = 1.40(4)$ \cite{CPV-02} 
and $\phi_4 = 0.145(7)$ \cite{CPV-02-4}
obtained using FT methods, and $\phi_2 = 1.387(30)$ \cite{Hu-01}
from MC simulations.

It has been argued that the O(5) universality class is relevant for the 
description of high-$T_c$ superconductivity. In the SO(5) theory
\cite{Zhang-97}, one considers a model with symmetry O(3)$\oplus$U(1) = 
O(3)$\oplus$O(2) with two order parameters: one is related to the 
antiferromagnetic order, the other one is associated with 
$d$-wave superconductivity. Neglecting the fluctuations of the electromagnetic field
\cite{HLM-74}, the most general Hamiltonian with symmetry 
O(3)$\oplus$O(2) describing the interactions of the 
two order parameters is 
\begin{eqnarray}
{\cal H}_{\rm eff} = \int d^3 x 
\Bigl[ 
\case{1}{2}( \partial_\mu \phi_1)^2  + \case{1}{2} (
\partial_\mu \phi_2)^2 + \case{1}{2} r_1 \phi_1^2  
 + \case{1}{2} r_2 \phi_2^2  
+ u_1 (\phi_1^2)^2 + u_2 (\phi_2^2)^2 + w \phi_1^2\phi_2^2 \Bigr],
\label{bicrH} 
\end{eqnarray}
where $\phi_1$ is the  three-component 
antiferromagnetic order parameter and $\phi_2$ is the two-component 
superconductivity order parameter. Such Hamiltonian has been 
extensively studied in the literature, see, e.g. Ref.~\cite{KNF-76},
since it describes a variety of multicritical phenomena. See also
Sec. \ref{LGW-multicritical}.

The main issue is whether the SO(5) symmetry can be realized at a bicritical
point where two critical lines, with symmetry O(3) and O(2) respectively,
meet. In RG terms, this can generally occur if 
the O(5) fixed point has only {\em two} relevant 
O(3)$\oplus$ O(2)-symmetric perturbations.
On the other hand, as shown in Ref. \cite{CPV-02-4}, 
(at least) {\em three} relevant perturbations exist:
beside the usual O(5)-symmetric interaction associated with the temperature,
there are two perturbations with O(5)-spin 2 and 4 respectively 
(explicit formulae are given in Sec. \ref{lsec-cubic} and 
in Ref. \cite{CPV-02-4}) that are relevant,
since $\phi_2 > 0$ and $\phi_4>0$ (see the estimates reported above).
The stable fixed point is expected to be the tetracritical 
decoupled fixed point. Indeed, using nonperturbative arguments 
one can show \cite{Aharony-02,Aharony-02-2} that the RG dimension $y_w$
associated with the perturbation $w\phi_1^2\phi_2^2$ of the 
decoupled fixed point is negative, i.e. the perturbation is irrelevant.
This follows from
\begin{equation}
y_w = {1\over 2}\left({\alpha_{XY}\over
\nu_{XY}} + {\alpha_{O(3)}\over \nu_{O(3)}}\right) < 0.
\end{equation} 
As a consequence, the SO(5) fixed point can only be reached by tuning an 
additional parameter. 

These conclusions are in apparent contrast with the 
MC results of Ref.~\cite{Hu-01}, that seem to favor the 
picture based on a stable bicritical O(5) fixed point.
On the other hand, as suggested by Aharony \cite{Aharony-02}, the MC
data of Ref.~\cite{Hu-01} may just show  a 
slow crossover towards either the stable tetracritical decoupled point
or a weak first-order transition.
This hypothesis is somehow
supported by the small value of the crossover exponent
$\phi_4$ at the O(5) fixed point, i.e. $\phi_4=0.145(7)$.

\section{The two-dimensional $XY$ universality class}
\label{XYd2}

\subsection{The Kosterlitz-Thouless critical behavior}
\label{KTcb}

The two-dimensional $XY$ universality class is characterized by the 
Kosterlitz-Thouless (KT) critical behavior~\cite{KT-73,Kosterlitz-74}
(see, e.g.,
 Refs.~\cite{It-Dr-book,Zinn-Justin-book} for reviews on this issue).
According to the KT scenario, the free energy has an essential
singularity at $T_c$ and the correlation length diverges as
\begin{equation}
\xi\sim \exp \left( {b/ t^{\sigma}} \right)
\label{xidiv}
\end{equation}
for $t\equiv T/T_c-1 \to 0^+$.
The value of the exponent is  $\sigma=1/2$ and $b$ is a 
nonuniversal positive constant. At the critical temperature, the
asymptotic behavior for $r\rightarrow\infty$ of the two-point
correlation function should be 
\begin{equation}
G(r)_{\rm crit} \sim {(\ln r)^{2\theta}\over r^\eta}
\,\left[ 1 + O\left( {\ln\ln r\over \ln r}\right)\right],
\label{gx}
\end{equation}
with $\eta=1/4$ and $\theta=1/16$. Near criticality,
i.e., for $0 < t \ll 1$, the behavior of the magnetic susceptibility
can be derived from Eq.~(\ref{gx}):
\begin{eqnarray}
\chi\sim \int_0^\xi  dr \,G(r)_{\rm crit}
\sim \xi^{2-\eta} \left( \ln \xi \right)^{2\theta}
\left[ 1 + O\left( {\ln\ln \xi\over \ln \xi}\right)\right]
\sim \xi^{2-\eta} t^{-2\sigma\theta}
\left[ 1 + O\left( t^{\sigma}\ln t \right)\right].
\label{chi}
\end{eqnarray}
In addition, the two-dimensional $XY$ model is 
characterized by a line of critical
points, starting from $T_c$ and extending to $T=0$,
with $\eta\sim T $ for $T \rightarrow 0$.
At criticality the two-dimensional $XY$ model corresponds to a conformal
field theory with $c=1$, see, e.g., Ref.~\cite{It-Dr-book}.

Transitions of KT type occur in a number of effectively 
two-dimensional systems with O(2) symmetry, such
as thin films of superfluid helium and 
of easy-plane magnetic materials.
In particular,
planar ferromagnets are realized by layered compounds 
such as K$_2$CuF$_4$ \cite{HY-79}
and Rb$_2$CrCl$_4$ \cite{HDJP-86,ANBHMIV-93} that effectively behave 
as two-dimensional systems.
The crossover from two-dimensional to three-dimensional behavior 
has been observed
in CoCl$_2$ intercalated in graphite \cite{WZS-94},
in Gd$_2$CuO$_4$ \cite{MRBSGV-00},
and in YBa$_2$Cu$_3$O$_{6+x}$ \cite{Mont-etal-98}.
We also mention the experimental results of Ref. \cite{BGMBB-01}
for iron films epitaxially grown on GaAs(001).

The KT critical scenario describes roughening transitions, i.e.
phase transitions from a smooth to a rough surface, 
which are, for example, observed
in the equilibrium structure of crystal interfaces. 
For a general introduction to roughening, see, e.g.,
Refs.~\cite{Abraham-86,vN-87,FLN-91,It-Dr-book}.
At a roughening transition, the large-scale interface behavior changes 
from being smooth at low temperature to being rough at high temperature.
This qualitative picture can be made quantitative,
for example by 
looking at the dependence of the interfacial width on the size $L$
of the interface:
in the smooth phase the interfacial width remains finite when $L\to\infty$,
while it diverges logarithmically in the rough phase.

\subsection{The roughening transition and solid-on-solid models}
\label{routr}

The interfacial thermodynamic behavior 
can be modeled by
solid-on-solid (SOS) models defined on two-dimensional lattices.
The partition function of a SOS model is
\begin{equation}
Z = \sum_{\{ h \} } \exp \Bigl[ -\sum_{\langle xy \rangle} V(h_x-h_y) \Bigr],
\end{equation}
where the Hamiltonian is a sum over nearest-neighbor pairs.
The variables $h_x$ can be interpreted as heights with respect to 
a certain base.
The summation is over equivalence classes of height configurations $\{ h \}$,
defined by identifying two configurations that differ only by a global 
vertical shift.
Several realizations of the SOS model have been proposed, see, e.g.,
Refs.~\cite{Abraham-86,vN-87,HP-97}.
We mention:
\begin{itemize}
\item[(i)]
The absolute-value SOS model (ASOS).
It can be considered
as the SOS approximation of a lattice plane interface of an Ising model 
on a simple
cubic lattice, and it is 
defined by the function
\begin{equation}
V_{\rm ASOS} = k | h_x - h_y|,
\end{equation}
where $h_x$ takes integer values.
For finite positive $k$ the Hamiltonian suppresses configurations with large
differences between nearest-nei\-gh\-bor sites. 
If $k$ is below a certain critical value,
the surface becomes rough, and the surface thickness diverges when the size
of the system goes to infinite.

\item[(ii)]
The body-centered SOS model (BCSOS). It represents a SOS approximation of
an interface in an Ising model on a bcc lattice \cite{vanBeijeren-77}. 

\item[(iii)]
The discrete Gaussian (DG) model defined by
\begin{equation}
V_{\rm ASOS} = k ( h_x - h_y )^2,
\end{equation}
where $h_x$ takes integer values.
It is dual to the
Villain formulation of the $XY$ model \cite{Savit-80}. 

\item[(iv)]
The dual of the standard $XY$ model,  
\begin{equation}
H_{XY} = - \beta \sum_{\langle xy\rangle} \vec{s}_x \cdot \vec{s}_y,  
\end{equation}
can be considered as a SOS model with partition function
\begin{equation}
Z = \sum_{ \{ h \} } \prod_{\langle xy\rangle} I_{|h_x-h_y|}(\beta),
\end{equation}
where $h_x$ are integer variables and $I_n$ are modified Bessel
functions. 
\end{itemize}

The BCSOS model  is the only one that
has been proved to undergo a KT transition \cite{vanBeijeren-77}.
It can be transformed into the $F$ model \cite{vanBeijeren-77}, 
which is a special six-vertex model
that can be solved exactly using transfer-matrix 
methods \cite{Lieb-67,LW-72,Baxter-82}.
The transitions in the other models, including the standard $XY$ model, 
are conjectured to be of the KT type as well, and
numerical evidences support this fact.

A much studied prototype of roughening is the Ising interface transition, which
occurs in the LT phase of the three-dimensional Ising model,
i.e. for $T_r < T_c$ where $T_c$ is the bulk critical temperature \cite{WGL-73,HP-97-XY}.
The SOS approximation amounts to neglecting overhangs of the Ising interface
and bubbles in the two phases separated by the interface. 
Lattice studies of the interfacial properties use the fact that
an interface of size $L$ can be realized on a $L^2\times T$ lattice by imposing
antiperiodic boundary conditions along the third direction.

Roughening transitions also occur in the lattice formulation of 
four-dimensional nonabelian gauge theories \cite{Wilson-74}, 
which, in the critical limit $T\rightarrow 0$, 
provide a nonperturbative definition of QCD,
the theory of strong interactions.
See, e.g., Refs.~\cite{Creutz-book,Mo-Mu-book} for an introduction to
lattice gauge theories.
An important issue concerning nonabelian gauge theories is related to
confinement. In the absence of quarks, the question of confinement 
is related to the behavior of the Wilson loop in the limit of large area.
Confinement requires a nonzero string tension $\sigma$.
In the HT region $\sigma$ is not zero. For $T\to\infty$ it behaves as 
$\sigma \sim \ln T$,  with $T^{-n}$ corrections
that can be systematically computed.
Theoretical arguments show that, independently of the 
gauge group, the string tension is affected by a weak singularity
associated with the roughening transition
\cite{Parisi-79,HHH-81,IPZ-80,Luscher-81,LMW-81,DZ-81,MW-81}
(see also Ref.~\cite{It-Dr-book}).
However, the roughening transition does not imply deconfinement, and 
the string tension should not vanish in the weak-coupling region, and
therefore in the continuum limit.
The change is essentially related to the fact that
at strong coupling the contributions to the string tension come
from smooth surfaces, while in the weak coupling region
the relevant surfaces become rough.

\subsection{Numerical studies}
\label{nust}

Numerical studies of the $XY$ model based on MC simulation techniques,
HT expansions, and CRG methods support the KT behavior 
\cite{BP-89,BG-92,JN-93,HMP-94,Olsson-94,SM-94,KI-95,Kim-95,Janke-96,HP-97,GW-01}.

In MC simulations, FSS investigations at criticality must be
very precise in order to pinpoint the logarithm in the two-point
Green's function, cf. Eq. \reff{gx}.  
On the other hand, if this logarithmic correction
is neglected, i.e. one sets $\theta=0$, 
a precise check of the prediction $\eta=1/4$
at $\beta_c$ may be quite hard. The relevance of such logarithmic
corrections and some of the consequences of neglecting them have been
examined in Refs.~\cite{KI-95,Janke-96}. 

HT expansion calculations \cite{BC-93,BC-94,CPRV-96-XY}
support the KT mechanism as well. 
In Ref.~\cite{CPRV-96-XY} the two-point correlation function
was  calculated on the square, triangular,
and honeycomb lattices 
respectively up to 21st, 15th, and 30th order. 
The results from all considered lattices
were consistent with universality and 
the KT exponential
approach \reff{xidiv}. The prediction
$\sigma=1/2$ has been confirmed with an uncertainty of few per
cent.  The prediction $\eta=1/4$ has been also substantially verified.

On the other hand, the value of $\theta$ predicted by the
KT theory, i.e. $\theta=1/16$,  has not yet got
a direct robust numerical confirmation. 
Some discrepancies have been noted in the results of MC simulations
\cite{KI-95,Janke-96} and HT expansions
\cite{CPRV-96-XY}.
But one should be cautious 
in considering these results as a real inconsistency,
since it is difficult to exclude a substantial
underestimate of the error.

The most accurate verification of the KT critical pattern was achieved in
Refs.~\cite{HMP-94,HP-97-XY}, by numerically matching the
RG trajectory of the dual of the $XY$ model with that
of the BCSOS model, which has been proven to
exhibit a KT transition.  The advantage of this strategy is that such
a matching occurs much earlier than the onset of the asymptotic
regime, where numerical simulations can provide quite accurate
results. Indeed, the authors argued that their method is subject to corrections
due to irrelevant operators, which are of order $L^{-\omega}$  with
$\omega = 2$, while standard MC results are affected by logarithmic 
corrections.
In Refs.~\cite{HMP-94,HP-97-XY} the same method was applied 
to the ASOS model, the DG model, and the interface in a simple-cubic Ising 
model, in order to demonstrate that they all belong to
the same universality class, and
therefore undergo a KT transition.

Finally, we mention the results
available for  the critical limit $g_4^+$ of
 zero-momentum four-point coupling $g_4$, defined
in Eq. \reff{grdef}, and the small-magnetization expansion of the
Helmholtz free energy.
In Table ~\ref{XYd2g4} we review the estimates of $g_4^+$ obtained
by various approaches:
high-temperature series (HT), Monte Carlo simulations (MC),
field-theoretical methods (FT), and the form-factor approach (FF).
The first few coefficients $r_{2j}$ of the small-magnetization
expansion of the scaling equation of state, cf. Eq.~(\ref{Fzdef}),
were estimated in Ref.~\cite{PV-00} by a constrained analysis
of their $O(\epsilon^3)$ series, obtaining $r_6=3.53(4)$ and
$r_8\approx 23$. 

Results concerning the small-momentum
behavior of the two-point function in the HT phase can be found
in Ref.~\cite{CPRV-96-XY}.

\begin{table*}
\caption{
Estimates of $g_4^+$ for the two-dimensional $XY$ universality class.
}
\label{XYd2g4}
\footnotesize
\begin{center}
\begin{tabular}{llccc}
\hline
\multicolumn{1}{c}{HT}& 
\multicolumn{1}{c}{MC}& 
\multicolumn{1}{c}{FT $d=2$ exp}&
\multicolumn{1}{c}{FT $\epsilon$ exp} &
\multicolumn{1}{c}{FF} \\
\hline  
13.65(8) \cite{PV-gr-98}  &  13.71(18) \cite{BNNPSW-00} & 13.57(23) \cite{OS-00} & 
13.7(2) \cite{PV-00} & 13.65(6) \cite{BNNPSW-01} \\
13.72(16) \cite{BC-96}&  13.3(3) \cite{Kim-95-2}& & & \\ 
\hline
\end{tabular}
\end{center}
\end{table*}

\section{Two-dimensional $N$-vector models with $N\geq 3$}
\label{sec-5.2}

Two-dimensional $N$-vector models with $N\ge 3$ are somewhat special since in this case 
there is no phase transition at finite values of $T$. The correlation length
is always finite and a critical behavior is observed only when 
$T\to 0$.\footnote{One can rigorously prove that systems with 
a vector order parameter do not have a magnetized LT phase in 
two dimensions \cite{MW-66}. This  does not exclude a LT
phase with algebraically decaying spin-spin correlation functions, 
as it happens 
for $N=2$. However, numerical and theoretical works indicate that this 
is not the case for $N\ge 3$. For a critical discussion of this issue, see 
Refs. \cite{PS-all,David,CEPS-96,NNW-97,AACP-99} and references therein.
Note also that the Mermin-Wagner theorem \cite{MW-66} excludes only
that the spin-spin correlation function be critical. Other correlation
functions may show a critical behavior for finite $T$,
see, e.g., Refs. \cite{MR-87,BGH-02,CP-02}.
} 

For $T\to 0$ the behavior of long-distance quantities is predicted 
by the perturbative RG applied to the 
$N$-vector model. One finds that 
these systems are asymptotically free with a nonperturbatively 
generated mass gap
\cite{Polyakov-75,BZ-76,BZL-76a,BLS-76}. Such a property is also present 
in QCD, i.e. the theory of strong interactions, 
and thus these two-dimensional theories are often used as  toy models
in order to understand nonperturbative properties and to test numerical 
methods that are of interest for the more complex theory of QCD
(see, e.g., Ref.~\cite{Polyakov-book}). 

The two-dimensional $N$-vector model is also important in condensed-matter
physics. For $N=3$ it describes the behavior for $T\to0$ of the two-dimensional 
spin-$S$ Heisenberg quantum antiferromagnet \cite{CHN-89}. 
Indeed, at finite temperature $T$ 
this quantum spin system is described by a (2+1)-dimensional 
$O(3)$ classical theory in which the Euclidean time direction 
has a finite extent $1/T$. In the critical limit
the relation $1/T\ll \xi$  is satisfied, therefore
 the system becomes 
effectively two-dimensional, and thus its critical behavior is  
described by the two-dimensional three-vector model. 

Moreover, the non-linear FT formulation of the
two-dimensional $N$-vector model with $N\geq 3$ is the starting point for the
$\varepsilon\equiv d-2$ expansion (see,
e.g., Ref.~\cite{Zinn-Justin-book} and references therein),
which provides information of the 
critical properties of $N$-vector models for $d\gtapprox 2$.
The $\varepsilon$ expansion of the critical exponents is known to
$O(\varepsilon^4)$ \cite{HB-78,Hikami-83,BW-86}. Contrary to
what happens for the $\epsilon=4-d$ expansion, $\varepsilon$-series 
have limited application and do not provide quantitative predictions
for physical three-dimensional
transitions, essentially because they are
not Borel summable \cite{HB-78} and 
it is not known how to resum them.
An attempt based on a variational
approach was presented in Ref.~\cite{Kleinert-00}.

Beside being asymptotically free, the two-dimen\-sio\-n\-al $N$-vector model
with $N\ge 3$ has another remarkable property. The corresponding quantum field theory
is integrable, in the sense that, assuming asymptotic freedom,  
one can establish the existence of nonlocal conserved 
charges and the absence of particle production 
\cite{Luescher-78,BL-79,BLR-86}. This implies the existence of a factorized 
$S$-matrix \cite{ZZ-79}  and allows the application of powerful 
techniques, such as the thermodynamic Bethe Ansatz and the form-factor
bootstrap approach (the latter has been 
applied only to the $N=3$ case). These approaches have led to exact predictions 
for several nonperturbative quantities 
\cite{HMN-90,HN-90,KW-78,LW-90,BaNi-97,BNNPSW-99,BNNPSW-00}.

\subsection{The critical behavior} \label{sec-5.2.1}

The critical behavior for $T\to 0$ can be computed in perturbation theory
for a particular class of $O(N)$ models that, in this context,
are usually called $\sigma$-models. On the lattice one may consider 
Hamiltonians 
\be 
{\cal H} = - {1\over T} \sum_{ij} K(i-j) \vec{s}_i\cdot \vec{s}_j,
\label{sigma-model}
\ee
where $K(x)$ is a short-range coupling and $|\vec{s}_i| = 1$. If only
nearest-neghbor spins are coupled, we reobtain the 
$N$-vector model \reff{NvectorHamiltonian}. It is also 
possible to add couplings that involve more than two spins without 
changing the universality class. The only important requirement 
is that the ground state of the system is ferromagnetically ordered. 

For this class of systems we can use spin-wave perturbation theory 
and the RG to obtain the behavior for $T\to 0$.
See, e.g., Ref.~\cite{Zinn-Justin-book} for a general reference.
For the true correlation length $\xi_{\rm gap}$ and the 
susceptibility $\chi$ one has
\begin{eqnarray}
\xi_{\rm gap}(T) &=& \xi_0 \left({\beta_0 T}\right)^{\beta_1/\beta_0^2}
\exp\left( {1\over \beta_0 T}\right)\, 
\exp\left[\int_0^T dt\left({1\over \beta(t)} +
{1\over \beta_0t^2} -
{\beta_1\over \beta_0^2 t}\right)\right] , 
\label{xi-2d-sigmamodel} \\
\chi(T) &=& \chi_0 
\left( {\beta_0 T}\right)^{2 \beta_1/\beta_0^2+\gamma_0/\beta_0} 
 \exp\left( {2\over \beta_0 T}\right) \times\nonumber \\
&&  \exp\left[\int_0^{T} dt\, \left({2\over \beta(t)} +
\, {2\over \beta_0 t^2} -\, {2\beta_1\over \beta_0^2 t}
-{\gamma(t)\over \beta(t)}
- \, {\gamma_0\over \beta_0 t}\right)\right], 
\label{chi-2d-sigmamodel}
\end{eqnarray}
where $\beta(T)$ is the $\beta$-function, defined by 
$\beta(T) = - a d T/da$ with $a$ being the lattice spacing, 
and $\gamma(T)$  is the anomalous dimension of the field. 
These functions describe how the temperature and the 
fundamental field should vary with the lattice spacing $a$ to keep
the renormalized Green's functions fixed.
They have a perturbative expansion of the form
\begin{eqnarray}
\beta(T) = - T^2 \sum_{n=0} \beta_n T^n, 
\qquad
\gamma(T)= T \sum_{n=0} \gamma_n T^n,
\end{eqnarray}
and are model-dependent. However, two particular 
combinations of the perturbative coefficients are universal,\footnote{
Often the claim is made that $\beta_0$, $\beta_1$, and $\gamma_0$ are 
universal. This is not strictly correct, since 
these quantities depend on the normalization of the temperature $T$. 
What it is usually meant 
is that, once a universal normalization is fixed, these constants are 
universal. The standard normalization of $T$ is such that at tree level
the two-point function is $\widetilde{G}^{-1}(q) = q^2/T$. Note that
the product $(\beta_0 T)$ is independent of the normalization of $T$.}
$\beta_1/\beta_0^2$ and 
$\gamma_0/\beta_0$, that appear as universal exponents in the critical behavior 
of $\xi$ and $\chi$. Explicitly: 
\begin{eqnarray}
{\beta_1 \over \beta_0^2} = {1\over N-2}, \qquad
{\gamma_1 \over\beta_0}   = {N-1\over N-2}.
\end{eqnarray}
The perturbative expansions of $\beta(T)$ and $\gamma(T)$ have been 
determined for several different models: to four-loop order for the 
continuum theory in the minimal subtraction scheme \cite{BW-86} 
and for the standard $N$-vector model \cite{CP-95,ACPP-99}, 
and to three-loop order for several other theories 
\cite{FT-86,CP-94,AP-00}.

The nonperturbative constants $\xi_0$ and $\chi_0$ are also nonuniversal. 
The ratio of $\xi_0$ (and also of $\chi_0$) in two different models 
can be computed in one-loop perturbation theory.
For the $N$-vector model with nearest-neighbour interactions, $\xi_0$
was computed exactly by means of the thermodynamic Bethe Ansatz
\cite{HMN-90,HN-90}:
\begin{eqnarray}
\xi_0 = \left({e\over 8}\right)^{1/(N-2)} 
    \Gamma\left({N-1\over N-2}\right) 2^{-5/2} 
    \exp\left(-{\pi\over 2 (N-2)}\right).
\label{xi0-HMN}
\end{eqnarray}
For $N=3$, there also exists a precise estimate of $\chi_0$ 
obtained by combining the form-factor approach and HT expansions
\cite{CPRV-97}, $\chi_0 = 0.01452(5)$.
Of course, one can consider other thermodynamic functions. In order to 
obtain the corresponding low-$T$ behavior, the anomalous dimensions 
are needed. For some composite operators, perturbative expressions 
can be found in Refs. \cite{BZL-76,CP-94,CMP-00}. 

The predictions \reff{xi-2d-sigmamodel} and \reff{chi-2d-sigmamodel} 
also apply to more general models in which the fields do not 
satisfy the condition $\vec{\phi}\cdot\vec{\phi}=1$, for instance to 
the $\phi^4$ theory \reff{latticephi4}. In this case, however, 
the functions $\beta(T)$ and $\gamma(T)$ cannot be determined analytically. 

The perturbative predictions have been checked in several different ways. 
The results for $\chi$ and $\xi$ have been checked in the large-$N$ 
limit, including the nonperturbative constant \reff{xi0-HMN}, 
to order $1/N$ \cite{BCR-90}. 
Several simulations checked the predictions for $N=3$, 4, and 8 
\cite{Wolff-90,EFGS-92,CRV-92,CEPS-95,CEMPS-96,MPS-96a,ABC-97,ACDG-99}
for the standard $N$-vector model and some other lattice versions in
the same universality class.
For $N=3$ there are significant discrepancies (of order 15-20\%) 
between MC results and 4-loop perturbative results at correlation lengths 
of order 100. This discrepancy decreases significantly at $\xi \approx 10^5$
where it is approximately 4\%. For larger values of $N$ the agreement is 
better and for $\xi \approx 100$ 4-loop perturbation theory differs from
MC results by 4\% ($N=4$) and 1\% ($N=8$). 
The agreement improves if one considers the so-called ``improved" 
perturbation theory \cite{Parisi-81,PR-81} in terms of effective temperature 
variables, which take somehow into account a large part of the 
perturbative contributions 
(see the discussions of Refs. \cite{CP-94,RV-94}).
The corresponding perturbative results are in 
better agreement with the numerical ones 
\cite{CEMPS-96,ABC-97,ACDG-99}.

It is interesting to mention 
that, at variance with the three-dimensional
case, the $1/N$ expansion is
quite predictive even for $N=3$ \cite{FL-91a,CPRV-96-gr}. 

As already mentioned, the above results can be extended to the 
antiferromagnetic quantum Heisenberg model on a square lattice, 
with Hamiltonian
\be 
  {\cal H} = {1\over T} \sum_{<xy>} \vec{s}_x \cdot \vec{s}_y,
\ee
where $\vec{s}_x$ is a spin operator satisfying the 
commutation relations 
$[s_{x,i}, s_{y,j}] = \delta_{xy} \epsilon_{ijk} s_{x,k}$ 
and the condition $s_x^2 = S(S+1)$. For $T\to 0$, the correlation length 
can be computed obtaining \cite{CHN-89,HN-91b,Hasenfratz-00}
\begin{eqnarray}
\xi_{\rm gap}
  = {e\over8} {c\over 2\pi \rho_s} \exp\left({2\pi \rho_s\over T}\right) 
   \left(1 - {T\over 4\pi \rho_s} + O(T^2)\right) f( \gamma),
\end{eqnarray}
where $\rho_s$ is the spin stiffness, $c$ the spin-wave velocity, 
$\gamma = 2 S/T$, and $f(\gamma)$ a function computed numerically 
in Ref. \cite{Hasenfratz-00} and which has the limit values: 
$f(\infty) = 1$, $f(\gamma) = e^{-\pi/2}/(8\gamma)$ for $\gamma\to 0$. 
Of course, for $S\to \infty$ we reobtain the low-$T$ 
behavior of the classical model.
These results have been compared with the experimental and the numerical
results obtained for the antiferromagnet. Similarly to the classical 
model, good agreement is observed only for very large values of the 
correlation length \cite{KLT-97,KT-98,BBGW-98,BCKM-00}.

\subsection{Amplitude ratios and two-point function} \label{sec-5.2.2}

The RG predictions \reff{xi-2d-sigmamodel} and \reff{chi-2d-sigmamodel}
are quite difficult to test because the neglected corrections 
are powers of $T$, thus corresponding to powers of 
$\log \xi$. As a consequence, numerical determinations of the 
nonperturbative constants $\xi_0$ and $\chi_0$ are extremely 
difficult. On the other hand, in the case of RG invariant quantities
scaling corrections are negative integer powers of $\xi$,
apart from logarithms.
Indeed, models with $N\ge 3$ are essentially Gaussian and thus 
the operators have canonical dimensions. 
For a generic RG invariant quantity we expect:
\bea
   R(T) = R^* + b {X^p\over \xi^2} 
    \left(1 + {a_{1}\over X} + {a_2 \log X\over X^2} + 
              {a_{3}\over X^2} +\ldots\right) 
   + O\left(\xi^{-4}\right),
\label{scaling-2d-Nge3}
\eea
where $X\equiv \log \xi$. Apart from logarithmic corrections,
the exponent associated with the leading scaling corrections
is given by $\omega=2$.
Such a behavior is verified explicitly in large-$N$ calculations, 
see, e.g., Ref. \cite{CCCPV-00}. This means that accurate estimates of the 
universal constants $R^*$ can be obtained 
by determining $R(T)$ at values of $T$ corresponding to 
relatively small $\xi$.

The approach to scaling can be somehow improved by considering 
models in which some scaling corrections vanish.
In the context of asymptotically free theories, 
one can determine analytically, using perturbation theory,
the Hamiltonian parameters that provide the cancellation 
of the leading logarithms associated with  each power correction
\cite{Symanzik-82,Symanzik-83}.
However, in practice the obtainable improvement using this approach
is rather modest: the scaling corrections
change only by powers of $\log \xi$. Several improved models of this type 
have been considered in Refs. \cite{Symanzik-83,HN-94,RV-96,CMP-99}.
On the other hand, 
a complete removal of the $O(\xi^{-2})$ scaling corrections would require
the tuning of an infinite numbers of parameters.

In Table~\ref{N3d2g4} we review the estimates 
of the four-point coupling constant $g_4^+$ for $N=3$
obtained by various approaches,
such as the form-factor approach (FF), Monte Carlo simulations (MC), 
high-temperature series (HT), and field-theoretical methods (FT).
The agreement is globally good. We only note that the HT results of
Refs.~\cite{PV-gr-98,CPRV-96-gr} are slightly smaller than
the precise FF and MC estimates. 
The HT estimates were taken at a value of the 
temperature corresponding to  a correlation length $\xi\approx 10$,
where the HT analysis was reliable. The difference with the other results 
should be essentially due to an underestimate of the scaling corrections.
Concerning the 
coefficients $r_{2n}$ that parametrize the equation of state for $M\to 0$,
we mention the estimates $r_6=3.33(2)$ and $r_8\approx 19$ for $N=3$, obtained 
by a constrained analysis
of their $O(\epsilon^3)$ series \cite{PV-00}.
For $N\ge 4$, estimates of $g_4^+$ and of the first few $r_{2n}$ 
can be found in Refs. \cite{CPRV-96-gr,BC-96,PV-gr-98,PV-00}.

\begin{table*}
\caption{
Estimates of $g_4^+$ for the two-dimensional O(3) universality class.
}
\label{N3d2g4}
\footnotesize
\begin{center}
\begin{tabular}{cllcc}
\hline
\multicolumn{1}{c}{FF} &
\multicolumn{1}{c}{MC}& 
\multicolumn{1}{c}{HT}& 
\multicolumn{1}{c}{FT $d=2$ exp}&
\multicolumn{1}{c}{FT $\epsilon$ exp} \\
\hline  
12.19(3) \cite{BNNPSW-99} & 12.21(4) \cite{BNNPSW-99,BNNPSW-00} & 11.82(6)
\cite{PV-gr-98} & 12.00(14) \cite{OS-00} & 12.0(2) \cite{PV-00} \\
&  11.9(2) \cite{Kim-95-2} & 11.9(2) \cite{CPRV-96-gr}& 11.99(11) \cite{FMPPT-83} & \\ 
\hline
\end{tabular}
\end{center}
\end{table*}

Finally, let us discuss the two-point function. Similarly to three-dimensional 
models, the scaling 
function $g^+(y)$ defined in Eq. \reff{scaling-G-HT} is well approximated 
by the Ornstein-Zernike behavior $g^+(y)\approx 1 + y$ up to $y\approx 1$. 
Indeed, the constants $c_n^+$ defined in Eq. \reff{gypiu} are very small .
For instance, for $N=3$
the HT analysis of Ref.~\cite{CPRV-97} obtained $c_2^+ = -0.0012(2)$,
$|c_3|\ltapprox 10^{-4}$,
$S_M^+=0.9987(2)$ and $S_Z^+=1.0025(4)$, cf. Eqs. (\ref{gypiu}) and (\ref{SMdef}). 
The most precise estimate of $S_M^+$ was obtained by
means of the form-factor approach, obtaining $S_M^+=0.998350(2)$
\cite{BNNPSW-99}.  
For $N=3$, $S_M^+$ has also been determined by means 
of MC simulations, 
obtaining $S_M^+=0.9985(12)$ \cite{Meyer} and $S_M^+=0.996(2)$ \cite{ABC-97}.
The overall agreement is good.

For large values of $q^2$, there are logarithmic deviations from the 
Ornstein-Zernike behavior. For $y\to \infty$ we have
\be 
    g^+(y) \sim y ( \ln y )^{-1/(N-2)}.
\ee
In $x$-space this implies, for $x\to 0$:
\be
  G(x) \sim (\ln 1/x  )^{N-1\over N-2}.
\ee
Finally, we report the result of Ref. \cite{BaNi-97} for the asymptotic 
behavior of $Z_{\rm gap}$, see Eq. \reff{Gx-largex-generale}.
For $T\to 0$ 
\begin{equation}
Z_{\rm gap} (T) = C(N) (\beta_0 T)^{\case{N-1}{N-2}} \left[ 1 +
O(T)\right].
\label{eq9b}
\end{equation}
The constant $C(N)$ is exactly known for $N=3$: 
$ C(3) = 3\pi^3$ \cite{BaNi-97}.

\section{The limit $N\to 0$, self-avoiding walks, and dilute polymers}
\label{n0}

In this section we discuss the limit $N\to 0$ of the $N$-vector model.
This is not an academic problem as it may appear at first. Indeed, in this 
limit, the $N$-vector model describes the statistical properties
of linear polymers in dilute solutions and in the good-solvent regime, 
i.e. above the $\Theta$ temperature 
\cite{DeGennes_book,desCloizeaux-Jannink_book,Flory_book,Freed_book}.
Note that FT methods can also be applied to describe the 
full crossover from the dilute to the semidilute regime, and, in 
particular, to compute the universal scaling behavior of the osmotic 
pressure in terms of the concentration 
\cite{desCloizeaux-Jannink_book,Freed_book,desCloizeaux-75,dCN-82,KSW-81}.

\subsection{Walk models} \label{sec.SAW}

Several walk models can be mapped into scalar theories with 
an $N$-vector field in the limit $N\to 0$. They 
belong to the same universality class of polymers in dilute solutions 
in the good-solvent regime, and thus their study provides quantitative 
predictions for the statistical behavior of long macromolecules. 

We begin by introducing the random walk (RW) and the self-avoiding walk (SAW). 
Given a regular lattice, an $n$-step lattice RW is a collection of 
lattice points $\{\omega_0,\ldots,\omega_n\}$ such that, for all 
$k$, $1\le k\le n$, $\omega_k$ and $\omega_{k-1}$ are lattice 
nearest neighbors. 
A lattice SAW is a nonintersecting lattice RW, i.e. a walk such that 
$\omega_i\not=\omega_j$ for all $i\not= j$. 
Analogously, 
an $n$-step self-avoiding (rooted) polygon 
(SAP) is defined as a closed $n$-step RW such that 
$\omega_0 = \omega_n$ and $\omega_i\not=\omega_j$ for all 
$0\le i\not= j\le n-1$.

Several quantities are usually introduced for SAW's and SAP's. We define:
\begin{itemize}
\item 
The number $c_n(x,y)$ of $n$-step SAW's going from $x$ to $y$ and 
the number $c_n$ of $n$-step SAW's starting from any given point (by translation
invariance it does not depend on the chosen point). 
\item 
The number $p_n$ of $n$-step SAP's starting at a given point.
\item
The squared end-to-end distance of a SAW $\omega$
\be
 R^2_e(\omega)  = (\omega_n - \omega_0)^2,
\ee
and its mean-value $R^2_e(n)$ over all $n$-step SAW's, i.e.
\begin{eqnarray}
R^2_e(n) =  {1\over c_n} \sum_{\{\omega\}} R^2_e(\omega) =
            {1\over c_n} \sum_{x} |x|^2 c_n(0,x).
\end{eqnarray}
\item 
The radius of gyration of a SAW or SAP $\omega$
\be
 R^2_g(\omega) = {1\over 2 (n+1)^2} \sum_{i,j=0}^n (\omega_i - \omega_j)^2,
\ee
and its mean value $R^2_g(n)$ over all $n$-step SAW's.
\item The end-to-end distribution function of $n$-step SAW's
\be
P_n(x) = {c_n(0,x)\over c_n}.
\ee
\item 
The form factor of $n$-step SAW's or SAP's
\begin{eqnarray}
H_n(q) = 
{1\over (n+1)^2} \left\<
  \sum_{i,j=0}^n \exp[iq\cdot(\omega_i-\omega_j)]\right\>,
\end{eqnarray}
where the average is over all $n$-step SAW's.
\item
The number $b_n$ of pairs of $n$-step SAW's $(\omega_1,\omega_2)$ such that 
$\omega_1$ starts at the origin, $\omega_2$ starts anywhere, and 
$\omega_1$ and $\omega_2$ have at least one point in common.
\end{itemize}
The quantities that we have defined have a natural interpretation. 
The end-to-end distance and the radius of gyration define the typical 
dimension of the walk and correspond to the correlation length in spin systems.
The end-to-end distribution function is the walk analogue of the 
two-point function in spin systems, while the form factor is the 
quantity that is relevant in scattering experiments. Finally, the 
number $b_n$ is related to the second virial coefficient $B_n = b_n/c_n^2$
that appears in the expansion of the osmotic pressure in powers of the 
concentration, see, e.g., Ref. \cite{desCloizeaux-Jannink_book}.
It measures the excluded volume between a pair of SAW's.
For $n\to\infty$, these quantities obey general scaling laws:
\begin{eqnarray}
&& c_n \sim \mu^n n^{\gamma - 1}, \qquad 
p_n \sim \mu^n n^{\alpha-2},   \qquad
b_n \sim \mu^{2n} n^{d\nu+2\gamma-2}, \nonumber \\
&&R^2_e(n)\approx a_e n^{2\nu}, \qquad 
R^2_g(n)\approx a_g n^{2\nu}.
\label{asymptotic-beh}
\end{eqnarray}
The constants $\gamma$, $\alpha$, and $\nu$ are critical indices. They are 
universal in the sense that any walk model which includes the 
basic property of SAW's, i.e. the local self-repulsion, has the same scaling 
behavior with exactly the same critical indices. 
The names of the indices were given by
analogy with the $N$-vector model. Indeed, as we shall discuss in the 
next section, $\gamma$, $\alpha$, and $\nu$ 
are respectively the exponents of the susceptibility, of the 
specific heat, and of the correlation length in the corresponding 
spin system in the limit $N\to0$. The quantities $\mu$, $a_e$, and $a_g$
are instead nonuniversal and depend on the specific model one is considering.
Some amplitude ratios are however universal, such as
\begin{eqnarray}
A = {a_g\over a_e} = \lim_{n\to\infty} {R^2_g(n)\over R^2_e(n)}, 
\qquad\qquad
\Psi = \left({d\over 12\pi}\right)^{d/2} 
            \lim_{n\to\infty} {b_n\over c_n^2 [R_g^2(n)]^{d/2} }.
\label{SAW-amplituderatios}
\end{eqnarray}
The universality of $A$ follows from the 
existence of a unique length scale at criticality. 
The universality of the interpenetration ratio $\Psi$ is more subtle, 
and it is related to a hyperscaling 
relation among critical indices---such a relation has already been used 
in writing the scaling of $b_n$ in Eq. \reff{asymptotic-beh}.

The SAW is a good model for dilute linear
polymers in a good solvent. Indeed, 
many properties of dilute polymers in the limit of a large number of 
monomers are universal and depend essentially on the following properties
\cite{DeGennes_book,desCloizeaux-Jannink_book,Flory_book,Freed_book}:
\begin{itemize}
\item The polymer can be considered as a continuous completely 
flexible chain above the persistence-length scale.
\item The interaction between monomers is strongly repulsive at small distances:
Clearly two monomers cannot occupy the same position in space.
\item Above the $\Theta$-temperature (good-solvent regime) the effective 
long-range attractive interaction between monomers can be neglected: The 
energetic attraction is much smaller than the entropic repulsion.
\end{itemize}
The SAW satisfies these properties, and therefore it belongs to the 
same universality class of real polymers. However, this is not the only 
model that one can consider, and indeed,
 many others have been introduced in the literature:
\begin{itemize}
\item {\em Domb-Joyce model}. This is again a lattice walk model, in which 
one considers RW's with Hamiltonian
\be
H(\omega) = w \sum_{i,j} \delta_{\omega_i,\omega_j}.
\ee
For $w\to\infty$, all intersections are suppressed and one recovers the SAW
model. The Domb-Joyce model interpolates
between the RW and the SAW model and offers
the possibility of constructing  an improved lattice walk model, i.e. a model 
in which the leading correction to scaling has a vanishing amplitude
\cite{BN-97}. The improved Domb-Joyce model corresponds to 
$\omega\equiv 1 - e^{-w} \approx 0.397$.
\item  {\em Off-lattice models}. The SAW and the Domb-Joyce models are defined
on a lattice. However, it is also possible to consider models of paths 
in continuous space. For instance, one can consider a model of off-lattice
SAW's defined as a collection of points $\{\omega_0,\ldots,\omega_n\}$, 
$\omega_i\in \mathbb{R}^d$, 
such that $|\omega_i-\omega_j|\ge 2a$ for all $i\not=j$,
where $a$ is the excluded radius. These walks are weighted with Hamiltonian
\be
H(\omega) = \sum_{i=0}^{n-1} V(\omega_i - \omega_{i+1}),
\ee
where $V(x)$ is an attractive potential. Two common choices are:
$V(x) = \delta(|x| - b)$ (stick-and-ball model) and $V(x) = {b\over 2} x^2$.
The stick-and-ball model is improved, i.e. the 
coefficient of the leading scaling correction vanishes,
for $a/b \approx 0.2235$ \cite{LK-99}.
\item{\em Continuous models}. 
For theoretical considerations it is useful to consider continuous 
paths, i.e. generic functions $r(s)$ with $0\le s \le n$, $n$ being 
the length of the path, with Hamiltonian \cite{Edwards_65}
\bea 
H[r(s)] = 
- {1\over4} \int_0^n ds \left({d{r}(s)\over ds}\right)^2 
- {w\over2} \int_0^n ds\,\int_0^n dt \,
   \delta[{r}(s) - {r}(t)]\; .
\label{twoparametermodel}
\eea
\end{itemize}

\subsection{$N$-vector model for $N\to0$ and self-avoiding walks}
\label{sec-SAW-Nvector}

In this section we derive the relation between SAW's and the limit $N\to 0$ of the 
$N$-vector model 
\cite{Daoud-etal_75,DeGennes_book,Emery-75}. Consider $N$-vector spin
variables $\vec{s}$, the lattice Hamiltonian 
\be
H = - N \sum_{\<ij\>} \vec{s}_i\cdot \vec{s}_j,
\ee
where the sum is extended over all nearest-neighbor pairs, and the partition
function
\be
Z = \int \prod_i d\Omega(s_i) e^{-\beta H},
\ee
where $d\Omega(s_i)$ is the invariant normalized measure on the sphere. 
The limit 
$N\to 0$ follows immediately from a basic result due to de Gennes
(see, e.g., Ref.~\cite{DeGennes_book}).
If
\be
I^{\alpha_1,\ldots,\alpha_n} = 
\lim_{N\to 0} N^{k/2} \int d\Omega(s) s^{\alpha_1} \ldots s^{\alpha_k},
\ee
then 
\be
I^{\alpha_1,\ldots,\alpha_n} =
 \cases{\delta_{\alpha_1\alpha_2} & \qquad if $k = 2$, \cr
        0                         & \qquad otherwise.}
\ee
Thus, for $N\to 0$, we can rewrite the partition function as 
\be
Z = \int \prod_i d\Omega(s_i) \prod_{\<ij\>} 
    (1 + N \beta \vec{s}_i\cdot \vec{s}_j) + O(N^2).
\ee
Then, expanding the product over the links and performing the integration,
we obtain a sum over a gas of self-avoiding loops. Each closed loop 
carries a power of $N$, so that 
\be
Z = 1 + N V \sum_n {p_n\over n}\beta^n + O(N^2),
\ee
where $V$ is the volume of the lattice, which is assumed to go to infinity.  
The factor $1/n$ is due to the fact that each loop can be considered as a 
SAP rooted in $n$ different points. 
Then, using the asymptotic behavior \reff{asymptotic-beh}, we obtain
for $\beta \to 1/\mu$, and $V\to \infty$
\be
\lim_{N\to 0} {1\over N V} \log Z = 
  \sum_{n=0}^\infty {p_n\over n} \beta^n \sim (\beta - 1/\mu)^{2-\alpha}.
\ee
We thus see that $\alpha$ is the specific-heat exponent 
and we may interpret $1/\mu$ as the critical temperature of the model.

Let us now compute the two-point function. Repeating the argument given above,
we obtain
\be
\lim_{N\to 0} G(x-y) = \lim_{N\to 0} \< \vec{s}_x \cdot \vec{s}_y \> = 
   \sum_n \beta^n c_n(x,y).
\ee
Then, using the asymptotic behavior \reff{asymptotic-beh},
for $\beta \to 1/\mu$ we obtain
\be
\lim_{N\to 0} \chi = \sum_n \beta^n c_n \sim (\beta - 1/\mu)^{-\gamma},
\ee
that shows that $\gamma$ is indeed the susceptibility exponent. 

Finally note that 
\bea
\lim_{N\to 0} \sum_x |x|^2 G(x) = 
   \sum_x \sum_n |x|^2 \beta^n c(0,x) = 
\sum_n c_n \beta^n R^2_{e}(n) \sim (\beta-1/\mu)^{-\gamma-2\nu},
\eea
where we have used the asymptotic formulae \reff{asymptotic-beh}.
Thus $\xi^2 \sim (\beta-1/\mu)^{-2\nu}$, which justifies the definition 
of $\nu$.

The interpretation of $b_n$ is less obvious. It can 
be also related to a quantity defined for a spin system in the 
the limit $N\to0$, but in this 
case two different fields must be considered 
\cite{desCloizeaux-Jannink_book,Muthukumar-Nickel_87}.

The above-reported discussion clarifies the equivalence between the 
$N$-vector model and the 
SAW model. However, also more general scalar models can be mapped in the 
limit $N\to0$ into 
walk models \cite{Fernandez-etal_book}. 
In general, one obtains a model of RW's with 
different weights that depend on the number of self-intersections. 
Finally, note that the continuous model \reff{twoparametermodel}
is equivalent to the usual $\phi^4$ theory in the continuum
\cite{DeGennes_72,desCloizeaux-Jannink_book,Oono_85}.

\subsection{Critical exponents and universal amplitudes}

In two dimensions exact results have been obtained for SAW's on the 
honeycomb lattice \cite{Nienhuis_82,Baxter_86}. In particular,  
the values of the exponents have been computed:
\be
\nu = {3\over4}, \qquad \gamma = {43\over 32}, \qquad 
\alpha = {1\over 2}.
\ee
These predictions have been checked numerically to a very high 
accuracy, see, e.g., Refs.  
\cite{Guttmann-Enting_88_89,Caracciolo-etal_90b,Guttmann-Wang_91,%
LMS-95,BennettWood-etal_98,Caracciolo-etal_99,Jensen-Guttmann_99}. 
In particular, accurate numerical works have shown  that the exponents do not depend 
on the lattice type---several nonstandard lattices, for instance 
the Manhattan lattice, have also been considered---and on the model---beside
SAW's, Domb-Joyce walks, trails (bond-avoiding walks), and 
neighbor-avoiding walks (note that the list is not exhaustive) 
have been considered.
In two dimensions 
conformal field theory provides exact predictions for some combinations 
of critical-amplitude  ratios
\cite{Cardy-Saleur_89,Caracciolo_etal_90,Cardy-Guttmann_93}. 
Results obtained using the form-factor approach can be found in Ref.~\cite{CM-93}.

\begin{table*}[t]
\caption{Estimates of the critical exponents for the $N=0$ universality class.
}
\label{N0exponents}
\footnotesize
\begin{center}
\begin{tabular}{rlllll}
\hline
\multicolumn{1}{c}{Ref.}& 
\multicolumn{1}{c}{Method}& 
\multicolumn{1}{c}{$\gamma$}& 
\multicolumn{1}{c}{$\nu$}& 
\multicolumn{1}{c}{$\eta$}&
\multicolumn{1}{c}{$\Delta$} \\   
\hline  
\cite{Prellberg-01} $_{2001}$  &  MC &  & 0.5874(2) &   &   \\
\cite{CCP-98} $_{1998}$  &  MC & 1.1575(6)  & &   &   \\
\cite{BN-97} $_{1997}$   &  MC &            & 0.58758(7)    & & 
                                         0.517(7)$^{+10}_{-0}$ \\
\cite{Grassberger-etal_97} $_{1997}$ & MC & 1.157(1) & 0.5872(5)& & \\
\cite{Pedersen-etal_96} $_{1996}$ & MC & & 0.5880 (18)& & \\
\cite{Eizenberg-Klafter_96} $_{1996}$ & MC & & 0.5885(9) & & \\
\cite{LMS-95} $_{1995}$  &  MC &            & 0.5877(6)    & & 0.56(3) \\
\cite{Grassberger_93} $_{1993}$ & MC & 1.608(3) & 0.5850(15) & & \\
\cite{Eizenberg-Klafter_93} $_{1993}$ & MC & & 0.591(1)      & & \\
\cite{Caracciolo-etal_91} $_{1991}$ & MC &  & 0.5867(13)    & & \\
\cite{Dayantis-Palierne_91} $_{1991}$ &  MC & & 0.5919(2)& & \\
\cite{Madras-Sokal_88} $_{1988}$ &  MC &    & 0.592(2)     & & \\
\cite{Rapaport_85} $_{1985}$ & MC&           & 0.592        & &  \\
\cite{MJHMJG-00} $_{2000}$ & HT sc  & 1.1585 & 0.5875    &  &   \\
\cite{BC-97-2} $_{1997}$ & HT sc  & 1.1594(8) & 0.5878(6)    &  &   \\
\cite{BC-97-2} $_{1997}$ & HT bcc & 1.1582(8) & 0.5879(6)    &  &   \\
\cite{MHKJ-92} $_{1992}$ & HT  & 1.16193(10)&     &  &    \\
\cite{Guttmann-89} $_{1989}$ & HT  & 1.161(2)& 0.592(3) &  &    \\
\cite{GZ-98}  $_{1998}$   & FT $d$=3 exp & 1.1596(20)  &  
                     0.5882(11)  & 0.0284(25)  & 0.478(10) \\
\cite{MN-91}  $_{1991}$   & FT $d$=3 exp & 1.1569(6)\{10\}&  
                     0.5872(4)\{7\} & 0.0297(9)\{6\} & \\
\cite{Muthukumar-Nickel_87} $_{1987}$ & FT $d$=3 exp & & 0.5886 & & 0.465 \\
\cite{LZ-77} $_{1977}$    & FT $d$=3 exp & 1.1615(20)  &  0.5880(15)  & 0.027(4)  &  0.470(24) \\
\cite{GZ-98} $_{1998}$ &  FT $\epsilon$ exp & 1.1575(60)  &  0.5875(25) & 0.0300(50) & 0.486(14)  \\
\cite{GZ-98} $_{1998}$ &  FT $\epsilon$ exp$|_{\rm bc}$    & 1.1571(30)  &  0.5878(11) & 0.0315(35) &  \\
\cite{NR-84}       $_{1984}$      & SFM  & 1.15(1)   &  0.585(5)   & 0.034(5) & 0.509(35) \\
\cite{GW-01} $_{2001}$ &  CRG (1st DE)   & 1.157  &  0.590 & 0.039 &  \\
%\cite{Cotton_80} $_{1980}$ & experiment& & 0.586 & & \\
\hline
\end{tabular}
\end{center}
\end{table*}

Extensive work has been performed in three dimensions to determine the critical 
exponents. 
A collection of results is reported in Table \ref{N0exponents}. 
At present, the most precise results are obtained from 
MC simulations. Indeed, SAW's and similar walk models are 
particularly easy to simulate. First, one works directly 
in infinite volume, and thus there are no finite-size systematic errors.
Second, there are very efficient algorithms, 
see, e.g., Refs. \cite{Madras-Sokal_88,Caracciolo-etal_92,Caracciolo-etal_99},
such that the autocorrelation time increases linearly with the length of the 
walk.\footnote{Translating into the standard language of spin systems, in three 
dimensions $\tau \sim \xi^{5/3}$ in SAW simulations, to be compared with 
$\tau \sim \xi^{3 + z}$ of spin systems. 
In practice, simulations with $\xi \sim 10^3$ 
(corresponding approximately to $n\sim 10^5$) are now feasible. 
This means that it is possible to obtain numerical estimates of critical
quantities with very good accuracy.} 
We mention that the older results obtained from MC simulations
and exact enumerations were providing larger estimates of $\nu$ and $\gamma$: 
typically $\nu \approx 0.592$, $\gamma \approx 1.160$.
This was mainly due to the fact that nonanalytic corrections
to scaling were neglected in the analysis (see the discussion in, e.g., Refs.
\cite{Caracciolo-etal_91,Dayantis-Palierne_94,LMS-95,Eizenberg-Klafter_96,%
MJHMJG-00}).

The FT results for $\nu$ are in good agreement 
with the latest lattice results, while $\gamma$ is slightly larger,
although the discrepancy is not yet significant, being at the level of 
one error bar. 

The experimental results for the critical exponents are quite old and, 
in general, not very precise
due to the difficulty to prepare monodisperse solutions. 
The analysis of the experimental data 
\cite{Yamamoto-etal_71,Fukuda-etal_74,Miyaki-etal_78}
for polystyrene in benzene
gives \cite{Cotton_80} $\nu \approx 0.586$, in good agreement with the 
theoretical estimates. 
The exponent $\nu$ has also been determined from the experimental
measurement of the second virial coefficient, obtaining $\nu \approx 0.582$
\cite{desCloizeaux-Jannink_book}. Another important class of experiments 
determines $\nu$ from the diffusion of
polymers \cite{Adam-Delsanti_76,Strazielle-Benoit_75,NMTK-84,VJP-86,RMMPW-01}, 
obtaining $\nu \approx 0.55$, which is lower than the 
theoretical estimates and the experimental results that we have presented above. 
As discussed in Refs.~\cite{Weill-desCloizeaux_79,SB-86,DRSK-02},
this is probably due to the 
fact that the quantity that is measured in these experiments
\cite{Kirkwood-54,DoiEdwards_book}, the 
hydrodynamic radius of the polymer, scales approximately as $n^\nu$ 
($n$ is here the number of monomers) only for values of $n$ that are much larger
than those of the macromolecules that are used 
(see also Refs. \cite{OF-82,DF-84a}). 
Therefore, 
these experiments only measure effective exponents, strongly affected 
by corrections to scaling.

There are also a few estimates of $\nu$ in two-dimensions: 
for two-dimensional polymer monolayers of polyvinylacetate \cite{VR-80}
$\nu \approx 0.79(1)$, while for DNA molecules confined in
fluid cationic liquid bilayers \cite{MR-99} $\nu \approx 0.79(4)$. 

Beside critical exponents, one can measure several amplitude ratios.
The ratio $A$  and the interpenetration ratio $\Psi$ defined in 
Eq. \reff{SAW-amplituderatios} are of particular interest. 
Theoretical and experimental estimates for linear polymers are reported in 
Table \ref{N0amplituderatios}. In Ref. \cite{BN-97}, 
the authors report $\Psi A^{3/2} = 0.0157965(7)$. Using 
the estimate $A = 0.15995(10)$ \cite{Grassberger-etal_97}, 
we obtain $\Psi = 0.2469(9)$.  All results are in good agreement, 
except the estimate of $\Psi$ of Ref. \cite{RF-96} that is 
apparently too low.  Another interesting ratio is 
\begin{equation}
 H = \lim_{n\to\infty}{\langle R^2_g(n)\rangle_{\rm ring} \over 
                        \langle R^2_g(n)\rangle_{\rm linear} },
\end{equation}
where in the ratio we consider two different conformations of the same polymer 
species. Experiments on cyclic and linear polydimethylsiloxane
give \cite{HDS-79} $H=0.526(50)$. One-loop $\epsilon$ expansion 
\cite{Prentis-84} gives $H\approx 0.568$, while a high-order 
perturbative analysis \cite{CPV-02-2} gives\footnote{\label{footN0} 
Ref. \cite{CPV-02-2} reports $\langle R^2_g(n)\rangle_{\rm ring}/
\langle R^2_e(n)\rangle_{\rm linear}$. To compute $H$,
we use $A= 0.15995(10)$ \cite{Grassberger-etal_97}.
} $H=0.556(19)$. By combining HT and MC results obtained in 
Refs. \cite{JEK-92,LMS-95,MJHMJG-00}, Ref. \cite{CPV-02-2}
obtained the numerical estimate $H\approx 0.519$. 
All results are in reasonable agreement. We should also mention results 
for universal combinations involving the hydrodynamic radius 
(see Ref. \cite{DRSK-02} and references therein), and for several 
invariant quantities characterizing the polymer shape 
(see Ref. \cite{ZP-01} and references therein).

\begin{table}
\caption{Estimates of the  amplitude ratios $A$ and $\Psi$ for 
linear polymers.
}
\label{N0amplituderatios}
\footnotesize
\begin{center}
\begin{tabular}{rlll}
\hline
\multicolumn{1}{c}{Ref.}& 
\multicolumn{1}{c}{Method}& 
\multicolumn{1}{c}{$A$}& 
\multicolumn{1}{c}{$\Psi$}\\ 
\hline  
\cite{Grassberger-etal_97} $_{1997}$ & MC & 0.15995(10) & \\
\cite{RF-96} $_{1996}$ & MC & & 0.2422(4) \\
\cite{LMS-95} $_{1995}$      & MC & 0.1599(2) & 0.2471(3) \\
\cite{Shanes-Nickel_94} $_{1994}$ & MC & 0.16003(3) & \\
\cite{Eizenberg-Klafter_93} $_{1993}$ & MC & 0.1596(2) & \\ 
\cite{Nickel_91} $_{1991}$ & MC & & 0.2465(12) \\
\cite{Madras-Sokal_88} $_{1988}$ & MC & 0.1603(4) & \\
\cite{Rapaport_85} $_{1985}$ & MC sc & 0.1597(3) & \\
\cite{Rapaport_85} $_{1985}$ & MC bcc& 0.1594(2) & \\
\cite{CL-02} $_{2002}$ & HT & 0.158(2) & \\
\cite{Shanes-Nickel_94} $_{1994}$ & FT $d$=3 exp & 0.16012(30) & \\
\cite{Benhamou-Mahoux_86} $_{1986} $ & FT $\epsilon$ exp & 0.158 &  \\
\cite{DF-84b} $_{1984}$ & FT $\epsilon$ exp & & 0.269 \\
\cite{OF-82} $_{1982}$ & FT $\epsilon$ exp & & 0.231 \\
\cite{desCloizeaux-81} $_{1981}$ & FT $\epsilon$ exp & & 0.249 \\
\cite{WS-78} $_{1978}$ & FT $\epsilon$ exp & & 0.268 \\
\cite{Yamakawa-etal_93} $_{1993}$ & experiment & & 0.245 \\
\hline
\end{tabular}
\end{center}
\end{table}

\subsection{Scaling functions}

We discuss now the distribution functions $P_n(x)$ and $H_n(q)$.
In $d$ dimensions, for $n\to\infty$, $|{x}|\to \infty$,
with $|{x}| n^{-\nu}$ fixed, the function $P_n({x})$ has the
scaling form \cite{Mazur_65,Fisher_66,McKenzie_76,%
McKenzie-Moore_71,DesCloizeaux_74_80}
\be
P_n({x}) \approx {1\over \xi^d_{n}} f(\rho),
\label{deffrho}
\ee
where $\rho = {x}/\xi_{n}$, and $\xi_n = R^2_e(n)/(2d)$.
For large $\rho$ it behaves as
\cite{Fisher_66,McKenzie-Moore_71,DesCloizeaux_74_80}
\be
f(\rho) \,\approx f_\infty \rho^\sigma
   \exp\left(-D \rho^\delta\right)\; ,
\label{flargerho}
\ee
where, in three dimensions, \cite{DesCloizeaux_74_80,Caracciolo-etal_00} 
\begin{eqnarray}
&f_\infty =  0.01581(2),\qquad
&\sigma = {6 \nu - 2 \gamma - 1\over 2 (1-\nu)}\, 
              =\, 0.255(2), 
\nonumber \\
&   D = 0.1434(2),\qquad
&   \delta = {1\over 1-\nu} \, =\, 2.4247(4).
 \label{delta} 
\end{eqnarray}
For $\rho\to 0$, we have \cite{McKenzie-Moore_71,DesCloizeaux_74_80}
\be
   f(\rho) \approx f_0 \left({\rho\over2}\right)^\theta,
\label{fsmallrho}
\ee
where \cite{DesCloizeaux_74_80,Caracciolo-etal_00}
\begin{eqnarray}
\theta = {\gamma - 1\over \nu} \, =\, 0.2680(10),
\qquad
f_0 =  0.019(3).     \label{theta} 
\end{eqnarray}
These theoretical predictions have been extensively checked, see
Refs. \cite{RF-96,Caracciolo-etal_00,TKC-02} and references therein. 
We refer to Ref. \cite{Caracciolo-etal_00} for a discussion of 
the two-dimensional case.

A phenomenological representation for the function $f(\rho)$
was proposed in Refs.
\cite{McKenzie-Moore_71,desCloizeaux-Jannink_book}:
\be
f(\rho) \approx {f}_{\rm ph}(\rho) =
   f_{\rm ph} \rho^\theta
    \exp\left(-{D}_{\rm ph} \rho^\delta\right).
\label{frhoapprox}
\ee
Here $\delta$ and $\theta$ are fixed by Eqs. \reff{delta} and \reff{theta},
while ${f}_{\rm ph}$ and ${D}_{\rm ph}$ are given by 
\be
D_{\rm ph} = 0.14470(14), \qquad 
f_{\rm ph} = 0.015990(8). 
\ee
The phenomenological representation is rather accurate, differing from 
the exact one by a few percent in the region where the distribution 
function is significantly different from zero, i.e. for $\rho \ltapprox 6$,
see Ref. \cite{Caracciolo-etal_00}.

The two-dimensional end-to-end distribution function 
has been recently measured in a system consisting 
of a linear chain of plastic spheres immersed in a planar fluid of 
self-propelled balls, a so-called granular polymer solution \cite{PS-02}.
The results are in good agreement with the theoretical predictions and allow
a direct estimate of $\nu$, $\nu=0.75(1)$.

Let us now consider the form factor $H_n(q)$.
For $n\to\infty$, $|q|\to 0$, with 
$|q| n^\nu$ fixed, the function $H_n(q)$ has the  scaling form 
$\widehat{H}(q R_g(n))$. For $q R_g(n)\ll 1$, we obtain the Guinier formula
\be
H_n(q) = 1 - {q^2 R^2_g(n)\over d} + O(q^4 R^4_g(n)),
\ee
which is often used in experimental determinations of the gyration radius.
For large $q R_g(n)$, we have 
\cite{Witten-Schafer_81,Ohta-etal_81_82,Witten_82,%
desCloizeaux-Duplantier_85,Duplantier_86_1,desCloizeaux-Jannink_book}
\be
H_n(q) \approx {h_\infty \over \left[ q^2 R^2_g(n)\right]^{1/(2\nu)} },
\label{HnlargeQ}
\ee
where $h_\infty$ is a universal constant. In three dimensions
we have 
\cite{desCloizeaux-Duplantier_85,Duplantier_86_1,desCloizeaux-Jannink_book}
$h_\infty \approx 1.0$ for linear polymers and 
\cite{CPV-02-2} $h_\infty \approx 0.63(9)$ for cyclic polymers. 
The form factor has been extensively studied numerically,
see, e.g., Refs. \cite{Pedersen-etal_96,PS-96,DLR-96,PS-99,TKC-02}.
Phenomenological expressions can be found 
in Refs. \cite{PS-96,FB-98} for linear polymers and in 
Ref. \cite{CPV-02-2} for cyclic polymers. 
Many experiments have measured the structure factor, although only
few of them \cite{Cotton-etal_74,Farnoux_76,Rawiso-etal_87} have 
been able to verify the large-$q$ behavior
(\ref{HnlargeQ}) with $\nu \approx 0.60$. Indeed, this requires to 
work with very long polymers, otherwise, after the initial Guinier 
regime one observes a $q^{-1}$ behavior due to the finite persistence 
length. In this case, the experimental data are better described by models
that do not take into account self-avoidance, but instead keep track of the 
finite persistence length and of the polymer transverse radius, 
see, e.g., Ref. \cite{PHBAP-00}.

\section{Critical crossover between the Gaussian and the Wilson-Fi\-sher
fixed point}
\label{crossover}

\subsection{Critical crossover as a two-scale problem}

Every physical situation of experimental relevance has at least two scales: 
one scale is intrinsic to the system, 
while the second one is related to experimental conditions.
In statistical mechanics the correlation length $\xi$ is 
related to experimental conditions (it depends on the temperature), 
while the interaction length (related to the Ginzburg parameter $G$) is intrinsic.
The opposite is true in 
quantum field theory: here the correlation length (inverse mass gap) 
is intrinsic, while the interaction scale (inverse momentum) 
depends on the experiment.
Physical predictions are functions of ratios of these two scales and
describe the crossover from the correlation-dominated ($\xi^2 G$ or $p/m$ large) 
to the interaction-dominated ($\xi^2 G$ or $p/m$ small) regime.
In a properly defined limit they are universal and define the unique flow
between two different fixed points. This universal limit is obtained when
these two scales become very large with respect to any other
microscopic scale. Their ratio becomes the universal
control parameter of the system, whose transition from $0$
to $\infty$ describes the critical crossover.

In this section we  consider the crossover between the Gaussian fixed
point where mean-field predictions hold (interaction-dominated regime) 
to the standard Wilson-Fisher fixed point (correlation-dominated
regi\-me). In recent years a lot of work has been devoted to understanding
this crossover, either experimentally 
\cite{Corti-Degiorgio_85,Dietler-Cannel_88,Anisimov-etal_95,%
BB-96,Anisimov-etal_96,Jacob-etal_98,AS-review}
or theoretically 
\cite{Thouless_69,Bagnuls-Bervillier_84,BB-85,BBMN-87,%
Fisher_86,Bagnuls-Bervillier_87,%
SD-89,Chen-etal_90,Anisimov-etal_92,%
Belyakov-Kiselev_92,MB-93,LBB-96,LBB-97,LB-98,PRV-98,PRV-99,CCPRV-99,%
LB-99,KWAS-99,Luijten-Maryland,OPF-00,AAS-01,BB-01,LK-01,PR-01,LK-02}.
For a recent review on crossover phenomena in polymer blends and solutions
see, e.g., Ref.~\cite{BLMWB-00}. 

The traditional approach to the crossover between the Gaussian and 
the Wilson-Fisher fixed point starts from the standard Landau-Ginzburg Hamiltonian. On a 
$d$-dimensional lattice, it can be written as 
\bea
  H= \sum_{x_1,x_2} \smfrac{1}{2}J({x_1}-{x_2})
\left(\phi_{x_1} - \phi_{x_2}\right)^2 
+ \sum_x \left[ \smfrac{1}{2}  r\phi_x^2 +
{u\over 4!} \phi_x^4 - h_x \cdot \phi_x\right] ,
\label{lham}
\eea
where $\phi_x$ are $N$-dimensional vectors, and
$J(x)$ is the standard nearest-neighbour
coupling. For this model the interaction scale is controlled by the 
coupling $u$ and the relevant parameters are the (thermal) Ginzburg number
$G$ \cite{Ginzburg_60} and its magnetic counterpart $G_h$ 
\cite{LBB-97,PRV-98} defined by:
\be
G_{\hphantom{h}} =\, u^{2/(4-d)}, \qquad
G_h =\, u^{(d+2)/[2(4-d)]}.
\ee
Under a RG transformation $G$ 
scales like the (reduced) temperature, 
while $G_h$ scales as the magnetic field. For 
$t \equiv r - r_c \ll G$ and $h\ll G_h$ one observes the standard critical 
behaviour, while in the opposite case the behaviour is classical.
The critical crossover limit corresponds to considering 
$t,h,u\to 0$ keeping 
\begin{equation}
\widetilde{t} \equiv  t/G, \qquad\qquad \widetilde{h} \equiv  h/G_h 
\label{ttilde}
\end{equation}
fixed. This limit is universal, i.e. independent of the detailed 
structure of the model: any Hamiltonian of the form \reff{lham} shows
the same universal behaviour as long as the interaction is 
short-ranged, i.e. for any $J(x)$ such that 
$\sum_{x}x^2\, J(x) < + \infty$. In the HT phase
the crossover functions can be related to the RG functions 
of the standard continuum $\phi^4$ theory if one 
expresses them in terms of the zero-momentum four-point renormalized 
coupling $g$ \cite{Bagnuls-Bervillier_84,BB-85,BBMN-87,SD-89}.
For the quantities  that are traditionally studied in statistical mechanics,
for instance the susceptibility or the correlation length, 
the crossover functions can be computed to high precision in
the fixed-dimension expansion in $d=3$ 
\cite{Bagnuls-Bervillier_84,BB-85,BBMN-87,PRV-99,BB-01}.

\subsection{Critical crossover functions in field theory} 
\label{sec2}

The critical crossover functions can be computed in the framework of
the continuum $\phi^4$  theory
\be
H = \int d^d x\left[ {1\over2} (\partial_\mu \phi)^2 +
                     {r\over2} \phi^2 +
                     {u\over 4!} \phi^4 \right],
\ee
where $\phi$ is an $N$-dimensional vector, by considering the limit
$u\to 0$, $t\equiv r - r_c\to 0$, 
with $\widetilde{t}\equiv t/G = t u^{-2/(4-d)}$
fixed.  In this limit, we have for the susceptibility $\chi$ and 
the correlation length $\xi$ 
\bea
\widetilde{\chi}  \equiv \chi\, G \longrightarrow \, F_\chi (\widetilde{t}), \qquad
\widetilde{\xi}^2 \equiv \xi^2\, G \longrightarrow \, F_\xi (\widetilde{t}).
\eea
The functions $F_\chi (\widetilde{t})$ and
$F_\xi (\widetilde{t})$ can be accurately computed by means of perturbative
FT calculations.
There are essentially two methods: (a) the
fixed-dimension expansion \cite{Bagnuls-Bervillier_84,%
BB-85}, which is at present the most precise
one since seven-loop series are available
\cite{BNGM-77,MN-91};
(b) the so-called minimal renormalization without $\epsilon$ expansion
\cite{Dohm-85,SD-89,KSD-90} that uses five-loop
$\epsilon$-expansion results \cite{CGLT-83,KNSCL-93}. In these two schemes
the crossover functions are expressed in terms of various
RG functions whose perturbative series can be resummed with high accuracy
using standard methods \cite{LZ-77,Zinn-Justin-book}. 
Here we shall consider the first approach 
although essentially equivalent results can be obtained using the second
method. 
For $F_\chi(\widetilde{t})$ and $F_\xi(\widetilde{t})$ 
one obtains \cite{BB-85}
\begin{eqnarray}
&& 
F_\chi(\widetilde{t}) = \chi^* 
   \, \exp\left[ - \int_{y_0}^g dx\, {\gamma(x)\over \nu(x) \beta (x)}\right], 
\label{Fchi-FT-fixedd}\\
&& 
F_\xi(\widetilde{t}) = \left(\xi^*\right)^2
   \, \exp\left[ - 2 \int_{y_0}^g dx\, {1\over \beta (x)}\right], 
\label{Fxi-FT-fixedd}
\end{eqnarray}
where $\widetilde{t}$ is related to the zero-momentum four-point
renormalized coupling $g$ by
\begin{eqnarray}
&&\widetilde{t} =
  - t_0 \, \int^{g^*}_g dx\, 
 {\gamma(x)\over \nu(x) \beta (x)} 
  \exp\left[ \int_{y_0}^x dz\, {1\over \nu(z) \beta (z)} \right],
\label{ttilde-FT-fixedd}\\
&&\nu(x) = \left( 2 - \eta_\phi(x) + \eta_t(x)\right)^{-1},\qquad
\gamma(x) = \left( 2-\eta_\phi(x)\right) \nu(x),
\end{eqnarray}
where $\eta_\phi(x)$, $\eta_t(x)$, and $\beta(x)$ are the standard RG
functions introduced in Sec. \ref{sec-2.4.1}.
We recall that $g^*$ is the critical value of $g$ defined by
$\beta(g^*) = 0$, and $\chi^*$, $\xi^*$, $t_0$, and $y_0$ are normalization
constants. 

The expressions \reff{Fchi-FT-fixedd}, \reff{Fxi-FT-fixedd}, and 
\reff{ttilde-FT-fixedd} are valid for any dimension $d<4$. The first
two equations are well defined, while Eq. \reff{ttilde-FT-fixedd}
has been obtained with the additional hypothesis that the integral
over $x$ is convergent when the integration is extended up to 
$g^*$. This hypothesis is verified when the system becomes critical 
at a finite value of $\beta$ and shows a standard critical behaviour. 
Therefore, Eq. \reff{ttilde-FT-fixedd} is well defined for 
$d>2$, and, in two dimensions, for $N\le2$. 
For $N> 2$, one can still define $\widetilde{t}$ by integrating up to an arbitrary point
$g_0$ \cite{PRV-99}. For these values of $N$, $\widetilde{t}$ varies between 
$-\infty$ and $+\infty$.

The functions $F_\chi(\widetilde{t})$ and $F_\xi(\widetilde{t})$ 
can be computed by using the perturbative results of 
Refs. \cite{BNGM-77,MN-91} and the resummation techniques
presented in Sec. \ref{sec-2.4.3}.
Explicit expressions can be found 
for $N=1,2,3$ and $d=3$ in Refs. \cite{Bagnuls-Bervillier_84,BB-85,BB-01},
and for the two-dimensional Ising model in Ref.~\cite{PRV-99}.
Large-$N$ results appear in Refs. \cite{PRV-98,PRV-99}.

The above results apply to the HT phase of the
model.  For $N=1$ the critical crossover can also be defined in the 
LT phase \cite{BBMN-87} and the crossover 
functions can be computed in terms of (resummed) perturbative 
quantities. Using the perturbative results
of Ref. \cite{BBMN-87}, 
$F_\chi(\widetilde{t})$ in the LT
phase of the three-dimensional Ising model 
has been calculated in Refs. \cite{PRV-98,PRV-99,BB-01}, showing 
a nonmonotonic behavior.

\subsection{Critical crossover in spin models with medium-range interaction}

The $\phi^4$ Hamiltonian relevant for
spin models with medium range interaction can be written as
\bea
  H =\sum_{x_1,x_2} \smfrac{1}{2}J({x_1}-{x_2})
\left(\phi_{x_1} - \phi_{x_2}\right)^2 
+ \sum_x \left[ \smfrac{1}{2}  r\phi_x^2 +
{u\over 4!} \phi_x^4 - h_x \cdot \phi_x\right] ,
\label{lhammr}
\eea
where $J(x)$ has the following form \cite{MB-93,LBB-97} 
\be
J({x})=\, \cases{ J & \qquad for ${x}\in {D}$, \cr
                  0 & \qquad for ${x}\not\in {D}$,
                    }
\label{defJ}
\ee
and $D$ is a lattice domain characterized by some scale $R$.
Explicitly, we define $R$ and the corresponding domain volume
$V_R$ by
\be
V_R  \equiv  \sum_{{x}\in D} 1,
\qquad
R^2  \equiv {1\over 2d\,V_R} \sum_{{x}\in D} x^2\; .
\label{defR}
\ee
The shape of ${D}$ is irrelevant in the critical crossover limit as 
long as $V_R\sim R^d$ for $R\to\infty$. The constant $J$ defines the 
normalization of the fields. Here we assume $J=1/V_R$, since this 
choice simplifies the discussion of the limit $R\to\infty$.
To understand the connection between the theory with medium-range
interactions and the short-range model, we consider the continuum 
Hamiltonian that is obtained by replacing in Eq. \reff{lham} the 
lattice sums with the corresponding integrals. Then, we perform a scale
transformation \cite{LBB-97} by defining new (``blocked") coordinates 
$y = x/R$ and rescaling the fields according to 
\be
\widehat{\phi}_y = R^{d/2} \phi_{R y}, \qquad
\widehat{h}_{y} = R^{d/2} h_{R y}.
\ee
The rescaled Hamiltonian becomes 
\bea
  \widehat{H}=
     \int d^d y_1 \, d^dy_2\, \smfrac{1}{2}\widehat{J}(y_1-y_2)
  \left(\widehat{\phi}_{y_1} - \widehat{\phi}_{y_2}\right)^2 
 +\int d^dy\, \left[ \smfrac{1}{2}  r\widehat{\phi}_y^2 +
{1\over 4!} {u\over R^d}\, \widehat{\phi}_y^4 - 
    \widehat{h}_y\cdot \widehat{\phi}_y\right] , 
\label{lham1}
\eea
where now the coupling $\widehat{J}(x)$ is of short-range type in the limit 
$R\to\infty$.  Being short-ranged, we can apply the 
previous arguments and define Ginzburg parameters: 
\begin{eqnarray}
&&   G_{\hphantom{h}} =  \left(u R^{-d}\right)^{2/(d-4)}
  =  u^{2/(d-4)} R^{-2d/(4-d)}, 
\\
&&   G_h = \; R^{-d/2} \left(uR^{-d}\right)^{(d+2)/[2(d-4)]} 
  = u^{(d+2)/[2(d-4)]}\, R^{-3d/(4-d)}.
\end{eqnarray}
Therefore, in the medium-range model, the critical crossover limit can 
be defined as $R\to\infty$, $t,h\to 0$, with
$\widetilde{t}\equiv t/G$,
$\widetilde{h}\equiv t/G_h$ fixed. 
The variables that are kept fixed are the same, but a different mechanism 
is responsible for the change of the Ginzburg parameters:
in short-range models we vary $u$ keeping the range $R$ fixed and finite,
while here we keep the interaction strength $u$ fixed and vary the 
range $R$.
The important consequence of the argument presented above is that the 
critical crossover functions defined using the medium-range Hamiltonian
and the previous limiting procedure agree with those computed in the
short-range model, apart from trivial rescalings.

If we consider the critical limit with
$R$ fixed, the Hamiltonian (\ref{lhammr})
defines a generalized $O(N)$-sym\-me\-tric
model with short-range interactions. If $d>2$, 
for each value of $R$ there is a critical point\footnote{In two dimensions a 
critical point exists only for $N\le 2$. Theories with $N\ge 3$ are 
asymptotically free and become critical only in the limit $\beta\to \infty$.
See Sec. \ref{sec-5.2}.}
$\beta_{c,R}$; for $\beta\to \beta_{c,R}$ the susceptibility and the 
correlation length have the standard behavior
\begin{eqnarray}
\chi_R(\beta) &\approx& A_\chi(R) t^{-\gamma} (1 + B_\chi(R) t^\Delta) ,
\label{chiRfisso} \\
\xi_R^2(\beta) &\approx& A_\xi(R) t^{-2\nu} (1 + B_\xi (R) t^\Delta),
\end{eqnarray}
where $t \equiv (\beta_{c,R} - \beta)/\beta_{c,R}$ and we have neglected 
additional subleading corrections.
The exponents $\gamma$, $\nu$, and $\Delta$ do not depend on $R$. 
On the other hand, the 
amplitudes are nonuniversal.
For $R\to\infty$, they behave as \cite{LBB-96,LBB-97}
\begin{eqnarray} 
A_\chi(R) \approx    A_\chi^\infty R^{2 d(1-\gamma)/(4 - d)} ,\qquad 
B_\chi(R) \approx    B_\chi^\infty R^{2 d \Delta/(4 - d)} ,
\label{eq3.4chi}
\end{eqnarray}
and 
\begin{eqnarray} 
A_\xi(R) \approx A_\xi^\infty   R^{4 (2 - d\nu)/(4-d)}, \qquad
B_\xi(R) \approx  B_\xi^\infty  R^{2 d \Delta /(4-d)} .
\label{eq3.4xi}
\end{eqnarray}
The critical point $\beta_{c,R}$ depends explicitly on $R$. 
The critical crossover limit is obtained
by taking the limit \cite{LBB-96,LBB-97}
$R\to\infty$, $t\to0$, with 
$R^{2d/(4-d)}t\equiv \widetilde{t}$ fixed.
It has been shown that
\begin{eqnarray}
&&
\widetilde{\chi}_R \equiv 
         R^{-2d/(4-d)} \chi_R(\beta) \longrightarrow f_\chi(\widetilde{t}) ,
\label{fchi} \\
&&
\widetilde{\xi}^2_R \equiv 
R^{-8/(4-d)} \xi^2_R(\beta) \longrightarrow f_\xi(\widetilde{t}) ,
\label{fxi}
\end{eqnarray}
where the functions $f_\chi(\widetilde{t})$ and 
$f_\xi(\widetilde{t})$ are universal apart from an overall 
rescaling of $\widetilde{t}$ and a constant factor, in agreement with the
above-presented argument.

There exists an equivalent way to define the 
cross\-ov\-er limit which is due to Thouless \cite{Thouless_69}. 
Let $\beta^{\rm (exp)}_{c,R}$ be the expansion of $\beta_{c,R}$ for 
$R\to\infty$ up to terms of order $R^{-2d/(4-d)}/V_R$, i.e. such that
\be
\lim_{R\to\infty} R^{2d/(4-d)} \beta_{c,R}^{-1}\, 
     (\beta_{c,R} - \beta^{\rm (exp)}_{c,R}) = 
      b_c,
\label{def-bc}
\ee
with $|b_c|<+\infty$. Then one may introduce 
\be
\widehat{t} =\, R^{2d/(4-d)} \beta_{c,R}^{{\rm (exp)}\,-1}\, 
              (\beta^{\rm (exp)}_{c,R} - \beta).
\label{def-that}
\ee
In the standard crossover limit 
$\widetilde{t} = \widehat{t} + b_c$. Therefore, the crossover limit 
can be defined considering the limit $R\to\infty$, 
$\beta\to\beta^{\rm (exp)}_{c,R}$ with $\widehat{t}$ fixed. The crossover 
functions will be identical to the previous ones apart from a 
shift. Thouless' definition of critical crossover has an important 
advantage. It allows us to define the critical 
crossover limit in models that do not have a critical point for finite values
of $R$: indeed, even if $\beta_{c,R}$ does not exist, one can define 
a quantity $\beta^{\rm (exp)}_{c,R}$ and a variable $\widehat{t}$ such that 
the limit $R\to\infty$ with $\widehat{t}$ fixed exists. This is the
case of two-dimensional models with $N\ge 3$
and of one-dimensional models with $N\ge 1$.

Medium-range spin models can be studied 
by a systematic expansion around mean field \cite{PRV-99}. 
In this framework one proves the equivalence between the crossover
functions computed starting from 
the continuum $\phi^4$ model and the results obtained in the 
medium-range model. Moreover, one can compute the nonuniversal
normalization constants relating the two cases,
so that the comparison of  the FT predictions with the numerical 
results obtained by simulating medium-range spin models 
\cite{LBB-96,LBB-97,LB-98,Luijten-FSS} can be done
without any free parameter. 

\begin{figure}[t]
\vspace*{-1cm} \hspace*{-1cm}
\begin{center}
\epsfxsize = 0.5\textwidth
\leavevmode\epsffile{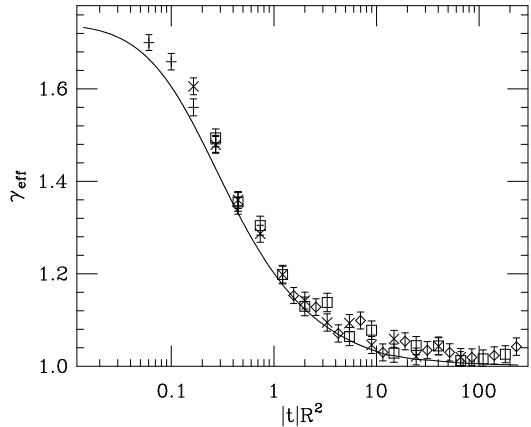}
%\quad \vspace{8cm}  %% TEMPORARY UNTIL FILE IS THERE
\end{center}
\vspace*{-1cm}
\caption{
Effective susceptibility exponent $\gamma_{\rm eff}(\widetilde{t})$ 
as a function of
$\widetilde{t}$ in the HT phase of the two-dimensional
Ising model. 
The points are the numerical results of Ref. \protect\cite{LBB-97}:
pluses, crosses, squares, and diamonds
correspond to data with $\rho^2=10$, 72, 140, and 1000 respectively.
The continuous curve was computed using FT methods in Ref.~\cite{PRV-99}.
In the mean-field limit $\gamma_{\rm eff}=1$, while for $\widetilde{t}\to 0$,
$\gamma_{\rm eff} = 7/4$.
}
\label{gammaeff2d}
\end{figure}

\begin{figure}
\vspace*{-1cm} \hspace*{-1cm}
\begin{center}
\epsfxsize = 0.5\textwidth
\leavevmode\epsffile{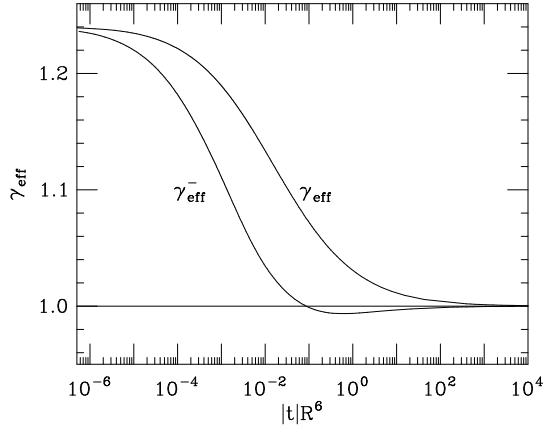}
%\quad \vspace{8cm}  %% TEMPORARY UNTIL FILE IS THERE
\end{center}
\vspace*{-1cm}
\caption{
Effective susceptibility exponent $\gamma_{\rm eff}(\widetilde{t})$
as a function of
$\widetilde{t}$ for the HT ($\gamma_{\rm eff} $) and LT
($\gamma_{\rm eff}^- $) phase of the three-dimensional
Ising model. In the mean-field limit $\gamma_{\rm eff} = 1$, while for 
$|\widetilde{t}|\to 0$, $\gamma_{\rm eff}\approx 1.237$.
Results from Ref. \cite{PRV-99}.
}
\label{gammaeff}
\end{figure}

In Figs. \ref{gammaeff2d} and \ref{gammaeff}
we report the graph of the effective magnetic susceptibility exponent 
$\gamma_{\rm eff}$, defined by
\begin{equation}
\gamma_{\rm eff} (\widetilde{t}) \equiv 
- {\widetilde{t}\over f_\chi(\widetilde{t})}
   {df_\chi(\widetilde{t})\over d\widetilde{t}},
\end{equation}
for the Ising model in two and three dimensions respectively.
They have been determined using the FT results discussed
in Sec. \ref{sec2} with the appropriate rescaling computed 
using the mean-field approach \cite{PRV-99}.

In two dimensions the results for 
$\gamma_{\rm eff}(\widetilde{t})$ can be compared with the numerical ones of 
Ref. \cite{LBB-97}, obtained by simulating the medium-range Ising model
\reff{lhammr}, \reff{defJ} with
\be
D =\, \left\{x: \sum_{i=1}^d x^2_i \le \rho^2\right\}.
\ee
They are shown in Fig. \ref{gammaeff2d}. The agreement is good,
showing nicely the equivalence of medium-range and FT
calculations.
In Fig. \ref{gammaeff} we report the three-dimensional results for 
$\gamma_{\rm eff}(\widetilde{t})$  in both phases. 
MC simulations of the three-dimensional models have
been reported in Ref.~\cite{LB-98}.
In the HT phase, $\gamma_{\rm eff}(\widetilde{t})$ agrees 
nicely with the MC data in the mean-field region, 
while discrepancies appear in the 
neighborhood of the Wilson-Fisher point. However,
for $\widetilde{t}\to 0$, only data with small values of $\rho$
are present, so that the differences that are observed should be 
due to corrections to the universal behavior. 
The LT phase 
shows a similar behaviour: good agreement in the mean-field region,
and a difference near the Wilson-Fisher point where again only
data with small $\rho$ are available.

The leading corrections to the universal scaling behavior are
studied analytically in Ref.~\cite{PRV-99},
showing that they are of order $R^{-d}$ in $d$ dimensions, provided one
chooses appropriately the scale $R$, i.e. as in Eq.~(\ref{defR}).
There one may find also a critical discussion of
the phenomenological methods used to described nonuniversal
crossover effects, such as those proposed in
Refs.\cite{Chen-etal_90,Chen-etal_90b,Anisimov-etal_92,Anisimov-etal_96,%
Luijten-Maryland},
which have been applied to many different experimental situations
\cite{Chen-etal_90,Chen-etal_90b,Anisimov-etal_95,Anisimov-etal_96,%
Melnichenko-etal_97,Jacob-etal_98}.

\subsection{Critical crossover in self-avoiding walk models with 
medium-range jumps}
\label{n0cr}

The models with medium-range interactions can be studied in the limit 
$N\to 0$. In this case, repeating the discussion presented in
Sec. \ref{n0}, one can write the $N$-vector model with 
medium-range interactions in terms of 
SAW's with medium-range jumps.
To be explicit, we define an $n$-step SAW with range $R$ as a 
sequence of lattice
points $\{\omega_0,\cdots,\omega_n\}$ with $\omega_0 = (0,0,0)$ and
$(\omega_{j+1} - \omega_j)\in D$, such that
$\omega_i\not=\omega_j$ for all $i\not= j$. 
Then, if $c_{n,R}$ is 
the total number of $n$-step walks and $R^2_e(n,R)$ is the end-to-end 
distance---see Sec. \ref{sec.SAW} for definitions---we have
\begin{eqnarray}
\lim_{N\to 0} \chi(\beta) &= &
     \sum_{n=0}^\infty \widehat{\beta}^n c_{n,R} \; ,
\label{chiRNeq0} \\
\lim_{N\to 0}  \xi^2(\beta) \chi(\beta) &= & {1\over 2d}
    \sum_{n=0}^\infty \widehat{\beta}^n c_{n,R} R^2_e({n,R}).
\label{xiRNeq0}
\end{eqnarray}
where $\widehat{\beta} = \beta/V_R$ and $\chi$ and $\xi^2$ are defined 
for the $N$-vector model with medium-range interaction. The proof 
is identical to that presented in Sec. \ref{sec-SAW-Nvector}.

Since $n$
is the dual variable (in the sense of Laplace transforms) of
$t$, the critical crossover is obtained by performing the limit 
$n\to\infty$, $R\to\infty$ with $\widetilde{n}\equiv n R^{-2d/(4-d)}$ fixed.
Using  Eqs. \reff{fchi} and  \reff{fxi} one infers the existence of the  following limits:
\begin{eqnarray}
&& \widetilde{c}_{n,R} \equiv \;
c_{n,R} \beta_{c}(R)^n \longrightarrow  g_c(\widetilde{n}), \nonumber \\
&& \widetilde{R}^2_e(n,R) \equiv \;
R^2_e(n,R) R^{-8/(4-d)} \longrightarrow  g_E(\widetilde{n}),
\end{eqnarray}
where the functions $g_c(\widetilde{n})$ and $g_E(\widetilde{n})$ are related
to $f_\chi(\widetilde{t})$ and
$f_\xi(\widetilde{t})$ by a Laplace transform. Explicitly
\begin{eqnarray}
f_\chi(t) &=& \int^\infty_0 du\,g_c(u) e^{-ut},  \nonumber \\
f_{\xi}(t) f_\chi(t) &=&  
{1\over 2d} \int^\infty_0 du\, g_c(u) g_E(u) e^{-ut}. 
\end{eqnarray}
Using perturbation theory, it is possible to derive predictions for 
$R^2_e(n,R)$ and $c_{n,R}$. For $R^2_e(n,R)$ we can write
\be
g_{E,PT}(\widetilde{n}) =\; a_E\, \widetilde{n}\, h_E(z),
\label{gEPT}
\ee
where $z = (\widetilde{n}/l)^{1/2}$. The function 
$h_E(z)$ has been computed in perturbation theory to six-loop order
\cite{Muthukumar-Nickel_84,Muthukumar-Nickel_87}. 
Resumming the series with a Borel-Leroy
transform, one finds that a very good approximation\footnote{
There exist several other expressions for the function
$h_E(z)$, see Refs. \cite{YT-67,DB-76,desCloizeaux-81,DF-84-85,dCCJ-85,%
Muthukumar-Nickel_87,CN-92,Schafer-94} and references therein.} 
is provided by \cite{BN-97}
\be
h_E(z) = (1 + 7.6118 z + 12.05135 z^2)^{0.175166}\; .
\label{hEresum}
\ee
Comparison with a detailed MC simulation of the
Domb-Joyce model indicates \cite{BN-97}
that this simple expression
differs from the exact result by less than 0.02\% for $z < 2$.

The constants $a_E$ and $l$ appearing in Eq. \reff{gEPT}
are nonuniversal. For the specific model considered here, they are given by
$a_E = 6$, and  $l =(4\pi)^3$.
Ref.~\cite{CCPRV-99} reports results of 
an extensive simulation  of medium-range SAW's, in which 
walks of length up to $N\approx 7 \cdot 10^4$ were generated. 
The domain $D$ was chosen
as follows:
\be
D = \; \left\{x:\, \sum_i |x_i|\le \rho\right\}\; .
\ee
In the simulation $\rho$ was varied between 2 and 12.

\begin{figure}
\vspace*{-1cm} \hspace*{-1.3cm}
\epsfxsize = 0.5\textwidth
\begin{tabular}{cc}
\leavevmode\epsffile{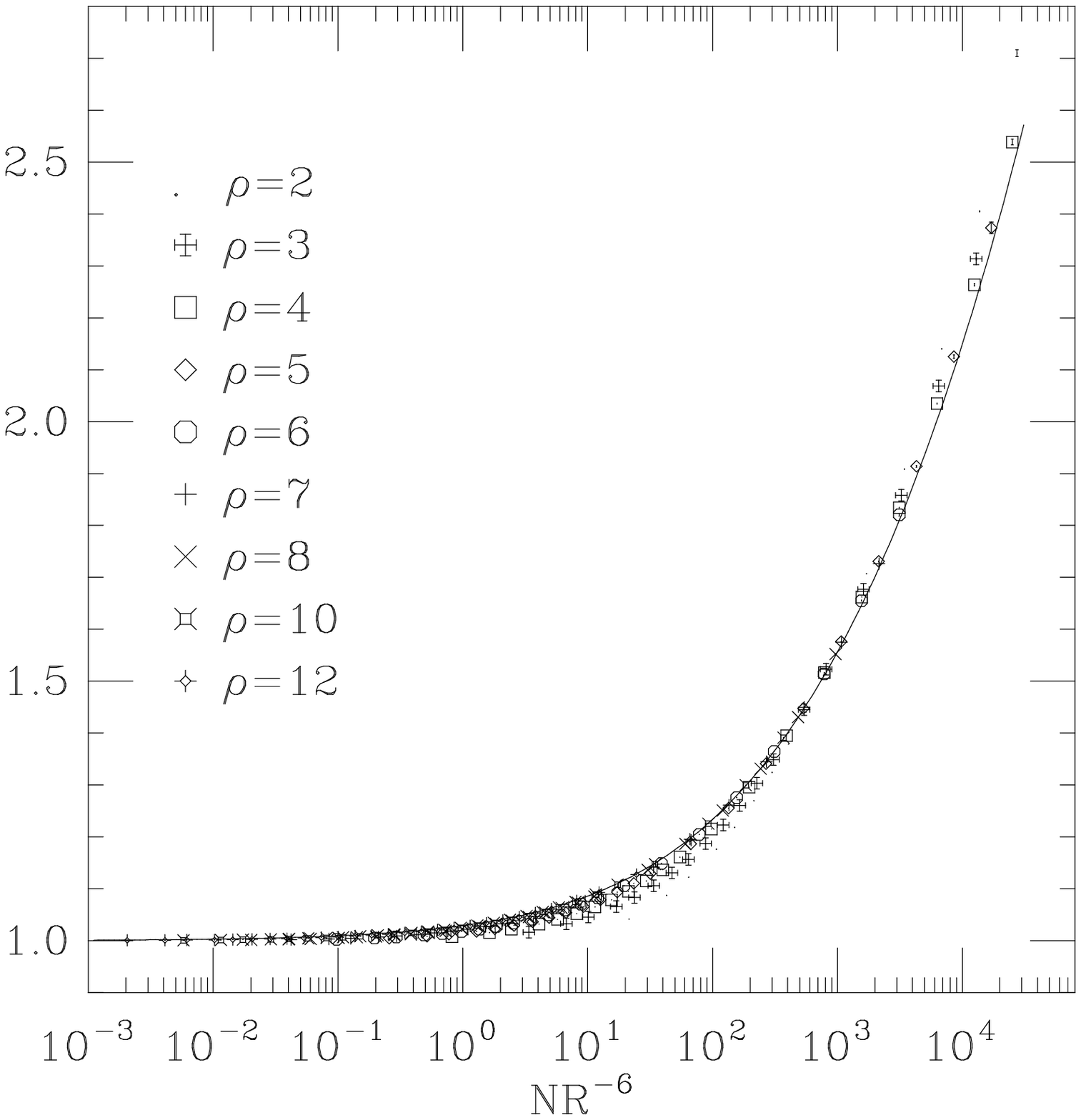} &
\leavevmode\epsffile{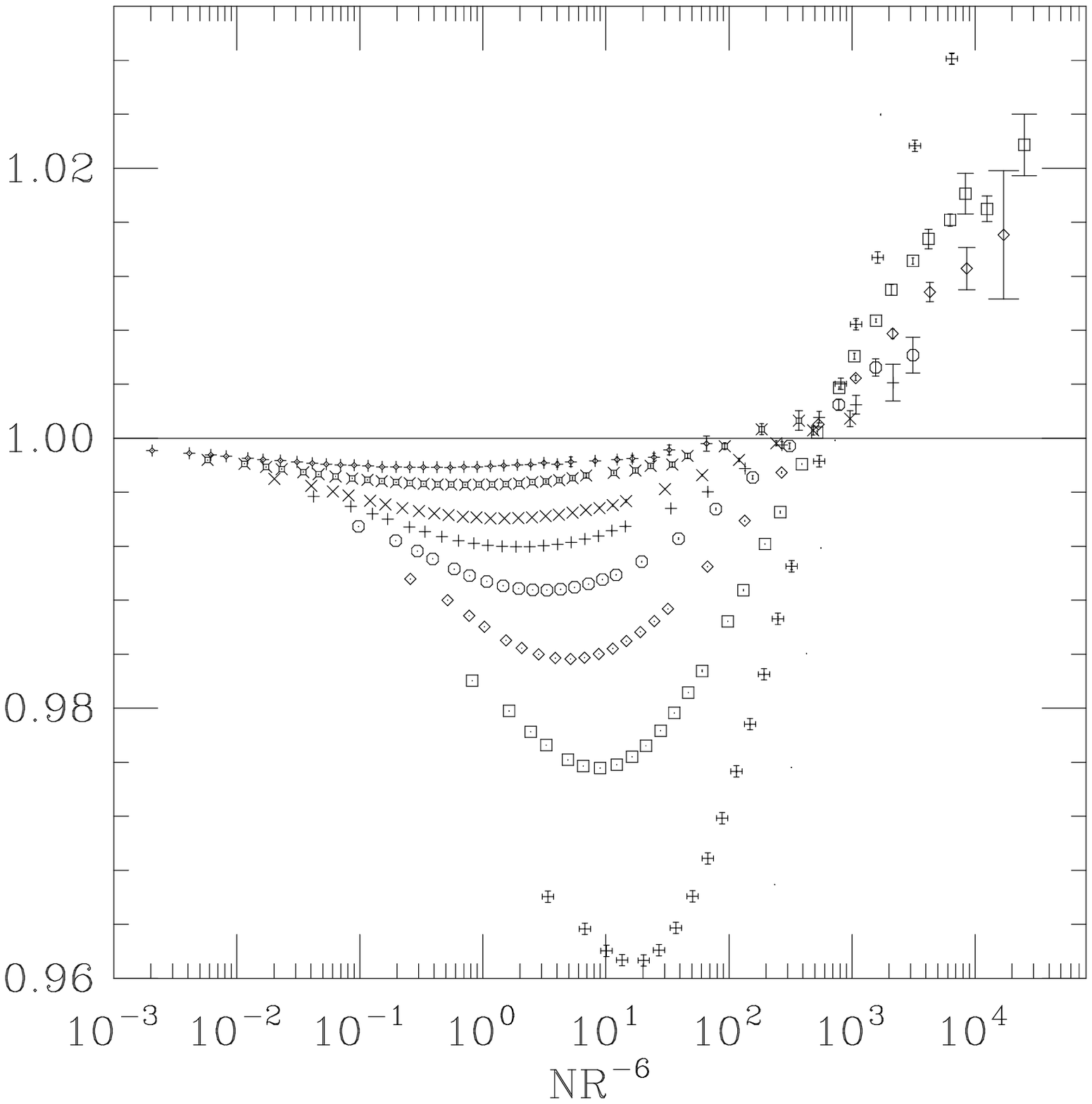}
\end{tabular}
%\quad \vspace{8cm}  %% TEMPORARY UNTIL FILE IS THERE
\vspace*{-1cm}
\caption{
Left: 
results for $\widetilde{R}^2_e(n,R)/(6\widetilde{n})$; the solid line is
the theoretical prediction (\ref{gEPT}), (\ref{hEresum}). 
Right: Results for $\widetilde{R}^2_e(n,R)/g_{E,PT}(\widetilde{n})$. 
Results from Ref. \cite{CCPRV-99}.
}
\label{figR2}
\end{figure}

%% \begin{figure}
%% \vspace*{-1cm} \hspace*{-1.3cm}
%% \begin{center}
%% \epsfxsize = 0.5\textwidth
%% \leavevmode\epsffile{n0cr2.ps}
%% %\quad \vspace{8cm}  %% TEMPORARY UNTIL FILE IS THERE
%% \end{center}
%% \vspace*{-1cm}
%% \caption{
%% Results for $\widetilde{R}^2_e(n,R)/g_{E,PT}(\widetilde{n})$. The same
%% symbols as in Fig. \ref{figR2} are used. Results from Ref. \cite{CCPRV-99}.
%% }
%% \label{figratio}
%% \end{figure}

\begin{figure}
\vspace*{-1cm} \hspace*{-1.3cm}
\begin{center}
\epsfxsize = 0.5\textwidth
\leavevmode\epsffile{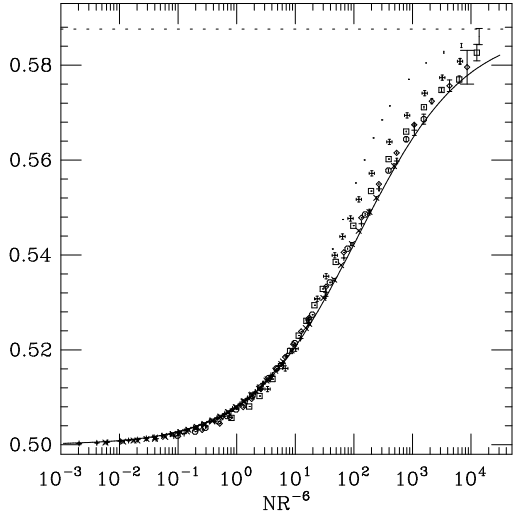}
%\quad \vspace{8cm}  %% TEMPORARY UNTIL FILE IS THERE
\end{center}
\vspace*{-1cm}
\caption{Effective exponent $\nu_{\rm eff}$ as a function of 
$\widetilde{n}$.
Symbols are defined as in Fig.~\ref{figR2}.
The dashed line is the self-avoiding walk value
$\nu=0.58758(7)$. The solid line is the theoretical prediction.
Results from Ref.~\cite{CCPRV-99}.
}
\label{fignu}
\end{figure}

In Fig.~\ref{figR2} the results for $\widetilde{R}^2_e(n,R)$ are
reported together
with the perturbative prediction $g_{E,PT}(\widetilde{n})$
defined in Eqs. (\ref{gEPT}) and (\ref{hEresum}).
The agreement is good, although one can see clearly the presence
of corrections to scaling, see the plot of 
the ratio $\widetilde{R}^2_e(n,R)/g_{E,PT}(\widetilde{n})$. 
As expected different points converge to 1 as $\rho$ increases.
A more detailed discussion of the
expected scaling corrections can be found in
Refs.~\cite{PRV-99,CCPRV-99}.
One may also define an effective exponent $\nu_{\rm eff}$
\be
\nu_{\rm eff} =\, {1\over 2\log 2}\; 
\log\left( {R^2_e(2n,R)\over R^2_e(n,R)} \right).
\ee
The corresponding curve  is reported in Fig. \ref{fignu}. It shows the expected
crossover behaviour between the mean-field value $\nu = 1/2$
and the self-avoiding walk value $\nu = 0.58758(7)$ 
\cite{BN-97}.

It is important to note that FT gives only the asymptotic behavior for 
$R\to\infty$, while corrections cannot be described by FT and one 
has to resort to phenomenological models. In Ref. \cite{CCPRV-99}
a very simple parametrization was proposed:
\begin{equation}
\widetilde{R}^2_e(n,R) = \widetilde{n} h_E(z) 
   \left[1 + {1\over R^d} {a + b z + c z^2\over 1 + d z + e z^2} \right],
\label{R2-phen}
\end{equation}
where $a$, $b$, $c$, $d$, $e$ are free parameters. Such a parametrization
describe accurately all data of Ref. \cite{CCPRV-99} with $\rho \ge 2$. 
Other phenomenological models are described in Refs. \cite{LK-01,LK-02,CN-92}.

The results of Ref. \cite{CCPRV-99} also apply to other polymer models. 
For instance, as in Ref. \cite{LK-01,LK-02}, one may model the polymer
as a sequence of rigidly-bonded hard spheres of diameter $\sigma$, with the 
bond length fixed to $\rho$. In this case, the relevant scale 
is $\rho/\sigma$ and the crossover limit is obtained for 
$\rho/\sigma\to\infty$, $n\to\infty$ with 
$\widetilde{n}\equiv n (\rho/\sigma)^{-2d/(4-d)}$ fixed. The limiting 
crossover curve should describe such chains already for 
$\rho/\sigma\gtrsim 3$, while for lower values of this ratio one could
use a parametrization of the form (\ref{R2-phen}).

For applications to polymers, 
the universal crossover functions for the radius of gyration and 
for the universal constant $\Psi$ are also of interest. 
In the critical crossover limit one finds \cite{BN-97}
\begin{eqnarray}
&&R^2_g (n,R) R^{-8/(4-d)} \longrightarrow  {a_E\over 6} \widetilde{n} 
\left(1 + 7.286 z + 9.51 z^2\right)^{0.175166}, \nonumber 
\\
&&\Psi(n,R) \longrightarrow 
z \left(1 + 14.59613 z + 63.7164 z^2 + 79.912 z^3\right)^{-1/3} .
\end{eqnarray}

\section{Critical phenomena described by Landau-Ginzburg-Wilson Hamiltonians}
\label{LGWHa} 

\subsection{Introduction}
\label{lsec-intro}

In the framework of the RG approach to critical phenomena,
a quantitative description of 
many continuous phase transitions  can be obtained
by considering an effective Landau-Ginzburg-Wilson (LGW) Hamiltonian, 
having an $N$-component fundamental field $\phi_i$
and containing up to fourth-order powers of the field components.
The fourth-degree polynomial form of the potential depends
on the symmetry of the system.
We have already discussed 
the O($N$)-symmetric $\phi^4$ theories, that describe 
the helium superfluid transition, 
the liquid-vapor transition in fluids,
the critical behaviors of many magnetic materials, and
the statistical properties of dilute polymers. We now consider 
other physically interesting systems 
whose continuous transitions are described by more complex $\phi^4$ theories. 

In general,
considering an $N$-component order parameter $\phi_i$, the 
Hamiltonian can be written as 
\bea
{\cal H} = \int d^d x \Bigl[ 
{1\over 2} \sum_i (\partial_\mu \phi_{i})^2 + 
{1\over 2} \sum_i r_i \phi_{i}^2  + 
{1\over 4!} \sum_{ijkl} u_{ijkl} \; \phi_i\phi_j\phi_k\phi_l \Bigr],
\label{Hcomplessa}
\eea
where the number of independent parameters $r_i$ and $u_{ijkl}$ 
depends on the symmetry group of the theory. An interesting 
class of models is characterized by the fact that $\sum_i \phi^2_i$ is the 
only quadratic polynomial that is invariant under the symmetry group of the 
theory. In this case, all $r_i$ are equal, $r_i = r$, and 
$u_{ijkl}$ satisfies the trace condition~\cite{BLZ-74-many}
\begin{equation}
\sum_i u_{iikl} \propto \delta_{kl}.
\label{tracecond}
\end{equation}
In these models, criticality is driven by tuning the single parameter
$r$, which physically may correspond to the temperature.

In the absence of a sufficiently large symmetry restricting the form of the
potential, many quartic couplings must be introduced---see, e.g., Refs.
\cite{BLZ-74,Mukamel-75,MK-75,NF-75,BLZ-76,Aharony-76,MK-76-1,%
MK-76-2,BM-76,GP-80,STJAJG-82,TMTB-85,B-etal-90,AS-94}---and
the study of the critical behavior may become quite complicated.
In the case of a continuous transition, 
the critical behavior is described by the infrared-stable fixed point (FP)
of the theory, which determines a universality class. 
The absence of a stable FP is instead 
an indication for a first-order phase transition.
Note that even in the presence of a stable FP,
a first-order phase transition may  still occur
for systems that are outside its attraction domain. 
For a discussion of this point see, e.g.,
Refs. \cite{BLM-77,KM-81} and references therein.

In this section we discuss several examples of 
systems whose critical behavior is 
described by $\phi^4$ LGW Hamiltonians with two or three quartic couplings,
such as the Hamiltonian with cubic anisotropy that is relevant for magnets,
the so-called $MN$ model whose limit $N\rightarrow 0$ is related to 
the critical behavior of spin models in the presence of quenched randomness,
the O($M$)$\times$O($N$)-symmetric Hamiltonian that (in the case $M=2$) 
describes the critical behavior
of $N$-component frustrated models with noncollinear order, and, finally,
the tetragonal Hamiltonian that is  relevant for some structural
phase transitions. 
We essentially review the results of FT studies, especially the most recent
ones. We then compare them with the results of other approaches 
and with the experiments. We will also briefly discuss the $\phi^4$ theory
with symmetry O$(n_1)\oplus$O$(n_2)$. This case is somewhat more complex 
since there are two independent $r_i$'s. Therefore, transitions where 
all components of the field are critical are only obtained by tuning 
two independent parameters.

\subsection{The field-theoretical method for generic $\phi^4$ theories with 
a single quadratic invariant}
\label{lsec-FT}

Let us consider the generic $\phi^4$ Hamiltonian \reff{Hcomplessa}
with a single parameter $r$, i.e. $r_i=r$ for all $r$. Correspondingly,
the quartic coupling satisfies the trace condition (\ref{tracecond}).

The perturbative series of the exponents are obtained by a trivial 
generalization of the method presented in Sec. \ref{sec-2.4.1}.
In the fixed-dimension FT approach one 
renormalizes the theory by introducing a set of zero-momentum conditions 
for the two-point and four-point correlation functions:
\begin{eqnarray}
&& \hskip -1truecm
\Gamma^{(2)}_{ij}(p) = \delta_{ij} Z_\phi^{-1} \left[ m^2+p^2+O(p^4)\right],
\label{ren1g}  \\
&& \hskip -1truecm
\Gamma^{(4)}_{ijkl}(0) = m\,Z_\phi^{-2} \,g_{ijkl},
\label{rencond}  
\end{eqnarray} 
which relate the mass $m$ and the zero-momentum
quartic couplings $g_{ijkl}$ to the corresponding Hamiltonian parameters
$r$ and $u_{ijkl}$.
In addition, one introduces the function $Z_t$ that is defined by the relation
\begin{equation}
\Gamma^{(1,2)}_{ij}(0) = \delta_{ij} Z_t^{-1},
\label{rencond2}
\end{equation}
where $\Gamma^{(1,2)}$ is the (one-particle irreducible)
two-point function with an insertion of $\case{1}{2}\sum_i \phi_i^2$.

The FP's of the theory are given by the common  
zeros $g_{abcd}^*$ of the $\beta$-functions
\begin{equation}
\beta_{ijkl}(g_{abcd}) =  
m \left. {\partial g_{ijkl}\over \partial m}\right|_{u_{abcd}} .
\label{bijkl}
\end{equation}
In the case of a continuous transition,
when $m\rightarrow 0$, the couplings $g_{ijkl}$ are driven toward an
infrared-stable zero $g_{ijkl}^*$ of the  $\beta$-functions.
The absence of stable FP's is usually considered 
as an indication of a  first-order transition.
The stability properties of the FP's are controlled  by the 
eigenvalues $\omega_i$ of the matrix 
\begin{equation}
\Omega_{ijkl,abcd} = { \partial \beta_{ijkl} \over \partial g_{abcd} } 
\end{equation}
computed at the given FP:
a FP is stable if all eigenvalues $\omega_i$ are positive.
The smallest eigenvalue $\omega$ determines the leading scaling corrections,
which vanish as $m^{\omega}\sim |t|^{\Delta}$ where $\Delta=\nu\omega$.
The critical exponents are then obtained by evaluating 
the RG functions
\begin{eqnarray} 
\eta_\phi(g_{ijkl}) = \left. {\partial \ln Z_\phi\over \partial \ln m}
      \right|_{u} , \qquad
\eta_t(g_{ijkl}) = \left. {\partial \ln Z_t \over \partial \ln m} 
      \right|_u
\end{eqnarray}
at the FP $g_{ijkl}^*$: 
\begin{eqnarray}
\eta = \eta_\phi(g_{ijkl}^*), \qquad
\nu  = [ 2 - \eta_\phi(g_{ijkl}^*) + \eta_t(g_{ijkl}^*)]^{-1}.
\end{eqnarray}
From the pertubative expansion of the correlation functions
$\Gamma^{(2)}$, $\Gamma^{(4)}$, and $\Gamma^{(1,2)}$ and 
using the above-reported relations, one derives the expansion of the
RG functions $\beta_{ijkl}$, $\eta_\phi$, and $\eta_t$
in powers of $g_{ijkl}$. 

In the $\epsilon$ expansion approach, it is convenient to renormalize the theory 
using the minimal-subtraction scheme (or the $\normalmsbar$ scheme). 
Then, one determines the zeroes $g_{ijkl}^*$
of the $\beta$-functions as 
series in powers of $\epsilon$, and the 
critical exponents by computing 
the RG functions $\eta(g_{ijkl}^*)$ and $\nu(g_{ijkl}^*)$.
In this approach one obtains a different
$\epsilon$ expansion for each FP.
A strictly related scheme is the so-called
minimal-subtraction scheme without 
$\epsilon$ expansion~\cite{SD-89} that we already mentioned in 
Sec. \ref{sec-2.4.2}.  

The resummation of the perturbative series can be performed by generalizing 
to expansions of more than one variable the methods of Sec. \ref{sec-2.4.3}. 
For this purpose, 
the knowledge of the large-order behavior of the coefficients 
is useful.
For some models with two quartic couplings, $u$ and $v$ say, 
the large-order behavior of the perturbative expansion in $u$ and $v$
at $v/u$ fixed is reported in Refs.~\cite{CPV-00,PV-00-r,PRV-00}.

\subsection{The LGW Hamiltonian with cubic anisotropy}
\label{lsec-cubic}

The magnetic interactions in crystalline solids with cubic symmetry
like iron or nickel are usually modeled using the 
$O(3)$-symmetric Heisenberg Hamiltonian. However, this is a 
simplified model, since other interactions are present.
Among them, the magnetic anisotropy that is induced by the lattice 
structure (the so-called crystal field) is particularly relevant 
experimentally \cite{Chikazumi-book,YT-57,Slonczewski-61}. In cubic-symmetric
lattices it gives rise to additional single-ion contributions, the 
simplest one being $\sum_i \vec{s}^{\ 4}_i$. 
These terms are usually not considered
when the critical behavior of cubic magnets is discussed.
However, this is strictly justified only
if these nonrotationally invariant interactions, that have the 
reduced symmetry of the lattice, are irrelevant in the RG
sense. 

Standard considerations
based on the canonical dimensions of the operators,
see, e.g., Ref.~\cite{Aharony-76}, 
indicate that 
there are two terms that one may add to the Hamiltonian and that are 
cubic invariant: a cubic hopping term $\sum_{\mu=1,3} (\partial_\mu\phi_\mu)^2$
and a cubic-symmetric quartic interaction term $\sum_{\mu=1,3} \phi_\mu^4$.
The first term was studied in 
Refs. \cite{Aharony-73-3,Bruce-74,NT-75,Aharony-76,Nattermann-77}.
A two-loop $O(\epsilon^2)$ calculation indicates that it is irrelevant 
at the symmetric point, although it induces slowly-decaying crossover effects.
In order to study the effect of the cubic-symmetric interaction, we consider
a $\phi^4$ theory with two quartic couplings \cite{Aharony-73,Aharony-76}:
\bea
{\cal H}_{\rm cubic} =  \int d^d x \Bigl\{ {1\over 2} \sum_{i=1}^{N}
      \left[ (\partial_\mu \phi_i)^2 +  r \phi_i^2 \right]  
+{1\over 4!} \sum_{i,j=1}^N \left( u_0 + v_0 \delta_{ij} \right)
\phi^2_i \phi^2_j  \Bigr\}.
\label{Hphi4cubic}
\eea
The cubic-symmetric term $\sum_i \phi_i^4$
breaks explicitly the O($N$) invariance of the model,
leaving a residual discrete cubic symmetry given by the reflections
and permutations of the field components.  

\begin{figure*}[tb]
\hspace{0cm}
\vspace{0cm}
\centerline{\psfig{width=12truecm,angle=-90,file=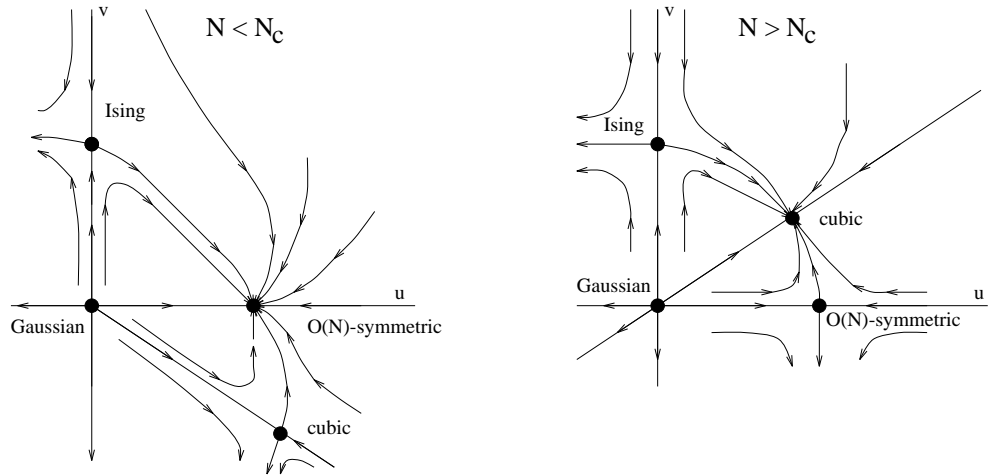}}
\vspace{-2cm}
\caption{
RG flow in the coupling plane $(u,v)$ for
$N<N_c$ and $N>N_c$ for magnetic systems with cubic anisotropy.
}
\label{cubicrgflow}
\end{figure*}

The theory defined by the Hamiltonian (\ref{Hphi4cubic}) has 
four FP's~\cite{Aharony-73,Aharony-76}:
the trivial Gaussian one, the Ising one in which the $N$ components of the 
field decouple,
the O($N$)-symmetric and the  cubic FP's. 
The Gaussian FP is always unstable, 
and so is the Ising FP for any number of components $N$. 
Indeed, at the Ising  FP one may interpret the cubic Hamiltonian
as the Hamiltonian of $N$ Ising systems coupled by the 
O($N$)-symmetric interaction. 
The coupling term $\int d^d x \,\phi_i^2\phi_j^2$ with $i\neq j$ scales
as the integral of the product of two operators $\phi^2$. Since
the $\phi^2$ operator has RG dimension
$1/\nu_I$---indeed, it is associated with the temperature---the combined
operator has RG dimension
$2/\nu_I - d = \alpha_I/\nu_I$, and
therefore the associated crossover exponent is given by $\phi=\alpha_I$,
independently of $N$ \cite{Sak-74,Aharony-76}. 
Since $\alpha_I>0$,
the Ising FP is unstable independently of $N$. 
On the other hand,
the stability properties of the O($N$)-symmetric and of the cubic FP's 
depend on $N$.
For sufficiently small values of $N$, $N<N_c$, the 
O($N$)-symmetric FP is stable and the cubic one is unstable.
For $N>N_c$, the opposite is true:
the RG flow is driven towards the cubic FP,
which now describes the generic critical behavior of the system.
Figure~\ref{cubicrgflow} sketches the 
flow diagram in the two cases $N<N_c$ and $N>N_c$.

Outside the attraction domain of the FP's, the flow goes away
towards more negative values of $u$ and/or $v$ and finally
reaches the region where the quartic interaction no longer 
satisfies the stability
condition. These trajectories should be related to
first-order phase transitions. Indeed, in mean-field theory,
the violation of the positivity conditions 
\begin{equation}
u+v > 0,\qquad Nu+v>0,
\end{equation}
leads to first-order transitions.\footnote{
RG trajectories leading to
unstable regions have been considered in the study of
fluctuation-induced first-order transitions
(see, e.g., Refs.~\cite{Rudnick-75,ASYZ-97,Tetradis-98}).}

It is worth mentioning that, for $v_0\to -\infty$, one obtains the model 
described in Refs. \cite{KL-75,KLU-75} in which the spins 
align along the lattice axes. A HT analysis on the 
face-centered cubic lattice indicates that these models have a 
first-order transition
for $N \gtapprox 2$. This is consistent with the above argument that 
predicts the transition to be of first order for any $v_0<0$ and 
$N>N_c$. More general models, that have Eq. (\ref{Hphi4cubic}) as their
continuous spin limit for $v_0\to -\infty$, were also 
considered in Ref. \cite{Aharony-77}. The first-order nature of the 
transition for negative (small) $v_0$ and large $N$ was also
confirmed in Ref. \cite{Wallace-73}.

If $N>N_c$, cubic anisotropy is relevant and therefore the critical
behavior of the system is not described by the O($N$)-symmetric theory.
In this case,
if the cubic interaction favors the alignment of 
the spins along the diagonals of the cube,
i.e. for a positive coupling $v$,
the critical behavior is controlled by the cubic
FP and the cubic symmetry is retained even at the 
critical point. On the other hand, if the system tends to 
magnetize along the cubic axes---this corresponds to a negative
coupling $v$---then the system undergoes a first-order phase 
transition \cite{Wallace-73,Aharony-76,Aharony-77,SD-99}.
Moreover, since the symmetry is discrete, 
there are no Goldstone excitations in the LT phase.
The longitudinal and the transverse
susceptibilities are finite for $T<T_c$ and $H\rightarrow 0$,
and diverge as $|t|^{-\gamma}$ for $t\propto T - T_c \rightarrow 0$.
For $N>N_c$, the O($N$)-symmetric FP is a tricritical point.

If $N < N_c$, the cubic term in the Hamiltonian is irrelevant, and therefore, 
it generates only scaling corrections $|t|^{\Delta_c}$ with
$\Delta_c>0$. However,
its presence leads to important physical consequences. 
For instance, the transverse susceptibility at the coexistence curve (i.e.
for $T<T_c$ and $H\rightarrow 0$), which is divergent in the 
O($N$)-symmetric case, is now finite and
diverges only at $T_c$ as $|t|^{-\gamma-\Delta_c}$ 
\cite{Wallace-73,KW-73,BA-75,Aharony-76,BLZ-76}. 
In other words, below $T_c$,
the cubic term is a ``dangerous'' irrelevant operator.
Note that for $N$ sufficiently close to $N_c$,
irrespective of which FP is the stable one,
the irrelevant interaction
bringing from the unstable to the stable FP gives rise to  very
slowly decaying corrections to the leading scaling behavior. 

In three dimensions, a simple argument 
ba\-s\-ed on the symmetry of the two-component cubic model~\cite{Korz-76}
shows that the cubic FP is unstable for $N=2$.
Indeed, for $N=2$, a $\pi/4$ internal rotation
maps ${\cal H}_{\rm cubic}$ into a new one of the same form 
but with new couplings $(u_0',v_0')$ given by
$u_0' = u_0+\case{3}{2}v_0$ and  $v_0' = -v_0$.
This symmetry maps the Ising FP onto the cubic one. Therefore,
the two FP's describe the same theory and have the same stability properties.
Since the Ising point is unstable, the cubic point is unstable too, 
so that the stable point is the isotropic one.
In two dimensions, this is no longer true. Indeed, one expects 
the cubic interaction to be truly marginal for $N=2$~\cite{JKKN-77,PN-76,NR-82} and relevant for 
$N>2$, and therefore  $N_c=2$ in two dimensions.

\begin{table*}
\caption{
Summary of the results in the literature. 
The values of the smallest eigenvalues $\omega_2$ 
of the stability matrix $\Omega$ refer to $N=3$.
The subscripts ``$s$" and ``$c$" indicate that it 
is related to the symmetric and to the cubic FP respectively.
}
\label{literature}
\footnotesize
\begin{center}
\begin{tabular}{rllll}
\hline
\multicolumn{1}{c}{Ref.}& 
\multicolumn{1}{c}{Method}& 
\multicolumn{1}{c}{$\omega_{2,s}$}& 
\multicolumn{1}{c}{$\omega_{2,c}$}& 
\multicolumn{1}{c}{$N_c$}\\
\hline  
\cite{FHY-01}  $_{2000}$ & $d=3$ exp: $O(g^6)$ &  & 0.015(2) & 2.862(5) \\
\cite{CPV-00}   $_{2000}$ & $d=3$ exp: $O(g^6)$ & $-0.013(6)$ & 0.010(4)& 2.89(4) \\
\cite{CPV-00}   $_{2000}$ & $\epsilon$ exp: $O(\epsilon^5)$ & $-0.003(4)$ & 0.006(4) & 2.87(5) \\
\cite{Varnashev-00}    $_{2000}$ & $d=3$ exp: $O(g^4)$ & $-0.0081$ & 0.0077 & 2.89(2) \\
\cite{SAS-97}    $_{1997}$ & $\epsilon$ exp: $O(\epsilon^5)$ & &  & 2.86 \\
\cite{KTS-97,KT-95}  $_{1995}$ & $\epsilon$ exp: $O(\epsilon^5)$ & $-0.00214$ & 0.00213 &  $N_c <3$ \\
\cite{KS-95}     $_{1995}$ & $\epsilon$ exp: $O(\epsilon^5)$  & &  & 2.958 \\
\cite{MSS-89}    $_{1989}$ & $d=3$ exp: $O(g^4)$  & & 0.008  & 2.91  \\
\cite{NKF-74}    $_{1974}$ & $\epsilon$ exp: $O(\epsilon^3)$  & & & 3.128  \\
\cite{TMVD-01}     $_{2002}$ & CRG & & & 3.1    \\
\cite{NR-82}     $_{1982}$ & scaling-field & & & 3.38    \\
\cite{YH-77}     $_{1977}$ & CRG  & $-0.16$ & & 2.3   \\
\cite{CH-98}     $_{1998}$ & MC & 0.0007(29)&  & $N_c \approx 3$ \\
\cite{FVC-81}    $_{1981}$ & HT exp. $O(\beta^{10})$ & 
$-0.89(14)$ &  & $N_c<3 $  \\
\hline
\end{tabular}
\end{center}
\end{table*}

The model (\ref{Hphi4cubic}) has 
been the object of several studies
\cite{GKW-72,Aharony-73,KW-73,NKF-74,BLZ-74,NT-75,Nattermann-77,YH-77,DR-79,FVC-81,NR-82,MS-87,%
Shpot-89,MSS-89,KS-95,KT-95,KTS-97,SAS-97,CH-98,MV-98-c,Varnashev-00,CPV-00,CC-01,FHY-01,PS-01,TMVD-01}. 
In the 70's several computations were done using the 
$\epsilon$ expansion \cite{Aharony-73,KW-73,NKF-74,BLZ-74}, and predicted
$3< N_c < 4$, indicating that cubic magnets
are  described by the O(3)-invariant Heisenberg model. 
More recent FT studies have questioned this conclusion, and
provided a robust  evidence that $N_c<3$, implying that the critical 
properties of cubic magnets are not described by the O(3)-symmetric
theory, but instead by the cubic model at the cubic FP. 
In Table \ref{literature} we report a summary of the results 
for $N_c$ and, in the physically interesting case $N=3$, for the 
smallest eigenvalue $\omega_2$ of the stability matrix $\Omega$
at the O(3)-symmetric and cubic FP's,
$\omega_{2,s}$ and $\omega_{2,c}$ respectively.
  
The most recent FT perturbative analyses are based on 
five-loop series in the framework of the $\epsilon$ expansion \cite{KS-95}
and on six-loop series in the fixed-dimension approach \cite{CPV-00}.
In Ref.~\cite{CPV-00} the analysis of the six-loop
fixed-dimension expansion was done 
exploiting Borel summability and the knowledge of the large-order behavior
of the expansion in $u$ and $v$ at $v/u$ fixed.
The same series were analyzed in Ref.~\cite{FHY-01}
using the pseudo $\epsilon$-expansion technique \cite{LZ-77}.
In the analysis of the five-loop 
$\epsilon$ expansion reported in Ref.~\cite{CPV-00}, beside the large-order behavior of the series, 
the exact two-dimensional result $N_c=2$ was used
to perform a constrained analysis (for the method, see 
Sec. \ref{sec-2.4.3}).

The results of the FT analysis, 
see Table \ref{literature}, show that in the Heisenberg case,
i.e. $N=3$,  the isotropic FP is unstable, while the cubic
one is stable. Indeed, the smallest eigenvalue $\omega_2$ is
positive at the cubic FP,
and negative at the symmetric one. 
The analyses of the six-loop fixed-dimension series give 
\cite{CPV-00} $\omega_{2,c}=0.010(4)$,
$\omega_{2,s}=-0.013(6)$, and
\cite{FHY-01} $\omega_{2,c} = 0.015(2)$.
The other eigenvalue
$\omega_1$ turns out to be much larger, i.e. $\omega_1=0.781(4)$ at the
cubic FP \cite{CPV-00}. 
The critical value $N_c$ is therefore smaller than three, but close to
it:  FT analyses show that $N_c \lesssim 2.9$. 
Note that the recent study \cite{TMVD-01}, based on CRG methods,
provides an apparent contradictory result, i.e. $N_c>3$,
probably due to the low level of approximation in the corresponding
derivative expansion.

For the physically relevant case
$N=3$,  the cubic  critical exponents differ very little 
from those of the Heisenberg universality class.
The analyses of Refs.~\cite{CPV-00} and \cite{FHY-01} give respectively
$\nu_c=0.706(6)$ and $\nu_c=0.705(1)$, $\gamma_c=1.390(12)$ and
$\gamma_c=1.387(1)$, $\eta_c=0.0333(26)$,
which should be compared with the Heisenberg
exponents reported in Sec.~\ref{expO3th}.
A more careful analysis of the six-loop fixed-dimension expansion \cite{CPV-02-3}
shows that there are peculiar cancellations in the differences 
between the cubic and Heisenberg exponents, obtaining 
\begin{eqnarray}
\nu_{c} - \nu_H = -0.0003(3) ,\qquad
\eta_{c} - \eta_H = -0.0001(1),\qquad
\gamma_{c} - \gamma_H = -0.0005(7) .\label{diffexp}
\end{eqnarray}
Note that these differences 
are much smaller that the  typical experimental error, see Sec.~\ref{expO3ex}.
Therefore, distinguishing the cubic and the Heisenberg universality class
should be very hard, taking also into account that
crossover effects decay as $t^{\Delta}$ with a very small
$\Delta$, i.e. $\Delta= \omega_{2,c}\nu_c = 0.007(3)$. 
These results 
justify the discussion of Sec.~\ref{O3d3}, where
we compared the experimental results with the theoretical predictions
for the O(3)-symmetric universality class,
neglecting the effects of the cubic anisotropy.
Using the most precise  estimates for the Heisenberg
exponents, i.e. those denoted by MC+IHT in Table~\ref{O3exponents},
and the differences (\ref{diffexp}), one obtains
$\nu_c=0.7109(6)$,  $\eta_c=0.0374(5)$  and
$\gamma_c=1.3955(12)$.

Large corrections to scaling appear also for $N=2$. Indeed,
at the XY FP (the stable one), the subleading exponent $\omega_2$ is 
given by $\omega_2 = 0.103(8)$ \cite{CPV-00}. Thus, even though the 
cubic-symmetric interaction is 
irrelevant, it induces strong scaling corrections behaving 
as $t^\Delta$, $\Delta=\omega_2\nu\approx 0.06$. 

Estimates of the critical exponents at the cubic FP 
for $N>3$ can be found in Ref. \cite{CPV-00}. For $N\to \infty$, 
keeping $Nu$ and $v$ fixed,
one can derive 
exact expressions for the exponents at the cubic FP.
Indeed, for $N\to\infty$ the system can be reinterpreted as 
a constrained Ising model~\cite{Emery-75}, leading to a 
Fisher renormalization of the Ising critical exponents~\cite{Fisher-68}.
One has~\cite{Aharony-73-2,Emery-75,Aharony-76}: 
\begin{eqnarray}
\eta = \eta_I+O\left( {1/N}\right),\qquad
\nu = {\nu_I\over 1-\alpha_I}+O\left( {1/N}\right),
\label{largen}
\end{eqnarray}
where $\eta_I$, $\nu_I$, and $\alpha_I$ 
are the critical exponents of the Ising model.

We mention that 
the equation of state for the cubic-symmetric critical theory
is known to $O(\epsilon)$ in the framework of the 
$\epsilon$ expansion \cite{Aharony-74}. 
Moreover, Ref.~\cite{PS-01} reports a study of the $n$-point
susceptibilities  in the HT phase using the fixed-dimension
expansion.

Using RG arguments, it has been argued  that the critical behavior
of the four-state
antiferromagnetic Potts model on a cubic lattice should be described
by the three-component $\phi^4$ theory 
(\ref{Hphi4cubic}) with cubic anisotropy and with negative
coupling $v$ \cite{BGJ-80}. 
Thus, as a consequence of the RG flow for $N>N_c$
shown in Fig.~\ref{cubicrgflow}, the system is expected to
undergo a first-order phase transition since the region  $v<0$ is outside
the attraction domain of the stable cubic FP.
Ref.~\cite{Itakura-99} presents a MC study
of the four-state antiferromagnetic Potts model.
The numerical results are however not conclusive on the nature of the 
transition.

It is worth noting that the computation of $\omega_2$ at the 
$O(N)$ FP directly gives the crossover exponent 
$\phi_4$ associated with the spin-4 perturbation of the 
$O(N)$ FP, see Sec. \ref{sec-1.5.8}. Indeed, the quartic
interaction can be written as
\be
u(\phi^2)^2 + v \sum_i\phi^4_i = 
    v \sum_i {\cal O}^{iiii}_4 + 
   \left(u + {3v\over N+2}\right) (\phi^2)^2 ,
\ee
where ${\cal O}^{ijkl}$ is the spin-4 operator defined in Sec. \ref{sec-1.5.8}.
The cubic perturbation is nothing but a particular combination
of the components of the spin-4 operator and therefore, 
$\phi_4 = -\omega_2 \nu$, where $\omega_2$ is the eigenvalue of 
the stability matrix $\Omega$ associated with the cubic-symmetric 
interaction at the $O(N)$ FP. Therefore, the results for the cubic 
theory, see, e.g., Ref. \cite{CPV-00}, can be used to compute $\phi_4$.
This correspondence implies that
the results for the stability of the O($N$)-symmetric FP
with respect to cubic perturbations  can be
extended to all spin-4 perturbations. In particular,
for $N>N_c$ (resp. $N<N_c$) the O($N$) FP is unstable (resp. stable) 
under any spin-4 quartic perturbation.

\subsection{Randomly dilute spin models}
\label{lsec-random}

The critical behavior of systems with quenched disorder is of considerable 
theoretical and experimental interest. 
A typical example is obtained by mixing an (anti)-ferromagnetic 
material with a nonmagnetic one, obtaining the so-called dilute 
magnets. These materials are usually described in terms of 
a lattice short-range Hamiltonian of the form
\begin{equation}
{\cal H}_x = - J\,\sum_{<ij>}  \rho_i \,\rho_j \;\vec{s}_i \cdot \vec{s}_j,
\label{latticeH}
\end{equation}
where $\vec{s}_i$ is an $M$-component spin and 
the sum is extended over all nearest-neighbor sites. The quantities
$\rho_i$ are uncorrelated random variables, which are equal to one 
with probability $x$ (the spin concentration) and zero with probability $1-x$
(the impurity concentration). The pure system corresponds to $x=1$.
One considers quenched disorder, since the relaxation time associated with 
the diffusion of the impurities is much larger than all other typical time 
scales, so that, for all practical purposes, one can consider the position
of the impurities fixed.
For sufficiently low spin dilution $1-x$, i.e. as long as one is above the 
percolation threshold of the magnetic atoms,
the system described by the Hamiltonian ${\cal H}_x$ undergoes a second-order 
phase transition at $T_c(x) < T_c(x=1)$
(see, e.g., Ref.~\cite{Stinchcombe-83} for a review).

The relevant question in the study of this class of systems is the effect
of disorder on the critical behavior. 
The Harris criterion~\cite{Harris-74} states that the addition of 
impurities to a system that undergoes a second-order 
phase transition does not change the critical behavior 
if the specific-heat critical exponent $\alpha_{\rm pure}$ of the pure 
system is negative. If $\alpha_{\rm pure}$ is positive, the transition
is altered.  Indeed the specific-heat exponent 
$\alpha_{\rm random}$ in a disordered system 
is expected to be negative~\cite{Ma-book,SS-81,CCFS-86,SF-88,PSZ-97,AHW-98}.
Thus, if $\alpha_{\rm pure}$ is positive, 
$\alpha_{\rm random}$ differs from $\alpha_{\rm pure}$, so that
the pure system and the dilute one have a different critical behavior.
In pure $M$-vector models with $M>1$, the specific-heat exponent 
$\alpha_{\rm pure}$ is negative; therefore, according to the Harris criterion, 
no change in the critical asymptotic behavior is expected
in the presence of weak quenched disorder. This means that in these systems
disorder leads only to irrelevant scaling corrections.
Three-dimensional Ising systems are more interesting, since 
$\alpha_{\rm pure}$ is positive.
In this case, the presence of quenched impurities leads
to a new random Ising universality class.

Theoretical investigations, using approaches based on  RG
\cite{HL-74,Emery-75,Lubensky-75,Khmelnitskii-75,AIM-76,GL-76,%
Aharony-76,GMM-77,JK-77,Shalaev-77,NR-82,Jug-83,Newlove-83,%
MS-84,PA-85,DG-85,Shpot-89,MSS-89,Mayer-89,Shpot-90,HS-92,BS-92,JOS-95,%
SAS-97,FHY-98,HY-98,Mayer-98,FHY-99,Varnashev-00,FHY-00,PS-00,TMVD-01,FHY-01},
and MC simulations 
\cite{Landau-80,MLT-86,CS-86,BS-88,BSHF-89,WC-89,WWMC-90,%
HF-90,Heuer-90,WHF-92,Heuer-93,Hennecke-93,WD-98,BFMMPR-98,Hukushima-00,MGI-00},
support the existence of a new random Ising FP describing the 
critical behavior along the $T_c(x)$ line: critical exponents are 
dilution independent
(for sufficiently low dilution) and different from those of the 
pure Ising model. We mention that, in 
the presence of an external magnetic field along the 
uniaxial direction,
dilute Ising systems present a different critical behavior, 
equivalent to that of the random-field Ising 
model~\cite{FA-79}, which is
also the object of intensive theoretical and experimental 
investigations (see, e.g., Refs.~\cite{Belanger-98,Nattermann-98,Belanger-00}).

Experiments confirm the theoretical picture. 
Crystalline mixtures of an Ising-like
uniaxial antiferromagnet (e.g., FeF${}_2$, MnF${}_2$) 
with a nonmagnetic material
(e.g., ZnF${}_2$)
provide a typical realization of the random Ising model (RIM)
(see, e.g., Refs. 
\cite{DG-81,BCSJBKJ-83,HCK-85,BKJ-86,Barret-86,MCYBUB-86,%
BKFJ-88,RKLHE-88,RKJ-88,TPBH-88,FKJ-91,%
BWSHNLRL-95,BWSHNLRL-96,HFHBRC-97,SBF-98,SB-98,SBF-99,SBF-01,MBSF-00}).
Some experimental results are reported in Table~\ref{experiments}. 
This is not a complete list, but it gives 
an overview of the experimental state of the art. 
Recent reviews of the experiments can be found in 
Refs.~\cite{Belanger-98,FHY-00,Belanger-00,FHY-01}.
The experimental estimates are definitely different from the values of the 
critical exponents for pure Ising systems. 
Moreover, they appear to be independent of concentration.

\begin{table*}
\footnotesize
\caption{
Experimental estimates of the critical exponents for systems in the 
RIM universality class, taken from Ref.~\protect\cite{FHY-00}.
}
\label{experiments}
\begin{center}
\begin{tabular}{rlcllll}
\hline
\multicolumn{1}{c}{Ref.}& 
\multicolumn{1}{c}{Material}& 
\multicolumn{1}{c}{$x$}& 
\multicolumn{1}{c}{$\gamma$}& 
\multicolumn{1}{c}{$\nu$}& 
\multicolumn{1}{c}{$\alpha$}& 
\multicolumn{1}{c}{$\beta$}\\
\hline  
\cite{MBSF-00}   $_{2000}$ & Mn${}_x$Zn${}_{1-x}$F${}_2$ & $0.35$ & $\approx 1.31$ & $\approx 0.69$ & & \\
\cite{SBF-99}   $_{1999}$ & Fe${}_x$Zn${}_{1-x}$F${}_2$ & $0.93$ & $1.34(6)$ & $0.70(2)$ & & \\
\cite{SB-98}      $_{1998}$& Fe${}_x$Zn${}_{1-x}$F${}_2$ & $0.93$ &                  &       & $-0.10(2)$& \\
\cite{HFHBRC-97}  $_{1997}$ & Fe${}_x$Zn${}_{1-x}$F${}_2$ & $0.5$  &&                  & & $0.36(2)$ \\
\cite{BWSHNLRL-96}  $_{1996}$ & Fe${}_x$Zn${}_{1-x}$F${}_2$ & $0.52$  &&                  & & $0.35$ \\
\cite{BWSHNLRL-95}  $_{1995}$ & Fe${}_x$Zn${}_{1-x}$F${}_2$ & $0.5$  &                  &                  & & $0.35$ \\
\cite{TPBH-88}    $_{1988}$ & Mn${}_x$Zn${}_{1-x}$F${}_2$ & $0.5$  &                  &                  & & $0.33(2)$ \\
\cite{RKJ-88}     $_{1988}$ & Mn${}_x$Zn${}_{1-x}$F${}_2$ & $0.40,0.55,0.83$  &                  &       & $-0.09(3)$& \\
\cite{RKLHE-88}   $_{1988}$ & Fe${}_x$Zn${}_{1-x}$F${}_2$ & $0.9$  &                  &                  & & $0.350(9)$ \\
\cite{MCYBUB-86}  $_{1986}$ & Mn${}_x$Zn${}_{1-x}$F${}_2$ & $0.75$ & $1.364(76)$ & $0.715(35)$ & & \\
\cite{Barret-86}  $_{1986}$ & Fe${}_x$Zn${}_{1-x}$F${}_2$    & $0.925-0.950$ &  &  & & 0.36(1) \\
\cite{BKJ-86}  $_{1986}$ & Fe${}_x$Zn${}_{1-x}$F${}_2$    & $0.46$ & $1.31(3)$ & $0.69(1)$ & & \\
\cite{BCSJBKJ-83}  $_{1983}$ & Fe${}_x$Zn${}_{1-x}$F${}_2$    & $0.5,0.6$ & $1.44(6)$ & $0.73(3)$ & $-0.09(3)$ & \\
\cite{DG-81}  $_{1981}$ & Mn${}_x$Zn${}_{1-x}$F${}_2$    & $0.864$ &  &  & & 0.349(8) \\
\hline
\end{tabular}
\end{center}
\end{table*}

Several experiments also tested the effect of disorder 
on the $\lambda$-transition of ${}^4$He that belongs to the 
$XY$ universality class, corresponding to $M=2$
\cite{KHR-75,FGWC-88,CBMWR-88,TCR-92,YC-97,ZR-99}. They studied
the critical behaviour of ${}^4$He completely filling the pores 
of porous gold or Vycor glass. The results indicate that the 
transition is in the same universality class of the 
$\lambda$-transition of the pure system, in agreement with the 
Harris criterion. 
Ref. \cite{YC-97} reports $\nu=0.67(1)$ and Ref. \cite{ZR-99} 
finds that the exponent $\nu$ is compatible with 2/3. These
estimates agree with the best results for the pure system 
reported in Sec.~\ref{expXY}. 
\footnote{Experiments 
for ${}^4$He in aerogels find larger values for
the exponent $\nu$ \cite{CBMWR-88,MMGA-91,CMR-96}. 
The current explanation of these results is that, in aerogels, 
the silica network is correlated to long distances, and therefore,
the Harris criterion and the model studied here do not apply.
A simple model describing these materials was studied in 
Ref. \cite{LT-89}.}

Experiments
on disordered magnetic materials of the  isotropic 
random-exchange type show that the critical exponents 
are unchanged by disorder, see Sec.~\ref{expO3ex},
confirming the theoretical expectation.

The randomly dilute Ising model (\ref{latticeH}) has been investigated by many 
numerical simulations (see, e.g., Refs.
\cite{Landau-80,MLT-86,CS-86,BS-88,BSHF-89,WC-89,WWMC-90,HF-90,%
Heuer-90,WHF-92,Heuer-93,Hennecke-93,WD-98,BFMMPR-98}).
The first simulations were apparently finding critical exponents 
depending on the spin concentration. Later, Refs.~\cite{Heuer-93,JOS-95}
remarked 
that this could be a crossover effect: the simulations 
were not probing the critical region and were computing effective exponents
strongly affected by corrections to scaling. Recently,
the critical exponents were computed 
using FSS techniques \cite{BFMMPR-98}. 
The authors found very strong corrections
to scaling, decaying with a rather small exponent 
$\omega = 0.37(6)$---correspondingly $\Delta=\omega\nu = 0.25(4)$---which 
is approximately a 
factor of two smaller than the corresponding pure-system
exponent. By taking into proper account the confluent corrections, 
they showed that the critical exponents are universal with 
respect to variations of
the spin concentration in a wide interval above the percolation point.
Their final estimates are \cite{BFMMPR-98} 
\begin{eqnarray}
\gamma =  1.342(10), \qquad
\nu = 0.6837(53), \qquad
\eta = 0.0374(45).\label{MCrest}
\end{eqnarray}

The starting point of the FT approach to the study of 
ferromagnets in the presence of quenched disorder
is the LGW Hamiltonian~\cite{GL-76}
\bea
{\cal H} = \int d^d x 
\left\{ {1\over 2}(\partial_\mu \phi(x))^2 + {1\over 2} r \phi(x)^2 
+ {1\over 2} \psi(x) \phi(x)^2  +
{1\over 4!} g_0 \left[ \phi(x)^2\right]^2 \right\},
\label{Hphi4ran}
\eea
where $r\propto T-T_c$, and $\psi(x)$ is a spatially uncorrelated 
random field with Gaussian distribution
\begin{equation}
P(\psi) = {1\over \sqrt{4\pi w}} \exp\left[ - {\psi^2\over 4 w}\right].
\end{equation}
We consider quenched disorder. Therefore, in order
to obtain the free energy of the system, one must 
compute the partition function $Z(\psi,g_0)$
for a given distribution $\psi(x)$, and then average the corresponding
free energy over all distributions with probability 
$P(\psi)$. Using the standard replica trick, it is possible to
replace the quenched average with an annealed one.
First, the system is replaced by $N$ non-interacting
copies with annealed disorder.
Then, integrating over the disorder, one obtains the
Hamiltonian~\cite{GL-76}
\bea
{\cal H}_{MN} = \int d^d x 
\Bigl\{ \sum_{i,a}{1\over 2} \left[ (\partial_\mu \phi_{a,i})^2 + 
         r \phi_{a,i}^2 \right]  
 + \sum_{ij,ab} {1\over 4!}\left( u_0 + v_0 \delta_{ij} \right)
          \phi^2_{a,i} \phi^2_{b,j} 
\Bigr\},
\label{Hphi4MN}
\eea
where $a,b=1,...M$ and $i,j=1,...N$.
The original system, i.e.   
the dilute $M$-vector model, is recovered 
in the limit $N\rightarrow 0$. Note that 
the coupling $u_0$ is negative (being proportional to minus the variance of
the quenched disorder), while the coupling $v_0$ is positive. 

In this formulation, 
the critical properties of the dilute $M$-vector model
can be investigated by studying the RG flow
of the Hamiltonian (\ref{Hphi4MN}) in the limit $N\rightarrow 0$, 
i.e. ${\cal H}_{M0}$. One can then apply
conventional computational schemes, such
as the $\epsilon$ expansion, the fixed-dimension $d=3$ expansion, 
the scaling-field method, etc.
In the RG approach, 
if the FP corresponding to the pure 
model is unstable and the RG flow moves
towards a new random FP, then the random system has a different
critical behavior.

\begin{figure*}[tb]
\hspace{-1.0cm}
%\hspace{-2cm}
\vspace{-0.5cm}
\centerline{\psfig{width=7truecm,angle=-90,file=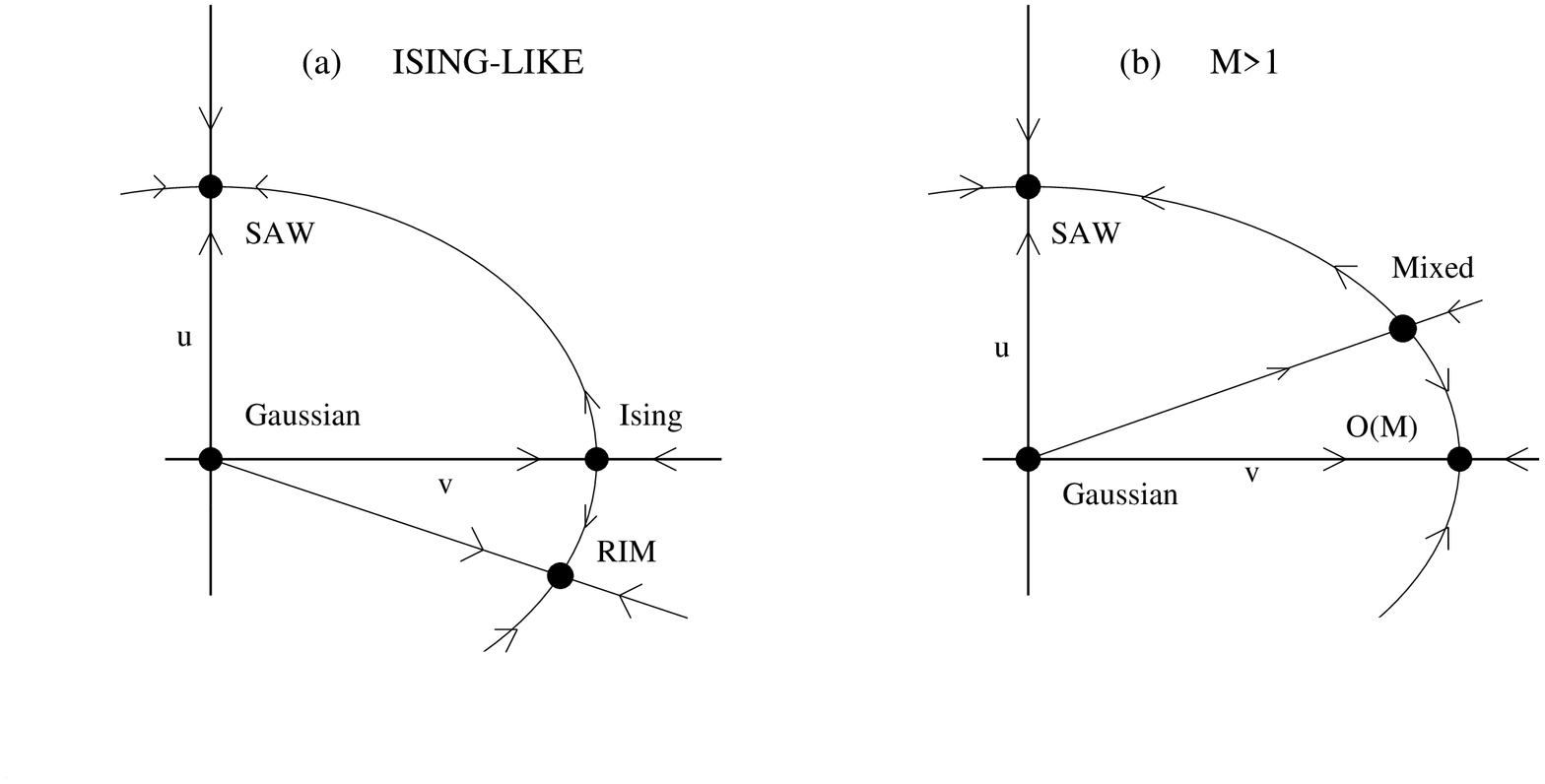}}
\vspace{-0.5cm}
\caption{
RG flow in the coupling plane $(u,v)$ for
(a) Ising ($M=1$) and (b) $M$-component ($M>1$) randomly dilute systems. 
}
\label{randomrgflow}
\end{figure*}

In the RG approach one assumes
that the replica symmetry is not broken.
In recent years, however, this picture has been
questioned~\cite{DHSS-95,DF-95,Dotsenko-99} on the ground that 
the RG approach may not take into account other  
local minimum configurations of the random Hamiltonian (\ref{Hphi4ran}), which
may cause the spontaneous breaking of the replica symmetry.
Arguments in favor of the stability of the critical behavior
with respect to replica-symmetry breaking are
reported in Ref.~\cite{PPF-01}.
They consider an appropriate effective Hamiltonian
allowing for possible replica-symmetry breaking terms and,
using
two-loop calculations, argue 
that the replica-symmetric FP is stable.

For generic values of $M$ and $N$, the Hamiltonian ${\cal H}_{MN}$ describes
$M$ coupled $N$-vector models 
and it is usually called $MN$ model~\cite{Aharony-76}.
Figure~\ref{randomrgflow} sketches the expected
flow diagram for Ising ($M=1$) and multicomponent ($M>1$) systems
in the limit $N\rightarrow 0$. There are four FP's: 
the trivial Gaussian one, an O($M$)-symmetric FP,
a self-avoiding walk (SAW) FP and a mixed FP.
We recall that 
the region relevant for
quenched disordered systems corresponds to negative values of 
the coupling $u$~\cite{GL-76,AIM-76}. 
The SAW FP is stable and
corresponds to the $(M N)$-vector theory for $N\rightarrow 0$, 
but it is not of interest for the critical 
behavior of randomly dilute spin models, since it is located in the region $u>0$.
The stability of the other
FP's depends on the value of $M$.
Nonperturbative arguments \cite{Sak-74,Aharony-76}
show that the stability of the O($M$) FP  
is related to the specific-heat critical exponent of the O($M$)-symmetric
theory. Indeed, ${\cal H}_{MN}$ at the O($M$)-symmetric FP can be interpreted 
as the Hamiltonian of $N$ $M$-vector systems coupled by the 
O($MN$)-symmetric term. Since this interaction is the sum of the products 
of the energy operators of the different $M$-vector models,
the crossover exponent associated with the O($MN$)-symmetric quartic
interaction is given by
the specific-heat critical exponent $\alpha_M$ of the $M$-vector model, 
independently of $N$. 
This implies that for $M=1$ (Ising-like systems) the pure 
Ising FP is unstable since $\phi = \alpha_I > 0$, 
while for $M>1$ the O($M$) FP is stable given that $\alpha_M<0$,
in agreement with the Harris criterion.
For $M>1$  the mixed FP is in the region of positive 
$u$ and is unstable \cite{Aharony-76}.
Therefore, the RG flow 
of the $M$-component model with $M>1$ is driven towards the pure O($M$) 
FP. Quenched
disorder yields corrections to scaling proportional to
the spin dilution and to $|t|^{\Delta_r}$ with 
$\Delta_r = - \alpha_M$.
Note that, for the physically interesting two- and three-component models,
the absolute value of $\alpha_M$ is very small: 
$\alpha_2\approx -0.014$ and $\alpha_3\approx  - 0.13$.
Thus, disorder gives rise to very slowly-decaying scaling corrections.
For Ising-like systems, the pure Ising FP is instead unstable, 
and the flow for negative
values of the quartic coupling $u$ leads to the stable mixed or 
random FP which is located in the region of negative values of $u$.
The above picture emerges clearly in the framework of the $\epsilon$ expansion,
although the RIM FP is 
of order $\sqrt{\epsilon}$~\cite{Khmelnitskii-75} rather than $\epsilon$.

The Hamiltonian ${\cal H}_{MN}$ has been
the object of several FT studies, especially for $M=1$ and $N=0$,
the case that describes the RIM. 
In Table~\ref{FTrexponents} we report a summary
of the FT results obtained for the RIM universality class.
Several computations have been done in the framework
of the $\epsilon$ expansion and of the fixed-dimension $d=3$ 
expansion. Other results have been obtained by nonperturbative methods (CRG
and scaling field) 
\cite{TMVD-01,NR-82},

\begin{table*}
\caption{FT 
estimates of the critical exponents for the 
RIM universality class. Here ``$d=3$ exp" denotes the massive scheme 
in three dimensions, ``$d=3$ MS" the minimal subtraction scheme 
without $\epsilon$ expansion. All perturbative results have been obtained 
by means of Pad\'e-Borel or Chisholm-Borel resummations, except the 
results of Ref. \protect\cite{Mayer-89} indicated by ``$\epsilon$W" 
obtained using the $\epsilon$-algorithm of Wynn and of those of 
Ref. \cite{PV-00-r}.
}
\label{FTrexponents}
\footnotesize
\begin{center}
\begin{tabular}{rlllllll}
\hline
\multicolumn{1}{c}{Ref.}& 
\multicolumn{1}{c}{Method}& 
\multicolumn{1}{c}{$\gamma$}& 
\multicolumn{1}{c}{$\nu$}& 
\multicolumn{1}{c}{$\eta$}& 
\multicolumn{1}{c}{$\alpha$}& 
\multicolumn{1}{c}{$\beta$}&
\multicolumn{1}{c}{$\omega$} \\   
\hline  
\cite{PV-00-r}   $_{2000}$ & $d=3$ exp $O(g^6)$ &
   1.330(17) & 0.678(10) & 0.030(3) & $-0.034(30)$ & 0.349(5) & 0.25(10) \\
\cite{PS-00}      $_{2000}$ & $d=3$ exp $O(g^5)$ &
        1.325(3) & 0.671(5)  & 0.025(10) & $-0.013(15)$ & 0.344(4) & 0.32(6) \\
\cite{FHY-00,FHY-99}  $_{2000}$   & $d=3$ MS $O(g^4)$ &
    1.318   & 0.675 & 0.049  &  $-0.025$ & 0.354 & 0.39(4)      \\
\cite{Varnashev-00}  $_{2000}$ & $d=3$ exp $O(g^4)$ & 
        1.336(2) & 0.681(12) & 0.040(11) & $-0.043(36)$ & 0.354(7) & 0.31 \\
%  & &   1.323(5) & 0.672(4)  & 0.034(10) & 0.33 \\
\cite{Mayer-89}  $_{1989}$   & $d=3$ exp $O(g^4)$ &    1.321   &
0.671  & 0.031 & $-0.013$ & 0.346 & \\
\cite{Mayer-89}  $_{1989}$   & $d=3$ exp $O(g^4)$ $\epsilon$W & 1.318   & 0.668
                   &  0.027 & $-0.004$ & 0.343 & \\
\cite{MSS-89}    $_{1989}$  & $d=3$ exp $O(g^4)$ &        1.326 &
                   0.670 & 0.034 & $-0.010$ &0.346 &  \\
\cite{TMVD-01}     $_{2002}$ & CRG   &    1.306   & 0.67  & 0.05 & $-0.01$&0.352 & \\
\cite{NR-82}     $_{1982}$ & scaling field   &    & 0.697  & & $-0.09$
& & 0.42 \\
\hline
\end{tabular}
\end{center}
\end{table*}

The analysis of the FT expansions is made difficult by 
the more complicated analytic structure of the field theory  
corresponding to quenched disordered models.
This issue has been investigated considering the free energy
in zero dimensions.
The large-order behavior of its double expansion 
in the quartic couplings $u$ and $v$,
$F(u,v) = \sum_{ij} c_{ij} u^i v^j$, 
shows that the expansion in powers of $v$, 
keeping the ratio $u/v$ fixed, is not Borel summable~\cite{BMMRY-87}. 
As shown in Ref.~\cite{McKane-94}, this 
a consequence of the fact that, because of the quenched average,
there are additional singularities corresponding to the 
zeroes of the partition function $Z(\psi,g_0)$ defined using the  
Hamiltonian (\ref{Hphi4ran}). The problem is  reconsidered 
in Ref.~\cite{AMR-99}. In the same context of the zero-dimensional model,
it is shown that a more elaborate resummation can provide the 
correct determination of the
free energy from its perturbative expansion. The procedure is still 
based on a Borel summation, which is performed in two steps:
first, one writes
$F(u,v) = \sum_i e_i(v)  u^i$ where $e_i(v) = \sum_j c_{ij} v^j$
and resums the coefficient functions $e_i(v)$
of the series in $u$;
then, one resums the resulting series in the coupling $u$. 
There is no proof that this procedure works also in higher dimensions, 
since the method relies on the fact that the zeroes of the partition function 
stay away from the real values of $v$. This is far from obvious 
in higher-dimensional systems.

The $MN$ model has been extensively 
studied in the framework of the $\epsilon$ expansion
\cite{HL-74,Lubensky-75,Khmelnitskii-75,AIM-76,GL-76,Aharony-76,GMM-77,JK-77,%
Shalaev-77,Newlove-83,PA-85,DG-85,Shpot-90,KS-95,SAS-97}. 
Several studies also considered the equation of state
\cite{GMM-77,Newlove-83,Shpot-90} and the two-point correlation function
\cite{GMM-77,PA-85}. In spite of these efforts,
studies based on the $\epsilon$ expansion
have not been able to go beyond a qualitative description of the physics of 
three-dimensional randomly dilute spin models. 
The $\sqrt{\epsilon}$ expansion \cite{Khmelnitskii-75} turns out not to be 
effective for a quantitative
study of the RIM (see, e.g., the analysis of the five-loop series done in 
Ref. \cite{SAS-97}).
The related minimal-subtraction renormalization scheme without 
$\epsilon$ expan\-sion \cite{SD-89} have been also considered.  
The three-loop \cite{JOS-95} and four-loop \cite{FHY-98,FHY-99,FHY-00}
results turn out to be  in reasonable agreement with the estimates obtained 
by other methods. At five loops, however, no random FP 
is found \cite{FHY-00} using this method. 
This negative result has been  
interpreted as a consequence of the 
non-Borel summability of the perturbative expansion. In this case,
the four-loop series might represent the ``optimal" truncation.

The most precise FT results have been obtained 
using the fixed-dimension expansion in $d=3$. 
Several quantities have been  computed: 
the critical exponents \cite{Jug-83,MS-84,Shpot-89,MSS-89,%
Mayer-89,HS-92,HY-98,FHY-99,Varnashev-00,PS-00,PV-00-r}, the equation of state
\cite{BS-92}, ratios of $n$-point susceptibilities in
the HT phase \cite{PSS-02}, 
and the hyperuniversal ratio $R^+_\xi$ \cite{BS-92,Mayer-98}.
The RG functions of the $MN$ model were  calculated to six-loops
in Ref.~\cite{PV-00-r}. In the case relevant for the RIM universality
class, i.e. $M=1$ and $N= 0$,
several methods of resummation have been  applied. 
In Ref. \cite{PV-00-r} 
the method proposed in Ref.~\cite{AMR-99} was applied to the
three-dimensional
series. The analysis
of the $\beta$-functions for the determination of the FP 
does not lead to a particularly  accurate estimate of the random FP. 
Nonetheless, 
the RG functions associated with the exponents
are largely insensitive to the 
exact position of the FP, so that
accurate estimates of the critical exponents can still be obtained, 
see Table \ref{FTrexponents}.
Earlier analysis of the RIM series up to 
five-loops were done using Pad\'e-Borel-Leroy approximants~\cite{PS-00},
thus assuming Borel summability.  
In spite of the fact that the series are not 
Borel summable, the results for the critical exponents 
turn out to be  relatively stable, depending very little  
on the order of the series and the details of the analysis. They
are in substantial agreement with the six-loop results
of Ref.~\cite{PV-00-r}.
This fact may be explained by the observation of Ref.~\cite{BMMRY-87}
that the Borel resummation applied in the standard way (i.e. at fixed $v/u$)
may give a reasonably accurate
result if one truncates the expansion 
at an appropriate point, i.e. for not too long series.

In conclusion, 
the agreement among FT results, experiments, and MC estimates is overall
good. The FT method appears to have a good predictive power,
in spite of the complicated analytic structure of the theory.

For $M\ge 2$ and $N=0$ the analysis of the corresponding
six-loop series shows
that no FP exists in the region $u<0$ and that the 
$O(M)$-symmetric FP is stable \cite{PV-00-r}, in agreement
with the Harris criterion.

Finally, we mention that the combined effect of cubic anisotropy 
and quenched uncorrelated impurities on multicomponent systems 
has been studied in Ref.~\cite{CPV-02-3}.

\subsection{Frustrated spin models with noncollinear order}
\label{lsec-frustrated}

\begin{figure}[tb]
\hspace{0cm}
\centerline{\psfig{width=7.5truecm,angle=0,file=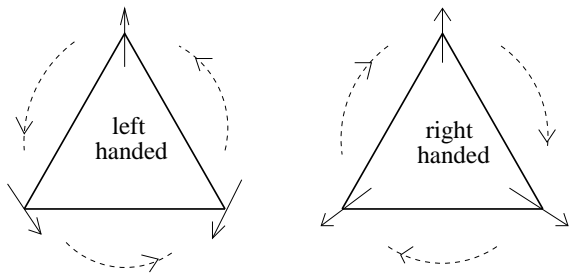}}
\vspace{0cm}
\caption{
The ground-state configuration of three $XY$ spins on a triangle
coupled antiferromagnetically. 
}
\label{chiralityfig}
\end{figure}

\subsubsection{Physical relevance} \label{sec.11.5.1}

The critical behavior of frustrated spin systems with noncollinear or 
canted order has been the object of intensive
theoretical and experimental studies
(see, e.g., Refs. \cite{Kawamura-98,CP-97,Kawamura-01} for 
recent reviews on this subject).  
Noncollinear order is due to frustration that may arise
either because of the special geometry of the lattice, or from the competition 
of different kinds of interactions.
Typical systems of the first type are 
three-dimensional
stacked triangular antiferromagnets (STA's), 
where magnetic ions are located at each site of 
a three-dimensional stacked triangular lattice.
Examples are some ${\rm ABX}_3$-type compounds,
where A denotes elements such as Cs and Rb, B is a magnetic
ion such as Mn, Ni, and V, and X stands for halogens as Cl, Br, and
I. Analogous behavior is observed in some BX$_2$ materials 
like VCl$_2$ and VBr$_2$. See Ref.~\cite{CP-97}
for a detailed description of the magnetic behavior of these materials.
Frustration due to the competition of interactions may 
be realized in helimagnets, in which 
a magnetic spiral is formed along a certain direction of the lattice.
The rare-earth metals Ho, Dy, and Tb provide physical examples of such systems.

All these systems are strongly anisotropic and the critical behavior is 
complex due to the competition between the $c$-axis coupling and the uniaxial
anisotropy. Effective Hamiltonians describing the phase diagram 
of these compounds in a magnetic field are discussed in 
Refs. \cite{PHC-88,PCH-89,PC-90,KCP-90,Kawamura-93}. 
Mean-field and RG analyses predict several transition lines and the 
appearance of tetracritical and bicritical points, in good agreement 
with experiments.

The main point under discussion is the nature of the critical behavior. In 
particular, the question is whether along some transition lines or at 
the multicritical points one should observe a new {\em chiral} 
universality class, as originally conjectured by Kawamura
\cite{Kawamura-85,Kawamura-86}.  On this question, there is still much 
debate, FT methods, MC simulations, and experiments providing 
contradictory results in many cases.

\subsubsection{Models} \label{sec.11.5.2}

According to RG theory, 
the existence of a new (chiral) universality class 
may be investigated in simplified models 
retaining the basic features of the physical systems. One considers 
a three-dimensional stacked triangular lattice, 
which is obtained by stacking two-dimensional triangular layers, 
and the Hamiltonian
\begin{equation}
{\cal H}_{\rm STA} = 
     - J\,\sum_{\langle vw\rangle_{xy}}  \vec{s}(v) \cdot \vec{s}(w) -
       J'\,\sum_{\langle vw\rangle_z}  \vec{s}(v) \cdot \vec{s}(w),
\label{latticeSTA}
\end{equation}
where $J<0$, the first sum is over nearest-neighbor pairs within the
triangular layers ($xy$ planes), and the second one is over 
orthogonal interlayer nearest neighbors.
The sign of $J'$ is not relevant, since there is no
frustration along the direction orthogonal to the triangular layers.
The variables $\vec{s}$ are $N$-dimensional unit spins defined on the sites 
of the lattice; of course, $N=2$ and $N=3$ are the cases of physical relevance.

Triangular antiferromagnets are frustrated. Nonetheless, for $N\ge 2$
they admit an ordered ground state. For instance, for $N=2$ the ground state
shows the 120$^o$ structure of Fig.~\ref{chiralityfig}. 
There are two chirally degenerate configurations, 
according to whether the noncollinear spin configuration is right- or
left-handed.
The chiral degrees of freedom are related to 
the local quantity~\cite{Kawamura-88}
\begin{equation}
C_{ij} \propto \sum_{<vw>\in\vartriangle} 
\left[s_i(v)s_j(w) - s_j(v) s_i(w)\right],
\label{chirality}
\end{equation}
where the summation runs over the three bonds of the given triangle.

Helimagnets can be modeled similarly. 
A simple model Hamiltonian is (see, e.g., Ref.~\cite{Kawamura-98})
\begin{equation}
{\cal H}_h = - J_1\,\sum_{\langle ij\rangle_{nn}}  \vec{s}_i \cdot \vec{s}_j -
J_2\,\sum_{\langle ij\rangle_{nnn,z}}  \vec{s}_i \cdot \vec{s}_j,
\label{latticeH2}
\end{equation}
where the first sum represents nearest-neighbor ferromagnetic interactions,
so that $J_1>0$, while the second one describes antiferromagnetic
next-nearest-neighbor interactions, i.e. $J_2<0$, 
along only one crystallografic axis  $z$. In the LT phase,
depending on the values of $J_1$ and $J_2$,
competition of ferromagnetic and 
antiferromagnetic interactions may lead 
to incommensurate helical structures along the $z$-axis.
The chiral degeneracy discussed in STA's is also
present in helimagnets.

On the basis of the structure of the ground state, one expects a breakdown of 
the symmetry from O$(N)$ in the HT phase to O$(N-2)$ in the LT phase. 
Therefore, the LGW Hamiltonian describing these systems must be
characterized by
a matrix order parameter. The determination of the effective Hamiltonian
goes through fairly standard steps. One starts from the spin model
(\ref{latticeSTA}), performs a Hubbard-Stratonovich
transformation that allows to replace the fixed-length spins
with variables of unconstrained length, expands around the
instability points, and  drops terms beyond quartic 
order \cite{Kawamura-88}. One finally obtains the $O(N)\times O(M)$
symmetric  Hamiltonian \cite{Kawamura-88,Kawamura-98}
\bea
{\cal H}  = \int d^d x
 \Bigl\{ {1\over2}
      \sum_{a} \Bigl[ (\partial_\mu \phi_{a})^2 + r \phi_{a}^2 \Bigr]
+ {1\over 4!}u_0 \Bigl( \sum_a \phi_a^2\Bigr)^2
+ {1\over 4!}  v_0
\sum_{a,b} \Bigl[ ( \phi_a \cdot \phi_b)^2 - \phi_a^2\phi_b^2\Bigr]
             \Bigr\},
\label{LGWH}
\eea
where $\phi_a$ ($1\leq a\leq M$) are $M$ sets of $N$-component
vectors. The case $M=2$ with $v_0>0$
describes frustrated systems with
noncollinear ordering such as STA's.
Negative values of $v_0$ correspond to simple ferromagnetic or
antiferromagnetic
ordering, and to magnets with sinusoidal spin structures~\cite{Kawamura-88}.

We mention a few other applications of the Hamiltonian (\ref{LGWH}).
The superfluid phase of liquid $^3$He can be described by a field
theory for complex 3$\times$3 matrices representing fermion pairs.
Due to the magnetic dipole-dipole interaction that couples orbital
momentum and spin, the superfluid order parameter is expected
to have O(3)$\times$U(1) symmetry, which is the symmetry
of the Hamiltonian (\ref{LGWH}) for $M=2$ and $N=3$.
According to Refs.~\cite{JLM-76,BLM-77},
in the absence of an external magnetic field and neglecting the strain
free-energy term, the transition from normal to planar superfluid is
described by the effective LGW Hamiltonian (\ref{LGWH})
with $v_0<0$.
The same LGW Hamiltonian, but with $v_0>0$,
should describe the transition from normal to
superfluid A$_1$ phase in the presence of a magnetic 
field~\cite{JLM-76,BLM-77}.
The model (\ref{LGWH}) can be also applied to
the superconducting phase transition
of heavy-fermion superconductors such as UPt$_3$~\cite{Joynt-93},
and to the quantum phase transition of certain Josephson-junction arrays in
a magnetic field~\cite{GK-90} (see also Ref.~\cite{AS-94}
for a discussion of these systems).
One may also consider more general O($N$)$\times$O($M$) models
with $M>2$~\cite{DR-89,Kawamura-90,RGB-92,MP-93,ADDJ-93,Kawamura-98,Loison-00,TDM-00-2}.
In particular, the principal chiral model with $N=M=3$ may be relevant for
magnets  with noncollinear noncoplanar spin  ordering.

In the following we only consider the $M=2$ case
that is relevant for frustrated models with noncollinear order.
In this case the LGW Hamiltonian (\ref{LGWH})
can also be written in terms of an $N$-component complex field $\psi$
as~\cite{AS-94}
\bea
{\cal H} = \int d^d x
\left[  {1\over 2} \left( \partial_\mu \psi^*\partial_\mu \psi
 + r \psi^* \psi  \right)
+{1\over 4!} y_0 \left( \psi^*\cdot \psi\right)^2 + {1\over 4!}  w_0
|\psi\cdot \psi |^2 \right].
\label{LGWHb}
\eea
The couplings of the models (\ref{LGWH}) and (\ref{LGWHb}) are related
by $y_0=u_0-v_0/2$ and $w_0=v_0/2$.
Note also that, for $N=2$, the transformation
\begin{eqnarray}
&&
\phi_{11} = {\phi'_{11}-\phi'_{22}\over \sqrt{2}},\quad
\phi_{12}  = {\phi'_{12}-\phi'_{21}\over \sqrt{2}},\quad 
\phi_{21} = {\phi'_{12}+\phi'_{21}\over \sqrt{2}},\quad
\phi_{22} = {\phi'_{11}+\phi'_{22}\over \sqrt{2}}, \nonumber \\
&& \qquad u'_0 = u_0 + v_0/2, \qquad v'_0 = - v_0
\label{symmetry}
\end{eqnarray}
maps the chiral Hamiltonian \reff{LGWH} into the Hamiltonian (\ref{Hphi4MN}) 
of the $MN$ model
with $M=2$, $N=2$.

\subsubsection{Theoretical results}
\label{lsec-frustrated-b}

\begin{table*}[t]
\caption{
Theoretical estimates of the critical exponents for 
two- and  three-component chiral systems. Results labeled MC have been
obtained by means of Monte Carlo simulations, those labeled FT from the 
analysis of six-loop perturbative field-theoretic expansions in $d=3$.
}
\label{chiralMC}
\footnotesize
\begin{center}
\begin{tabular}{ccclllll}
\hline
\multicolumn{1}{c}{}& 
\multicolumn{1}{c}{Ref.}& 
\multicolumn{1}{c}{Method}& 
\multicolumn{1}{c}{$\gamma$}& 
\multicolumn{1}{c}{$\nu$}& 
\multicolumn{1}{c}{$\beta$}&
\multicolumn{1}{c}{$\alpha$} &
\multicolumn{1}{c}{$\eta$}\\   
\hline  
$N=2$  
& \cite{PRV-00} $_{2000}$ & FT & 1.10(4) &  0.57(3) & 0.31(2) &
0.29(9) & 0.09(1) \\
& \cite{ZIT-01} $_{2001}$ & MC    & 1.074(13) & 0.514(7) & & &\\
& \cite{BLD-96} $_{1996}$ & MC & 1.15(5) & 0.48(2) & 0.25(2) & 0.46(10) &\\
& \cite{PM-94}  $_{1994}$ & MC & 1.03(4) & 0.50(1) & 0.24(2) & 0.46(10) &\\
& \cite{Kawamura-92} $_{1992}$ & MC & 1.13(5) & 0.54(2) & 0.253(10) & 0.34(6)& \\
\hline
$N=3$  
& \cite{PRV-00} $_{2000}$ & FT & 1.06(5) & 0.55(3) & 0.30(2) & 0.35(9)
& 0.10(1) \\
& \cite{BBLJ-94} $_{1994}$ & MC & 1.176(20) & 0.585(9) & 0.289(10) & & \\
& \cite{MPC-94} $_{1994}$ & MC & 1.185(3) & 0.586(8) & 0.285(11) & & \\
& \cite{LD-94} $_{1994}$ & MC & 1.25(3) & 0.59(1) & 0.30(2) & & \\
& \cite{Kawamura-92} $_{1992}$ & MC & 1.17(7) & 0.59(2) & 0.30(2) & 0.24(8)& \\
\hline
\end{tabular}
\end{center}
\end{table*}

The critical behavior of
frustrated systems with noncollinear order is quite controversial, since
different theoretical methods, such as MC, CRG, 
and perturbative FT approaches provide 
contradictory results.
Since all these approaches rely on  
different approximations and assumptions, their comparison and consistency 
is essential before considering the issue substantially understood.

Frustrated models with noncollinear order
have been much studied using FT RG methods
\cite{BM-76,GP-76,BLM-77,BW-82,YD-85,Kawamura-86-2,Kawamura-88,ADDJ-93,%
Zumbach-94,AS-94,ASV-95,JD-96,LSDASD-00,TDM-00,PRV-00,CP-01,PRV-01,%
TDM-01,PRV-02}. 
Two different expansion schemes have been used: the $\epsilon$ expansion
and the fixed-dimension $d=3$ expansion. 

A detailed discussion of the  $\epsilon$-expansion 
results is presented in Ref.~\cite{Kawamura-98}.
Near four dimensions, 
the $\epsilon$ expansion predicts four regimes.
For $N>N_+$, there are four FP's:
the Gaussian FP,
the $O(2N)$ FP, the XY FP and a mixed FP. The latter is the stable one and 
can be identified with the chiral FP.  
For $N_-< N < N_+ $, only the Gaussian and the Heisenberg
O($2N$)-symmetric FP's are present, and none of them is stable.
For $N_H < N < N_-$, there are again four FP's,
but none of them belongs to the physically 
relevant region $v_0 > 0$. 
For $N < N_H$, there are four FP's,
and the Heisenberg O$(2N)$-symmetric FP is the stable one. 
Three-loop calculations \cite{ASV-95}  give  
\begin{eqnarray}
&&N_+ = 21.80 - 23.43 \epsilon + 7.09 \epsilon^2 + O(\epsilon^3),
\nonumber \\
&&N_- = 2.20 - 0.57 \epsilon + 0.99 \epsilon^2 + O(\epsilon^3),
\nonumber \\
&&N_H = 2 - \epsilon + 1.29 \epsilon^2 + O(\epsilon^3).
\end{eqnarray}
Therefore, according to a ``smooth extrapolation'' of this scenario
to three dimensions, 
the existence of chiral universality classes
for $N=2,3$ requires $N_+ < 3$ in three dimensions.
The analysis of the $\epsilon$ expansion of $N_+$
\cite{AS-94,ASV-95,PRV-00} shows that $N_+\approx 5$ in three dimensions.
This estimate of $N_+$ is confirmed by CRG calculations,
giving $N_{+} \approx 4$ \cite{TDM-00} and
$N_+ \approx 5$ \cite{TDM-01}. 
Therefore, there is a rather robust indication that
$N_+ > 3$ in three dimensions, so that the stable chiral FP found near $d=4$ is not
relevant for the three-dimensional physics of these systems.

\begin{figure}
\vspace*{0truecm} \hspace*{-0.2cm}
\centerline{\psfig{width=6truecm,angle=0,file=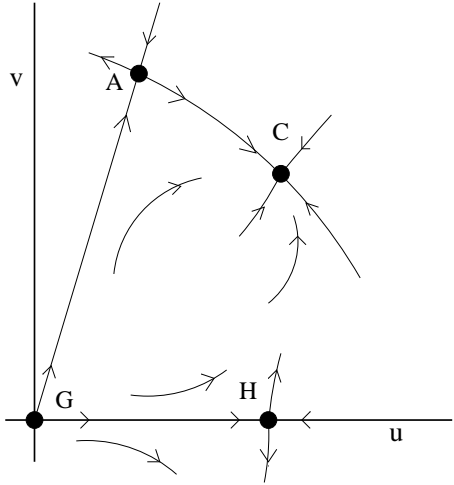}}
\caption{RG flow in the $(u,v)$ plane for $N=2,3$.
}
\label{fig-diagramma-di-flusso}
\end{figure}

On the other hand, one cannot exclude the existence of FP's
that are not smoothly connected with the FP's 
described by the $\epsilon$ expansion.
The investigation of such a possibility requires
a strictly three-dimensional scheme.
For both $N=2$ and $N=3$ cases,
high-order calculations in the framework of the
fixed-dimension $d=3$ expansion support
the existence of a stable FP corresponding to the conjectured
chiral universality class, and the RG flow diagram
drawn in Fig. \ref{fig-diagramma-di-flusso}.
Indeed, the six-loop analysis of Ref.~\cite{PRV-00} provides a rather robust evidence
of their existence,  contradicting earlier
FT results based on three-loop series \cite{AS-94,LSDASD-00}.
In Ref.~\cite{CPS-02}, on the basis of the six-loop
fixed-dimension series, it has been argued that the stable chiral FP
is actually a focus, essentially because the eigenvalues of 
its stability matrix turn out to be complex
(only the positivity of their real part is required for the
stability of the FP).
The exponents at the chiral FP are given in Table \ref{chiralMC}.
The major drawback of this computation is that 
the chiral FP lies in a region where the perturbative
expansions are not
Borel summable,  although it is still within the region in which 
one can take into account the leading large-order behavior
by a standard analysis based on a Borel transformation.
Nevertheless, the observed stability of the results
with the order of the series, from four to six loops,
appears quite robust.
We also mention that in the fixed-dimension approach,
no FP is found  for $5\ltapprox N \ltapprox 7$,
while for $N\gtapprox 7$, a stable chiral FP is again present.
These results may be interpreted as follows.
The stable FP found for $N\gtapprox 7$ is smoothly
related to the large-$N$ and small-$\epsilon$ chiral FP. Such a FP 
disappears for $5\ltapprox N \ltapprox 7$, so that we can identify
$5\ltapprox N_+ \ltapprox 7$, in agreement with the above-reported estimates.
According to the $\epsilon$-expansion scenario,
for $N< N_+$ no stable FP's should be found.
However, the existence of a stable chiral FP for $N=2,3$
indicates that the situation is more 
complex in three dimensions: another value 
$3 < N_{d3} < N_+$ exists such that, 
for $N<N_{d3}$, the system shows again a chiral critical behavior 
with a FP unrelated to the small-$\epsilon$ and large-$N$
chiral FP. 

The new chiral FP's found for $N=2,3$
should describe the apparently continuous
transitions observed in $XY$ and Heisenberg chiral systems.
Note that the presence of a stable FP
does not exclude the possibility that some systems undergo
a first-order transition. Symmetry arguments are not sufficient
to establish the order of the transition.
Indeed, within the RG
approach, first-order transitions are still possible for systems that 
are outside the attraction domain of the chiral FP.
In this case, the RG flow would run away to a 
first-order transition. This means that, even if 
some systems show a universal continuous transition
related to the presence of a stable FP, 
other systems may exhibit a first-order transition.
The different behavior of these
systems is not due to the symmetry, but arises from the particular values
of the microscopic parameters, which may be or not be in the
attraction domain of the stable FP.  

Studies based on approximate solutions of 
continuous RG equations (CRG)~\cite{Zumbach-94,TDM-00,TDM-01,TDM-01-2} 
favor a weak first-order transition, since no evidence of 
stable FP's has been found.
In this scenario,
the transition should be weak enough to effectively appear
as continuous in experimental works. The weakness of the
transition is somehow supported by the observation of a  
range of parameters in which the RG flow appears very slow,
with effective critical exponents 
close to those found in experiments, for instance
$\nu = 0.53$, $\gamma = 1.03$, $\beta = 0.28$ for $N = 3$. 
Note however that,
as already discussed in Sec.~\ref{CRGth},
the practical implementation of CRG
methods requires approximations and/or truncations of the effective action.
Ref.~\cite{Zumbach-94} employed  a local potential approximation (LPA);
Ref.~\cite{TDM-00} used a more refined approximation that 
allows for an anomalous scaling of the field and
therefore for a nontrivial value of $\eta$ (ILPA);
finally, Ref.~\cite{TDM-01-2} mentioned some attempts 
for a partial first-order derivative expansion approximation.
These approximations are essentially limited to the lowest
orders of the derivative expansion, so that their results 
may not be conclusive.

Also MC simulations (see, e.g., Refs.~\cite{Kawamura-85,Kawamura-86,Diep-89,%
Kawamura-92,KZ-93,MPC-94,PM-94,LD-94,BBLJ-94,DD-95,BLD-96,Loison-00,LS-98,%
LS-00,Itakura-01,ZIT-01,ZIT-02}) 
have not been conclusive in setting the question. 
Most simulations of the 
STA Hamiltonian observe second-order phase transitions.
Some results are reported in Table \ref{chiralMC}.
We observe small differences among the results of the MC simulations
and the FT approach. 
Moreover, some MC results are not consistent with general exponent 
inequalities. Indeed, one must have $\eta\ge 0$, which follows 
from the unitarity of the
corresponding quantum field theory~\cite{PP-79,Zinn-Justin-book}
(one may show that the model (\ref{latticeH}) is
reflection positive and thus the corresponding field theory
is unitary).  Using $\gamma = (2-\eta)\nu$ and 
$\beta = {1\over2}\nu(1+\eta)$, we obtain the inequalities
$\gamma \le 2\nu$ and $\beta \ge {1\over2}\nu$. As it can be seen
from the results of Table \ref{chiralMC}, the first inequality is 
not satisfied by the results of Refs. \cite{BLD-96,LD-94,ZIT-01}, while the 
second one is barely satisfied by the results of 
Ref. \cite{Kawamura-92}. 
This fact has been interpreted as an additional indication
in favor of the first-order transition hypothesis \cite{LS-00,LSDASD-00}.
But, it may also be explained by sizeable scaling corrections,
that are neglected in all these numerical studies.
Ref.~\cite{Itakura-01} reports simulations of various
systems, and in particular STA spin systems.
The results favor a first-order transition,
although the evidence that 
the asymptotic critical behavior has been probed is not clear.
We should also say that distinguishing a weak first-order transition
from a continuous one is in general a hard task in numerical
simulations. 
First-order transitions have been clearly observed in
MC investigations~\cite{LS-98,LS-00} 
of modified lattice spin systems that, according to 
general universality ideas,
should belong to the same universality class of the 
Hamiltonian (\ref{latticeSTA}).
But, as we already said,
this does not necessarily contradict the existence of a stable FP.
Indeed, mean-field arguments suggest a first-order transition
for such modified systems \cite{Kawamura-98}.

Also higher values of $N$ have been studied,
although they are not of physical interest. For $N=6$, 
MC simulations \cite{LSDASD-00} and CRG calculations \cite{TDM-00} 
provide evidence for a second-order phase transition,
showing also a good agreement in the estimates of the 
critical exponents. 

We finally mention that
in the many-component limit $N\rightarrow\infty$ at fixed $M$, the O($M$)$\times$O($N$) theory
can be expanded in powers of $1/N$ \cite{Kawamura-88,Kawamura-98}. 
In the $1/N$-expansion the transition in the noncollinear
case, i.e. for $v>0$, is continuous, and the exponents have been
computed to $O(1/N^2)$ \cite{PRV-01,Kawamura-88}. 
For $d=3$ and $M=2$ the critical exponents are given by
\begin{eqnarray}
&& \hskip -0.8truecm
\nu = 1 - {16\over \pi^2} {1\over N} - \left( {56\over \pi^2}-{640\over 3 \pi^4} \right) {1\over N^2}
+O\left( {1\over N^3}\right), \nonumber \\ 
&& \hskip -0.8truecm
\eta = {4\over \pi^2} {1\over N} - {64\over 3 \pi^4} {1\over N^2}
+O\left( {1\over N^3}\right).
\label{nuln}
\end{eqnarray}

\subsubsection{Experimental results} \label{sec.11.5.4}

\begin{table*}[t]
\caption{
Experimental estimates 
of the critical exponents for two- and  three-component systems. 
We report results for stacked triangular antiferromagnets (STA) 
and helimagnets (HM).
}
\label{chiralexper}
\footnotesize
\begin{center}
\begin{tabular}{clllll}
\hline
\multicolumn{1}{c}{}& 
\multicolumn{1}{c}{Material}&
\multicolumn{1}{c}{$\gamma$}& 
\multicolumn{1}{c}{$\nu$}& 
\multicolumn{1}{c}{$\beta$}&
\multicolumn{1}{c}{$\alpha$} \\   
\hline  
$N=2$  
&CsMnBr$_3$ & 1.10(5) \cite{ref25} & 0.57(3) \cite{ref25} & 
              0.25(1) \cite{ref25} & 0.39(9) \cite{ref26}\\
STA   
 &          & 1.01(8) \cite{ref24b} & 0.54(3) \cite{ref24b} & 
              0.21(2) \cite{ref24b} & 0.36(4) \cite{ref26} \\
    &   & & & 0.24(2) \cite{ref39}  & 0.40(5) \cite{ref27} \\
    &   & & & 0.22(2) \cite{ref24a} & \\
& RbMnBr$_3$ & & & 0.28(2) \cite{KAAKII-95} & \\
& CsNiCl$_3$ & & & 0.243(5) \cite{ref93} & 0.37(8)\cite{ref91}   \\
&            & & & &  0.342(5)\cite{ref92}   \\
& CsMnI$_3$ & & & & 0.34(6)\cite{ref91}  \\
\hline
$N=2$ 
& Ho & 1.24(15) \cite{LMP-70} & 0.57(4) \cite{GHC-88} & 
             0.37(10) \cite{THGHGS-93,THHGGS-94} & 0.34(1) \cite{JCS-85} \\
HM 
&    & 1.14(10) \cite{GHC-88} & 0.54(4) \cite{THHGGS-94} &
             0.327 \cite{TASH-94} & 0.27(2) \cite{JCS-85} \\
& & & &      0.41(4) \cite{HHTGKH-94} & 0.10(2) \cite{ref26} \\
& & & &      0.39(4) \cite{HHTGKH-94} & 0.22(2) \cite{ref26} \\ 
& & & &      0.39(2) \cite{DPVB-95} & \\
& & & &      0.38(1) \cite{PSBKGRV-01} & \\
& Dy & 1.05(7) \cite{GHC-88} & 0.57(5) \cite{GHC-88} &
             0.39(1) \cite{BDP-88} & 0.18(2) \cite{LS-74} \\
& & & &      0.38(2) \cite{DPVB-95} & 0.16(1) \cite{JCS-85} \\
\hline
$N=3$  
& VCl$_2$   
& 1.05(3) \cite{ref29} & 0.62(5) \cite{ref29} & 0.20(2) \cite{ref29} &   \\
STA 
& VBr$_2$   & & & & 0.30(5) \cite{ref30}   \\
& RbNiCl$_3$& & & 0.28(1) \cite{ref95} &  \\
& CsNiCl$_3$& & & 0.28(3) \cite{ref93} & 0.25(8) \cite{ref91}  \\
  &        & & & & 0.23(4) \cite{ref92}  \\
& CsMnBr$_3$ & & & & 0.28(6) \cite{ref138}  \\
& & & & & 0.44 \cite{ref92} \\
  & CsMn(Br$_{0.19}$I$_{0.81}$)$_3$ & & & & 0.23(7) \cite{BWLOT-01}  \\
\hline
\end{tabular}
\end{center}
\end{table*}

For a critical discussion of the experimental results we refer to
Refs.~\cite{CP-97,DPVB-95,Kawamura-98}.
As already mentioned,
on the basis of symmetry,
one expects two classes of systems to have a similar behavior: 
STA's and helimagnets. Apparently, all these systems show 
continuous phase transitions with the exception of 
CsCuCl$_3$.\footnote{
For this material the transition is of first order
\cite{WWWLS-96}. Note however that CsCuCl$_3$ is a peculiar 
material (see, e.g., Refs.~\cite{CP-97,Kawamura-98}), since the triangular 
crystal structure is distorted, probably due to an 
additional Dzyaloshinsky-Moriya interaction $\sum \vec{D}_{ij} \cdot 
(\vec{s}_i \times \vec{s}_j)$, where $\vec{D}_{ij}$ is a vector pointing 
slightly off the $z$-axis. This interaction breaks the chiral symmetry
and thus the chiral universality class is expected to describe only
pretransitional behavior as observed experimentally 
\cite{CP-97,Kawamura-98}. 
Some experiments on Ho also found some evidence of 
a first-order transition. For a critical discussion of these studies, 
see Ref. \cite{DPVB-95}.}  
Experimental results are reported in Table \ref{chiralexper}.
It is not a complete list, but it gives an overview of the
experimental state of the art.
Additional results are reported in Refs.~\cite{CP-97,DPVB-95,Kawamura-98}.

Overall, experiments on STA's favor a continuous transition
belonging to a new chiral universality class.
The measured 
critical exponents are in satisfactory agreement with 
the theoretical results of Table~\ref{chiralMC}. However, as 
some MC results, the experimental 
estimates do not apparently satisfy the inequality $\beta\ge {1\over2} \nu$.
This fact could be explained by the presence of scaling corrections
that are not considered in most of the experimental analyses.
Of course, another possible 
explanation \cite{Zumbach-94,TDM-00,TDM-01,TDM-01-2}  is that no chiral
universality class exists, so that the transitions are weakly 
first-order ones and the 
measured exponents are simply effective.

The behavior of helimagnets is even more controversial.
The estimates of the exponent $\beta$ are substantially larger
than the experimental results for STA's and also than
the theoretical results of Table~\ref{chiralMC}. 
But, as discussed in Ref.~\cite{Kawamura-98},
special care should be taken in extracting information on the
asymptotic critical behavior of rare-earth metals,
essentially due to the more complicated physical mechanism that
gives rise to the effective model (\ref{latticeH2}) for
helimagnets. 
Apart from the explanation in terms of a weak first-order transition,
it is also possible that experiments have not really probed the  
asymptotic regime.
For a discussion, see, e.g., Ref.~\cite{Kawamura-98}.
Another possibility is that the current
modelling of these systems becomes invalid near the critical point. 
There could be other interactions that are quantitatively small, 
but still change the asymptotic critical behavior of these systems. 
In both cases, one would be observing a crossover between 
different regimes.

\subsubsection{Chiral crossover exponents}

In the standard $O(N)$ model there is only one crossover exponent
at quadratic order, which is associated with the spin-2 operator 
defined in Sec. \reff{sec-1.5.8}. In the $O(M)\times O(N)$ model, there are 
four different quadratic operators 
\cite{Kawamura-88,Kawamura-98}. Two of them are particularly relevant, 
those associated with chirality and with the uniform anisotropy. 
Correspondingly, we define chirality exponents $\phi_c$, $\gamma_c$, 
and $\beta_c$, and anisotropy exponents 
$\phi_a$, $\gamma_a$, and $\beta_a$. These exponents are not 
independent: they are related by the relations \reff{expcross}.

The chirality exponents are associated with the operator \reff{chirality}, 
or in the FT framework, with the operator
\begin{equation}
C_{cd,kl}(x) = \phi_{ck}(x) \phi_{dl}(x) - \phi_{cl}(x) \phi_{dk}(x).
\label{chiralop}
\end{equation}
For $N=2$, there are several theoretical estimates. The analysis
of six-loop perturbative series in the framework of the fixed-dimension
expansion \cite{PRV-02} gives $\phi_c =  1.43(4)$, $\beta_c =  0.28(10)$.
MC simulations give:
$\beta_c=0.45(2)$, $\gamma_c=0.77(5)$, $\phi_c=1.22(6)$ 
\cite{Kawamura-92};
$\beta_c=0.38(2)$, $\gamma_c=0.90(9)$, $\phi_c=1.28(10)$
\cite{PM-94}; 
$\gamma_c = 0.81(3)$ \cite{ZIT-02}. The agreement 
is satisfactory, keeping into account the different systematic 
errors of the various approaches. 
Such exponents have been recently measured in Refs.~\cite{PKVMW-00,PSBKGRV-01}. 
For the $XY$ STA CsMnBr$_3$, it was found  \cite{PKVMW-00}
$\phi_c=1.28(7)$, $\beta_c=0.44(2)$, 
measured respectively in the HT and LT phases.  
These results are in reasonable agreement with the theoretical ones 
for the XY chiral universality class. On
the other hand, for the helimagnet holmium it was found \cite{PSBKGRV-01}
$\beta_c = 0.90(3)$, $\gamma_c = 0.68(6)$, which are sensibly
different from the theoretical results. Again, the reason for this 
discrepancy is unclear.

For $N=3$ only theoretical estimates are available: 
perturbative FT gives $\phi_c =  1.27(4)$, $\beta_c =  0.38(10)$ 
\cite{PRV-02}, while MC simulations give
$\beta_c=0.55(4)$, $\gamma_c=0.72(8)$, $\phi_c=1.27(9)$
from Ref.~\cite{Kawamura-92}
and $\beta_c=0.50(2)$ $\gamma_c=0.82(4)$ and $\phi_c=1.32(5)$ 
from Ref.~\cite{MPC-94}. 

An important question is the relevance of the $\mathbb{Z}_2$ chiral 
symmetry for the critical behavior of these systems. 
The experimental results of Ref.~\cite{PKVMW-00} show that chiral order and
spin order occur simultaneously. Still, one may wonder whether the absence of 
the $\mathbb{Z}_2$-symmetry changes the critical behavior of these systems. 
In this respect, the results of Ref.~\cite{ZIT-01} are interesting.
They  considered two-dimensional spins on a stacked 
triangular lattice and the biquadratic Hamiltonian 
\be
{\cal H} = - J \sum_{\<ij\>} \left( \vec{s}_i\cdot \vec{s}_j\right)^2,
\ee
with $J<0$. Because of the $\mathbb{Z}_2$ gauge symmetry 
$\vec{s}_i\to -\vec{s}_i$, chirality is identically zero. 
Nonetheless, the system shows a continuous transition with critical 
exponents $\gamma = 1.072(9)$, $\nu = 0.520(3)$, that are clearly 
compatible with the $XY$ chiral exponents,
but again they do not satisfy the relation $\gamma < 2\nu$.
These results would suggest that 
frustration, not chirality, is the relevant ingredient 
characterizing the phase transition.

The anisotropy exponent $\phi_a$ describes 
the crossover near multicritical points in the presence of a 
magnetic field. Experimental results are discussed 
in Ref. \cite{Kawamura-98}.

\subsection{The tetragonal Landau-Ginzburg-Wilson Hamiltonian}
\label{lsec-tetragonal}

In this section we study the critical behavior of statistical systems
 that are described by the three-coupling LGW Hamiltonian 
\bea
{\cal H} = \int d^d x 
\Bigl\{ \sum_{i,a}{1\over 2} 
\left[ (\partial_\mu \phi_{a,i})^2 + r \phi_{a,i}^2 \right]  
+
 \sum_{ij,ab} {1\over 4!}\left( u_0 + v_0 \delta_{ij} + w_0 \delta_{ij}\delta_{ab} \right)
\phi^2_{a,i} \phi^2_{b,j} \Bigr\},
\label{Hphi4tetra}
\eea
where $a,b=1,...M$ and $i,j=1,...N$.
Note that, as particular cases, one may recover
the $MN$ model, for $w_0=0$, the $(M\times N$)-component model with cubic anisotropy for $v_0=0$,
and $N$ decoupled $M$-component cubic models for $u_0=0$.
The models with $M=2$ are physically interesting: They should
describe the critical properties in some structural and
antiferromagnetic phase transitions
\cite{Mukamel-75,MK-75,NF-75,Aharony-76,MK-76-1,MK-76-2,BM-76,GP-80,STJAJG-82,TMTB-85}.
Therefore,
we will restrict ourselves to the case $M=2$. In the following the Hamiltonian
(\ref{Hphi4tetra}) with $M=2$ will be named tetragonal. 

We mention that in the literature 
the tetragonal Hamiltonian is also written in terms of a $2N$-component
vector field $\varphi_i$:
\bea
{\cal H} = \int d^d x 
\Bigl\{ {1\over 2}\sum_{i=1}^{2N}  
\left[ (\partial_\mu \varphi_{i})^2 + r \varphi_{i}^2 \right] 
+{1\over 4!} z_1 ( \sum_{i=1}^{2N} \varphi_i^2 )^2 +
{1\over 4!} z_2 \sum_{i=1}^{2N} \varphi_i^4 
+ {1\over 4!} 2 z_3 
\sum_{j=1}^N \varphi_{2j-1}^2 \varphi_{2j}^2 \Bigr\}.
\eea
The relations between the two sets of couplings are
$z_1 = u_0$, $z_2 = v_0 + w_0$, and $z_3 = v_0$.

Note that 
the tetragonal Hamiltonian is symmetric under the transformation~\cite{Korz-76}
\bea
(\phi_{1,i}\;,\;\phi_{2,i}) &\longrightarrow&  {1\over \sqrt{2}} 
(\phi_{1,i}+\phi_{2,i}\;,\;\phi_{1,i}-\phi_{2,i}),
\nonumber \\
(u_0,v_0,w_0) &\longrightarrow& (u_0,v_0+\case{3}{2}w_0,-w_0).
\label{sym1}
\eea

Many physical systems are expected to be described by
the tetragonal Hamiltonian.
Indeed, for $N=2$ the tetragonal Hamiltonian should be relevant
for the  structural phase transition in NbO$_2$ 
and, for $w_0=0$,  for the antiferromagnetic transitions in 
TbAu$_2$ and DyC$_2$. 
The case $N=3$ describes the antiferromagnetic phase transitions in the
K$_2$IrCl$_6$ crystal and, for $w_0=0$, those in TbD$_2$ and Nd.
Experimental results show continuous phase transitions in all the 
above-mentioned cases (see, e.g., Ref.~\cite{TMTB-85} and references therein).

\begin{figure*}[tb]
\vspace{-2cm}
\hspace{0cm}
\centerline{\psfig{width=12truecm,angle=-90,file=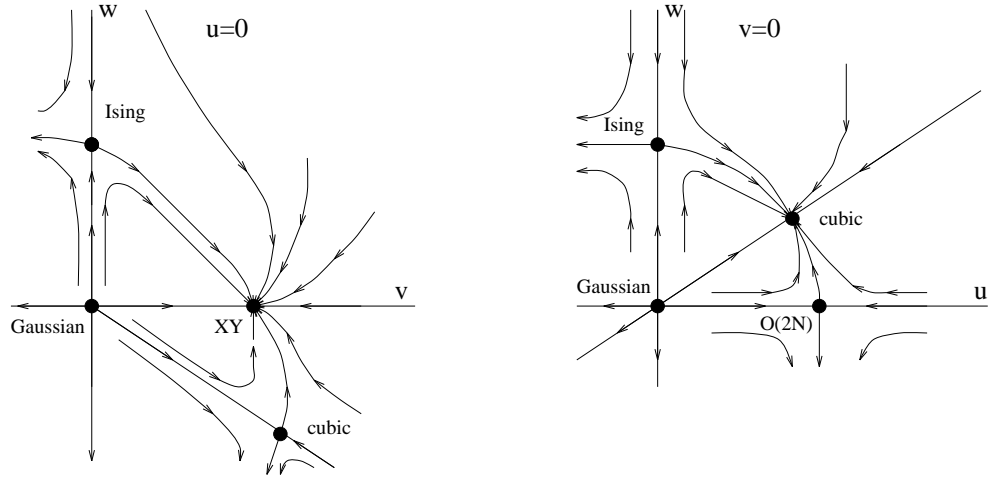}}
\vspace{-3cm}
\caption{
RG flow in the planes $u=0$ and $v=0$.
}
\label{tetrargflowuv0}
\end{figure*}

The $\epsilon$ expansion analysis of the tetragonal Hamiltonian
indicates the presence
of eight FP's~\cite{Mukamel-75,MK-75,MK-76-1}.
In order to understand their physical properties,
we begin by discussing 
the special cases when one of the couplings is zero.
As already mentioned, for $u=0$ the model is equivalent to 
$N$ decoupled cubic models with  two-component spins, 
while for $v=0$  the model is equivalent to a cubic model 
with $2N$-component spins.
Since $N$ is supposed to be larger than one, 
using the results reported in  Sec.~\ref{lsec-cubic}, we conclude that
in the plane $u=0$ the stable FP is the $XY$
one, and the cubic and the Ising FP's are equivalent because
they can be related through the symmetry (\ref{sym1}).
On the other hand, in the plane $v=0$ the stable FP is the cubic one.
Figure~\ref{tetrargflowuv0} shows sketches of the 
flow diagram in the two planes $u=0$ and $v=0$.

\begin{figure*}[tb]
\vspace{-2cm}
\hspace{0cm}
\centerline{\psfig{width=12truecm,angle=-90,file=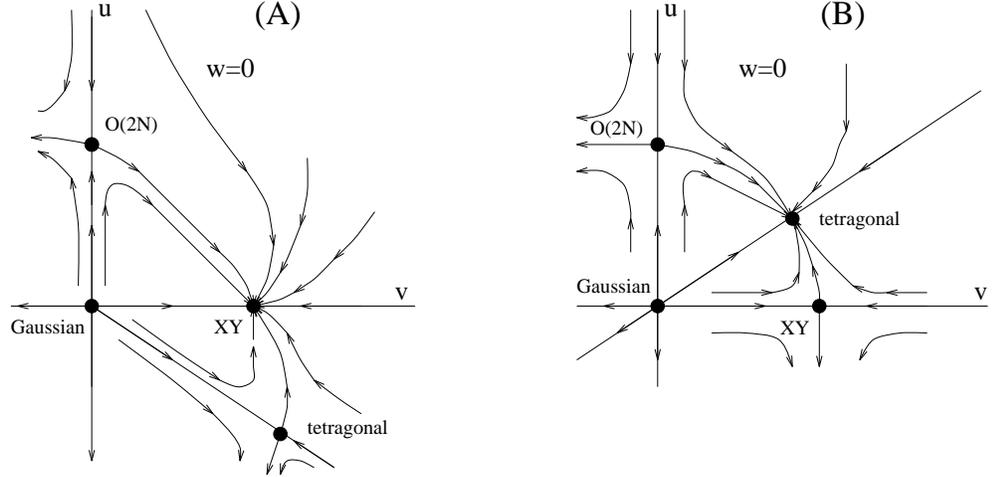}}
\vspace{-3cm}
\caption{
Two possibilities for the  RG flow in the plane $w=0$.
}
\label{tetrargfloww0}
\end{figure*}

In the case $w=0$ the tetragonal Hamiltonian describes $N$ coupled
$XY$ models. Such theories have four FP's~\cite{BLZ-74-many,Aharony-76}:
the trivial Gaussian one, the $XY$ one where the $N$ $XY$ models decouple,
the O($2N$)-symmetric and the  mixed tetragonal  FP's.
The Gaussian one is again never stable.
One can argue that, at the $XY$ FP,
the crossover exponent related to the O($2N$)-symmetric interaction 
is given by $\phi=\alpha_{XY}$~\cite{Sak-74,Aharony-76,CB-78},
where $\alpha_{XY}$ is the specific heat exponent of the $XY$ model.
This result is again based on the observation that
when $w=0$ the tetragonal Hamiltonian describes $N$ interacting $XY$ models,
and the O($2N$)-symmetric interaction can be represented as the product of two energy
operators of the $XY$ subsystems~\cite{Sak-74}. 
Since $\alpha_{XY}$ is negative,
the $XY$ FP should be stable with respect to 
the O($2N$)-symmetric interaction. 
In turn, one expects that the O($2N$)-symmetric and the tetragonal FP's are
unstable. The resulting sketch of the RG flow in the plane $w=0$ 
is given by the case (A) of Fig.~\ref{tetrargfloww0}.

We have so identified  seven out of eight FP's.
The eighth one can be  obtained by applying the transformation
(\ref{sym1}) to the cubic FP lying in the $v_0=0$ plane.

Therefore, the above-reported analysis leaves us with three possible 
stable FP points: the cubic one in the $v=0$ plane and its symmetric 
counterpart, and the $XY$ FP with $u=w=0$. 
The cubic FP, which is stable in the $v=0$ plane, 
turns out to be unstable with respect to the quartic
interaction associated with the coupling $v$. This is clearly seen
from the analyses of both the $\epsilon$ and the
fixed-dimension expansions. Of course, also its symmetric counterpart is 
unstable and therefore,
the $XY$ FP is the only---at least among the FP's predicted by the 
$\epsilon$ expansion---stable FP 
of the tetragonal theory, independently of the value of $N$.
Thus, systems described by the tetragonal Hamitonian are expected to have
$XY$ critical behavior.

The global stability of the $XY$ FP 
has been apparently contradicted by FT studies.
The analysis of the two-loop 
$\epsilon$ expansion \cite{Mukamel-75,MK-75,MK-76-1}
predicts a globally stable tetragonal FP, 
which is the one in the plane $w=0$,
and an unstable $XY$ FP. In the plane $w=0$, the predicted RG
flow is given by case (B) of Fig.~\ref{tetrargfloww0}. 
This fact should not come unexpected  
because $\alpha_{XY} = \epsilon/10 + O(\epsilon^2)$, 
so that, according
to the arguments of Refs.~\cite{Sak-74,Aharony-76,CB-78},
sufficiently close to $d=4$
the FP describing $N$ decoupled $XY$ models is unstable
and the tetragonal FP
dominates the critical behavior.
However, in order to obtain reliable results in three dimensions 
from the $\epsilon$ expansion,  
higher-order calculations with a proper resummation of the series 
are necessary.

The RG flow (B) of Fig.~\ref{tetrargfloww0} is further supported by 
recent higher-loop calculations.
The stability of the tetragonal FP has been  confirmed  
by  calculations up to $O(\epsilon^4)$ in the framework of the
$\epsilon$-expansion \cite{MV-00,MV-98,DG-85}.
The same result has been obtained by a Pad\`e-Borel analysis of the three-loop
series in the framework of the fixed-dimension expansion~\cite{SV-99,Shpot-89}.
However, we mention that
the authors of  Ref. \cite{SV-99}, noting the closeness of the apparently
stable tetragonal and unstable $XY$ FP's,
argued that the respective stability-instability may be 
a misleading effect of the relatively few terms of the series.

In order to clarify this issue,
we have extended the fixed-dimension expansion of the tetragonal Hamiltonian
to six loops.
Note that, since the tetragonal model for $w_0=0$ is 
nothing but the $MN$ model with $M=2$, the results of Sec.~\ref{lsec-frustrated}
show that, at least for $N=2$, there is another stable 
FP in the region
$v<0$, whose presence is not predicted by the $\epsilon$ expansion.
In the following we will not investigate this issue, 
although it would be worthwhile to perform a more systematic study,
but we will only focus on the stability properties of the $XY$ FP.

The tetragonal FT theory is renormalized by introducing 
a set of zero-momentum conditions for the one-particle irreducible two-point 
and four-point correlation functions, such as Eq.~(\ref{ren1g}) and
\bea
\Gamma^{(4)}_{ai,bj,ck,dl}(0) = m
Z_\phi^{-2}  
\left( u A_{ai,bj,ck,dl} 
+ v B_{ai,bj,ck,dl} 
+ w C_{ai,bj,ck,dl} \right)
\label{ren2g}  
\eea
where, setting $\delta_{ai,bj} \equiv \delta_{ab}\delta_{ij}$,
\begin{eqnarray}
&&A_{ai,bj,ck,dl} = 
 \case{1}{3} 
\left(\delta_{ai,bj}\delta_{ck,dl} + \delta_{ai,ck}\delta_{bj,dl} 
+ \delta_{ai,dl}\delta_{bj,ck} \right),
\nonumber \\
&& B_{ai,bj,ck,dl} = 
\delta_{ij}\delta_{ik}\delta_{il}\,\case{1}{3} 
\left(\delta_{ab}\delta_{cd} + \delta_{ac}\delta_{bd} + \delta_{ad}\delta_{bc} \right),\nonumber\\
&&C_{ai,bj,ck,dl} = 
\delta_{ij}\delta_{ik}\delta_{il}\,\delta_{ab}\delta_{ac}\delta_{ad}.
\label{tensors}
\end{eqnarray}
The mass $m$, and the zero-momentum
quartic couplings $u$, $v$, and $w$ are related 
to the corresponding Hamiltonian parameters
by 
\bea
u_0 = m u Z_u Z_\phi^{-2}, \qquad
v_0 = m v Z_v Z_\phi^{-2}, \qquad
w_0 = m w Z_w Z_\phi^{-2}.
\eea
The FP's of the theory are given by 
the common  zeros of the $\beta$-functions $\beta_u(u,v,w)$, $\beta_v(u,v,w)$,
and $\beta_w(u,v,w)$, associated with the couplings $u$, $v$,
 and $w$ respectively.
Their stability properties are controlled  by the matrix 
\begin{equation}
\Omega =
\left(\matrix{
\frac{\partial \beta_u}{\partial u} & \frac{\partial \beta_u}{\partial v} &
\frac{\partial \beta_u}{\partial w} \cr
\frac{\partial \beta_v}{\partial u} & \frac{\partial \beta_v}{\partial v} &
\frac{\partial \beta_v}{\partial w} \cr
\frac{\partial \beta_w}{\partial u} & \frac{\partial \beta_w}{\partial v} &
\frac{\partial \beta_w}{\partial w} 
}\right) \; .
\label{omegatetra}
\end{equation}

We have computed the perturbative expansion of the two-point and four-point
correlation functions to six loops. The diagrams contributing to this calculations
are approximately 1000.
We handled them with a symbolic manipulation program, which  generates the diagrams 
and computes the symmetry and group factors of 
each of them. We did not calculate the integrals associated with each diagram,
but we used the numerical results compiled in Ref.~\cite{NMB-77}.
Summing all contributions, we determined the RG functions
to six loops.
We report our results in terms of the rescaled couplings
\begin{equation}
u \equiv {16 \pi\over 3} R_{2N} \bar{u}, \qquad 
v \equiv   {16 \pi\over 3} R_2  \bar{v} ,\qquad
w \equiv   {16 \pi\over 3} \bar{w} ,\label{resc}
\end{equation}
where $R_K = 9/(8+K)$. The resulting series are
\begin{eqnarray}
&&\beta_{\bar{u}}  =
-\bar{u} + \bar{u}^2 + {4\over 5} \bar{u} \bar{v} + {2\over 3} \bar{u} \bar{w} 
- {2 (95+41N)\over 27(4+N)^2} \bar{u}^3 - {80 \over 27(4+N)} \bar{u}^2 \bar{v}
- {200 \over 81(4+N)} \bar{u}^2 \bar{w}\nonumber \\ 
&& - {92 \over 675}\bar{u}\bar{v}^2 
- {92 \over 729 }\bar{u}\bar{w}^2 
- {92\over 405} \bar{u}\bar{v}\bar{w} 
+\bar{u} \;(\sum_{i+j+k\geq 3} b^{(u)}_{ijk} \;\bar{u}^i \bar{v}^j \bar{w}^k ),
\label{bu}\\
&&\beta_{\bar{v}}  =  -\bar{v} + \bar{v}^2 + 
{6\over 4+N} \bar{u} \bar{v} + {2\over 3}\bar{v}\bar{w} 
 - {272 \over 675} \bar{v}^3 
- {724 \over 135  (4+N)} \bar{u}\bar{v}^2 - {2(185+23N) \over 27(4+N)^2} \bar{u}^2 \bar{v} 
\nonumber \\ &&
- {40 \over 81} \bar{v}^2 \bar{w} 
- {92 \over 729} \bar{v}  \bar{w}^2  
- {308\over 81 (4+N)} \bar{u}\bar{v}\bar{w} 
+\bar{v} \;(\sum_{i+j+k\geq 3} b^{(v)}_{ijk} \;\bar{u}^i \bar{v}^j \bar{w}^k ),
\label{bv}\\
&&\beta_{\bar{w}}  =  -\bar{w} + \bar{w}^2 + 
+ {6\over 4+N }\bar{u}\bar{w} + {6\over 5}\bar{v}\bar{w}  
- {308 \over 729} \bar{w}^3 
- {416 \over 81 (4+N)} \bar{u}\bar{w}^2 
- {416 \over 405} \bar{v}\bar{w}^2 \nonumber \\ 
&&
- {2(185+23N) \over 27(4+N)^2} \bar{u}^2 \bar{w}
- {416 \over 675} \bar{v}^2 \bar{w} 
- {832\over 135 (4+N)} \bar{u}\bar{v}\bar{w} 
+\bar{w} \;(\sum_{i+j+k\geq 3} b^{(w)}_{ijk} \;\bar{u}^i \bar{v}^j \bar{w}^k ).
\label{bw}
\end{eqnarray}
The coefficients 
$b^{(u)}_{ijk}$, $b^{(v)}_{ijk}$, $b^{(w)}_{ijk}$,
with $3\leq i+j+k\leq  6$ 
are reported in the Tables~\ref{betauc}, \ref{betavc}, and \ref{betawc},
respectively.

We report the RG functions $\eta_\phi$ and $\eta_t$
to two loops only 
(the complete six-loop series are available on request), since we will not use
them in our analysis. They are 
\begin{eqnarray}
\eta_\phi &=&
{4 (1 + N)\over 27(4+N)^2 } \bar{u}^2 + {16 \over 135 (4+N) } \bar{u} \bar{v} 
+{8 \over 81 (4+N) } \bar{u} \bar{w} + 
{8 \over 675 } \bar{v}^2 
+ {8 \over 405 } \bar{v} \bar{w} + {8 \over 729 } \bar{w}^2 
+ ...
%\sum_{i+j+k\geq 3} e^{(\phi)}_{ijk} \;\bar{u}^i \bar{v}^j \bar{w}^j,
\label{etaphi}\\
\eta_t &=& -{ 1+N \over 4+N } \bar{u} -{2\over 5 } \bar{v} -{1\over 3 } \bar{w} 
+{ 1+N \over (4+N)^2 } \bar{u}^2 
+{ 4\over 5 (4+N) } \bar{u} \bar{v}
+{ 2\over 25 } \bar{v}^2 
+{ 2\over 3 (4+N) } \bar{u} \bar{w}\nonumber \\ 
&&+{ 4\over 30 } \bar{v}\bar{w}
+{ 2\over 27 } \bar{w}^2  + ...
%+\sum_{i+j+k\geq 3} e^{(t)}_{ijk} \;\bar{u}^i \bar{v}^j \bar{w}^k.
\label{etat}
\end{eqnarray}

In the following we limit ourselves to check the stability of the
$XY$ FP, whose coordinates are
$\overline{u}^*=0$, $\overline{v}_{XY}^*=1.402(4)$ \cite{CHPRV-01,GZ-98},
and $\overline{w}^*=0$.
One can easily see that the eigenvalues of the stability matrix 
(\ref{omegatetra}) at the $XY$ FP are given simply by
\begin{eqnarray}
\omega_1 = {\partial \beta_{\overline{u}} \over \partial\overline{u}}
  (0, \overline{v}_{XY}^*,0),\qquad
\omega_2 = {\partial \beta_{\overline{v}} \over \partial\overline{v}}
  (0, \overline{v}_{XY}^*,0),\qquad
\omega_3 = {\partial \beta_{\overline{w}} \over \partial\overline{w}}
  (0, \overline{v}_{XY}^*,0). 
\end{eqnarray}
Note that $\omega_i$ are $N$-independent, as it can be checked
by looking at the corresponding series.
According to the nonpertubative argument reported above, 
the $XY$ FP is stable, and the smallest eigenvalue
of the stability matrix $\Omega$ should be given by
\begin{equation}
\omega_1= -{\alpha_{XY}\over \nu_{XY}}, 
\label{omega1}
\end{equation}
where $\alpha_{XY}$ and $\nu_{XY}$ are the
critical exponents of the $XY$ model. 
In the analysis we exploit the knowledge of the 
large-order behavior of the series, 
which is determined by the
$XY$ FP only and therefore it is the same as the one of the
O(2)-symmetric theory. We skip the details, since
the analysis is identical to that performed in Ref.~\cite{PV-00-r} to study 
the stability of the O($M$)-symmetric FP in the $MN$ model.
Our estimate is
\begin{equation}
\omega_1 = 0.007(8).
\end{equation}
The stability of the $XY$ FP is substantially confirmed,
although the apparent error of the analysis does not completely
exclude the opposite sign for $\omega_1$.
The estimate of $\omega_1$
turns out to be substantially consistent with the value one obtains 
using Eq.~(\ref{omega1}).
Indeed, $\alpha_{XY}/\nu_{XY} = - 0.0217(12)$
using the recent estimates of the $XY$ critical 
exponents of Ref.~\cite{CHPRV-01},  
and $\alpha_{XY}/\nu_{XY} = -0.016(7)$ and $\alpha_{XY}/\nu_{XY} = -0.010(9)$ 
using the estimates respectively of Ref.~\cite{GZ-98} and \cite{LZ-77} 
that were obtained by 
a more similar technique, i.e. from the
analysis of the fixed-dimension expansion of the O(2)-symmetric model.
It is easy to see that 
$\omega_3$ is  equal to the smallest eigenvalue of stability matrix of 
the two-component cubic model at the $XY$ FP,
see Sec.~\ref{lsec-cubic}, thus $\omega_3=0.103(8)$. 
The eigenvalue $\omega_2$ is the one determining 
the leading scaling corrections in the $XY$ model and it is given by
$\omega_2=0.789(11)$ ~\cite{GZ-98}.

In conclusion the analysis of the six-loop fixed-dimension expansion
turns out to be substantially consistent with the nonperturbative
prediction indicating that the $XY$ FP is globally stable
independently of $N$.

\subsection{LGW Hamiltonian with symmetry 
O$(n_1)\oplus$O$(n_2)$ and multicritical phenomena}
\label{LGW-multicritical}

We now consider an $N$-component system with symmetry 
O$(n_1)\oplus$O$(n_2)$ with $n_1 + n_2 = N$. The most general Hamiltonian
containing up to quartic terms is given
by \cite{PJF-74,Fisher-75,KNF-76} 
\begin{eqnarray}
{\cal H} = \int d^3 x \Bigl[ 
\case{1}{2} ( \partial_\mu \phi_1)^2  + \case{1}{2} (
\partial_\mu \phi_2)^2 + \case{1}{2} r_1 \phi_1^2  
 + \case{1}{2} r_2 \phi_2^2  
+ u_1 (\phi_1^2)^2 + u_2 (\phi_2^2)^2 + w \phi_1^2\phi_2^2 \Bigr],
\label{bicrHH} 
\end{eqnarray}
where $\phi_1$, $\phi_2$ are $n_1$-, $n_2$-component fields 
with $n_1+n_2=N$.
We are interested in the critical behavior at the 
multicritical point, where
two critical lines with O$(n_1)$ and 
O$(n_2)$ symmetry meet. 
For this purpose, one must analyze the FP's of the theory when both 
$r_1$ and $r_2$ are tuned to their critical value. 
According to the $O(\epsilon)$ analysis of Ref.~\cite{KNF-76} 
(see also Ref.~\cite{Aharony-02-2}) the model has six FP's. 
Three of them, i.e. the Gaussian, the O($n_1$) and the O($n_2$) FP's,  
are always unstable. The other three FP's are
the bicritical $O(N)$-symmetric FP, 
and the tetracritical decoupled and biconal FP's. The stability of these FP's 
depends on $n_1$ and $n_2$. For the decoupled FP, one can use 
nonperturbative arguments to establish its stability properties 
with respect to the $w$-interaction \cite{Aharony-02-2}. The RG dimension 
$y_w$ of the operator $w\phi_1^2\phi_2^2$ that couples the 
two fields $\phi_1$ and $\phi_2$ is given by
\be
y_w = {\alpha_1\over 2\nu_1} + {\alpha_2\over 2\nu_2},
\ee
where $\alpha_i$ and $\nu_i$ are the critical exponents of the 
O$(n_i)$ theory. For $n_1 \ge 2$ and $n_2 \ge 2$, we have 
$\alpha_i < 0$, so that $y_w < 0$. Therefore, the perturbation is 
irrelevant and the decoupled FP is stable. For $n_1 = 1$,
the perturbation is irrelevant for $\nu_2 > \nu_I/(3 \nu_I - 1) \approx 
0.7077(2)$, where we have used the estimate of $\nu_I$ of Ref. \cite{CPRV-02}.
Therefore, the decoupled FP is stable for $n_2 \ge 3$ and unstable for 
$n_2 = 1,2$. 

In order to study the stability properties of the bicritical O($N$) FP,
we  consider generic
O($n_1$)$\oplus$O($n_2$) invariant perturbations ${\cal P}_{ml}$ 
at the O($N$)-summetric FP, where $m$ is
the power of the fields
and $l$ the spin of the representation of the O($N$) group \cite{CPV-02-4}.
For $m=2$ (resp. 4), the only possible values of $l$ are 
$l=0,2$ (resp. $l=0,2,4$). Explicitly, 
\begin{eqnarray}
&&{\cal P}_{2,0}= \Phi^2,\qquad  \qquad  
{\cal P}_{2,2}= \sum_{i=1}^{n_1} {\cal O}_2^{ii} = 
                \phi_1^2-{n_1\over N} \Phi^2 , \nonumber \\
&&{\cal P}_{4,0}= (\Phi^2)^2,\qquad  \qquad  
{\cal P}_{4,2}= \Phi^2 {\cal P}_{2,2} ,\nonumber \\
&&{\cal P}_{4,4}=  \sum_{i=1}^{n_1} \sum_{j=n_1+1}^{n_2} {\cal O}_4^{iijj} = 
\phi_1^2 \phi_2^2- {\Phi^2 (n_1 \phi_2^2+n_2 \phi_1^2)\over N+4}+
{n_1 n_2 (\Phi^2)^2 \over (N+2)(N+4)}, 
\end{eqnarray}
where $\Phi$ is the $N$-component field $(\phi_1,\phi_2)$, 
and ${{\cal O}_2^{ij}}$, ${\cal O}_4^{ijkl}$ are respectively the 
spin-2 and spin-4 operators defined in Sec. \ref{sec-1.5.8} and expressed 
in terms of the field $\Phi$.
The perturbations ${\cal P}_{2,0}$ and ${\cal P}_{2,2}$
are always relevant. They must be tuned to approach a multicritical point.  
As discussed in Sec. \ref{lsec-cubic}, any spin-4 perturbation---therefore,
${\cal P}_{4,4}$ too---of the O($N$) FP is relevant for $N>N_c$ with 
$N_c\ltapprox 2.9$. Therefore, the O($N$) FP is unstable for $N\ge 3$. 
Note that,
for $N=3$, the associated crossover exponent is very small, i.e. $\phi_4 = 0.009(4)$.

The Hamiltonian \reff{bicrHH} has been used to describe a variety of 
multicritical phenomena. We should mention the critical 
behavior of uniaxial antiferromagnets in a magnetic field parallel to 
the field direction \cite{KNF-76}---in this case $n_1=1$ and $n_2=2$---and 
the SO(5) theory of high-$T_c$ superconductors \cite{Zhang-97},
corresponding to $n_1=3$ and $n_2=2$, that was already 
discussed in Sec.~\ref{N5case}.
Note that the instability of the O(3) FP implies that uniaxial
antiferromagnets should not show a O(3)-symmetric bicritical transition point. 
Since the decoupled FP is also unstable, 
the multicritical behavior should be controlled by the 
biconal FP \cite{CPV-02-4}, which, however,
is expected to be close to the O(3) FP,
so that the critical exponents should be very close to the Heisenberg ones.
Thus, differences should be hardly distinguishable in experiments.

\begin{table*}
\caption{
The coefficients $b^{(u)}_{ijk}$, cf. Eq. (\ref{bu}).
}
\label{betauc}
\tiny
\begin{center}
\begin{tabular}{cl}
\hline
\multicolumn{1}{c}{$i,j,k$}& 
\multicolumn{1}{c}{$R_{2N}^{-i} R_2^{-j} b^{(u)}_{ijk}$}\\ 
\hline
3,0,0 &$ 0.27385517 + 0.15072806\,N + 0.0074016064\,{N^2}$\\
2,1,0 &$ 0.903231 + 0.072942424\,N$\\
1,2,0 &$ 0.60730385 + 0.0068245729\,N$\\
0,3,0 &$ 0.13854816$\\
2,0,1 &$ 0.67742325 + 0.054706818\,N$\\
1,1,1 &$ 0.91095577 + 0.010236859\,N$\\
0,2,1 &$ 0.31173336$\\
1,0,2 &$ 0.4154565 + 0.0051184297\,N$\\
0,1,2 &$ 0.27646528$\\
0,0,3 &$ 0.090448951$\\
4,0,0 &$ -0.27925724 - 0.1836675\,N - 0.021838259\,{N^2} + 
   0.00018978314\,{N^3}$\\
3,1,0 &$ -1.2584488 - 0.22200749\,N + 0.0032992093\,{N^2}$\\
2,2,0 &$ -1.4679273 - 0.029437397\,N$\\
1,3,0 &$ -0.65789001 - 0.0052472383\,N$\\
0,4,0 &$ -0.11873585$\\
3,0,1 &$ -0.94383662 - 0.16650561\,N + 0.002474407\,{N^2}$\\
2,1,1 &$ -2.201891 - 0.044156096\,N$\\
1,2,1 &$ -1.4802525 - 0.011806286\,N$\\
0,3,1 &$ -0.35620754$\\
2,0,2 &$ -0.96497888 - 0.024920289\,N$\\
1,1,2 &$ -1.3045513 - 0.010625658\,N$\\
0,2,2 &$ -0.47070809$\\
1,0,3 &$ -0.42331874 - 0.0035418858\,N$\\
0,1,3 &$ -0.30532865$\\
0,0,4 &$ -0.075446692$\\
5,0,0 &$ 0.35174477 + 0.26485003\,N + 0.045288106\,{N^2} + 
   0.00043866975\,{N^3} + 0.000013883029\,{N^4}$\\
4,1,0 &$ 2.0278677 + 0.51868097\,N + 0.0059085942\,{N^2} + 
   0.00033898434\,{N^3}$\\
3,2,0 &$ 3.4214862 + 0.18321912\,N + 0.0024313106\,{N^2}$\\
2,3,0 &$ 2.4773377 + 0.034202317\,N$\\
1,4,0 &$ 0.92541748 + 0.0039821066\,N$\\
0,5,0 &$ 0.1462366$\\
4,0,1 &$ 1.5209008 + 0.38901073\,N + 0.0044314456\,{N^2} + 
   0.00025423826\,{N^3}$\\
3,1,1 &$ 5.1322293 + 0.27482868\,N + 0.0036469659\,{N^2}$\\
2,2,1 &$ 5.5740098 + 0.076955213\,N$\\
1,3,1 &$ 2.7762524 + 0.01194632\,N$\\
0,4,1 &$ 0.54838726$\\
3,0,2 &$ 2.2073347 + 0.13067265\,N + 0.00142597\,{N^2}$\\
2,1,2 &$ 4.8290973 + 0.066949343\,N$\\
1,2,2 &$ 3.5983005 + 0.016143463\,N$\\
0,3,2 &$ 0.94562264$\\
2,0,3 &$ 1.5315693 + 0.021353803\,N$\\
1,1,3 &$ 2.2741668 + 0.010775585\,N$\\
0,2,3 &$ 0.89377961$\\
1,0,4 &$ 0.56035196 + 0.0026938962\,N$\\
0,1,4 &$ 0.44016041$\\
0,0,5 &$ 0.087493302$\\
6,0,0  &$  -0.5104989 - 0.4297050\,N - 0.09535750\,{N^2} - 
   0.0040017345\,{N^3} + 0.00003226842\,{N^4} + 
   0.00000141045\,{N^5} $\\ 
5,1,0  &$  -3.5978778 - 1.20182\,N - 0.057714496\,{N^2} + 
   0.00061831126\,{N^3} + 0.000043766306\,{N^4} $\\ 
4,2,0  &$  -8.0310329 - 0.80074836\,N - 0.0021586547\,{N^2} + 
   0.00029396543\,{N^3} $\\ 
3,3,0  &$  -8.2478554 - 0.20470296\,N + 0.00055966931\,{N^2} $\\ 
2,4,0  &$  -4.7055601 - 0.028798289\,N $\\ 
1,5,0  &$  -1.4837563 - 0.0061256278\,N $\\ 
0,6,0  &$  -0.20437244 $\\ 
5,0,1  &$  -2.6984083 - 0.90136504\,N - 0.043285872\,{N^2} + 
   0.00046373344\,{N^3} + 0.00003282473\,{N^4} $\\ 
4,1,1  &$  -12.046549 - 1.2011225\,N - 0.0032379821\,{N^2} + 
   0.00044094814\,{N^3} $\\ 
3,2,1  &$  -18.557675 - 0.46058167\,N + 0.0012592559\,{N^2} $\\ 
2,3,1  &$  -14.11668 - 0.086394867\,N $\\ 
1,4,1  &$  -5.5640863 - 0.022971104\,N $\\ 
0,5,1  &$  -0.91967597 $\\ 
4,0,2  &$  -5.1135549 - 0.53538355\,N - 0.0025247004\,{N^2} + 
   0.00015530699\,{N^3} $\\ 
3,1,2  &$  -15.854479 - 0.41049007\,N + 0.00078044444\,{N^2} $\\ 
2,2,2  &$  -18.106633 - 0.11544087\,N $\\ 
1,3,2  &$  -9.5137746 - 0.03966586\,N $\\ 
0,4,2  &$  -1.964478 $\\ 
3,0,3  &$  -4.9317312 - 0.13514942\,N + 0.00011311235\,{N^2} $\\ 
2,1,3  &$  -11.278684 - 0.075967076\,N $\\ 
1,2,3  &$  -8.8867987 - 0.0375632\,N $\\ 
0,3,3  &$  -2.4446492 $\\ 
2,0,4  &$  -2.754683 - 0.019167341\,N $\\ 
1,1,4  &$  -4.3412061 - 0.01856332\,N $\\ 
0,2,4  &$  -1.7904475 $\\ 
1,0,5  &$  -0.86229463 - 0.003712664\,N $\\ 
0,1,5  &$  -0.71141747 $\\ 
0,0,6  &$  -0.1179508 $\\ 
\hline
\end{tabular}
\end{center}
\end{table*}

\begin{table*}
\caption{
The coefficients $b^{(v)}_{ijk}$, cf. Eq. (\ref{bv}).
}
\label{betavc}
\tiny
\begin{center}
\begin{tabular}{cl}
\hline
\multicolumn{1}{c}{$i,j,k$}& 
\multicolumn{1}{c}{$R_{2N}^{-i} R_2^{-j} b^{(v)}_{ijk}$}\\ 
\hline
3,0,0 &$ 0.64380517 + 0.11482552\,N - 0.0068647863\,{N^2}$\\
2,1,0 &$ 1.97782 - 0.000039427734\,N$\\
1,2,0 &$ 1.5893912$\\
0,3,0 &$ 0.43198483$\\
2,0,1 &$ 1.2813995 - 0.0062019643\,N$\\
1,1,1 &$ 1.8846568$\\
0,2,1 &$ 0.73213007$\\
1,0,2 &$ 0.56468457$\\
0,1,2 &$ 0.42057493$\\
0,0,3 &$ 0.090448951$\\
4,0,0 &$ -0.76706177 - 0.17810933\,N + 0.00016284548\,{N^2} - 
   0.00070068894\,{N^3}$\\
3,1,0 &$ -3.2372708 - 0.11004576\,N - 0.0010508505\,{N^2}$\\
2,2,0 &$ -4.2003729 + 0.017778007\,N$\\
1,3,0 &$ -2.3041302$\\
0,4,0 &$ -0.48457321$\\
3,0,1 &$ -2.117322 - 0.066630894\,N - 0.0014713371\,{N^2}$\\
2,1,1 &$ -5.1751293 + 0.019981131\,N$\\
1,2,1 &$ -4.0752638$\\
0,3,1 &$ -1.1078678$\\
2,0,2 &$ -1.6595755 + 0.00505486\,N$\\
1,1,2 &$ -2.4989894$\\
0,2,2 &$ -0.98989917$\\
1,0,3 &$ -0.5483926$\\
0,1,3 &$ -0.42686062$\\
0,0,4 &$ -0.075446692$\\
5,0,0 &$ 1.0965348 + 0.31582586\,N + 0.0094338525\,{N^2} - 
   0.00049177077\,{N^3} - 0.000086193996\,{N^4}$\\
4,1,0 &$ 5.9292953 + 0.40901849\,N - 0.010102522\,{N^2} - 
   0.00031594921\,{N^3}$\\
3,2,0 &$ 10.739283 - 0.0048408111\,N + 0.00072672306\,{N^2}$\\
2,3,0 &$ 9.0637188 - 0.058086499\,N$\\
1,4,0 &$ 3.8277762$\\
0,5,0 &$ 0.66233546$\\
4,0,1 &$ 3.9073944 + 0.26207189\,N - 0.0068664349\,{N^2} - 
   0.00031886052\,{N^3}$\\
3,1,1 &$ 13.511919 - 0.017385016\,N + 0.00047240741\,{N^2}$\\
2,2,1 &$ 16.458953 - 0.10118022\,N$\\
1,3,1 &$ 9.0245987$\\
0,4,1 &$ 1.9145972$\\
3,0,2 &$ 4.4690201 - 0.0072615868\,N - 0.000071990874\,{N^2}$\\
2,1,2 &$ 10.482169 - 0.060013651\,N$\\
1,2,2 &$ 8.4121105$\\
0,3,2 &$ 2.3394333$\\
2,0,3 &$ 2.3860591 - 0.012232105\,N$\\
1,1,3 &$ 3.7649897$\\
0,2,3 &$ 1.5529231$\\
1,0,4 &$ 0.6859313$\\
0,1,4 &$ 0.56304585$\\
0,0,5 &$ 0.087493302$\\
6,0,0 &$ -1.774553 - 0.6080863\,N - 0.03773523\,{N^2} + 
   0.0005359509\,{N^3} - 0.0001051598\,{N^4} - 0.00001200996\,{N^5}$\\
5,1,0 &$ -11.741739 - 1.2643992\,N + 0.011113177\,{N^2} - 
   0.0013773749\,{N^3} - 0.000070170026\,{N^4}$\\
4,2,0 &$ -27.537411 - 0.39161313\,N - 0.0024557833\,{N^2} - 
   0.000052719756\,{N^3}$\\
3,3,0 &$ -31.89832 + 0.25340217\,N - 0.0042556584\,{N^2}$\\
2,4,0 &$ -20.465063 + 0.12482592\,N$\\
1,5,0 &$ -7.0723339$\\
0,6,0 &$ -1.0395295$\\
5,0,1 &$ -7.7827796 - 0.8288997\,N + 0.0078564644\,{N^2} - 
   0.0010207165\,{N^3} - 0.000063719952\,{N^4}$\\
4,1,1 &$ -35.145003 - 0.47099755\,N - 0.0052695631\,{N^2} - 
   0.00017352674\,{N^3}$\\
3,2,1 &$ -59.094258 + 0.45356273\,N - 0.0074832062\,{N^2}$\\
2,3,1 &$ -49.340075 + 0.28869185\,N$\\
1,4,1 &$ -20.9357$\\
0,5,1 &$ -3.6425627$\\
4,0,2 &$ -11.831866 - 0.15794071\,N - 0.0024381429\,{N^2} - 
   0.00010567985\,{N^3}$\\
3,1,2 &$ -38.56208 + 0.27559695\,N - 0.0045217676\,{N^2}$\\
2,2,2 &$ -47.222342 + 0.26270351\,N$\\
1,3,2 &$ -26.292061$\\
0,4,2 &$ -5.6513079$\\
3,0,3 &$ -8.9681447 + 0.05667789\,N - 0.00095212462\,{N^2}$\\
2,1,3 &$ -21.590806 + 0.11456243\,N$\\
1,2,3 &$ -17.822421$\\
0,3,3 &$ -5.0667675$\\
2,0,4 &$ -3.9823089 + 0.020527147\,N$\\
1,1,4 &$ -6.5311036$\\
0,2,4 &$ -2.7738504$\\
1,0,5 &$ -1.0205971$\\
0,1,5 &$ -0.8660073$\\
0,0,6 &$ -0.1179508$\\
\hline
\end{tabular}
\end{center}
\end{table*}

\begin{table*}
\caption{
The coefficients $b^{(w)}_{ijk}$, cf. Eq. (\ref{bw}).
}
\label{betawc}
\tiny
\begin{center}
\begin{tabular}{cl}
\hline
\multicolumn{1}{c}{$i,j,k$}& 
\multicolumn{1}{c}{$R_{2N}^{-i} R_2^{-j} b^{(w)}_{ijk}$}\\ 
\hline
3,0,0 &$ 0.64380517 + 0.11482552\,N - 0.0068647863\,{N^2}$\\
2,1,0 &$ 2.2471073 + 0.0081904303\,N$\\
1,2,0 &$ 2.2552977$\\
0,3,0 &$ 0.75176591$\\
2,0,1 &$ 1.6853305 + 0.0061428227\,N$\\
1,1,1 &$ 3.3829466$\\
0,2,1 &$ 1.6914733$\\
1,0,2 &$ 1.3138294$\\
0,1,2 &$ 1.3138294$\\
0,0,3 &$ 0.3510696$\\
4,0,0 &$ -0.76706177 - 0.17810933\,N + 0.00016284548\,{N^2} - 
   0.00070068894\,{N^3}$\\
3,1,0 &$ -3.6514455 - 0.13125033\,N - 0.00013991835\,{N^2}$\\
2,2,0 &$ -5.7009462 + 0.026692512\,N$\\
1,3,0 &$ -3.7828358$\\
0,4,0 &$ -0.94570894$\\
3,0,1 &$ -2.7385841 - 0.098437749\,N - 0.00010493876\,{N^2}$\\
2,1,1 &$ -8.5514193 + 0.040038768\,N$\\
1,2,1 &$ -8.5113805$\\
0,3,1 &$ -2.8371268$\\
2,0,2 &$ -3.3477204 + 0.015083679\,N$\\
1,1,2 &$ -6.6652735$\\
0,2,2 &$ -3.3326368$\\
1,0,3 &$ -1.8071874$\\
0,1,3 &$ -1.8071874$\\
0,0,4 &$ -0.37652683$\\
5,0,0 &$ 1.0965348 + 0.31582586\,N + 0.0094338525\,{N^2} - 
   0.00049177077\,{N^3} - 0.000086193996\,{N^4}$\\
4,1,0 &$ 6.6487314 + 0.46860779\,N - 0.011049797\,{N^2} - 
   0.00020675107\,{N^3}$\\
3,2,0 &$ 14.201957 + 0.0086575882\,N + 0.0015502926\,{N^2}$\\
2,3,0 &$ 14.309604 - 0.097439028\,N$\\
1,4,0 &$ 7.1060826$\\
0,5,0 &$ 1.4212165$\\
4,0,1 &$ 4.9865485 + 0.35145584\,N - 0.0082873478\,{N^2} - 
   0.0001550633\,{N^3}$\\
3,1,1 &$ 21.302936 + 0.012986382\,N + 0.0023254389\,{N^2}$\\
2,2,1 &$ 32.19661 - 0.21923781\,N$\\
1,3,1 &$ 21.318248$\\
0,4,1 &$ 5.329562$\\
3,0,2 &$ 8.3645284 + 0.0079241123\,N + 0.00085452488\,{N^2}$\\
2,1,2 &$ 25.291106 - 0.17118531\,N$\\
1,2,2 &$ 25.119921$\\
0,3,2 &$ 8.373307$\\
2,0,3 &$ 6.8946012 - 0.046174799\,N$\\
1,1,3 &$ 13.696853$\\
0,2,3 &$ 6.8484264$\\
1,0,4 &$ 2.8857918$\\
0,1,4 &$ 2.8857918$\\
0,0,5 &$ 0.49554751$\\
6,0,0 &$ -1.774553 - 0.6080863\,N - 0.03773523\,{N^2} + 
   0.0005359509\,{N^3} - 0.0001051598\,{N^4} - 0.00001200996\,{N^5}$\\
5,1,0 &$ -13.10644 - 1.4235988\,N + 0.011751068\,{N^2} - 
   0.0013937944\,{N^3} - 0.000055380116\,{N^4}$\\
4,2,0 &$ -35.75223 - 0.54684266\,N - 0.00034126582\,{N^2} + 
   0.000073209725\,{N^3}$\\
3,3,0 &$ -48.800935 + 0.40885839\,N - 0.0070450255\,{N^2}$\\
2,4,0 &$ -36.538548 + 0.23920711\,N$\\
1,5,0 &$ -14.519736$\\
0,6,0 &$ -2.4199561$\\
5,0,1 &$ -9.8298296 - 1.0676991\,N + 0.0088133007\,{N^2} - 
   0.0010453458\,{N^3} - 0.000041535087\,{N^4}$\\
4,1,1 &$ -53.628345 - 0.82026399\,N - 0.00051189873\,{N^2} + 
   0.00010981459\,{N^3}$\\
3,2,1 &$ -109.8021 + 0.91993139\,N - 0.015851307\,{N^2}$\\
2,3,1 &$ -109.61564 + 0.71762134\,N$\\
1,4,1 &$ -54.449012$\\
0,5,1 &$ -10.889802$\\
4,0,2 &$ -21.073538 - 0.33257393\,N - 0.000059310729\,{N^2} + 
   0.000035990819\,{N^3}$\\
3,1,2 &$ -86.32735 + 0.71521067\,N - 0.012400534\,{N^2}$\\
2,2,2 &$ -129.28253 + 0.84571708\,N$\\
1,3,2 &$ -85.62454$\\
0,4,2 &$ -21.406135$\\
3,0,3 &$ -23.569724 + 0.19143373\,N - 0.00335616\,{N^2}$\\
2,1,3 &$ -70.60619 + 0.46125161\,N$\\
1,2,3 &$ -70.144939$\\
0,3,3 &$ -23.381646$\\
2,0,4 &$ -14.927998 + 0.097362599\,N$\\
1,1,4 &$ -29.661271$\\
0,2,4 &$ -14.830635$\\
1,0,5 &$ -5.1298717$\\
0,1,5 &$ -5.1298717$\\
0,0,6 &$ -0.74968893$\\
\hline
\end{tabular}
\end{center}
\end{table*}

\end{document}